\pdfoutput=1

\documentclass[a4paper,12pt,oneside]{book}

\usepackage{nicefrac}
\usepackage[latin1]{inputenc}
\usepackage{graphicx}
\usepackage{amssymb,amsmath}
\usepackage{epstopdf}
\usepackage{mdwlist}
\usepackage[a4paper,top=3cm,bottom=3.5cm,left=3.5cm,right=3cm,heightrounded,bindingoffset=5mm]{geometry}
\usepackage{setspace}
\usepackage[bookmarks=true,bookmarksopen=true,pdfhighlight=/I,pdfpagemode=UseOutlines,linktocpage,plainpages=false]{hyperref}  
\usepackage{amsfonts}
\usepackage[usenames,dvipsnames]{color}
\usepackage[all]{xy}
\hypersetup{colorlinks=true,urlcolor=Sepia,linkcolor=BlueViolet,citecolor=OliveGreen}
\usepackage{array} 
\usepackage{multirow}
\usepackage{acronym}
\def\listofsymbols{\begin{tabbing}\label{listofsymbols}

$AQFT$~~~~~~~~~~~~~~\=\parbox{4in}{Algebraic Quantum Field Theory  \hfill \textit{\pageref{symbol:AQFT}}}\\
\addsymbol BV: {Blasone-Vitiello}{symbol:BV}
\addsymbol CAR: {Canonical Anti-commutation Relations}{symbol:CAR}
\addsymbol CCR: {Canonical Commutation Relations}{symbol:CCR}
\addsymbol CP: {Charge-parity}{symbol:CP}
\addsymbol DE: {Dark Energy}{symbol:DE}
\addsymbol DM: {Dark Matter}{symbol:DM}
\addsymbol FLO: {Flavour ladder operators}{symbol:FLO}
\addsymbol FRW: {Friedman-Robertson-Walker}{symbol:FRW}
\addsymbol MLO: {Massive ladder operators}{symbol:MLO}
\addsymbol MSSM: {Minimal Supersymmetric Standard Model}{symbol:MSSM}
\addsymbol MSW: {Mikheyev-Smirnov-Wolfenstein}{symbol:MSW}
\addsymbol NFO: {Neutrino flavour oscillations}{symbol:NFO}
\addsymbol QFT: {Quantum Field Theory}{symbol:QFT}
\addsymbol SM: {Standard Model}{symbol:SM}
\addsymbol SUSY: {Supersymmetry}{symbol:SUSY}
\\
\addsymbolnoref \mbox{iff}: {if and only if}
\addsymbolnoref c.c.: {complex conjugate (of the previous term)}
\addsymbolnoref h.c.: {hermitian conjugate  (of the previous term)}
\addsymbolnoref	\mbox{vev}: {vacuum expectation value}
\addsymbolnoref	\mbox{\textit{f}-vev}: {flavour vacuum expectation value}
\\
\addsymbol \slot: {empty slot}{symbol:slot}
\addsymbolnoref \slot^{T}: {transpose}
\addsymbol \slot^*: {complex/hermitian conjugate}{symbol:cc}
\addsymbol \slot^{\dagger}: {$(\slot^*)^T$}{symbol:dagger}
\addsymbolnoref ;:	{such that}
\addsymbolnoref \sim:	{of the order of}
\addsymbolnoref \approx:	{approximately equal to}
\addsymbolnoref \equiv:	{defined as}
\addsymbolnoref \propto:	{proportional to}
\addsymbolnoref \propto\!\!\!\!\!/:	{not proportional to}
\addsymbolnoref	\field{N}_0: {Nonnegative Integer Numbers}
\addsymbolnoref diag\{\slot\}: {diagonal matrix}
\addsymbolnoref \{X_1 \dotsc X_n \mid Y\}: {set of elements $X_1 \dotsc X_n $, with condition $Y$}
\\
\addsymbol \mathfrak{H}: {non-separable Hilbert space for physical states}{symbol:nonseparable}
\addsymbol \mathfrak{F}_{\slot}: {separable Hilbert space (Fock space)}{symbol:separable}
\addsymbol \mathfrak{F_0}: {Fock space for massive states}{symbol:massfock}
\addsymbol \mathfrak{F_f}: {Fock space for flavour states}{symbol:flavourfock}
\addsymbolnoref \vk,\vp,\vq: {momentum}
\addsymbolnoref K: {momentum cutoff}
\addsymbolnoref \w_i(k):	{energy of a particle state: $\sqrt{\vk^2+m_i^2}$}
\addsymbolnoref 	\w_\pm:	{$\w_1\pm\w_2$}
\addsymbolnoref a^{(\dagger)}_{\slot}(\vk): {annihilation (creation) operator}
\addsymbolnoref a^{(\dagger)}_i(\vk): {MLO ($i=1,2$)}
\addsymbolnoref a^{(\dagger)}_\iota(\vk): {FLO ($\iota=A,B$)}
\addsymbol a^\mp_{\slot}:	{$a^\dagger_{\slot} (-\vk)$}{symbol:aplusminus}
\addsymbol \rmv: {vacuum for MLO}{symbol:rmv}
\addsymbol \rfv: {vacuum for FLO (\textit{flavour vacuum})}{symbol:rfv}
\addsymbolnoref \langle \slot \rangle:	{vev: $\lmv \slot \rmv$}
\addsymbolnoref _f\langle \slot \rangle_f:	{\textit{f}-vev: $\lfv \slot \rfv$}
\addsymbolnoref {:\slot:}:	{normal ordering: $\slot-\lmv \slot \rmv$}
\addsymbol \fnormal{f(\varphi_1,\varphi_2)}: {\small $f(\ct \varphi_1-\st \varphi_2,\st\varphi_1+\ct\varphi_2)-%
f(\varphi_1,\varphi_2)$\normalsize}{symbol:fnormal}
\addsymbol G_{\theta}: {mapping operator between massive and flavour states}{symbol:G}
\\
\addsymbolnoref \lag: {Lagrangian}
\addsymbolnoref \varphi: {generic field}
\addsymbolnoref \phi,S: {scalar field}
\addsymbolnoref \pi_{\slot}: {conjugate momentum of $\phi_{\slot}$: $\dot{\phi}_{\slot}$}
\addsymbolnoref \psi: {spin-\nicefrac{1}{2} field}
\addsymbolnoref \gamma^{\mu}:	{gamma matrices}
\addsymbolnoref	\diracpartial: {Dirac notation: $\gamma_\mu \partial^\mu$}
\addsymbolnoref x p:	{$x_\mu p^\mu=x^\mu p^\nu \e_{\mu\nu}$}
\\
\addsymbol T_{\mu\nu}: {stress-energy tensor}{symbol:set}
\addsymbolnoref T_{jj}: {space-space components: $T_{11},T_{22},T_{33}$}
\addsymbol \rho_{\slot}: {energy density}{symbol:energy}
\addsymbol \pressure_{\slot}: {pressure}{symbol:pressure}
\addsymbol w: {equation of state}{symbol:eqofstate}
\addsymbol \Lambda: {cosmological constant}{symbol:cosmcos}
\addsymbol G:	{Newton's gravitational constant}{symbol:newton}
\addsymbol M_{\odot}:	{solar mass}{symbol:solarmass}
\\
\addsymbolnoref \e_{\mu\nu}: {metric in flat spacetime (Minkowski)}
\addsymbolnoref g_{\mu\nu}: {metric in curved spacetime}
\addsymbolnoref \slot_\flat:	{expression in \textit{flat} spacetime}
\addsymbolnoref D_{\mu}: {covariant derivative}
\addsymbol \slot_{;\nu}:	{$D_\nu \slot$}{symbol:;cov}
\addsymbol a(t): {scale factor}{symbol:scalefactor}
\addsymbol \e: {conformal time}{symbol:conftime}
\addsymbol \Ce: {conformal scale factor}{symbol:confscalefactor}
\addsymbol \we:	{$\sqrt{\vp^2+m^2\Ce}$}{symbol:we}
\\
\addsymbol X^n[Y]:	{$\underbrace{[X,[X,[X,...}_{n},Y\underbrace{]]...]}_{n}$}{symbol:Xtothenth}
\addsymbolnoref \otto{\slot}:	{interchange of the indices 1 and 2}
\addsymbolnoref A\Big|_{x\rightarrow y}:	{$x$ is replaced by $y$ in the expression $A$}
\addsymbol A\stackrel{\mbox{\tiny{M}}}{\slot }B:	{$A$ has been reduced to $B$ via \textit{Mathematica}}{symbol:overscoreM}
\addsymbol A\stackrel{\mbox{\tiny{(n)}}}{=}B:	{$A=B$ follows from equation ($n$)}{symbol:overscore}
\end{tabbing} \clearpage}
\def\addsymbol #1: #2#3{$#1$ \> \parbox{4in}{#2 \hfill \textit{\pageref{#3}}}\\}
\def\addsymbolnoref #1: #2{$#1$ \> \parbox{4in}{#2 \hfill}\\}
\def\newnot#1{\label{#1}} 
\usepackage[export,reorder]{biblio_and_other_files/mysplitbib} 

\begin{category}{Neutrino Physics}
\SBentries{Strumia:2006db,Beuthe:2001rc,Bernardini:2010zba,Kayser:1981ye,Giunti:1993se,Giunti:1991cb,Hossenfelder:2010zj,Bilenky:2010zza,Pontecorvo:1957qd,Pontecorvo:1967fh,Gribov:1968kq,Maki:1962mu,MSW1,MSW2}
\end{category}
\begin{category}{Blasone \& Vitiello Formalism}
\SBentries{Capolupo:2004av,Blasone:1995zc,Fujii:1998xa,Blasone:1999jb,Fujii:2001zv,Blasone:2001sr,Blasone:1998hf,Blasone:2001du,Blasone:2002jv,Ji:2001yd,Blasone:2004hr,Blasone:2004yh,Blasone:2010zn,Capolupo:2006et,Capolupo:2006re,Capolupo:2008rz,Blasone:2007iq,Blasone:2008rx,Blasone:2006ch,Blasone:2008ii,Blasone:2006jx,Hannabuss:2000hy,Ji:2002tx,Capolupo:2007hy,Blasone:2007jm,Blasone:2003hh,Giunti:2003dg,Blasone:2005ae,Blasone:2001qa}
\end{category}
\begin{category}{String/Brane Theory}
\SBentries{Maartens:2003tw,Csaki:2004ay,Mavromatos:2006yy,Mavromatos:2005bu,Mavromatos:2008bz,Ellis:2004ay,Ellis:2005ib,Kogan:1996zv,Mavromatos:2007sp,Ellis:1999jf,Ellis:1999uh,Ellis:2000sx,Ellis:1999sd,Ellis:1999sf,Ellis:2008gg,Li:2009tt,Ellis:2003sd,Ellis:2003if,Mavromatos:1998nz,Kogan:1995df,Barenboim:2004ev,Barenboim:2006xt,mavrosarkar,Polchinski:1995mt,Horava:1995qa,Rubakov:1983bb,Akama:1982jy,Randall:1999vf,Schwarz:1999xj,Polchinski:1998rr,Kane:2010zz}
\end{category}
\begin{category}{Quantum Field Theory}
\SBentries{Umezawa:1982nv,Streater:1989vi,Rugh:2000ji,Barton:1963zi,Haag:1955ev,Lupher2,Lupher1,Bohr,weyl1950theory,blank2008hilbert,bratteli2003operator,Itzykson:1980rh,Peskin:1995ev,Borne:2002,Lehmann:1954rq,calzetta2008nonequilibrium,Bytsenko:2003tu}
\end{category}
\begin{category}{Cosmology}
\SBentries{Riess:1998cb,Perlmutter:1998np,Peebles:1994xt,Komatsu:2010fb,Weinberg:1988cp,Gurwich:2010gb,Peacock:1999ye,Padmanabhan:2002ji,Peebles:2002gy,Frieman:2008sn,Seljak:2004xh,Tegmark:2006az,Eisenstein:2005su,Copeland:2006wr,Kolb:2005da,Rau:2011fu,Garrett:2010hd,Einasto:2009zd,Bertone:2004pz,Jungman:1995df,Liddle:2009zz,amendola,BertoneBook,waldGR,weinberg}
\end{category}
\begin{category}{Quantum Field Theory in Curved Spacetime}
\SBentries{Hawking:1974sw,Unruh:1976db,Wald:2006ty,Ford:1997hb,Mukhanov:2007zz,Wald:1995yp,Kay:2006jn,Birrell:1982ix,Weldon:2000fr,Parker:1969au,Parker:1971pt}
\end{category}
\begin{category}{Supersymmetry}
\SBentries{Ellis:2010kf,Sohnius:1985qm,Coleman:1967ad,Haag:1974qh,Wess:1992cp,FigueroaO'Farrill:2001tr,Aitchison:2005cf,deWit:2002vz,Dine:2007zp,Aad:2011hh,Wess:1973kz}
\end{category}

\newcommand{\be}{\begin{equation}}
\newcommand{\ee}{\end{equation}}
\newcommand{\bea}{\begin{IEEEeqnarray}{rCl}}
\newcommand{\eea}{\end{IEEEeqnarray}}
\newcommand{\nn}{\nonumber\\}
\newcommand{\ba}{\begin{array}}
\newcommand{\ea}{\end{array}}

\newcommand{\w}{\omega}

\newcommand{\we}{\omega(p,\eta)}
\newcommand{\intw}{\int{\we d\e}}
\newcommand{\C}{\mathcal C}
\newcommand{\Ce}{\mathcal{C}(\eta)}
\newcommand{\Cpe}{\mathcal{C}'(\eta)}
\newcommand{\Cp}{\mathcal{C}'}
\newcommand{\rfv}{|0\rangle_{f}}
\newcommand{\lfv}{_{f}\langle 0 |}
\newcommand{\rmv}{| 0 \rangle}
\newcommand{\lmv}{\langle 0 |}
\newcommand{\bpsi}{\bar{\psi}}
\newcommand{\vp}{\vec{p}}
\newcommand{\vk}{\vec{k}}
\newcommand{\vq}{\vec{q}}
\newcommand{\vx}{\vec{x}}
\newcommand{\vy}{\vec{y}}
\newcommand{\e}{\eta}

\newcommand{\lag}{\mathcal L}
\newcommand{\g}{g^{\mu \nu}}
\newcommand{\st}{\sin \theta}
\newcommand{\sst}{\sin^2 \theta}
\newcommand{\ssst}{\sin^3 \theta}
\newcommand{\ct}{\cos \theta}
\newcommand{\cct}{\cos^2 \theta}
\newcommand{\ccct}{\cos^3 \theta}

\newcommand{\diracpartial}{\partial\!\!\!/}

\newcommand{\Gt}{G_{\theta}(t)}

\newcommand{\Gti}{G_{\theta}^{-1}(t)}
\newcommand{\dg}{\dagger}

\newcommand{\field}[1]{\mathbb{#1}}

\newcommand{\hilbert}[1]{$\mathfrak{#1}$}
\newcommand{\ket}[1]{| #1 \rangle}
\newcommand{\bra}[1]{\langle #1 |}

\definecolor{lightGray}{RGB}{220,220,220}
\newcommand{\slot}{\textcolor{lightGray}{\blacksquare}}
\definecolor{MyDarkBlue}{rgb}{0,0.08,0.45}

\definecolor{MyDarkGray}{RGB}{140,140,140}

\newcommand{\tgamma}{\tilde{\gamma}}
\newcommand{\bes}{\begin{equation}\begin{split}}
\newcommand{\ees}{\end{split}\end{equation}}
\newcommand{\pa}{\phi_A}
\newcommand{\pb}{\phi_B}
\newcommand{\po}{\phi_1}
\newcommand{\pt}{\phi_2}

\definecolor{lightGray}{RGB}{220,220,220}
\definecolor{MyDarkGray}{RGB}{140,140,140}
\newcommand{\via}[1]{\stackrel{\mbox{\scriptsize{(\ref{#1})}}}{=}}
\newcommand{\doublevia}[2]{\stackrel{\stackrel{\mbox{\scriptsize{(\ref{#1})}}}{\mbox{\scriptsize{(\ref{#2})}}}}{=}}
\newcommand{\triplevia}[3]{\stackrel{\stackrel{\stackrel{\mbox{\scriptsize{(\ref{#1})}}}{\mbox{\scriptsize{(\ref{#2})}}}}{\mbox{\scriptsize{(\ref{#3})}}}}{=}}
\newcommand{\viam}{\stackrel{\mbox{\scriptsize{M}}}{=}}
\newcommand{\otto}[1]{\underline{#1}_{1\stackrel{\rightarrow}{\leftarrow}2}}

\newcommand{\cdel}{\begin{center}
\------------------------------------------
\end{center}}
\newcommand{\com}[2]{ \left[#1,#2\right] }
\definecolor{MyDarkBlue}{rgb}{0,0.08,0.45}
\newcommand{\U}{\mathcal U}
\newcommand{\V}{\mathcal V}
\newcommand{\fvev}[1]{\lmv \because #1 \because \rmv}
\newcommand{\fnormal}[1]{\because #1 \because}
\newcommand{\subsectionItalic}[1]{\subsection{\textit{#1}}}

\newcommand{\mfA}{\mathfrak{A}}
\newcommand{\mfB}{\mathfrak{B}}
\newcommand{\mfC}{\mathfrak{C}}
\newcommand{\mfD}{\mathfrak{D}}
\newcommand{\mfa}{\mathfrak{a}}
\newcommand{\mfb}{\mathfrak{b}}
\newcommand{\mfc}{\mathfrak{c}}
\newcommand{\mfd}{\mathfrak{d}}

\newcommand{\Ku}{\mathfrak{A}_{rs}}
\newcommand{\Kd}{\mathfrak{B}_{rs}}
\newcommand{\Kt}{\mathfrak{C}_{rs}}
\newcommand{\Kq}{\mathfrak{D}_{rs}}
\newcommand{\ku}{\mathfrak{a}_{rs}}
\newcommand{\kd}{\mathfrak{b}_{rs}}
\newcommand{\kt}{\mathfrak{c}_{rs}}
\newcommand{\kq}{\mathfrak{d}_{rs}}

\newcommand{\uad}{u^{(a,d)}(\vp)}
\newcommand{\uadd}{u^{(a,d)}(\vp)^{\dagger}}
\newcommand{\uade}{u^{(a,d)}(\vp,\e)}

\newcommand{\VN}{V^2_N(p)}
\newcommand{\VM}{V^2_M(p)}

\newcommand{\pressure}{\mathsf{P}}
\newcommand{\trho}{\tilde{\rho}}

\newlength{\eqboxstorage}
\newcommand{\eqbox}[1]{
\setlength{\eqboxstorage}{\fboxsep}
\setlength{\fboxsep}{6pt}
\boxed{#1}
\setlength{\fboxsep}{\eqboxstorage}
}

\newenvironment{abstract}{ 
    \chapter*{\centering Abstract}%
}{}
\newenvironment{acknowledgments}{ 
    \section*{\centering Acknowledgments}%
}{}
\newenvironment{declaration}{ 
    \section*{\centering Declaration}%
}{}

\usepackage{biblio_and_other_files/IEEEtrantools}

\usepackage{xspace}

\newcommand{\myName}{{\scshape walter tarantino}\xspace}
\newcommand{\myTitle}{
{\scshape Flavour Condensate and\\ the Dark Sector of\\ the Universe}\xspace}
\newcommand{\mySubTitle}{{\scshape A thesis submitted in partial fulfilment
						for the degree of Doctor of Philosophy}\xspace}

\newcommand{\myGroup}{King's College London\xspace}
\newcommand{\myUrl}{{\sffamily \href{mailto:walter.tarantino@kcl.ac.uk}{walter.tarantino@kcl.ac.uk}}\xspace}
\newcommand{\myTime}{May 2011\xspace}
\begin{document}

\begin{titlepage}
\pdfbookmark[1]{Titlepage}{Titlepage}
\null\vfill
\begin{center}
\large

\bigskip

{\large \myName} \\

\bigskip
\bigskip

{\huge  \myTitle \\
}

\bigskip
    
\vspace{9cm}

\begin{tabular} {cc}
\parbox{0.4\textwidth}{\includegraphics[width=4cm]{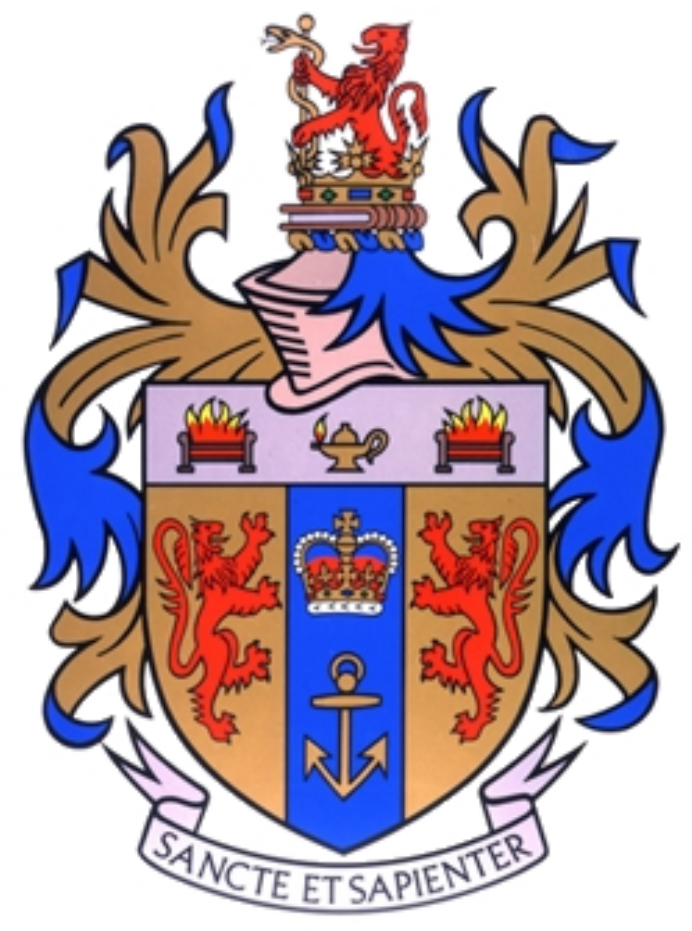}}
&
\parbox{0.6\textwidth}{{\large \mySubTitle } \\ 

					{\large
				
					 \myGroup \\
					{\normalsize \myUrl }\\
					\myTime}}
			\end{tabular}

\end{center}
\vfill
\end{titlepage}

\setcounter{page}{2}
\thispagestyle{empty}
\declaration{This thesis is the result of my own work undertaken between October 2007 and April
2011 at the Physics Department of King's College London, under the supervision of
Professor Nikolaos E. Mavromatos.\\
\\I certify that this thesis contains only my own work except where otherwise indicated.
This thesis has not been previously submitted in whole or in part for any degree or diploma
at this or any other university.\\
\\Some of the material presented in this thesis contributed to the argument of a published paper
(\href{http://dx.doi.org/10.1103/PhysRevD.80.084046}{Phys.\ Rev.\  {\bf D80 } (2009)  084046}),
a conference proceedings paper
(\href{http://dx.doi.org/10.1002/andp.201010423}{Annalen Phys.\  {\bf 19 } (2010)  258-262}),
and a manuscript currently under review
(preprint \href{http://arxiv.org/pdf/arXiv:1010.0345}{arXiv:1010.0345 [hep-th]}).
\vspace{2cm}
\begin{flushright}
Walter Tarantino\\
London, May 2011
\end{flushright}}
\vfill

\acknowledgments{
This thesis would not have been possible without the guidance and the help of my supervisors, Professor Mavromatos and Professor Sarkar,
to whom I am sincerely grateful.
}
\abstract{This thesis is devoted to the development of a non-perturbative quantum field theoretical approach to flavour physics,
with special attention to cosmological applications.
Neutrino flavour oscillation is nowadays a fairly well-established experimental fact.
However, the formulation of flavour oscillations in a relativistic field theoretical framework
presents non-trivial difficulties. A non-perturbative approach for building flavour states
has been proposed by Blasone, Vitiello and coworkers.
The formalism implies a non-trivial physical vacuum (called \textit{flavour vacuum}),
which might act as a source of Dark Energy. 
Furthermore, such a vacuum has been recognized as the effective vacuum state arising
in the low energy limit of a string theoretical model, \textit{D-particle Foam Model}. 
Developed in the Braneworld scenario, the model is characterized by a 4-dimensional brane,
representing our universe, embedded in a higher dimensional bulk space, which is punctured by
zero-dimensional topological defects, such as D0-branes.
In the attempt of probing the observable phenomenology of the D-particle foam model,
a simple toy model (two scalars with mixing \textit{\`a la} Blasone \& Vitiello on a adiabatically expanding background)
has been studied, proving
that the flavour vacuum might behave as Dark Energy under certain assumptions.\\
The first work presented in this thesis represents a development of this approach.
A more realistic model is considered, which includes two flavoured Dirac fermions
on a generic Friedmann-Robertson-Walker universe. In this framework we show that the flavour vacuum presents different
features, which are incompatible with Dark Energy.
Motivated by this discrepancy, we next embark on the analysis of a simple supersymmetric model in flat
spacetime (free Wess-Zumino), proving that the bosonic component of flavour vacuum acts as Dark Energy,
whereas the fermionic as a source of Dark Matter.
Finally we develop a new method of calculation that open the way to a non-perturbative extension
of these results for interactive theories.}

\tableofcontents
\chapter*{INTRODUCTION}\addcontentsline{toc}{chapter}{INTRODUCTION}
\pagestyle{myheadings}
\markright{\textit{INTRODUCTION}}

\section*{Introduction and Motivation}\label{sIM}\addcontentsline{toc}{section}{Introduction and Motivation}

Neutrino physics has gained attention and interest in recent years,
giving us the first chance of a glimpse beyond the Standard Model\newnot{symbol:SM} (SM).
Despite the majority of experiments in high energy physics being devoted to enhancing the precision of measurements
of the parameters of the SM, or the strength of bounds on new physics, data collected during the last decade on neutrinos
seems to require an extension or a modification of the SM itself:
in particular, the phenomenon of neutrino flavour oscillations\newnot{symbol:NFO} (NFO) finds a natural explanation
if neutrinos are massive particles,
while in the SM they are treated as massless particles.

\textit{Flavour oscillation} (or \textit{flavour mixing}) occurs when a particle created with a specific lepton flavor
(electron, muon or tau, for neutrinos) is later detected with a different flavor.
At the moment, strong experimental evidence for the neutrino flavour mixing comes mainly from two sources \cite{Strumia:2006db}:
\begin{itemize}
	\item \textit{solar evidence}: historically, the first evidence of NFO; the mechanism of \textit{flavour mixing}
	was suggested as a solution
	of the ``solar neutrino problem'' (a deficiency of electronic neutrinos, coming from the Sun,
	as expected from the Standard Solar Model), now proven by various experiments;
	\item \textit{atmospheric evidence}: disappearance of muon neutrinos in cosmic showers, which allowed
	SuperKamiokande	to give the first clear proof of the NFO mechanism, and was then investigated and confirmed
	by MAcro and other experiments.
\end{itemize}

Besides its use in neutrino physics, particle \textit{flavour oscillation}
has represented an important tool in a wide range of particle physics experiments as an interference phenomenon:
the mechanism has been used, for instance, in measuring charge-parity\newnot{symbol:CP} (CP) violation, meson mass differences,
unknown parameters of the Standard Model in its minimal extension,
Einstein-Podolsky-Rosen correlations and Bell inequalities \cite{Beuthe:2001rc,Capolupo:2004av}.

Despite the importance of the \textit{flavour mixing} in experimental particle physics,
there is not full agreement among theoreticians on its formalization.
Although the procedure of formulating a theory with flavour mixing
in the framework of the Quantum Field Theory is well-known
(Cabibbo-Kobayashi-Maskawa matrix in the SM is such an example),
the derivation of its phenomenology in the context of flavour oscillations
(\textit{i.e.} the probability for a particle of changing from a specific flavour to another in a given time)
is still a matter of debate.

Usual formalism developed for scattering processes relies on the assumption of well-defined asymptotic states,
that is not adequate for describing flavour-oscillating particles.
For a long time \cite{Beuthe:2001rc,Bernardini:2010zba}
it has been thought that one-particle quantum mechanics was convenient enough for describing
flavour oscillation, and a simple approach based on plane waves has been known since the end of '60s.
Such an approach is still sufficient for fitting current observational data.
However, the increasing interest in experiments devoted to flavour oscillation phenomenology
in the last years demanded a deeper understanding of the theoretical background.
Already in  1981 Kayser \cite{Kayser:1981ye} showed that the quantum mechanical approach based on plane waves
was inherently inadequate for describing neutrino oscillations, and proposed a model based on wave packets.
A field-theoretical approach was suggested by Giunti, Kim, Lee and Lee later on \cite{Giunti:1993se},
which overcomes some drawbacks of Kayser's model.
Far from being the conclusive solution of the theoretical problem,
many other approaches have been developed in more recent years in the attempt
of providing a coherent relativistic framework for oscillation phenomenology
(see for instance \cite{Beuthe:2001rc} and references therein).

Among others, an interesting non-perturbative field-theoretical approach has been suggested and developed
by Blasone, Vitiello and coworkers \cite{Blasone:1995zc}.
The aim of their formalism is to solve one of the most important open problems:
the definition of sensible \textit{flavour states}, in the framework of the Second Quantization.	
As extensively explained later on,
in this attempt, they are led by the established formalism to consider a Fock space for the \textit{flavour states}
that is different from the usual Fock space ordinarily used in Quantum Field Theory\newnot{symbol:QFT} (QFT)
in which the states of the basis have a \textit{well defined mass}.

Surprisingly, these results opened the door to physical implications far beyond the neutrino oscillations.
As a consequence of the different Fock space considered, it has been claimed that the ground state of this new
Fock space (the so called \textit{flavour vacuum}), differing from the standard vacuum, might have non-zero energy
and provide a source for \textit{Dark Energy} (DE) \cite{Blasone:2004yh,Capolupo:2008rz}.

\textit{Dark Energy} is the hypothetical fluid which fills all the observable Universe homogeneously, that has been suggested
in order to explain the present acceleration of the expansion of the Universe.
Many attempts have been made in literature to find a suitable candidate for this exotic fluid,
but so far no satisfactory answer has been found.
From this point of view, Blasone \& Vitiello (BV) formalism
could represent an important connection between particle physics and astrophysics,
providing an elegant solution for one of the most relevant problems in astrophysics,
through a better understanding of the Theory of Quantum Fields.

Moreover, the formalism has found an interesting application in the context of Quantum Gravity too.
A pioneering work in this direction was published in 2007 \cite{mavrosarkar}:
the connection between the \textit{flavour vacuum} and a toy model inspired by the Brane World scenario \cite{mavrosarkar},
was explored. Features of this model had already been studied extensively in literature.
In this work, the authors considered a microscopic (at the string scale) mechanism,
that can explain the emergence of flavour mixing at macroscopic scales (for which ordinary QFT applies).
In their analysis, they suggested that the BV formalism could provide the correct description
for the effective vacuum state in the low energy limit of their stringy model.
An effective toy model including the BV formalism was examined with specific attention to cosmological phenomenology.
Although the analysis was worked out by means of several heuristic assumptions in order to simplify calculations and to
supply details of the brane model not yet understood, preliminary results were promising.

This thesis moves a step forward in this same direction.
As we will see, we will consider two field-theoretical models and focus on the phenomenology
of the \textit{flavour vacuum} at cosmological scales.

BV formalism for particle flavour mixing might not be a mature theory yet.
Despite its current problems, it is an interesting and promising theory, though. From one side,
it offers a completely non-perturbative tool that might enlighten some aspects of the field-theoretical
formulation of the flavour mixing mechanism; on the other side,
there is a chance that the ultimate version of the formalism might be the first brick
of a new bridge connecting particle physics and cosmology in a novel as well as elegant way.

Neutrino physics has already been proven to represent a window into new physics.
It is believed that it might offer chances of getting deeper insights on particle physics
and cosmology, or even on the thorny problem of Quantum Gravity (see \cite{Hossenfelder:2010zj} and references therein).
In a time in which research in such a field finds it difficult to establish connections between theory and experiments,
it is our belief that such an opportunity cannot be neglected.


\section*{Overview}\label{sO}\addcontentsline{toc}{section}{Overview}

As mentioned in the previous section, the formalism proposed by Blasone and Vitiello (the \textit{BV formalism})
provides a quantum field-theoretical tool for constructing
eigenstates of the flavour charge, in a free theory with two or more flavours.
This thesis is meant to be a development of the formalism, paying special attention to its cosmological implications.
The chain that connects the BV formalism and this work of thesis counts two links in between:
the work done by Capolupo, Blasone, Vitiello and collaborators
and the work done by Mavromatos and Sarkar.
The basic ideas underlying these studies are the following:
\begin{enumerate}
	\item\label{BVlink} Blasone and Vitiello showed that the formalism implies a non-trivial structure of the ground state
	(the so called \textit{flavour vacuum}), giving rise to a \textit{vacuum condensate} \cite{Blasone:1995zc};
	the features of this condensate depend on the model considered (number of flavours, spin of the particles involved, and so on).
	\item\label{BVClink} Capolupo, Blasone, Vitiello \textit{et al.} suggested that such a structured vacuum might contribute to the
	Dark Energy budget of the universe \cite{Blasone:2004yh,Capolupo:2008rz}.
	\item\label{MSlink} Mavromatos and Sarkar embedded BV formalism in a model inspired
	by the \textit{Brane World} scenario \cite{mavrosarkar}. As a low-energy limit of their stringy toy model,
	they studied a simple QFT model involving two real scalar fields with flavour mixing \textit{\`a la} BV
	on a curved $1+1$ dimensional spacetime background.
\suspend{enumerate}
The thesis itself consists of three more links of the chain:
\resume{enumerate}
	\item\label{WTlinkCURVED} An extension of the work of \ref{MSlink}., in which a more realistic model (Dirac fermions on a $3+1$
	dimensional curved spacetime), is considered.
	\item\label{WTlinkSUSY} Motivated by differences in the results of \ref{MSlink}. and \ref{WTlinkCURVED}.,
	a new set-up for BV formalism that includes Supersymmetry (on flat spacetime) is considered.
	\item\label{WTlinkNEW} A first step towards the study of interactive models is moved (all existing analyses being on free theories),
	thanks to a method of calculation that has been fruitfully developed for the supersymmetric model of \ref{WTlinkSUSY}.
\end{enumerate}

Our first analysis (link \ref{WTlinkCURVED}.) has a twofold importance: on one hand, it applies ideas developed in \ref{MSlink}.
in a more realistic context; on the other hand, a step forth in defining BV formalism on curved backgrounds is moved.
In particular, analogies between BV formalism and the formulation of QFT on curved backgrounds are enlightened,
and special care is dedicated to the definition of the \textit{flavour vacuum} in such contexts.
Although the model considered (and therefore its phenomenology)
reflects specific features of the microscopic model in which \textit{BV formalism} is embedded,
the discussion about the definition of the  \textit{flavour vacuum} in curved spacetime is of generic valence.

The second analysis (\ref{WTlinkSUSY}.) adds the new ingredient of Supersymmetry to the usual BV formalism in flat spacetime.
Besides new possible insights from the theoretical side (Supersymmetry is broken by the flavour vacuum state,
even if the symmetry is not \textit{spontaneously broken}), this analysis is of great importance for its phenomenology:
the \textit{flavour vacuum} arising from the model seems to offer a source both for \textit{Dark Energy} and \textit{Dark Matter},
the elusive matter species required by gravitational analysis of astrophysical observations, but not yet directly detected.
It should be remarked that the model, although initially motivated by the analysis of our previous model of \ref{WTlinkCURVED}.,
does not present any distinctive features imposed by a possible underlying theory, besides Supersymmetry.
Its phenomenology can be regarded therefore as a consequence of BV formalism only, when implemented on a supersymmetric theory.

The third link represents a first attempt to characterize the phenomenology of the \textit{flavour vacuum}
in presence of interactions, in a completely non-perturbative way.
An exhaustive definition of the \textit{flavour vacuum} in interactive theories is beyond the aims of the present work.
However, some technical tools developed for the analysis of \ref{WTlinkSUSY}. seem to provide some insights
on interactive models, independently of specific schemes of perturbative analysis, such as regularization and renormalization.


\section*{Outline}\addcontentsline{toc}{section}{Outline}

\paragraph{Chapter \ref{cB}} We shall guide the reader through the works that motivated this thesis:
BV formalism (Section \ref{sBVF}), the possible connection between the flavour vacuum and Dark Energy (Section \ref{sFVDE})
and the first attempt to embed  the formalism in a quantum gravity model (Section \ref{sSFV}).
Furthermore, we shall introduce some general knowledges of modern cosmology, useful for
our forthcoming analyses (Section \ref{sDSU}).
\paragraph{Chapter \ref{cAFVICS}} We shall present our first original analysis, in which BV formalism for Dirac spinors
will be implemented in curved spacetime. Special attention will be dedicated to the definition
of the \textit{flavour vacuum} on curved backgrounds.
\paragraph{Chapter \ref{cAFVIASM}} Our second analysis, in which BV formalism will be implemented in a supersymmetric theory
(Wess-Zumino model), will be presented. The idea that the \textit{flavour vacuum} might provide a source both for Dark Energy
and Dark Matter will then be introduced and discussed.
\paragraph{Chapter \ref{cANMOF}} Remarks on a specific method of calculation developed for the latter model will be presented.
We shall then emphasize the importance of this tool in further analyses on interactive theories,
by showing preliminary non-perturbative results.
\paragraph{Chapter \ref{cC}} Conclusions will be drawn.

\paragraph{Appendix \ref{aC23}} Details about calculations behind results in Chapter \ref{cAFVICS} will be provided
in this first appendix.
\paragraph{Appendix \ref{aC4}}  All relevant results of BV formalism for real spin-0 and spin-\nicefrac{1}{2} fields
(already present in literature) will be entirely rederived.
Furthermore, (original) calculations that lead to results of Chapter \ref{cAFVIASM} will be presented.
This appendix is a self contained piece of work which might be useful for future references on BV formalism and
\textit{flavour vacuum}.
\paragraph{Appendix \ref{aC5}} The method presented in Chapter \ref{cANMOF} will be used to rederive results of the previous
analysis in a much more direct way.


\section*{Notation}\addcontentsline{toc}{section}{Notation}\label{section:notation}

The analysis of model in curved spacetime, that will be presented in Chapter \ref{cAFVICS}, and the analysis of the supersymmetric model discussed
in Chapters \ref{cAFVIASM} and \ref{cANMOF} are completely independent. Moreover, the background literature
from which our analysis started is different:
\begin{itemize}
	\item Chapter \ref{cAFVICS} mainly refers to works in \cite{mavrosarkar,Birrell:1982ix,Parker:1971pt,Blasone:1995zc},
	which provided us both with the formalism for QFT in curved backgrounds and BV formalism for Dirac fermions.
	\item Chapters \ref{cAFVIASM} and \ref{cANMOF}, on the other hand, refer to
	\cite{Sohnius:1985qm,FigueroaO'Farrill:2001tr,Capolupo:2004av,Blasone:2003hh},
	in which one can find the realization of the WZ-model here used, and BV formalism for neutral scalars and Majorana fermions.
\end{itemize}
By virtue of this fact, we thought to keep the notation of the corresponding literature. The obvious disadvantage is that
throughout the present work different notations are used.
It follows that in Chapter \ref{cAFVICS}
the notation of \cite{mavrosarkar,Birrell:1982ix,Parker:1971pt,Blasone:1995zc} is adopted,
whereas in Chapters \ref{cAFVIASM} and \ref{cANMOF}
the notation of \cite{Sohnius:1985qm,FigueroaO'Farrill:2001tr,Capolupo:2004av,Blasone:2003hh} has been used.\footnote{To
be more precise: the cited works do not share the same \textit{exact} notation. In our work, we tried to
mediate between them and choose a notation that differs as little as possible from the respective references.}

In particular, it should be notices that:
\begin{itemize}
	\item for Chapters \ref{cB} and \ref{cAFVICS} Minkowski metric reads 
	\begin{equation*}\e_{\mu\nu}=diag\{-1,+1,+1,+1\},\end{equation*}
	whereas for Chapters \ref{cAFVIASM} and \ref{cANMOF} it reads 
	\begin{equation*}\e_{\mu\nu}=diag\{+1,-1,-1,-1\};\end{equation*}
	\item the symbol $\psi$ for spin-\nicefrac{1}{2} fields it refers to \textit{Dirac spinors} in Chapter \ref{cAFVICS},
	whereas in Chapters \ref{cAFVIASM} and \ref{cANMOF} it refers to \textit{Majorana spinors};
	\item the following symbols: $T_{\mu\nu}$ (stress-energy tensor), $\lag$ (Lagrangian),
	$\rho$ (energy density), $\pressure$ (pressure), and $w$ (equation of state),
	in different models have different expressions (in terms of fields or coordinates);
	\item $a_{\slot}^{(\dagger)}$ (ladder operators) refer to: scalar particles (Chapter \ref{cB}), Dirac particles in curved spacetime
	(Chapter \ref{cAFVICS}), Majorana particles in flat spacetime (Chapter  \ref{cAFVIASM}).
\end{itemize}
In spite of these apparent complications,
the two works here presented (the one in curved spacetime and the supersymmetric one) are perfectly distinct
and do never overlap (besides in the final comparison of the equations of state, which, however, is a feature of the model that
does not depend on a particular choice of the notation). Therefore we preferred to keep the notation
of the corresponding background literature, since it would facilitate the comparison with
already existing results and analyses. The counterpart in adopting a uniform nomenclature would have been a \textit{doubled} notation,
symbols overloaded with indices and not very intuitive.
For sake of clarity, each time a symbol will be introduced we shall recall its meaning; moreover, recurrent symbols are listed
on page \pageref{listofsymbols}.

Throughout the thesis, Planck units are used ($G=c=\hbar=1$, with $G$ Newton's gravitational constant,
$c$ the speed of light in vacuum, and $\hbar$ the reduced Planck constant).

\pagestyle{headings}
\chapter{BACKGROUND}\label{cB}

As mentioned in the introduction, this Chapter is dedicated to the background of the work of thesis.
In particular,
in Section \ref{sBVF} we will present and critically review the BV formalism for flavours states,
complementing original ideas of the group of Vitiello with discussions
about the interpretation of the \textit{flavour vacuum} as a \textit{vacuum condensate}.
In Section \ref{sFVDE} we will clarify how the \textit{flavour vacuum} might behave as a Dark Energy source,
and we will briefly report about existing analyses on the topic.
In Section \ref{sSFV} we will present the stringy model used in \cite{mavrosarkar}
and its connection with the \textit{flavour vacuum};
we will report about the results of that preliminary analysis and we will emphasize
the importance of the approach in the general framework of the study of the formalism.

The following presentation does not proceed in accordance with the historical developments of the formalism,
neither does it aspire to be an exhaustive compendium from a technical point of view.
Readers interested in further mathematical details may refer to \cite{Blasone:1995zc,Ji:2002tx,Umezawa:1982nv}.

\section{BV FORMALISM}\label{sBVF}\newnot{symbol:BV}

The formalism developed from Blasone, Vitiello and co-workers for the particle flavour mechanism is an attempt
to define a Fock space for flavour eigenstates. A first version was proposed in 1995 \cite{Blasone:1995zc}, but some
inconsistencies in the derivation of oscillation formulae
were noticed \cite{Fujii:1998xa,Blasone:1999jb,Fujii:2001zv,Blasone:2001sr} shortly after;
a revisited version, in which these discrepancies
were clarified and removed, was suggested and developed \cite{Blasone:1998hf,Blasone:2001du,Ji:2001yd,Ji:2002tx,Blasone:2002jv}
later on. More recently, a newer formulation of the mixing has been proposed \cite{Blasone:2010zn}.
In this latter work, the authors try to overcome one of the main drawbacks of the BV formalism:
the breaking of Lorentz invariance, induced by the explicit time-dependency of the \textit{flavour vacuum}, as defined in
\cite{Blasone:1995zc}.
In the present work we will consider the (amended) first version as stated in
\cite{Blasone:1995zc,Blasone:1998hf,Capolupo:2004av,Blasone:2004yh,Blasone:2003hh}.
The applicability of the analysis presented here on the recently suggested reformulation has not been studied yet.

As we already mentioned, despite the richness and relevance of phenomenology of the particle flavour mixing,
field-theoretical aspects of the mechanism
have been investigated only in recent times.
The work of Blasone and Vitiello \cite{Blasone:1995zc}
is an attempt to provide a quantum field-theoretical framework for free flavoured particles:
more specifically, they build a Fock space for states that are eigenvalues of the flavour charge operator.

In what follows we try to summarize the basic ideas underneath the BV formalism.
We will guide the reader through the quantization \textit{\`a la} BV of the simplest theory with flavour mixing:
by considering two free massive fields with flavour mixing\footnote{We will consider scalar fields for simplicity.
A generalization to more complicated systems is straightforward in principle, but might be tedious in practice because
of the complex notation. For more details, see Appendix \ref{aC4} or \cite{Blasone:2003hh,Capolupo:2004av}},
we will try to clarify how the \textit{flavour vacuum} arises from the theory, and what features we should expect it to have.
In the present Section we will concentrate on the aspects of the formalism that lead us to interpret the ground state of the theory
(the \textit{flavour vacuum}) as a \textit{vacuum condensate}. For more details about other aspects of the formalism
and the connection with NFO phenomenology, readers may refer to \cite{Blasone:2002jv}.

\subsectionItalic{Standard Neutrino Oscillation Formalism}\label{ssSNOF}

As anticipated in the introduction, a simple quantum-mechanical formalism is completely sufficient to account for current observational
data.\footnote{Such a formalism is due to the pioneering works of Pontecorvo, that first suggested a possible mixing mechanism for massive neutrinos,
 Maki, Nagawa and Sakata in the 60s \cite{Pontecorvo:1957qd,Pontecorvo:1967fh,Gribov:1968kq,Maki:1962mu}.}

In the simple case of two flavours $A$ and $B$, mixing between flavour states $\ket{\nu_A}$ and $\ket{\nu_B}$ is introduced by writing
them as a linear combination of eigenstates of the Hamiltonian $\ket{\nu_1}$ and $\ket{\nu_2}$ \cite{Bilenky:2010zza}:
\bea
\ket{\nu_A}&=&\ct \ket{\nu_1} +\st \ket{\nu_2}\nn
\ket{\nu_B}&=&-\st \ket{\nu_1} +\ct \ket{\nu_2}.
\eea
More specifically, $\ket{\nu_1}$ and $\ket{\nu_2}$ are massive eigenstate of the free Hamiltonian, and their time evolution is described by
\bea
e^{-iHt}\ket{\nu_1}&=&e^{-i\w_1t}\ket{\nu_1}\nn
e^{-iHt}\ket{\nu_2}&=&e^{-i\w_2t}\ket{\nu_2}
\eea
with $H$ the Hamiltonian operator and $\w_{\slot}$ the energy of the state.\footnote{The
symbol $\slot$\newnot{symbol:slot} will denote a generic completion of the expression, as deducible from the context.}

Therefore, the probability that the state $\ket{\nu_A}$ produced at time $t_0=0$ is then found to be $\ket{\nu_B}$ at a
generic time $t$ is given by
\bea\label{pontecorvoformula1}
\wp_{A\rightarrow B}&=&|\langle \nu_B\ket{\nu_A(t)}|^2=|\bra{\nu_B}e^{-iH t}\ket{\nu_A}|^2\nn
									&=&|\left(-\st \bra{\nu_1} +\ct \bra{\nu_2}\right)\left(\ct e^{-i\w_1t} \ket{\nu_1} +\st e^{-i\w_2t} \ket{\nu_2}\right)|^2=\nn
									&=&2\sst \cct (1-\cos(\w_1t-\w_2t))=\nn
									&=&\sin^2(2\theta) \sin^2\left(\frac{\w_1 t-\w_2 t}{2}\right)
\eea
since $\sin(2\theta)=2\st \ct$ and $\cos(2\theta)=1-2\sst$.
This formula is usually further simplified by considering
\be
\w_i=\sqrt{\vec{p}^{\;2}+m_i^2}\approx \vec{p}^{\;2}+\frac{m_1^2}{2|\vec{p}\;|}\approx \w+\frac{m_i^2}{2 \w}
\ee
with $\w\approx |\vec{p}\;|$ denoting the approximate energy of $\ket{\nu_A}$, and $t\approx L$, with $L$ the distance that
the particle traveled in the time $t$, giving rise to
\be\label{pontecorvoformula}
\wp_{A\rightarrow B}\approx \sin^2(2\theta)\sin^2\left(\frac{m_1^2-m_2^2}{4 \w}L\right).
\ee
Since formula (\ref{pontecorvoformula}) is phenomenologically well established, any field theoretical models that attempt
to extend this simple formalism must reproduce (\ref{pontecorvoformula}) in the high energy limit.

\subsectionItalic{Quantization of the Flavour Lagrangian}\label{ssQFL}

Before introducing the BV formalism, we shall start by showing the standard quantization for a \textit{field}
theory with flavour mixing.
We start by considering two classic real scalar fields: $\phi_A(x)$ and $\phi_B(x)$.
A Lagrangian \cite{Blasone:2003hh}
with \textit{flavour mixing} is built by grouping the two fields in a vector
\be\label{doublet}
\Phi_f\equiv \begin{pmatrix} \phi_A(x) \\ \phi_B(x) \end{pmatrix}
\ee
and then writing
\be\label{flavourlagrangian}
\lag =\partial^\mu \Phi_f^{\dg} \partial_\mu \Phi_f-\Phi_f^{\dg} M_f \Phi_f.
\ee
with $M_f$ a certain matrix: if the matrix was of the form
\be
\begin{pmatrix}
m^2_A & 0\\ 0 & m^2_B
\end{pmatrix}
\ee
the theory would simply describe two massive free scalar fields;
in order to have \textit{flavour mixing} we require that
\be
M_f=\begin{pmatrix}
m^2_A & m^2_{AB}\\ m^2_{AB} & m^2_B
\end{pmatrix}.
\ee
with
$m_{\slot}\in \field{R}$.
In this way cross-terms, such as $m^2_{AB} \phi_A \phi_B$, are present.
A Lagrangian in this form is said to describe two \textit{flavour fields}, $\phi_A(x)$ and $\phi_B(x)$.

In order to study the equations of motion, and therefore the dynamics of the theory,
we will to rewrite the Lagrangian in terms of two new fields $\phi_1(x)$ and $\phi_2(x)$, such that
the cross-terms disappear.
This is always possible by introducing the parameters $\theta$, $m_1$ and $m_2$ and writing
\bea\label{flavourandmassive}
\phi_A(x)&=& \phi_1 (x) \ct+\phi_2 (x) \st\nn
\phi_B(x)&=& -\phi_1 (x) \st+\phi_2 (x) \ct
\eea
and
\bea
m_A^2&=&m_1^2 \cct+m_2^2 \sst\nn
m_B^2&=&m_1^2 \sst+m_2^2 \cct\nn
m_{AB}^2&=& (m_2^2-m_1^2) \st \ct.
\eea
In terms of a vector
\be
\Phi_m\equiv \begin{pmatrix} \phi_1(x) \\ \phi_2(x) \end{pmatrix}
\ee
and a matrix
\be
M_m\equiv\begin{pmatrix}
m^2_1 & 0\\ 0 & m^2_2
\end{pmatrix}.
\ee
we can write (\ref{flavourlagrangian}) as
\be\label{linearized}
\lag =\partial^\mu \Phi_m^{\dg} \partial_\mu \Phi_m-\Phi_m^{\dg} M_m \Phi_m
\ee
that synthesizes the usual Lagrangian for two free scalar fields $\phi_1(x)$ and $\phi_2(x)$.

Once we have the Lagrangian written in terms of the ordinary fields $\phi_1(x)$ and $\phi_2(x)$,
we proceed with the quantization of the theory in the simplistic way here briefly summarized
\cite{Birrell:1982ix}:
\begin{itemize}
	\item we promote the fields from being complex functions to being operators (acting on a Hilbert space yet to be defined);
	\item we require that fields and their conjugate momenta satisfy canonical equal-time commutation rules:
		\be\label{commfield}
		[\phi_i(t,\vx),\pi_j(t,\vy)]=i\delta_{ij}\delta^3(\vx-\vy)\;\;\;\;\;\;\;{i,j=1,2}
		\ee
		with $[\phi,\phi]$ and $[\pi,\pi]$ vanishing;
	\item thanks to the equations of motion, we can decompose the fields in plane waves
		\be\label{fielddecomposition}
		\phi_i(x)=\int{\frac{d^3k}{(2 \pi)^{3/2}}\frac{1}{\sqrt{2 \w_i(k)}} \left( a_i(\vk)e^{-i\w_i(k)t}+ a_i^{\dg}(-\vk) e^{i\w_i(k)t} 					\right) e^{i \vk \cdot \vx}  }
		\ee
		with $\w_i(k)\equiv \sqrt{k^2+m_i^2}$ and $k=|\vk|$;
	\item we verify that the operators $a_i$ and $a_i^{\dg}$ satisfy the algebra
		\be\label{ladderalgebra}
		[a_i(\vk),a^\dg_j(\vq)]=\delta_{ij} \delta^3(\vk-\vq)
		\ee
		with $[a,a]$ and $[a^\dg,a^\dg]$ vanishing;
	\item we then build vectors of a space on which the operators act,
		using the so called \textit{ladder} operators $a_i$ and $a_i^\dg$:
		first we denote with $\rmv$\newnot{symbol:rmv} the vector such that $a_i(\vk)\rmv=0$ for all $\vk$,
		and second we act repeatedly on this vector with $a_i^\dg$ in order to get the other vectors:
		\be\label{fockspace}
		\begin{array}{c}
		a_i^\dg(\vk)\rmv\\
		a_i^\dg(\vk)a_i^\dg(\vk')\rmv\\
		a_i^\dg(\vk)a_i^\dg(\vk')a_i^\dg(\vk'')\rmv\\
		...
		\end{array}
		\ee
	\item we recognize that the set of all the vectors $\{ a_i^\dg(\vk)a_i^\dg(\vk')...a_i^\dg(\vk^{(n)})\rmv \}$
		(including $\rmv$) is the basis for a Fock space $\mathfrak{F}_i$: a Hilbert (\textit{i.e.} infinite-dimensional vector)  space
		that can represent physical states with an arbitrary (but finite) number of \textit{quanta};
	\item since $i=1,2$, and ladder operators with different indices commute, we can consider as total vector space the tensorial
		product $\mathfrak{F}_{TOT}=\mathfrak{F}_1 \otimes \mathfrak{F}_2$;
	\item observable operators are written in terms of the field operators; thanks to (\ref{fielddecomposition}), (\ref{ladderalgebra})
		and (\ref{fockspace}), we know how a generic operator would act on a vector of the space of physical states;
	\item it turns out that these vectors are eigenstates of the Energy and Momentum operators, defined from the Lagrangian;
	\item we are therefore led to interpret the vector $a_i^\dg(\vk)a_i^\dg(\vk')...a_i^\dg(\vk^{(n)})\rmv$
	as the physical state containing $n$ particles with momenta $\vk^{(i)}$ and mass $m_i$.
\end{itemize}
To come back to the \textit{flavour fields}, they are still defined through (\ref{flavourandmassive}), which now
has to be read as a relation between field operators. The dynamics of $\phi_A(x)$ and $\phi_B(x)$ are fully determined
by the dynamics of $\phi_1(x)$ and $\phi_2(x)$.

Therefore, we can say that from the perspective of the fields, everything is quite clear and neat.
But, what if we want to create a vector space that represents a particle of definite \textit{flavour} instead of definite \textit{mass}?
Because of the simple relation between $(\phi_A,\phi_B)$ and $(\phi_1,\phi_2)$, one would expect a simple relation
between a hypothetical \textit{flavour eigenstate} and a state of the space defined by the basis (\ref{fockspace}).
As Blasone and Vitiello showed, it turns out that things are not that easy.

\subsectionItalic{A Fock Space for the Flavour States}\label{ssFSFS}

First of all, we shall clarify why we need and what we mean with ``flavour eigenstate''.
Since in experiments we have to deal with (basically we produce and/or the detect) particles of well-defined flavour,
we wish to have an object in our theory that represents these kinds of particle. Such an object
would then be an \textit{eigenstate} of the \textit{flavour charge} operators, associated with a $SU(2)$ transformation
of the vector (doublet) $\Phi_f$ of (\ref{doublet}):
\be
\Phi_f'=e^{i \alpha_j \tau_j}\Phi_f
\ee
with $\alpha_j$ ($j=1,2,3$) real constants, $\tau_j=\sigma_j/2$, and $\sigma_j$ the Pauli matrices.
When flavour mixing is present, the Lagrangian is not invariant under this transformation and,
in the context of a theory for complex fields\footnote{A more delicate treatment is required for real fields, as we shall explain
later on.}, it leads to
\be
\delta \lag=-i\alpha_j \Phi^\dagger_f [M_f,\tau_j]\Phi_f
\ee
which allows to identify the currents
\be
J^\mu_{j}(x)=i\Phi^\dg_f\tau_j \stackrel{\leftrightarrow}{\partial^\mu}\Phi_f
\ee
where $\stackrel{\leftrightarrow}{\partial^\mu}\equiv \stackrel{\rightarrow}{\partial^\mu}-\stackrel{\leftarrow}{\partial^\mu}$
(thanks to the equations of motion) and the charges
\be
Q_{j}(t)\equiv \int d^3x J^0_j (x)
\ee
which correctly fulfill the $SU(2)$ algebra. Flavour charge operators are then defined \cite{Blasone:2001qa} as
\bea\label{flavourcharges}
Q_A(t)&\equiv& \frac{1}{2}Q+Q_{3}(t)\nn
Q_B(t)&\equiv& \frac{1}{2}Q-Q_{3}(t)\nn
\eea
with $Q$ the total charge of the system
\be
Q\equiv \int d^3x I^0(x)=\int d^3x \;i \Phi^\dg_f \stackrel{\leftrightarrow}{\partial_t}\Phi_f
\ee
which is a conserved charge, associated with the invariance of the Lagrangian under $U(1)$ transformations
\be
\Phi_f'=e^{i \alpha}\Phi_f \;\;\;\;\;\alpha\in \field{R}.
\ee
Here the flavour charges are time dependent, but in absence of mixing they would be separately conserved for each flavour.

Going back to our problem, we first point out that a generic procedure to find eigenstates for a given operator
is not known, at least at a non-perturbative level.
Before showing the solution found by Blasone and Vitiello, an important point needs to be emphasized:
it is easy to show that states such as
\bea\label{AeB}
\ket{A}&=&\ct a_1^\dg(\vk)\rmv+\st a_2^\dg(\vk)\rmv\nn
\ket{B}&=&-\st a_1^\dg(\vk)\rmv+\ct a_2^\dg(\vk)\rmv
\eea
are not eigenstates of the \textit{flavour operators} \cite{Blasone:2006jx,Blasone:2008ii},
although, in analogy with the rotation of the fields (\ref{flavourandmassive}),
one might be led to interpret them as the one-particle flavour states.
Indeed one finds \cite{Blasone:2005ae}
\begin{multline}\label{noponte}
\bra{A} :Q_A(t_0): \ket{A}=\\
\cos^4\theta+\sin^4\theta+2 \sin^2\theta \cos^2\theta \frac{1}{2}\left(\sqrt{\frac{\w_1(k)}{\w_2(k)}}+\sqrt{\frac{\w_2(k)}{\w_1(k)}}\right)
\end{multline}
where $:\slot:\equiv \slot-\lmv \slot \rmv$, $t_0=0$ for simplicity,
$\vk$ is the momentum of the particle and $\w_i\equiv\sqrt{\vk^2+m_i^2}$.
Provided that $\theta\neq0$, $m_1\neq m_2$, and $\vk\neq0$, the right hand side of (\ref{noponte})
is always less then $1$,
the flavour quantum number that one would have expected to be associated to $\ket{A}$.
It is difficult to underestimate the importance of this result: these states (\ref{AeB}) are the field-theoretical equivalent
of the quantum-mechanical states $\ket{\nu_A}$ and $\ket{\nu_B}$ introduced in Section \ref{ssSNOF} and
commonly used in the treatment of the neutrino oscillation: this is the reason why a field-theoretical approach
to the problem needs more sophisticated tools.

Starting from the operators $\phi_A(x)$ and $\phi_B(x)$, and recovering the procedure above explained for $\phi_1(x)$
and $\phi_2(x)$, it is not possible to build an equivalent Fock space for the \textit{flavour states}:
as soon as we start we are stopped by the fact that flavour fields cannot be decomposed into plane waves, just like
in (\ref{fielddecomposition}). This problem prevents us from building proper
\textit{flavour ladder} operators that follow the algebra (\ref{ladderalgebra}) (\textit{cf} \cite{Giunti:1991cb}).

A solution to this dilemma was found by Blasone, Vitiello and coworkers: they were actually able to define operators that satisfy
the algebra (\ref{ladderalgebra}), and using them they built a Fock space whose basis is a set of eigenstates of the
\textit{flavour charge} operators.
Starting again from
\bea\label{flavourandmassive1}
\phi_A(x)&=& \phi_1 (x) \ct+\phi_2 (x) \st\nn
\phi_B(x)&=& -\phi_1 (x) \st+\phi_2 (x) \ct
\eea
they introduced an operator $G$\newnot{symbol:G}, such that
\bea\label{G1WT}
\phi_A(x)&=& G^{-1}_{\theta} \phi_1 (x) G_{\theta}\nn
\phi_B(x)&=& G^{-1}_{\theta} \phi_2 (x) G_{\theta}
\eea
Thanks to this definition, they found an explicit expression for the operator $G_\theta$, learning that
it depends on time (not on space coordinates, therefore from now on we will denote it as $G_{\theta}(t)$) and is unitary
($G^{-1}_{\theta}(t)=G_{\theta}^{\dg}(t)$). In our simple model, $G_\theta(t)$  is given by
\be
G_\theta(t)=e^{i \theta \int d^3 x(\dot{\phi}_2(x)\phi_1(x)-\dot{\phi}_1(x)\phi_2(x))}
\ee
as shown in Appendix \ref{asusyBS}.
Using this operator, they defined \textit{new operators}
\bea\label{flavlad}
a_A(\vk,t)&\equiv& \Gti a_1(\vk) \Gt\nn
a_B(\vk,t)&\equiv& \Gti a_2(\vk) \Gt
\eea
and \textit{new vectors}
\be\label{flavourstates}
\begin{array}{c}
\rfv\equiv\Gti\rmv\\
\\
a_A^{\dg}(\vk,t)\rfv,\;\;a_A^{\dg}(\vk,t)a_A^{\dg}(\vk',t)\rfv,\;\;...\\
\\
a_B^{\dg}(\vk,t)\rfv,\;\;a_B^{\dg}(\vk,t)a_B^{\dg}(\vk',t)\rfv,\;\;...\\
\\
a_A^{\dg}(\vk,t)a_B^{\dg}(\vk',t)\rfv,\;\;...
\end{array}
\ee
and so on (by just iteratively applying operators $a_A^\dg$ and/or $a_B^\dg$ on the vector $\rfv$\newnot{symbol:rfv}).
The interesting properties of these operators and vectors are \cite{Blasone:2003hh}:
\begin{itemize}
	\item the vectors are eigenstates of the \textit{flavour charge} operators,
	\item the operators (\ref{flavlad}) and their hermitian conjugate follow the canonical equal-time commutation algebra:
		\be
		[a_\iota(\vk),a^\dg_\kappa (\vq)]=\delta_{\iota \kappa} \delta^3(\vk-\vq)\;\;\;\;\;\;\;\;\iota,\kappa=A,B
		\ee
		with $[a,a]$ and $[a^\dg,a^\dg]=0$ vanishing,
	\item the vector $\rfv$ is annihilated by all $a_\iota(\vk)$, with $\iota=A,B$:
		\be
		a_A(\vk)\rfv=a_B(\vk)\rfv=0\;\;\;\;\forall \vk\in \field{R}^3.
		\ee
\end{itemize}
In analogy with the Fock space for free particles of well defined momentum and mass, built in the previous section,
the authors suggested to interpret the set
\be
\big\{a_\iota^\dg (\vk) a_{\kappa}^\dg (\vk')... a_{\lambda}^\dg (\vk^{(n)})\rfv \mid
\iota,\kappa,...,\lambda=A,B \big\}
\ee
as a basis for a Fock space of states with an arbitrary number of \textit{flavour particles}:
the vector $a_A^{\dg}(\vk,t)\rfv$ would represent the state of a free particle with well defined \textit{flavour} $A$
and \textit{momentum} $\vk$, and so on. The vector $\rfv$ was therefore named the \textit{flavour vacuum},
whereas the operators in (\ref{flavlad}) are called \textit{flavour ladder} operators.
The feature of the \textit{flavour mixing} emerges quite naturally:
the \textit{flavour states} (\textit{flavour vacuum} included)
are \textit{not} eigenstates of the Hamiltonian that one derives from (\ref{linearized}),
therefore the physical system
(\textit{i.e.} the free particle created with a well defined flavour) might ``jump'' from a \textit{flavour state}
to the other (\textit{i.e.} the particle \textit{oscillates} between different flavours), while evolving with time.
The \textit{flavour mixing} mechanism is therefore recovered in this formalism and
field-theoretical \textit{exact} formulae for the oscillations can be deduced \cite{Blasone:2003hh}.
In particular, recalling our discussion in Section \ref{ssSNOF},
the probability that a particle of flavour $A$ at time $t_0=0$ is found to be of flavour $B$ at a generic time $t$ is given by
\cite{Blasone:2003hh}
\begin{multline}\label{NOBV}
\wp_{A\rightarrow B}=\sin^2(2\theta)\left(U_+^2(k)\sin^2\left(\frac{\w_2(k)-\w_1(k)}{2}t\right)+\right.\\
											\left.+U_-^2(k)\sin^2\left(\frac{\w_2(k)+\w_1(k)}{2}t\right)\right)
\end{multline}
with
\be
U_{\pm}\equiv \frac{1}{2}\left(\sqrt{\frac{\w_1(k)}{\w_2(k)}}\pm\sqrt{\frac{\w_2(k)}{\w_1(k)}}\right)
\ee
and $\w_i(k)=\sqrt{\vec{k}^2+m_i^2}$,
that correctly reduces to (\ref{pontecorvoformula1}) in the limit $|\vec{k}|\gg m_i$.

Such a result can also be obtained in a different way: in terms of the flavour ladder operators (\ref{flavlad}),
flavour charge operators defined in (\ref{flavourcharges}) are written as \cite{Blasone:2005ae}
\bea\label{flavourchargessimply}
::Q_A(t)::&=&\int d^3k\left(a^\dg_A(\vk,t)a_A(\vk,t)-b^\dg_A(-\vk,t)b_A(-\vk,t)\right)\nn
::Q_B(t)::&=&\int d^3k\left(a^\dg_B(\vk,t)a_B(\vk,t)-b^\dg_B(-\vk,t)b_B(-\vk,t)\right)
\eea
where $::\slot::$ denotes the normal ordering with respect to the flavour vacuum $::\slot::\equiv \slot-\lfv \slot \rfv$
(required in order to have $\lfv ::Q_\iota:: \rfv=0$, $\iota=A,B$) and $b^{(\dg)}$ are ladder operators for antiparticles.
The probability (\ref{NOBV}) is exactly recovered
when one considers the expectation value of the $B$-flavour charge operator with respect to the state $a^\dg_A\rfv$
\cite{Blasone:2001du,Capolupo:2004av}:
\begin{multline}\label{chargeexp}
\lfv a_A(\vk,t)::Q_B(t)::a^\dg_A (\vk,t)\rfv=\\
\sin^2(2\theta)\left(U_+^2(k)\sin^2\left(\frac{\w_2(k)-\w_1(k)}{2}t\right)+U_-^2(k)\sin^2\left(\frac{\w_2(k)+\w_1(k)}{2}t\right)\right).
\end{multline}
From (\ref{flavourchargessimply}) it is also simple to infer that the eigenvalues of the flavour charge operators
are given by all possible integers (positive as well as negative, thanks to the contribution of antiparticles).\footnote{In
case of real fields, flavour charge operators, as defined in this section, identically vanish. This can be easily seen in
formulae (\ref{flavourchargessimply}) when antiparticles are set to be equal to particles ($a^{(\dg)}=b^{(\dg)}$).
It is therefore necessary to find other operators which enable us to distinguish flavour states. 
In \cite{Capolupo:2004av} it has been suggested that $T^{0j}_A\equiv G^\dg_{\theta}T^{0j}_1 G_{\theta}$ and 
$T^{0j}_B\equiv G^\dg_{\theta}T^{0j}_2 G_{\theta}$ (with $T^{\mu\nu}_i$ the stress energy tensor for a free field of mass $m_i$)
may play this role, since it correctly reproduces flavour oscillations
when its expectation value (correctly normalized) is taken with respect to a flavour state,
in analogy of (\ref{chargeexp}) and (\ref{NOBV}).}

Formulae for realistic cases deduced within the BV formalism are consistent with the ordinary quantum-mechanical
analysis. However, corrections induced to ordinary formulae seem not be observable with current experimental data
\cite{Beuthe:2001rc,Capolupo:2004av}.
Nonetheless,
the phenomenology of the formalism might be richer than just correcting well-known formulae to inaccessible scales of energy.

A legitimate question might be raised here: provided with the definition of the flavour states in (\ref{flavourstates}),
can we write those states $\ket{\slot}_f$ in terms of the usual states $\ket{\slot}$
in a simpler way? In other words, being the linear space generated by linear combinations of
\be\label{massbasis}
\big\{a_i^\dg (\vk) a_{j}^\dg (\vk')... a_{k}^\dg (\vk^{(n)})\rmv \mid
i,j,...,k=1,2 \big\}
\ee
the space of \textit{all} many\footnote{yet finite
in total}-particle states, we would expect to be able to express a vector $\ket{\alpha}_f$ as a linear combination of vectors in
(\ref{massbasis}).
Moreover, what is the relationship between the ``flavour'' vacuum $\rfv$ and $\rmv$, the ordinary vacuum?
Do they coincide (up to some phase) or is the flavour ``vacuum'' is not actually empty? What would that mean \textit{physically}?
Surprisingly, a simple answer to these questions fostered a long discussion on unexpected physical implications of the formalism.

As shown in several works (by BV first\cite{Blasone:1995zc} and by other independent groups then \cite{Hannabuss:2000hy,Ji:2002tx}),
\textit{it is impossible to express the} flavour vacuum \textit{as a linear combination
of states belonging to (\ref{massbasis})}. Furthermore, not one of the flavour states (\ref{flavourstates}) can be expressed in such a way.
To understand the implications of the previous statements we should step back in our discussion
on the formulation of the \textit{flavour mixing} within the framework of QFT and recall some basic notions about QFT itself.

\subsectionItalic{Physical States, Hilbert Space and Fock Space}\label{ssPSHSFS}\newnot{symbol:nonseparable}\newnot{symbol:separable}

In a field-theoretical approach\footnote{This and the next Sections are based on \cite{Umezawa:1982nv,Lupher1,Lupher2}},
a many-particle state is described by a vector of a Hilbert space, that we will denote with \hilbert{H}.
In order to specify this Hilbert space, we assume that a single-particle state can be classified by
a discrete set of states, labeled by the index $i=1,2,3,...$, that we will call \textit{classifying-states} or
\textit{c-states}.\footnote{For instance, a wave packet can be expressed as a sum of harmonic oscillator wave functions,
classified by the \textit{principal quantum numbers}.}
Therefore we say that a vector representing a many-particle state can be identified by the number $n_i$
of particles occupying the $i$-th c-state, and we denote it with $\ket{n_1,n_2,n_3,...}$
(the so called \textit{occupation number representation}).
In the case of bosons $n_i\in \field{N}_0$, whereas for fermions $n_i=0$ or $1$.
This allows us to treat not only states with a finite number of particles, but also states with an \textit{infinite}
number of particles. For instance, the vector $\ket{1,1,1...}$ counts infinite particles, each of them in a different
c-state. It can be shown that the set $\{\ket{n_1,n_2,n_3,...}\}$ is
\textit{uncountable}\footnote{A
set is said to be \textit{countable} iff there exists a one-to-one correspondence between the elements of the set and
the natural numbers $\field{N}$ (formally the symbol $\aleph_0$ is used to denote such a \textit{countable} infinity);
it is said to be \textit{uncountable} iff its elements are in a one-to-one correspondence with real numbers $\field{R}$.
To understand why $\{\ket{n_1,n_2,n_3,...}\}$ is \textit{uncountable},
one can start by considering the fermionic case: the generic vector $\ket{n_1,n_2,n_3,...}$ has
a \textit{countable} infinite,  $\aleph_0$, number of ``slots'', denoting the c-states;
these slots can be filled with one of the two options: 0 or 1;
therefore, the total number of possible vectors is given by $2\cdot2\cdot2\cdot...=2^{\aleph_0}$,
that corresponds to an \textit{uncountable} infinity.
Similarly, in the case of bosons, the \textit{countable} infinite slots can be filled by a \textit{countable} number of particles;
therefore the total number of possible vectors corresponds to $\aleph_0 ^{\aleph_0}$, that is again an \textit{uncoutable} infinity
\cite{Lupher1}.
}
and therefore it cannot be used as a basis for a \textit{separable} Hilbert space.\footnote{A
Hilbert space is separable if it admits a \textit{countable} basis such that any element of the space is
either a linear combination of the basis or the limit of a series of linear combinations.
If such a basis exists but is \textit{uncountable}, the Hilbert space is said \textit{non-separable}.
}

In particle physics, we usually work on a space that is a subset of \hilbert{H}.
This subset is a Hilbert space itself and we will call it \hilbert{F_0}\newnot{symbol:massfock}.
Moreover, its basis is a \textit{countable} subset of $\{\ket{n_1,n_2,n_3...}\}$, therefore \hilbert{F_0} is \textit{separable},
a desirable property \cite{Streater:1989vi}. The countable basis that we consider is given by
all vectors of $\{\ket{n_1,n_2,n_3...}\}$, in which the total sum $\sum_i n_i$ is finite:
\be
\Big\{\ket{n_1,n_2,n_3,...} \mid \sum_i^\infty n_i=N,\forall N \in \field{N}_0\Big\}
\ee
For the free theory described by the Lagrangian (\ref{linearized}),
the space \hilbert{F_0} spanned by this basis is the ordinary \textit{Fock space} that we introduced in Section \ref{ssQFL}:
it includes all states with a \textit{finite}, though arbitrarily large,
total number of free particles
(the vector $\ket{1,1,1...}$, for instance, does not belong to this space).
The space \hilbert{F_0} of free particles is also used in interactive theories: in fact,
it is sufficient in perturbation theory
to account for scattering processes involving an arbitrary number of (incoming or outcoming) particles,
to an arbitrarily high order in perturbation theory.
Furthermore, \hilbert{F_0} and the creation and annihilation operators ($a_i$ and $a_i^\dagger$) acting on it realize
a representation\footnote{A representation is a realization of a certain algebra via linear operators acting on a suitable vector space.}
of the algebra of Canonical Commutation Relations.\footnote{In case of bosonic particles we have the Canonical Commutation Relations \newnot{symbol:CCR}
(CCR), as stated in (\ref{ladderalgebra}), whereas for fermions we have Canonical Anti-commutation Relations\newnot{symbol:CAR} (CAR),
in which the anti-commutator replaces the commutator.
Although we will mention just CCR, all what we say can be extended for fermionic particles and CAR representations.}

\subsectionItalic{Unitary Inequivalent Representations}\label{ssUIR}

\begin{figure}[t]
\centering
\includegraphics[width=6.5cm]{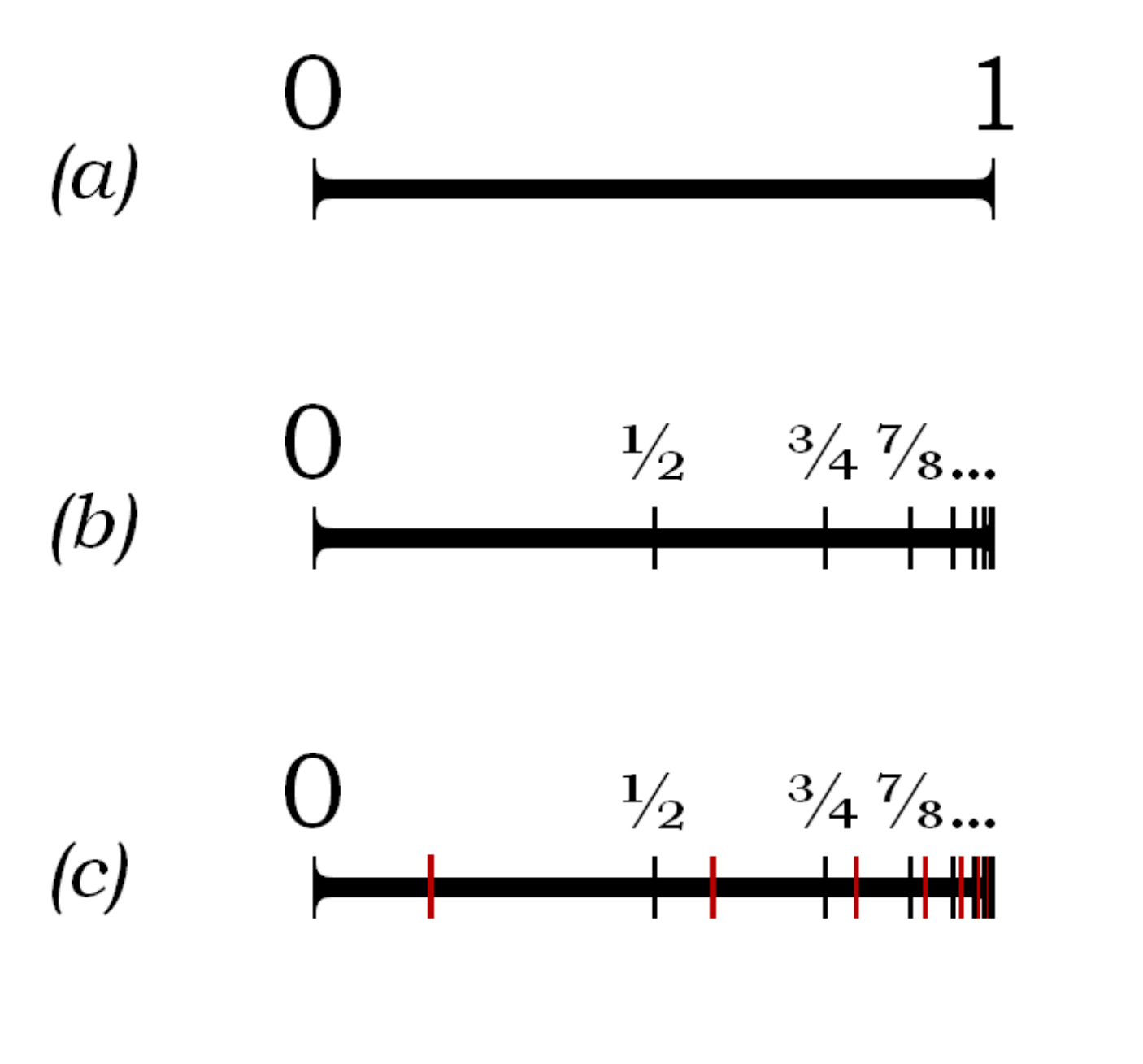} \hfill
\caption{\small{\textit{(a)}
In the fermionic case, it is easy to create a correspondence between the
vectors $\{\ket{n_1,n_2,n_3,...}\}$ (the basis of \hilbert{H}) and the numbers in the interval $[0,1)$.
In the \textit{binary} notations, a number of this interval is expressed as $0.\slot$,
where $\slot$ is a specific sequence of $0$'s and $1$'s; conversely, any such a sequence corresponds to a number of the interval,
and any sequence also corresponds to a vector of $\{\ket{n_1,n_2,n_3,...}\}$,
since each $n_i$ can be either $0$ or $1$. The one-to-one correspondence between the vectors
$\{\ket{n_1,n_2,n_3,...}\}$ and the numbers of the interval $[0,1)$ follows.
\textit{(b)}
The numbers in the interval $[0,1)$ are an uncountable infinity, and therefore the basis of \hilbert{H} is uncountable.
If we want to choose a \textit{countable} set of vectors, for building a \textit{separable} Hilbert space,
we can choose, for instance, the point $0$ (corresponding in the map above stated to the vector $\ket{0,0,0,...}$),
the point $1/2$ (corresponding to $\ket{1,0,0,...}$), the point $1/2+1/4$ ($\ket{1,1,0,0,...}$), and so on.
\textit{(c)}
A second countable set of points, and therefore a second basis for a new separable Hilbert space,
can be selected by choosing the first point between $0$ and $1/2$, the second point between $1/2$ and $1/4$, and so on.
Since there is a \textit{uncountable} infinity of points between $0$ and $1/2$, there are \textit{uncountable}
infinite ways to choose a second set, and therefore a second basis.
Such a basis will not share any vector with the previous basis,
and therefore the Hilbert space built of out it will be orthogonal to the one built out of the first basis of \textit{(b)}.}}
\label{fig:segment}
\end{figure}

But what happens if we choose a different subset of \hilbert{H}?
In that case, it might happen that the space spanned by the new basis (we call it \hilbert{F_1})
would not overlap with the previous one (see Figure \ref{fig:segment}).
This is exactly the case if both \hilbert{F_0} and \hilbert{F_1} are representations of the CCR algebra:
here, it is impossible to write vectors of one space as a superposition of basis vectors of the other space.
Intuitively, we can say that \hilbert{F_0} and \hilbert{F_1} are two \textit{orthogonal} subspaces of \hilbert{H}.

This is a consequence of a fundamental theorem of Quantum Mechanics, due to Stone and von Neumann\footnote{In
the following we will focus on the consequences of the theorem for our particular case, neglecting
technical details, at the expense of mathematical rigor. The interested reader
can find more details in \cite{weyl1950theory,blank2008hilbert}.}:
in the case of systems with a \textit{finite number} of degrees of freedom\footnote{The
	number of the degrees of freedom must not be
	confused with the dimensionality of the Hilbert space used to represent the phenomenon:
	we need a two-dimensional linear space to describe the spin of an electron (a single degree of freedom);
	but we need an infinite-dimensional Hilbert space for describing the energy of a particle in a box (characterized
	again by one degree of freedom) \cite{Lupher2}.},
given any two representations $S$ and $S'$ of CCR on a vector space $V$,
there exists a \textit{unitary} operator that maps the two representations over the
same vector space, in formulae:
\be\label{proper}
\exists\;\;\; U:V\rightarrow V
\ee
such that (unitary condition)
\be\label{unitary}
U^* U=U U^*=\field{I}
\ee
(where $\field{I}$ is the identity operator), and
such that
\be\label{map}
S=US'U^*.
\ee
The representations are said to be \textit{unitarily equivalent}.
This is actually the case for simple Quantum Mechanics,
in which context the theorem had great historical relevance, being crucial in the proof that
the Heisenberg's formulation in terms of infinite matrices was equivalent to Schr�dinger's wave function formalism.

In Quantum Field Theory the degrees of freedom are infinite\footnote{In
	fact, they are a \textit{countable} infinity. Since
	fields are defined on every point of the spacetime, one might be led to conclude that
	QFT should have \textit{uncountable} infinite degrees of freedom. However, a quantum field defined at spacetime
	points $\phi(x)$ is a mathematically ill-defined object. To give it sense, one considers a \textit{countable}
	set of test functions $f_i(x)$ in order to ``smear'' it spatially: $\phi(f_i,t)=\int dx \phi(x,t) f_i(x)$.
	The countability of the set $\{f_i(x)\}$ leads therefore to the countability of the degrees of freedom of $\phi$.
	The pathological behaviour of $\phi(x)$ reflects the unphysical idealization
	of the possibility of measuring the field with infinite precision (\cite{Bohr},
	mentioned in \cite{Rugh:2000ji}).},
and Stone-von Neumann theorem no longer applies: two different representations of CCR are \textit{unitarily inequivalent}.
More specifically, this can be realized as follows:
if two representations $S$ and $S'$ are mapped by the mean of unitary operator $U$, in accordance with (\ref{unitary}) and (\ref{map}),
the operator $U$ will no longer satisfy (\ref{proper}):
\be
\exists \;U:V\rightarrow V'\;\;\;\mbox{but}\;\;\;V\cap V'= \emptyset
\ee
in other words, the operator $U$, acting on a vector of $V$, gives as a result a vector of the space $V'$ ,
that is \textit{orthogonal} to $V$, and vice versa.
In this case, we say that the operator is \textit{improper} \cite{Barton:1963zi}, meaning that it maps orthogonal spaces.
The transformation between the two representations leaves the
CCR for the operators unchanged, but it does not transform vectors of one representation into other vectors of the same Hilbert space.

In this sense we can now understand the nature of the Fock space for the flavour states built in the BV formalism.
This space, provided with the \textit{flavour ladder operators}, is a representation of the CCR algebra
and, therefore, its intersection with \hilbert{F_0} is empty. Both \hilbert{F_0} and \hilbert{F_f} (that from now on will denote the
Fock space for flavours\newnot{symbol:flavourfock}) are subsets of \hilbert{H}, the space of linear combinations of all physical states, but
do not share any vector: it is impossible to write \textit{flavour states} as a linear combination
of states of \hilbert{F_0}, the usual Fock space, and vice versa.
In the case of the simple model presented in Section \ref{sBVF},
the role of $U$ is played by the operator $G_{\theta}$, defined by the relations in (\ref{G1WT}):
it is a \textit{unitary} operator, well defined on \hilbert{H},
but it also \textit{improper}, since it maps the vectors of \hilbert{F_0} into vectors of \hilbert{F_f}, and vice versa.\footnote{Since
	$G_\theta$ depends on time, the flavour Fock space defined at a chosen time is actually different than
	the flavour Fock space defined at a different time: being $\mathfrak{F_f}(t)$ defined as the space spanned by the basis
	$ \{ G_\theta (t) \ket{\alpha}|\ket{\alpha} \in \mathfrak{F_0} \}$, it follows that
	$\mathfrak{F_f}(t_1)\neq \mathfrak{F_f}(t_2)$ if $t_1 \neq t_2$. This gives rise to the complex structure of the
	oscillation formulae (the probability of one flavour of jumping into another state at different times). In this work we will
	not deal with the phenomenology of oscillations. The interested reader may refer to \cite{Capolupo:2004av} and works cited therein.}

The Fock space for flavours is just one of the (uncountable) infinite ways
to choose a subset of \hilbert{H} (Figure \ref{fig:segment}).
In fact, any time that we transform the ladder operators in new operators that follow the correct CCR algebra, we can
construct, by means of the new operators, a new Fock space that is orthogonal to \hilbert{F_0}.
Familiar examples \cite{Umezawa:1982nv} are the \textit{Bogoliubov transformations}, for which
the operators $\alpha(k)$ and $\beta(k)$ (with $[\alpha(k),\alpha^\dagger(l)]=[\beta(k),\beta^\dagger(l)]=\delta(l-k)$
and all other commutators being zero) are transformed in the new operators
\bea\label{bogoexample}
a(k)&=&c_k \alpha(k)-d_k \beta^\dagger(-k),\nn
b(k)&=&c_k \beta(k)-d_k \alpha^\dagger(-k)
\eea
with $c_k^2-d_k^2=1$,
and the simple \textit{c-number shift} (also called \textit{field translation}), from $\alpha(k)$ to
\be
a(k)=\alpha(k)+c
\ee
with $c\in \field{C}$, or, slightly more generally, to
\be
a(k)=\alpha(k)+c_k\;\;\;\;\;\;\;\;\;\;\;\;\mbox{with}\;\;\;\int d^3k |c_k|^2=\infty.
\ee
As we will see in Section \ref{ssTFFS}, the relations (\ref{flavlad}) that define the flavour ladder operators can be further
simplified and reduced to expressions resembling Bogoliubov transformations.

\subsectionItalic{Two Interpretations}\label{ssTI}

As already mentioned, vectors out of the Fock space \hilbert{F_0} are not used in the ordinary (perturbative) theory of scattering,
and consequently in the everyday calculations of particle physicists.
Nevertheless, they play a fundamental role in statistical quantum mechanics \cite{bratteli2003operator}:
modeling a gas, described by an uniform nonzero density of particles on a spatially infinitely extended region, requires
vectors that contain an \textit{infinite} total number of particles and therefore do not belong to \hilbert{F_0}.
From this, the first interpretation of \hilbert{F_f} comes naturally: rather than describing states with a finite number of \textit{flavour
particles}, vectors of \hilbert{F_f} can be regarded as states for a \textit{gas} with infinite particles \textit{of well-defined mass}.
Further details about this interpretation will be provided in the next Section, when we will discuss the role of
the ground state of \hilbert{F_f}, the \textit{flavour vacuum}, as a \textit{vacuum condensate}.

Without invoking infinite-particle systems, Haag taught us that inequivalent representations necessarily occur whenever
an interactive theory is considered from a non-perturbative point of view \cite{Haag:1955ev}.
Intuitively speaking, we could say:
in the perturbative expansion of an interactive theory, a state with two interactive particles can be written as a finite
series of vectors of \hilbert{F_0} (a finite number of free particle creations and annihilations occurs);
but, as we consider higher order perturbations, we add terms to this series (we consider more and more particles
being created or annihilated): the (infinite) limit of the series (the non-perturbative theory) is a vector that is not part of \hilbert{F_0}
(an infinite number of created and annihilated particles are considered).
From a more formal point of view, we can notice that the \textit{free} theory described by the Lagrangian (\ref{linearized})
can actually be regarded as an \textit{interactive} theory for the states of \hilbert{F_f}:
indeed, cross-terms $m^2_{AB} \phi_A \phi_B$ in the starting
Lagrangian (\ref{flavourlagrangian}) can be considered as terms of interaction between $\phi_A$ and $\phi_B$.


\section[FLAVOUR VACUUM AND DARK ENERGY]{FLAVOUR VACUUM AND DARK ENERGY}\label{sFVDE}

In the previous Section, we introduced the BV formalism for flavour states.
In attempting to find states of well-defined flavour, in the context of Second Quantization for QFT,
the formalism leads to a Fock space \hilbert{F_f} for flavour states that is orthogonal to the usual Fock space \hilbert{F_0}
for particles with well-defined mass, as commonly used in particle physics.
Nevertheless, \hilbert{F_f}, which describes states with a \textit{finite} number of particles with
well-defined \textit{flavour}, also makes sense as a collection of states with an \textit{infinite}
number of particles with well-defined \textit{mass}.
In the present Section we report on preliminary studies on the cosmological consequences of these ideas:
we will clarify the interpretation of the ground state of \hilbert{F_f} (the \textit{flavour vacuum})
as a \textit{vacuum condensate} (Section \ref{ssVC}).
This interpretation might lead to interesting phenomenology at cosmological scales, since the energy
of such a condensate might have gravitational effects relevant for the expansion of the universe (Section \ref{ssTP}
and \ref{ssCCP}). The first analyses on the cosmological consequences of the formalism are finally reported in Section \ref{ssFVDE}.

\subsectionItalic{Vacuum Condensate}\label{ssVC}

Perhaps the most important consequence of the orthogonality of \hilbert{F_0} and \hilbert{F_f} is that the vacuum $\rfv$ defined in
\hilbert{F_f} is not a trivial rotation of the vacuum $\rmv$ of \hilbert{F_0}.
They are both ``vacua'' in the sense that they are annihilated by all annihilation operators of the respective representation
($\rfv$ being annihilated by $a_f(\vk)$  and $\rmv$ by $a(\vk)$, for all values of $\vk$).
But only $\rmv$ represents the state with no particles at all: the state that in the \textit{occupation number representation}
is denoted by $\ket{0,0,0...}$.
This means that the state $\rfv$ represents a state with some sort of \textit{structure} (determined by the operator $G_\theta$
of (\ref{G1WT})), a state that in literature is called \textit{vacuum condensate}\footnote{Borrowing
	concepts and terminology from condensed matter physics,
	in QFT the vacuum is sometimes visualized as the low energy state of a material, and therefore, when
	an observable presents a non-zero expectation value in the vacuum state,
	it is costum to refer to it as to a \textit{condensate}.} \cite{Umezawa:1982nv}.
The two applications of Unitary Inequivalent Representations
mentioned earlier might help us in finding intuitive interpretations of this
\textit{vacuum condensate}: from one point of view, we could say that the \textit{flavour vacuum} represents the vacuum
of an interactive theory, in the sense of its \textit{lowest energy} state.
As in the case of other better-known interactive theories, the lowest energy state is subject to quantum fluctuations,
due to the possibility of creation and annihilation of particles.
Intuitively speaking, a single flavoured particle traveling in space interacts with the condensate at any point of its path,
and as a result the oscillation of its flavour occurs.

On the other hand, in this theory, particles are naturally defined through the representation of CCR
with well-defined flavour (the \textit{flavour states}) but not well-defined mass; nevertheless, it is possible to rewrite
the theory in terms of states that represent particles with well-defined mass
(as we did in (\ref{linearized})).
From this second point of view, the \textit{flavour vacuum} is not \textit{empty} but is populated
by an infinite number of particles (with well-defined \textit{mass}): the energy density is non-zero and uniform\footnote{As
	we will show in Chapter \ref{cAFVIASM}.}
over all the space.
And also a state with well-defined \textit{flavour} needs to be viewed as a collection of particles with well-defined \textit{mass}:
the interference between them causes the change of the flavour of the state.

In order to clarify the latter interpretation, we can count the particles of one representation, per unit volume,
that are present in the vacuum of the other representation.
More specifically, we can ask what is the number of particles of well-defined mass, per unit volume,
that are present in the \textit{flavour vacuum} state.
As we shall see (Appendix \ref{qab}), for the simple model introduced in the previous section with two free scalar fields,
such a number is given by
\be\label{number}
\frac{\sst}{(2 \pi)^2}\int_0^\infty k^2 \frac{(\w_1(k)-\w_2(k))^2}{\w_1(k) \w_2(k)}dk
\ee
with $\w_i(k)=\sqrt{k^2+m_i^2}$.
The integrand is clearly a positive function of the momentum $k$ and the masses $m_1$ and $m_2$; moreover,
it does not have singularities and it behaves well for large momenta, being
\be
k^2 \frac{(\w_1(k)-\w_2(k))^2}{\w_1(k) \w_2(k)}\approx \frac{1}{k^2} \;\;\;\;\;\;\mbox{for}\;\;\;k\rightarrow \infty
\ee
Therefore, we can say that the number of particles with well-defined mass per unit volume in the \textit{flavour vacuum}
is a \textit{finite positive} number. It is clear now that this state, which in the limit of \textit{infinite} volume contains
an \textit{infinite} number of particles, cannot be part of the usual Fock space for particles of well-defined  mass.

From a physical point of view, it is reasonable to think that the \textit{flavour vacuum} must be used as the \textit{physical} vacuum,
since in our experiments we detect particles of well-defined flavour.
Saying that a state is \textit{empty} means that \textit{no flavour particles are present},
and therefore it must be represented by the \textit{flavour vacuum}.
From a mathematical point of view, the \textit{flavour vacuum} arises naturally when one constructs a basis of
eigenstates of the flavour charge: the choice of the vacuum is inherent the definition of the \textit{flavour ladder operators},
being $\rfv$ the only state that is annihilated by all \textit{flavour annihilation operators}.
Moreover, Blasone, Vitiello and Henning \cite{Blasone:1998hf}
have shown that the attempt of using states such as $a_f^\dg \rmv$ (a \textit{flavour ladder operator} acting on the usual vacuum)
for describing flavour states leads to an inconsistency:
since the probability of a particle to conserve its flavour in the elapse of time $t$ would be written as
\bea
\wp_{A\rightarrow A}(t)&=&|\lmv a_A(\vec{k},t) a^\dagger_A(\vec{k},0)\rmv|^2\nn
			&=&\left|\cct+\sst e^{-i(\w_1-\w_2)t}\frac{(\w_1-\w_2)^2}{4\w_1\w_2}\right|^2
\eea
with $\w_i\equiv \sqrt{\vec{k}^2+m_i^2}$, in the limit $t\rightarrow 0$ we would have:
\be
\lim_{t\rightarrow 0}\wp_{f\rightarrow f}(t)=\left|\cct+\sst \frac{(\w_1-\w_2)^2}{4\w_1\w_2}\right|^2
\ee
that for generic values of $m_1$, $m_2$ and $\vec{k}$ can be different than $1$. However,
this is not compatible with the initial assumption that the particle at $t=0$ \textit{is} $a_A^\dagger\rmv$,
and therefore $\wp_{\slot}(0)=1$.

Adopting BV formalism implies therefore the assumption of a non trivial vacuum, as the lowest energy state.
But what is the energy of this state?
It is important to remark that once we fix $\rmv$ as the zero-point for the energy of our system (since $\rmv \equiv \ket{0,0,...}$),
the \textit{flavour vacuum} $\rfv$ must have \textit{nonzero} energy: a zero energy for both vacua would imply
the trivial identity $\rfv=\rmv$ \cite{Umezawa:1982nv}.
Not surprisingly, a direct calculation of the energy density per unit volume
\be
\rho=\lfv \mathcal H \rfv
\ee
diverges ($\mathcal H$ being the Hamiltonian density);
unfortunately, this divergence does not disappear even after normal ordering:
\be\label{normalordering1}
\rho=\lfv :\mathcal H: \rfv \equiv \lfv \mathcal H \rfv-\lmv \mathcal H \rmv
\ee
in which one subtracts the contribution of $\rmv$.
Again, the reader should not be surprised:
if the theory is regarded as an interactive theory (Section \ref{ssTI}), apart from (\ref{normalordering1})
further renormalizations were expected.

At the moment, non-perturbative tools have not been developed or applied to the problem.
On the other hand, as a first step,
conventional perturbative tools have been used: more precisely,
since the divergent quantity in (\ref{normalordering1}) is an integral in the momentum space,
a \textit{cutoff} for momenta has been introduced as a \textit{regulator} of the theory, and
preliminary analyses have been performed with the regularized theory, as we will soon see in Section \ref{ssFVDE}.

\subsectionItalic{From Theory to Physics}\label{ssTP}

So far we have presented the formalism in a rather abstract way.
Although we started from the analysis of a well-established phenomenon, the oscillation of neutrino flavour,
important consequences of the theory, such as the vacuum condensate,
have been analyzed only from a theoretical point of view.
We discussed a Fock space for flavour states that seems possible only in the presence
of a vacuum state that is not really empty, but presents some structure.
But how realistic is the model?
To answer to this question we have to step away from Particle Physics and direct our attention to
recent results in Astrophysics.

Recalling basic notions of General Relativity, a nonzero energy contribution of a vacuum condensate
distributed in all the space would affect the geometry of the spacetime itself:
if such a contribution would exist, it might be detectable via astronomic observations, in principle at least.
Indeed, in the past this has been the preferred way to investigate such a phenomenon,
and strict upper limits were found\footnote{We are here talking about the energy that the vacuum would have \textit{in the present};
a nonzero energy density of the vacuum has been considered in various models for the early universe inflationary era \cite{Peebles:1994xt}.}
\cite{Peebles:1994xt}.
Until a few years ago, the general attitude was to consider it to be negligible or even exactly zero, via a not-yet-known
mechanism.\footnote{Supersymmetry models are known to have a zero energy ground state,
without the need of renormalization, as we will see in Section \ref{secSUSY}. However supersymmetry is not a symmetry of Nature.}
Quite surprisingly in 1998 data from supernovae sources
\cite{Riess:1998cb,Perlmutter:1998np}
showed a different situation:
it was the first evidence of a current accelerated regime of the expansion of the Universe,
explicable via an uniform nonzero energy filling all the space.
One possibility is that this energy is inherent to the vacuum state itself.
Because of its unknown origin, this contribution to the total energy of the universe (causing its accelerated expansion)
was called Dark Energy. Many other alternatives to this elusive fluid have been proposed in recent years:
nonetheless, the accelerated expansion of the universe remains at the moment one of the most urgent problems in cosmology\footnote{A
dedicated section, with more details about DE and the acceleration of the universe will follow (Section \ref{ssDE})}.

A candidate for Dark Energy should behave as a perfect relativistic fluid that uniformly permeates all space.
Moreover, another special feature characterizes such a hypothetical fluid: it presents
\textit{negative} pressure. This condition is essential for
generating an accelerating expansion of the universe (see Section \ref{ssDE}).
Particle and interactive fields of the Standard Model (the only kind of field so far experimentally proven to exist)
do not satisfy the special features that Dark Energy is supposed to have.
On the other hand, other objects until now only theoretically conjectured might show the correct behaviour.
The vacuum condensate suggested by the BV formalism is one of them, as already mentioned.
Year by year, new astronomical data enhance boundaries on the two parameters that characterize Dark Energy:
the \textit{energy density} per unit volume, estimated to be $\rho_\Lambda=3\times 10^{-11} eV^4$,
and the \textit{equation of state}
$w$, that expresses the ratio between the pressure and the energy density $w=p/\rho$, being compatible with $-1$ according to
the most stringent estimates \cite{Komatsu:2010fb}.

\subsection{\textit{The} Cosmological Constant Problem}\label{ssCCP}

The existence of Dark Energy poses a rather important problem for QFT.
As we know, QFT does not consider gravitational effects and it usually considers only differences in energy;
but when we query the theory about the \textit{absolute} energy of the vacuum state,
the answer is ambiguous not to say problematic.

In a free theory, the direct calculation of the energy of the vacuum gives an \textit{infinity} as result.
This divergency arises from the integral
\be
\int d\vk \sqrt{\vk^2+m^2}
\ee
which corresponds to the \textit{un-}normal ordered vacuum expectation value of the energy operator,
in simple theory for one free massive particle.

If we consider QFT to be valid up to the Planck scale (where gravitational effects are expected and our concept of
spacetime as a continuum might break down, roughly $10^{27} eV$),
we can set a cutoff in the theory up to that energy. The result of the previous
calculation is therefore finite, although it is \textit{enormous}: $10^{108} eV^4$.

If we consider realistic interactive theories
and a cutoff that covers at least the regime of energies tested in laboratories
(for which we know that the theory works) we get a lower result $10^{48} eV^4$ (assuming that the SM works up to $1\;TeV$),
but still very high.

Another way to overcome the initial infinity is to \textit{redefine} the energy operator:
the theory allows us to add a constant to such an operator, with no changes in the expectation values of other
observable quantities. This reflects the fact that in field-theoretical models, only differences in energies
matter and therefore it is possible to set to zero the contribution of the vacuum state
(na\"ively speaking, by subtracting the infinite contribution of the expectation value of the vacuum from the energy operator).
The formal procedure is part of the \textit{renormalization} techniques and is the conventional way to perform calculations
in particle physics.

Progressing this argument further, one might argue that the arbitrary choice of the zero-point of the energy
is a sign that QFT is not adequate to reveal the nature of the vacuum.

Quite remarkably, no one of these answers is in accordance with what we know \textit{experimentally}. In particular,
the huge difference between the estimated value ($10^{108} eV^4$ or $10^{48} eV^4$) and the measured value ($10^{-11}eV^4$)
is the core of what in literature has been called \textit{the cosmological constant problem} \cite{Weinberg:1988cp}:
the energy scale of Dark Energy is far away from all natural scales provided by the SM via the particle masses.

Only one fundamental scale is known to be comparable with the Dark Energy one: the scale of
\textit{neutrino physics}.\footnote{Boundaries on total masses of neutrinos show that they are much
lighter then all other particles:	astrophysical data indicate that $\sum m_\nu<0.58\;eV$, with $95\%$ of confidence \cite{Komatsu:2010fb}
(the sum runs over all possible species - possibly more then three - that where present in the early Universe, and relies on some
cosmological assumptions, that will be partially addressed in Section \ref{ssMCAI}).
Moreover, direct observations on solar and atmospheric neutrinos show that
$\Delta m^2_{12}\approx 8 \cdot 10^{-5}\;eV^2$ and $\Delta m^2_{23}\approx 2.5 \cdot 10^{-3}\;eV^2$ (being
$m_i^2-m_j^2\equiv \Delta m_{ij}^2$, \textit{cf} \cite{Bilenky:2010zza} and  references therein).
These mass scales $10^{-1}\div 10^{-2}eV$ have to be compared with the scale $10^{-3}eV$,
that one obtains from $\rho_{\Lambda}=3\times10^{-11}eV^4$.}

The idea that Dark Energy and neutrinos might be connected gave rise to various works in the past years
(see \cite{Gurwich:2010gb} and references therein)
and BV formalism has to be considered as an encouraging step in this direction.
Besides, it also sheds new light on the issues with QFT just listed:
the formalism seems to suggest that the ambiguity of QFT about the absolute value of the energy of the vacuum
is solved by requiring that the state $\rmv$ is empty and has zero energy, as assumed ordinarily in QFT;
nevertheless, the \textit{physical} vacuum is \textit{not} represented by $\rmv$, but by $\rfv$ and,
being a \textit{vacuum condensate}, the \textit{physical vacuum} has nonzero energy.

Still, the absolute value of this energy is a matter of debate and subject to studies,
since, as we said, in the na\"ive calculation it diverges.

\subsectionItalic{Flavour Vacuum and Dark Energy}\label{ssFVDE}

What are the physical features of the flavour vacuum? Are they in accordance with the observations?
The group of Vitiello tried to explore the connection between the observed
vacuum energy and its mathematical construct, the flavour vacuum, in a series of works
\cite{Blasone:2004hr,Blasone:2004yh,Capolupo:2006et,Capolupo:2006re,Capolupo:2008rz,Capolupo:2007hy,Blasone:2007iq,Blasone:2008rx,Blasone:2007jm,Blasone:2006ch},
in a purely phenomenological spirit: the cutoff of the momenta in the regularized theory is considered as a physical cutoff,
in the sense that its value is determined by physical constraints.

The analysis has been performed for a simple model that includes three generations of free Dirac
particles.\footnote{They considered neutrinos but they
did not exclude a further contribution from quarks, the other particle of SM exhibiting flavour mixing;
as we will see, their contribution is excluded in the context of \cite{mavrosarkar} (Section \ref{ssDPFM}), for reasons dictated
by the microscopic model.}
To account for the current data about the ratio $w=-1$ they suggested that the vacuum condensate might have had a dynamical
evolution during different eras of the universe.
The Standard Cosmological model
distinguishes three different regimes of the expansion of the universe \cite{Liddle:2009zz}:
at very early times the Inflation occurred, causing an extremely rapid expansion,
in a regime of very high acceleration; after that, the matter and radiation content of the universe drove the expansion,
causing a deceleration; $8$ billions years ago,  the universe started a second accelerated expansion, although
a more moderate one.

In the analysis of the flavour vacuum, the authors of \cite{Capolupo:2008rz} (the most recent and complete analysis)
suggested that the vacuum condensate (for a theory with three generations of free Dirac spinors)
could have presented two different regimes:
1) $0<w<1/3$ (the range is spanned for values of the cutoff $K$ going from $0$ to $+\infty$),
in which a violation of Lorentz symmetry is present\footnote{As we will see in Section \ref{ssDE},
Lorentz symmetry requires that $w=-1$ exactly.};
2) $w\approx -1$, when Lorentz invariance is conserved or the violation is very small.
The first regime could have characterized the early universe, in which Lorentz invariance might have been violated;
the second one might have occurred afterworlds and might represent the present status of the condensate.
Moreover, provided with the value of the Dark Energy density $\rho_{\Lambda}\approx 10^{-11}eV^4$, its equation of state $w\approx 0.98$
and an estimate of the parameters of the Lagrangian
($m_1=4.6\times 10^{-3}eV$, $m_2=1\times 10^{-2}eV$, $m_3=5\times 10^{-2}eV$, $\sin^2\theta_{12}=0.31$, $\sin^2\theta_{23}=0.44$,
$\sin^2\theta_{13}=0.009$), one is able to extract an estimate for the cutoff, in terms of the masses of the particles:
$K\approx3\times10^{-3}eV$, that turns out to be of the same order of the neutrino masses.

These results are the first attempts to accomodate the supposed existence of the vacuum condensate with the actual data available.
Still, it seems that more effort is needed to accomplish this.\footnote{For instance, these results are achieved by taking into
account only certain terms of the full expression for the stress-energy tensor. However, the model does not explain \textit{why}
only those terms ought to be relevant at different stages of the evolution of the Universe.}
On the other hand, we will see in the present work that new elements in the simple model considered here may change
the behaviour of the condensate significantly, offering different perspectives on the problem.


\section{STRINGY FLAVOUR VACUUM}\label{sSFV}

After the works of the group of Vitiello,
the phenomenology of the flavour vacuum has been under examination in a quite different context: working
in the framework of String Theory, the authors of \cite{mavrosarkar}
suggested that the BV formalism might provide the low energy effective theory for a specific
model based on the Brane World approach \cite{Maartens:2003tw}, and extensively studied in literature
(\cite{Mavromatos:2006yy,Mavromatos:2005bu,Mavromatos:2008bz,Ellis:2004ay,Ellis:2005ib} to mention a few).

As defined in BV formalism, the flavour vacuum breaks Lorentz symmetry,
being explicitly time dependent, as mentioned in Section \ref{ssFSFS}, formula (\ref{flavourstates}).
At first, this feature has been marked as an undesirable drawback of the formalism.
However, the authors of \cite{mavrosarkar} claimed that the seeds of this violation could have been found
in an underlying microscopic theory; furthermore,
they suggested that the formalism itself would be consistent \textit{only} if embedded in an underlying microscopic theory
in which an explicit mechanism for Lorentz violation is provided.
The Lorentz symmetry breaking induced by the flavour vacuum would be regarded, from this perspective, as
a \textit{sign} of physics \textit{beyond} our current knowledges.

In a concrete attempt to test these ideas,
they tried to consistently embed BV formalism in a specific model,
coming from the framework of Brane/String Theory.
In this model, called \textit{D-particle foam model},
our universe is depicted as a $3+1$ dimensional manifold embedded in a higher dimensional bulk.
Moreover, in the bulk a cloud (or \textit{foam}) of point-like topological defects (\textit{D-particles}),
through which the manifold moves, is present.
The relative motion of the manifold and the cloud picks a preferred direction for the interaction of D-particles and
ordinary particles living on the manifold, which leads, form the point of view of an observer on the manifold,
to a violation of Lorentz symmetry.
Furthermore, the interaction of the foam with ordinary flavoured particles on the manifold can lead to a change in their flavours.
Such an effect might be well treated, in the low energy limit, via BV formalism.

Sections \ref{ssBW} and \ref{ssDPFM} we will be dedicated to a more detailed overview on the \textit{D-particle foam model}.
In Section \ref{ssEFV} we shall briefly present the effective toy model, used to describe the low energy limit
of the \textit{D-particle foam model}, in which BV formalism enters.
More details about this effective toy model will be provided in the next chapter,
since it will represent the ground of our first analysis.

In order to be used in this new stringy-inspired context, the formalism needed to be re-adapted, to better fit
the prerequisites of the microscopic model.
Nevertheless, the contribution of this work has general relevance for two reasons:
it represents a step further in the direction indicated by the existing literature,
since it includes a \textit{curvature} of the spacetime background, often invoked,
but never systematically included until then;\footnote{First attempts of extending
the formalism in curved spacetime can be found in \cite{Blasone:2004hr,Blasone:2004yh}.
However, after then a complete treatment was often invoked
\cite{Capolupo:2006et,Capolupo:2006re,Capolupo:2008rz,Capolupo:2007hy,Blasone:2007iq,Blasone:2008rx}
but never accomplished, or at least explicitly presented \cite{Blasone:2007jm,Blasone:2006ch}.}
and it presents a new perspective on the formalism, by considering it as naturally emergent from a more fundamental theory.

\subsectionItalic{Brane World}\label{ssBW}

In the mid 1990s string-theoreticians realized \cite{Polchinski:1995mt,Horava:1995qa}
the existence of non-perturbative objects in their theories, and called them \textit{branes}\footnote{\textit{Branes}
are implied by the theory, but are not ``visible'' from a perturbative point of view
(the usual approach for studying simple string theories).}:
just like a string is a one-dimensional object, a ``brane'' can be regarded as an extended object with arbitrary dimensionality
(up to the dimensionality of the space in which the brane is embedded). In this sense,
a string is nothing more than a one-dimensional brane.
A certain kind of brane, so called \textit{D-brane}, played in the last years a fundamental role in building cosmological models.
The special feature of such a brane is that in specific theories strings are either closed or they have their ending points attached
on a D-brane.
Recalling some old ideas \cite{Rubakov:1983bb,Akama:1982jy},
people started speculating about the possibility that our universe was nothing but a D-brane embedded in
higher dimensional space. Several models have been proposed since then (see for instance \cite{Csaki:2004ay}
and references therein), giving rise to the \textit{Brane world} scenario.
A simplistic way to explain how the approach works is the following:
first, a three-dimensional brane in a higher dimensional space (called \textit{bulk} space) and
strings attached on the D-brane are considered. Then,
one notices that, in accordance with the properties of the strings who they belong to, the end points might interact between themselves
mimicking the behaviour of real particles. Moreover, those end points are confined in a three dimensional space,
since they exist as intersecting points of the strings and the three-dimensional D-brane.
Therefore, one is led to think that the particles that are present in our universe are just the attaching points of strings
that live in a higher dimensional space.

The importance of the approach relies on the fact that it might provide a simple answer
to the question ``where are the extra-dimensions predicted by String Theory?'',
without involving the problematic \textit{compactification} procedure.

\subsectionItalic{D-Particle Foam Model}\label{ssDPFM}

\begin{figure}[ht]\label{fig:foam}
\centering
\includegraphics[width=6.5cm]{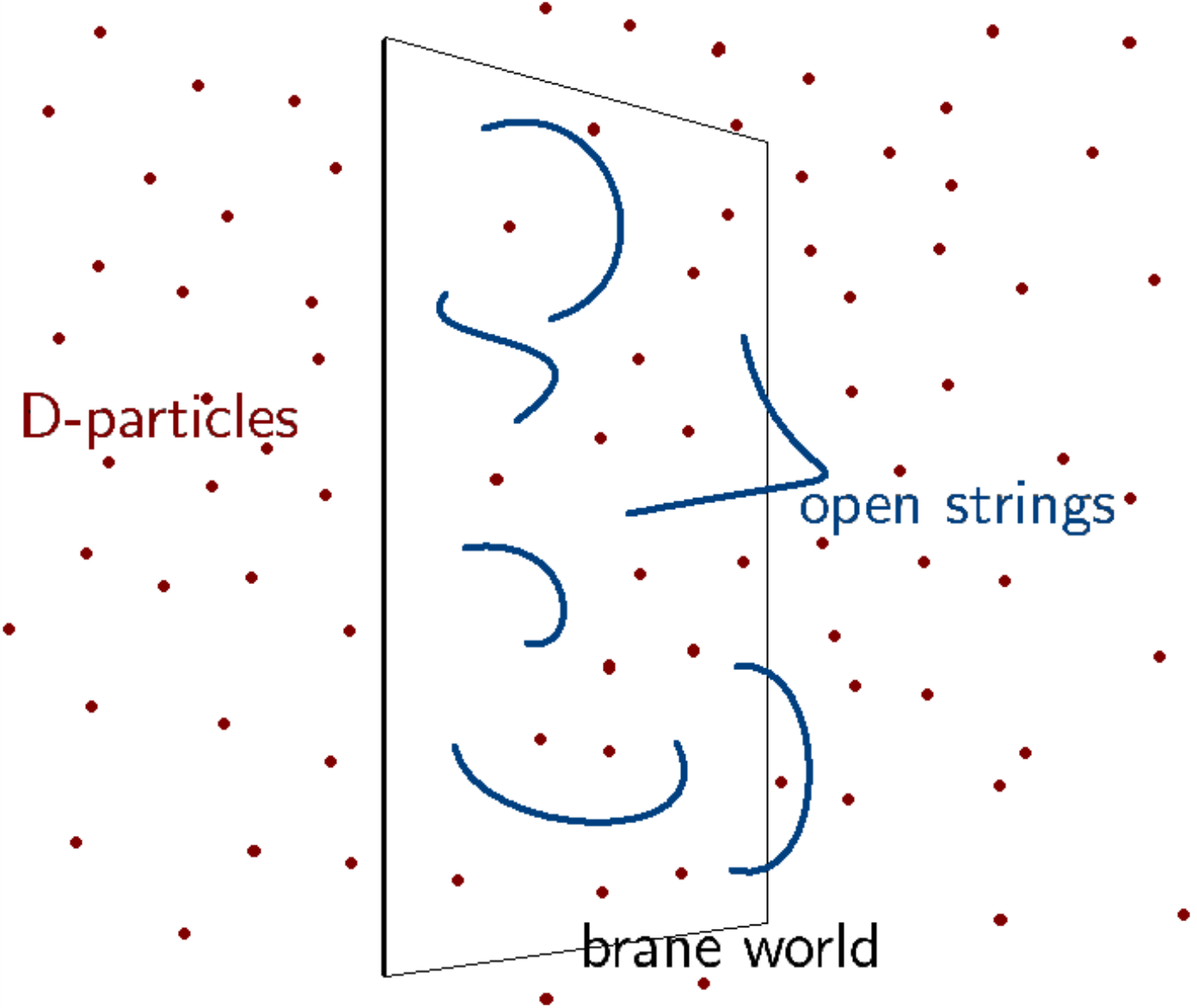} \hfill
\caption{Pictorial representation of the \textit{D-particle Foam Model}.}
\end{figure}

Although the Brane World scenario suffers from quite important unsolved problems
(for instance, it is unclear how to quantize branes with dimensionality
greater then one), a wide range of cosmological models have been studied \cite{Maartens:2003tw}.
One of those is the \textit{D-particle Foam Model} \cite{Mavromatos:2006yy,Mavromatos:2005bu,Mavromatos:2008bz,Ellis:2004ay,Ellis:2005ib}.
In this model, in addition to the standard set-up of a three-dimensional D-brane (that we will refer to as the ``D3-brane'')
and strings attached on it, other zero-dimensional objects are present in the bulk (Figure \ref{fig:foam}).
They can either be D-branes with zero dimensionality
(called ``D-particles'') or three dimensional D-branes wrapped up around small three-cycles so as
to resemble small spheres (giving rise to ``effective D-particles'').\footnote{The two possibilities, effective or
proper D-particles, do not coexists in the same theory: proper D-particles (zero-dimensional branes) occur in the context of
Type-IA (a T-dual of Type-I \cite{Schwarz:1999xj})
and Type-IIA string theory \cite{Ellis:2004ay,Ellis:2008gg} whereas effective D-particles (wrapped three dimensional branes)
are present in the context of Type-IIB string theory \cite{Li:2009tt}.}
The collection of D-particles forms a \textit{foam} in the bulk, and, in a dynamical picture, the brane-world
floats through this foam. When this happens, both the strings and the D3-brane interact with the foam in a highly non trivial way.
Na\"ively speaking, the interaction between a string and a single D-particle can be described as a ``capture'' of the
string by the D-particle, and a subsequent releasing. As a result of this interaction, the released string
might have different flavour than the incoming one.
On the other hand, when a single D-particle crosses the D3-brane, it causes a ``deformation'' of the brane itself.
From the point of view of an observer confined to the D3-brane (as real observes are regarded in this picture),
a deformation of the brane is viewed as a deformation of the metric
of the three dimensional space on which the observer lives.
Although the deformation is localized in the region where the D-particle interacts,
deformations induced by different D-particles sum up all together: from the macroscopic
point of view of the observer, a \textit{uniformly averaged} deformation of the metric due to the foam, viewed as a whole, occurs.

An analysis at the string level of the model has been possible only for single aspects of the process of interaction,
see for instance \cite{Mavromatos:1998nz,Kogan:1995df}
for the capture of the string and \cite{Kogan:1996zv,Mavromatos:2007sp,Ellis:1999jf} for the deformation of the metric.
Nevertheless, an interesting phenomenology at astrophysical scales has been widely studied during the last years
\cite{Ellis:1999jf,Ellis:1999uh,Ellis:2000sx,Ellis:1999sd,Ellis:1999sf,Ellis:2008gg,Li:2009tt}.

In one of the developments of the approach, strings with different flavours have been included in the model \cite{mavrosarkar}.
As a result of the interaction of the foam and the strings, it has been argued that
a change of the flavour of the captured string can occur.
Recalling that the attaching points of the strings are interpreted as the particles present in our universe,
the change of flavours of strings causes a change in flavour of the particles.
Moreover, since only electrically neutral particles are affected by the interaction with the microscopic foam,
the flavour mixing occurs only for \textit{electrically neutral particles}\footnote{This is the reason why in this
context only neutrinos are considered (see Section \ref{ssFVDE}).} \cite{Ellis:2003sd,Ellis:2003if}.
A preliminary investigation on the phenomenology of the model has been carried out by means of an effective field theoretical
toy model.

\subsectionItalic{Effective Flavour Vacuum}\label{ssEFV}

Since a formulation of the quantum theory of the \textit{D-particle foam model}
at all scales is not available, the authors of \cite{mavrosarkar} considered the simplest effective QFT model
that could catch the expected features in the low energy limit.
In the spirit of weak coupling string theory, the effects of the D-particle foam
on brane and strings are regarded as ``vacuum defects'' from the point of view of a (macroscopic) observer on the brane.
Therefore, the effective vacuum that describes the theory in the QFT regime should account both for
the particle flavour mixing and for the deformation of the metric.\footnote{The foam is also responsible
for an \textit{effective mass} of the particles, as we will explain in the next Chapter.
In fact, we can regard this effect as emerging from the interaction of the particles and the gravitational field inherent
in the background. Therefore, we can consider it as a ``side effect'' of the gravitational field itself.}
As we will see, this is achieved by a self-consistent inclusion of the \textit{flavour vacuum} and the curvature of the spacetime,
in a (otherwise) empty space.
The adopted toy model includes\footnote{To
seek simplicity, we omit here technical details, as an exhaustive discussion about the model will be provided in
the following Chapter.}:
\begin{enumerate}
	\item a $1+1$ dimensional spacetime,
	\item two real scalars with flavour mixing (reflecting the interaction between the foam and the strings),
	\item a specific FRW metric, describing a monotonically expanding universe,
				that interpolates smoothly between two asymptotically flat spacetimes (reflecting the interaction between
				the foam and the D3-brane).
\end{enumerate}
The curvature of the background is introduced in the effective toy model via
the standard techniques of \textit{QFT in Curved Spacetime}:
a topic that has been widely studied since the early Seventies, as a possible bridge between QFT and a yet-to-be discovered
quantum theory of gravity. Following its prescriptions,
the authors of \cite{mavrosarkar} formulated a quantum theory on a non-flat spacetime background, where
the curvature is regarded as \textit{classical} and \textit{fixed}:
it enters as an external field in the Lagrangian and it is not quantized (\textit{classical});
as such, they do not minimize the action with respect to it, they rather consider it as given and dependent just on spacetime
coordinates (\textit{fixed}).
In this way one clearly considers quantum effects of gravity as negligible and ignores Einstein's equation,
which dynamically connects the curvature of spacetime with the distribution of energy and matter in the spacetime itself.
Nevertheless, this kind of approach led in the past to notable results such as the Hawking radiation \cite{Hawking:1974sw}
and the Unruh effect \cite{Unruh:1976db}.

Just like the flat case, a direct evaluation of the expectation value of the stress-energy tensor
with respect to the flavour vacuum diverges.
To give sense to this expression, as we mentioned in Section \ref{sFVDE}, in the \textit{flat} case it has been used in the following procedure:
one first removes the contribution of the ordinary vacuum (the usual \textit{normal ordering}),
and then one sets a cutoff in the momentum space.
BV formalism does not prescribe any cutoff, and a way suggested to choose it is to constrain the value of expectation value of the energy
of the vacuum condensate with experimental data (Section \ref{ssFVDE}).

In the brane-inspired toy model in curved spacetime, the high energy theory (the brane model)
imposes conditions both on the normal ordering and the cutoff.
It should be stressed that the particular normal ordering introduced in this work differs from the standard \textit{normal ordering}:
as we will see in the relevant Section \ref{ssRANO},
it is very specific for the model and it strongly affects the behaviour of the condensate.

As a result, the flavour vacuum condensate behaves in these conditions as a perfect fluid with positive energy, negative pressure and
an equation of state $w=-1$, exactly and independently of the choice of the cutoff.
Although it is also possible to work out the absolute value of the energy density (in terms of the cutoff),
it is not possible to make a numeric estimate of it, since its value relies strongly on the unknown value of some parameter
that has been introduced to account for the features of the brane model.

Nevertheless, such an encouraging result led the authors of \cite{mavrosarkar} to keep working on the toy model,
and the present thesis can be considered as an extension of that work.

As already mentioned, this preliminary investigation had a twofold importance: from one side it was first introduced
a (classical) curvature of the background; from the other side, again for the first time, the BV formalism has been regarded
as the low energy limit of a Quantum Gravity theory. It is believed
\cite{Capolupo:2006et,Capolupo:2006re,Capolupo:2008rz,Capolupo:2007hy,Blasone:2007iq,Blasone:2008rx}
that treating the flavour mixing in curved spacetime
is a necessary step also in the study of the BV formalism \textit{per se},
even when no deeper theories are invoked. Such a conviction is clearly motivated by the fact that
the BV formalism is being used to enlighten a cosmological conundrum,
such as the accelerated expansion of the universe.
As we will see in more detail later on, gravitational effects due to the curved background can in general
affect very much the behaviour of our vacuum condensate, therefore it seems sensible to consider them
as a key ingredient that cannot be neglected.
From this perspective, the BV formalism acquires a potential role in the quest for Quantum Gravity:
as shown by the authors of \cite{mavrosarkar}, it is possible that the connection between the vacuum condensate and the structure of the
spacetime itself might be deep enough to disclose some characteristics of an underlying and more fundamental theory.
In \cite{mavrosarkar} this theory is a string theory, but the door is clearly open to other approaches.


\section{THE DARK SECTOR OF THE UNIVERSE}\label{sDSU}

In previous Sections we traced the path that from the formulation of flavour mixing \textit{\`a la} BV
\cite{Blasone:1995zc} led to the suggestive speculation that the formalism might actually reveal a profound
connection between neutrino physics, Dark Energy \cite{Blasone:2004yh} and even quantum gravity physics \cite{mavrosarkar}.
We tried to explain how a quantum field theory with flavour mixing might actually require a ground state
characterized by a very complex structure, behaving as a source of Dark Energy.
Moreover, such a structure, because of its explicit time dependency, might represent a first hint of
a deeper Lorentz symmetry violation at the Planck scale, as the one that occurs in the brany model just discussed.

Before moving to our first analysis, which extend results of \cite{mavrosarkar}, we shall recall a few notions
of Modern Cosmology (Section \ref{ssMCAI}), discuss more the concept of Dark Energy (Section \ref{ssDE})
and introducing Dark Matter (Section \ref{ssDM}).
Furthermore, we shall introduce basic technical tools, such as the \textit{conformal scale factor}, describing the
expansion of the universe, and the \textit{equation of state} of a perfect relativistic fluid,
that will be necessary for our upcoming discussions.
Indeed, in the next chapters we will mainly be interested in the study of the equation of state
that characterizes the \textit{flavour vacuum} in different models. As we shall soon see,
the knowledge of its equation of state
will enable us to establish important aspects of its phenomenology at cosmological scales.

\subsectionItalic{Modern Cosmology: an Introduction}\label{ssMCAI}

Neutrino oscillations do not represent the only problem posed by astrophysics to particle physics that find no answer in the SM.
Indeed, current models of cosmology explicitly require the existence of fields and mechanisms that particle physicists
have not being able to explain so far.
What is the ``inflaton''? Is baryogenesis a dynamical mechanism? What is Dark Matter made of? What is Dark Energy?
For many, an answer to these questions relies on physics \textit{beyond} the SM.

In particular, the problem concerning Dark Matter and Dark Energy is of specific interest for our work, and
we shall briefly review it in the present section.\footnote{For completeness: \textit{inflaton} is the name given to the unknown field invoked for explaining \textit{inflation},
the very short period ($\sim10^{-34}$ sec) of exponential growth of the universe (of a factor at least $10^{78}$ in volume)
occurred at very early times; \textit{baryogenesis} is the term that indicates the yet unknown mechanism
that produced an asymmetry between baryons and anti-baryons at early times,
resulting in the conspicuous abundance of matter and relatively lack of antimatter in the visible universe \cite{Liddle:2009zz}.}
In order to understand such a problem, we shall recall a few notions of cosmology modeling,
and introduce the basic tools that will be used in the rest of work, such as the \textit{conformal scale factor}
and the \textit{equation of state} for a perfect relativistic fluid.

Modern cosmological models
are built on the fundamental assumption of spatially homogeneity (translational invariance)
and isotropy (no preferred directions) of the universe (the so called \textit{Cosmological Principle}) \cite{weinberg}.
Such an assumption is valid at very large scales (from $10^8-10^9$ light years, large enough to contain many clusters of galaxies).
However, it is a very powerful tool that allows us to make fruitful use of the limited data provided to cosmology by observational astronomy.

In a universe obeying the Cosmological Principle, it is always possible to choose coordinates $t,r,\theta$ and $\phi$ for which
the metric takes the form
\be\label{RWmetric}
ds^2=-dt^2+a^2(t)\left(\frac{dr^2}{1-k r^2}+r^2 d\theta^2+r^2\sin^2(\theta)d\phi^2\right)
\ee
with $k=+1,0,$ or $-1$ (in a suitable choice of units for $r$) and $a(t)$ a function of time.
The value of $k$ determines the spatial geometry of the universe: it decides the sign of the three-dimensional curvature scalar,
that is given by $K(t)=k a^{-2}(t)$.
The most generic metric meeting the requests of the Cosmological Principle is represented by the above formula,
for a generic $a(t)$. Such a function is called \textit{scale factor}\newnot{symbol:scalefactor}, and its physical interpretation is easily recognized,
when considering the distance $d(t)$ between two galaxies at the time $t$, in terms of the distance $d_0$ at the time $t_0$:
\be
d(t)=a(t)d_0
\ee
assuming that $a(t_0)=1$: as the name suggests, $a(t)$ behaves as factor of rescaling for spatial distances.
The metric (\ref{RWmetric}), called Friedman-Robertson-Walker\newnot{symbol:FRW} (FRW) metric,
indeed describes a universe that expands and/or contracts, depending on the specific choice of the function $a(t)$.

For sake of convenience, in the following work we will express the FRW metric in a different coordinate system $(\e,r,\theta,\phi)$,
for which
\be\label{conformalFRWmetric}
ds^2=\Ce \left(-d\e^2+\frac{dr^2}{1-k r^2}+r^2 d\theta^2+r^2\sin^2(\theta)d\phi^2\right)
\ee
related with (\ref{RWmetric}) via the definition of the \textit{conformal time}\newnot{symbol:conftime}
\be
\e=\int^t \frac{dt'}{a(t')}
\ee
and a\newnot{symbol:confscalefactor} \textit{conformal scale factor}\footnote{The nomenclature ``conformal'' is here due to the fact that FRW metric
is related to the flat metric $\e_{\mu\nu}$ via the transformation
$\e_{\mu\nu}\rightarrow \Ce \e_{\mu\nu}$. This is a special case of \textit{conformal transformations},
that in General Relativity correspond to $g_{\mu\nu}(x)\rightarrow f(x)g_{\mu\nu}(x)$, with $f(x)$ an arbitrary spacetime
function \cite{Peacock:1999ye}.}
\be
\Ce =a^2(t(\e)).
\ee

The Cosmological Principle can also be invoked to restrict the kind of distribution that cosmic matter can assume in space.
In particular, homogeneity and isotropy imply that the matter content of the universe must distribute as a perfect fluid\footnote{A
fluid is said to be ``perfect'' if it has no shear stresses, viscosity, or heat conduction.},
at the rest in the coordinate system $r,\theta,\phi$, for which (\ref{RWmetric}).
The stress-energy tensor $T_{\mu\nu}$\newnot{symbol:set} for such a distribution is given by
\be\label{tmunurhop}
T_{\mu\nu}=(\rho+\pressure)U_\mu U_\nu+\pressure g_{\mu\nu}
\ee
with
\be\label{velocityfourvector}
U^0\equiv 1,\;\;\;U^i\equiv 0.
\ee
The expression (\ref{tmunurhop}) describes the stress-energy tensor for a perfect fluid,
with energy density $\rho$\newnot{symbol:energy} and pressure $\pressure$\newnot{symbol:pressure}. $U^\mu$ represents the local value of $dx^\mu/dt$ for a comoving fluid element;
the relation (\ref{velocityfourvector}) tells us that the fluid is at rest in the coordinate system $r,\theta,\phi$.
It is easy to show (Appendix \ref{aCS})
that in the coordinate system $\e,r,\theta,\phi$, the stress-energy tensor (\ref{tmunurhop})
becomes
\be\label{conformalTmunu}
T_{\mu\nu}=\Ce \begin{pmatrix}						\rho&0&0&0\\
													0&\pressure&0&0\\
													0&0&\pressure&0\\
													0&0&0&\pressure
					\end{pmatrix}.
\ee

In summary, the Cosmological Principle constrains the geometry of the universe to be described by the FRW metric,
((\ref{RWmetric}) or (\ref{conformalFRWmetric}) equivalently),
and the matter therein contained to be distributed as for a perfect fluid ((\ref{tmunurhop}) or (\ref{conformalTmunu})).
However, it does not prescribe the specific form of the scale factor $a(t)$, appearing in the metric,
leaving undetermined the evolution of the universe itself.
In order to build a \textit{dynamics} for the geometry of the universe, we invoke the
Einstein equations, that relate the \textit{curvature} of the universe (in the form of the Einstein tensor $G_{\mu\nu}$)
with its matter content (encoded in the stress-energy tensor $T_{\mu\nu}$):
\be\label{einstein}
G_{\mu\nu}+g_{\mu\nu}\Lambda=8\pi T_{\mu\nu}
\ee
with $\Lambda$ a constant, called \textit{cosmological constant}\newnot{symbol:cosmcos}.

When the FRW metric (\ref{RWmetric}) and the stress-energy tensor for a perfect fluid described by (\ref{tmunurhop})
and (\ref{velocityfourvector}) are considered,
equations (\ref{einstein}) reduce to
\be\label{fried1}
\left(\frac{\dot{a}}{a}\right)^2=\frac{8\pi}{3}\rho-\frac{k}{a^2}+\frac{\Lambda}{3}
\ee
and
\be\label{fried2}
\frac{\ddot{a}}{a}=-\frac{4\pi}{3}\left(\rho+3\pressure\right)+\frac{\Lambda}{3}
\ee
in which the explicit dependency on time of $a$, $\rho$ and $\pressure$ has been omitted.
Moreover, since the left hand side of (\ref{einstein}) is divergenceless, a conservation law must hold for $T_{\mu\nu}$,
namely $T^{\mu\nu}_{\;\;\;\; ;\nu}$, that in our case leads to
\be\label{Tdivergence}
\frac{d}{da}(\rho a^3)=-3\pressure a^2.
\ee
In order to solve (\ref{fried1})	and (\ref{Tdivergence}), an equation of state relating $\rho$ to $\pressure$ is required.
Once such a relation is known, one can use (\ref{Tdivergence}) to determine $\rho$ as a function of $a$, and then
(\ref{fried1}) to determine $a$ as a function of time.
In general, the equation of state is expressed in the form\newnot{symbol:eqofstate}
\be
w=\frac{\pressure}{\rho}.
\ee
For simple cases, the parameter $w$, itself called \textit{equation of state}, is constant in time and space.
A gas of electromagnetic radiation or any relativistic particle follows roughly $w=1/3$, whereas for non-relativistic particles
$w\approx 0$ holds.\footnote{In this context, ``relativistic'' refers to particles for which the energy is roughly given
by the absolute value of the momentum; ``non-relativistic'' on the other hand refers to particles with very small momentum
and for which the energy is roughly equal to the rest mass.}
Rearranging equation (\ref{fried2}), we can write
\be\label{wlessminus13}
\frac{\ddot{a}}{a}=-\frac{4\pi }{3}\left(1+3w\right)\rho
\ee
in which we have neglected the term $\Lambda/3$. Being the energy density $\rho$ always positive,
it is clear that the acceleration or deceleration of the universe is determined by the value of $w$:
if the energy/matter content of the universe was described by a fluid for which $w>-1/3$, the universe would
gradually decrease its expansion rate and eventually start contracting;
on the other hand, in case of $w<-1/3$, the fluid would enforce the expansion leading to a never-ending expansion.
If $\rho=\pressure=0$, equation (\ref{fried2}) implies that for any positive values of $\Lambda$ the expansion of the universe
gets accelerated, while negative values lead to a deceleration.

Equations (\ref{fried1}), (\ref{Tdivergence}) and the knowledge of the equation of state univocally identify, up to an integration constant,
the scale factor $a(t)$, and therefore the evolution of the geometry of the universe.
For instance, in the simple case $\Lambda=k=w=0$ (corresponding to a spatially flat universe, with vanishing cosmological constant
and filled by a pressureless fluid) we have that (\ref{Tdivergence}) implies $\rho(t)a^3(t)=constant$
and therefore $a(t)\propto t^{2/3}$, from (\ref{fried1}).

Most recent astrophysical data are rather well fitted in a very simple model, called $\Lambda$CDM model,
that is regarded as the \textit{standard cosmological model} (\textit{cf} \cite{Komatsu:2010fb} and references therein).
The model considers our universe as homogeneous and isotropic on large scales and spatially flat.
The ``$\Lambda$'' in the name puts emphasis on a non-vanishing cosmological constant, whereas ``CDM'' stands for
``Cold Dark Matter'', a non-barionic non-relativistic unknown kind of matter, that fills the universe along with ordinary
matter. Looking at the right hand side of equations (\ref{fried1}) and (\ref{fried2}),
gravitational effects due to the cosmological constant and Dark Matter accounts for $95\%$ of the total (roughly $72\%$ and
$23\%$, respectively), leaving a tiny $5\%$ for ordinary matter \cite{Komatsu:2010fb}.

\subsectionItalic{Dark Energy}\label{ssDE}\newnot{symbol:DE}

The non-vanishing cosmological constant term in the Einstein equations accounts for the observed acceleration of
the expansion of our Universe. After first data from Supernovae Type Ia \cite{Riess:1998cb,Perlmutter:1998np} back in 1998,
over a decade of other independent observations (temperature anisotropies
of Cosmic Microwave Background \cite{Komatsu:2010fb}, large-scale clustering of galaxies \cite{Tegmark:2006az,Seljak:2004xh}
and baryonic acoustic oscillations
\cite{Eisenstein:2005su})
confirmed the unexpected discovery that the growth of the Universe has been accelerating in the last few billions of years.

Although such a phenomenon can be easily accounted via the above mentioned non-vanishing cosmological constant term,
this solution requires an enormous fine tuning,
that can be expressed in terms of the dimensionless quantity $\Lambda G$  \cite{Padmanabhan:2002ji},
with $G$\newnot{symbol:newton} Newton's gravitational constant\footnote{In all previous formulae $G$ was set $G=1$, since we work in natural units.
However, we now keep it explicitly to emphasize the fact that $\Lambda G$ is a dimensionless pure number.}:
rather than being exactly zero (or of the order of the unit), observations suggest that $\Lambda G$ is of the order of
$10^{-123}$.
Although neither known symmetry mechanisms nor invariance principles prevent $\Lambda G$ being so small,
this fine tuning has been deemed as highly unsatisfactory from a theoretical point of view \cite{Padmanabhan:2002ji}.
Moreover, this value, that seems to be so tiny for no reason, is however such that the contribution to the curvature of the Universe
of the cosmological constant is of the same order of the curvature induced by its matter content.
This relation between two apparently unrelated quantities, known in literature as the \textit{``coincidence problem''} \cite{Padmanabhan:2002ji},
allows to speculate that the reason for $\Lambda G\sim \mathcal O (10^{-123})$ finds justification in an deeper unknown theory.
The last decade witnessed a strenuous effort in overcoming these theoretical issues, by means of a wide range of
models spanning from modified theories of gravity to exotic quantum fields.\footnote{For recent overviews on the theoretical
landscape one may refer to \cite{amendola,Frieman:2008sn,Copeland:2006wr}, and references therein.}

The vast majority of these approaches can be sorted in two distinct classes. In order to introduce them,
we must look back at the Einstein equations:
\be\label{einstein2}
G_{\mu\nu}+g_{\mu\nu}\Lambda=8\pi T_{\mu\nu}.
\ee
Instead of considering lambda as a geometric term on the left-hand side,
one can write
\be
G_{\mu\nu}=8\pi (T_{\mu\nu}+T^\Lambda_{\mu\nu})
\ee
with
\be\label{Tlambda}
T^\Lambda_{\mu\nu}=-(8\pi)^{-1}\Lambda g_{\mu\nu}.
\ee
The same equations, written in this form, are now open to a different interpretation:
the role of the cosmological constant can be regarded as being played by a fluid for which $w=-1$, characterized by the energy density
\be
\rho_{\Lambda}=T_{00}\C^{-1}(\eta)=-\frac{\Lambda}{8\pi} g_{00}\C^{-1}(\eta)=\frac{\Lambda}{8\pi}
\ee
and pressure
\be
\pressure_{\Lambda}=T_{jj}\C^{-1}(\eta)=-\frac{\Lambda}{8\pi} g_{jj}\C^{-1}(\eta)=-\frac{\Lambda}{8\pi},
\ee
in which the FRW metric $g_{\mu\nu}=\Ce \e_{\mu\nu}$ has been considered.
Such a fluid is usually called \textit{Dark Energy}.\footnote{In
discussing the example (\ref{wlessminus13}) of the previous section, we already anticipated the possibility that the acceleration
of the expansion might be caused by a fluid, whenever this presents $w<-1/3$, although we pointed out that
there is no known particle that can contribute to such a fluid.}

The first class of models covers all attempts to find a suitable fluid that acts as Dark Energy.
As mentioned in Section \ref{ssCCP}, a natural candidate for this fluid is the quantum vacuum state:
if we request that all freely moving inertial observers see the same vacuum state, the only possibility for
its stress-energy tensor is to be
\be
T^{vacuum}_{\mu\nu}=\rho_{vacuum}g_{\mu\nu}
\ee
with $\rho_{vacuum}$ a constant in a general coordinate labeling \cite{Peebles:2002gy},
that is in the same form of (\ref{Tlambda}).
However, as already mentioned in Section \ref{ssCCP}, with our current understanding of QFT we are not able to make
a sensible prediction of the value of $\rho_{vacuum}$ (\textit{cf} also \cite{Padmanabhan:2002ji}).

More sophisticated approaches belong to the first class, including scalar fields with slowly varying potentials (\textit{Quintessence} models),
or with non-canonical kinetic terms (\textit{k-essence} models), or even more exotic objects such as Chaplygin gases,
Ghost Condensates and Phantom Dark Energy, to mention a few (\textit{cf} \cite{Copeland:2006wr} and references therein).
Many of them predict an equation of state that might differ to $w=-1$ (even changing with time), therefore a compelling target
for observations is to constrain the estimated value of $w$,
that at the moment is believed to be $w=-1.10 \pm 0.14$ with $68\%$ of confidence \cite{Komatsu:2010fb}.
However, the chances to test those models are weakened by the numerous degrees of freedom that each model introduces and leaves undetermined.

The other class of approaches covers models that modify the geometric side of Einstein equations,
namely the left-hand side of (\ref{einstein2}).
If models in the former class were trying to solve the problem via a suitable new form of matter field, in this latter the solution is sought in new physics beyond General Relativity. It includes: $f(R)$ models, in which the Lagrangian $R-2 \Lambda$
with $R$ the Ricci scalar (that gives rise to Einstein equations), is replaced by a more complicated
non-linear function of the Ricci scalar, denoted precisely by $f(R)$; scalar-tensor models, in which the Ricci scalar couples with a scalar field $\phi$
via terms as $F(\phi)R$, for some function $F(\slot)$, braneworld models coming from String Theory and others (\textit{cf}
\cite{Copeland:2006wr,Frieman:2008sn} and references therein).
Although at the moment these models lack of self-consistency \cite{Frieman:2008sn}, it seems that they are more constrained
by cosmological observations and gravitational experiments then models belonging to the first class \cite{amendola}.

Apart those two classes, more conventional approaches have been explored, in which no new physics is required.
More specifically, it has proposed that dropping the assumption of spatial homogeneity at large scales might lead to models
fitting all current data \cite{Kolb:2005da}. However, they require the assumption that our galaxy occupies a privileged location in space,
that seems neither theoretically nor experimentally well-justified \cite{Frieman:2008sn}.

To summarize, available observational data indicate an acceleration in the expansion of the Universe, that is
commonly explained either with a non-vanishing cosmological constant in the Einstein equations,
or by means of a perfect fluid with equation of state $w=-1$, called Dark Energy, whose nature remains unrevealed.

\subsectionItalic{Dark Matter}\label{ssDM}

Beside Dark Energy, the other pillar of the standard cosmological model that eludes our understanding is Dark Matter\newnot{symbol:DM} (DM):
astrophysical observations clearly indicate a discrepancy between \textit{visible} matter and the total matter budget required
from gravitational analyses.

While the basic evidence for the existence of a Dark Energy fluid is provided by the accelerated expansion for the Universe,
the existence of Dark Matter is pointed out by many independent astronomical data and numerical
simulations.\footnote{Useful overviews of the topic are \cite{Bertone:2004pz,Jungman:1995df}, to which the present section
refers unless otherwise specified.}
The most compelling evidence for DM comes from observations on \textit{galactic scales}:
a great amount of data indicates that
luminous matter in spiral galaxies is not sufficient to account for their observed rotation curves (the graph traced by
the circular velocity of stars and gas as a function of their distance from the center of the galaxy).
Going to higher scales, a strong component of Dark Matter in \textit{galactic clusters} is requested to explain data on
motion of cluster member galaxies, gravitational lensing and X-ray gas temperatures.
Finally, at \textit{cosmological scales}, the Cosmic Microwave Background provides us with strong indications
that the \textit{total} amount of non visible matter in the Universe must be higher than ordinary matter (about five times bigger,
as mentioned in a previous Section).

Apart from first hints in the 30s,
the necessity of Dark Matter became evident in late 70s.\footnote{For
a historical perspective one can refer to \cite{Garrett:2010hd,Einasto:2009zd}.}
Since then, different features of DM have been clarified.
Numerical simulations of structure formation have shown that ``hot'' (relativistic) particles
cannot explain the observed structures at galactic scales, therefore we expect Dark Matter to be made out of fairly massive
and ``cold'' (non-relativistic) particles.\footnote{To be precise: ``hot'' denotes particles
that traveled at relativistic velocities for at least some fraction of the lifetime of the Universe \cite{Liddle:2009zz}.}
Big-Bang nucleosynthesis limits on the average baryonic content of the Universe exclude that
(the majority of) Dark Matter is made out of ordinary baryonic matter (\textit{i.e.} atoms).
Furthermore, although ``dark'', in the sense that does not emit nor absorb light (\textit{i.e.} electromagnetically neutral),
Dark Matter, might couples to ordinary matter in other ways (besides gravity); however, arguments on
its density and thermal production at early times imply that such a coupling must be weak.\footnote{In this context
``weak'' is used in the generic sense, and is not related with the \textit{Weak Interaction},
mediated by W and Z bosons of the Standard Model.}

Both astrophysics and particle physics have been proposing suitable candidates for Dark Matter
through the last three decades, giving rise to an enormous wealth of choice.\footnote{An
idea of the variety of candidates can be given by looking at their masses:
the full list spans from light particles called axions with $m\approx 10^{-71}M_{\odot}$\newnot{symbol:solarmass},
to black holes of mass $m\approx 10^4 M_{\odot}$ \cite{Jungman:1995df}.}

In particular, within the frame of Standard Model of particle physics, the only candidate is provided by neutrinos:
they are electrically neutral and almost collisionless. However, upper bounds on their total
relic density show that they are not abundant enough to be a dominant component of Dark Matter.
Moreover, being relativistic particles, they would just contribute to a possible but tiny \textit{hot} component.

The simplest supersymmetric extension of the Standard Model (Minimal Supersymmetric Standard Model)
offers the most studied candidate, called \textit{neutralino}.
In the model, the superpartners of the photon, the gauge boson $Z^0$, and Higgs bosons mix together
to give rise to four Majorana spinors mass eigenstates, called neutralinos.
The lightest of them should to be stable and it would provide a very suitable candidate for Dark Matter,
fulfilling the requests of weak interaction, heavy mass and stability.
Apart from \textit{neutralino}, Supersymmetry offers other candidates, such as \textit{sneutrinos},
\textit{gravitinos} and other superpartners of particles of the SM \cite{Ellis:2010kf}.
However, a definite proof of the existence of supersymmetric companions is left to particle experiments,
such as LHC (Large Hadron Collider).\footnote{More details on Supersymmetry will be provided in Chapter \ref{cAFVIASM}.}

Another well studied candidate is provided by a hypothetical particle, called \textit{axion},
that was originally introduced to solve the problem that no CP-violation is
observed in strong interactive processes.
Further particle candidates are: \textit{sterile} (neutral under Weak Interaction) \textit{neutrinos}, \textit{Kaluza-Klein particles}
(arising from compactified extra dimensions), \textit{cryptons} (thought to exist in the low energy limit of some string
theories), and others.

At the moment the strongest experimental effort is aimed at finding evidence of
particle Dark Matter, in particular in the form of WIMPs (Weakly Interactive Massive Particles),
either via indirect detection through annihilation products
or direct detection via interaction with baryonic matter \cite{Rau:2011fu}.

Nonetheless, for the time being the nature of
what we believe is the most abundant species of matter in the observable Universe remains
a matter of conjectures.

\chapter{A FLAVOUR VACUUM IN CURVED SPACETIME}\label{cAFVICS}
In the previous chapter we presented the BV formalism for flavour particles. In particular, we emphasized the role
of the \textit{flavour vacuum} state, which, despite just representing the physical state where no flavoured particles are present,
is characterized by a rich physical structure.

The flavour vacuum has been also recognized as the ground state of an effective theory that might emerge in the low energy limit
of a string theory model, the D-particle foam model. A first attempt to study the features of the flavour vacuum
from this perspective has been made in \cite{mavrosarkar}, in which the BV formalism has been implemented
for the first time on curved spacetime. In that work, two bosonic fields with flavour mixing in a $1+1$ dimensional spacetime
with a specific curvature has been considered. Speculative arguments seemed to suggest that the flavour
vacuum in such a context might provide a suitable candidate for Dark Energy.

In this Chapter we  would like to make a step further in that direction, by looking at a more realistic model: we shall consider
two free Dirac fields in a $3+1$ dimensional FRW universe.

Since a full formulation of the stringy model is not available, as explained in Section \ref{sSFV},
in order to study its low energy limit one considers a toy model that includes the most relevant features expected
from considerations in the high energy regime.
In Section \ref{sSUTM1} we shall discuss such expected features, as a complementary clarification
to the explanations provided in \cite{mavrosarkar}.

As mentioned in Section \ref{sSFV}, the D-particle foam model considers a non-trivial interaction between a brane world configuration
(a three dimensional brane with strings attached on it, representing our universe) and a ``foam'' of D-particles (zero dimensional
branes) present in the bulk (a higher dimensional spacetime, in which D-particles, strings and brane are embedded).
The low energy limit of the theory should reflect the interaction between the foam and the brane world.
On one hand, the interaction between the three dimensional brane and the foam leads to a deformation of the brane itself,
that from the point of view of an observer on the brane can be regarded as a curvature of the metric of the three dimensional
spacetime in which (s)he lives. Therefore our effective toy model will be formulated in the framework of \textit{QFT in curved spacetime}.
On the other hand, the interaction between the strings and the foam leads to
two effects for the matter fields in the low energy limit that can be described
by \textit{flavour mixing} between different fields and an \textit{Mikheyev-Smirnov-Wolfenstein gravitational effect}.
The flavour mixing will enter in our model via the BV formalism, extensively discussed in the previous Chapter.
An extension of the formalism in curved spacetime will be exposed in Section \ref{sSUTM1}.
The MSW gravitational effect and the way it will be incorporated in our model will be discussed in Section \ref{ssMGE}.

Collecting all these ingredients, in Section \ref{sFOAFFV} we will summarize the distinctive features of our model,
concluding the Chapter (Section \ref{sDAC}) with an analysis of the final results and a comparison with the work of \cite{mavrosarkar}.

\section{SETTING UP THE MODEL}\label{sSUTM1}

\subsectionItalic{QFT in Curved Spacetime}\label{ssQICS}

The subject of Quantum Field Theory in curved spacetime concerns the study of quantum field theories in which
gravitational effects enter as a classical curvature of the background.
Just like in the early days of quantum theories, the study of quantized matter in presence of a classical
electromagnetic external field anticipated some important features of QED, in absence of a full quantum
theory of gravity, QFT in curved spacetime has been regarded as a privileged tool for
catching significant aspects of the interplaying between gravity and matter, at a quantum level.
The approach is expected to describe phenomena in which the quantum nature of matter and gravitational effects are
both relevant, but the quantum nature of gravity itself does not play a fundamental role.
Therefore it has been used to examine the behaviour of quantum fields in presence of highly curved backgrounds:
more precisely, phenomena occurring in the very early universe and in proximity of a black hole have been studied.
And in this latter context, QFT in curved spacetime founded its most important application: using this approach,
Hawking was able to deduce that a black hole behaves as a black body, emitting a thermal radiation \cite{Hawking:1974sw}.
Such a result is now considered a true cornerstone in any attempts of building a Quantum Gravity theory.

Since the early investigations in the 70's that led to Hawking radiation and other promising results (\textit{e.a.}
the Unruh effect \cite{Unruh:1976db}),
QFT in curved spacetime has been developing in two directions \cite{Wald:2006ty}: on one hand, cosmologists interested in
the phenomenology of the young universe applied those techniques to the study of structure formation
and particle creation by gravitational fields \cite{Mukhanov:2007zz}; on the other hand, theoreticians interested in
generalizing those preliminary results tried to provide a general and coherent framework for the subject,
with the help of more sophisticated mathematical tools borrowed from Algebraic Quantum Field Theory\newnot{symbol:AQFT}
 (AQFT)
\cite{Wald:1995yp,Kay:2006jn}.

One of the major difficulties addressed is the ambiguity in choosing a vacuum state.
Closely related with our investigations, the problem will be presented in more details in the next section.
Such an ambiguity is strictly related with the notion of ``particle''\footnote{For instance, in the previous Chapter,
we \textit{defined} the vacuum state as the state with \textit{no particles} (\textit{cf} Section \ref{ssPSHSFS}).},
and in the modern approach of AQFT it is solved by considering ``fields'' as more fundamental objects.
More specifically,
the approach concentrates on the algebraic properties of fields, rather then implementing them
in specific representations \cite{Wald:2006ty}. In this way, the theory is formulated without the need of defining Fock spaces,
avoiding problems related with inequivalent representations of the same algebra (Section \ref{ssUIR}).

In the study of astrophysical implications of the BV formalism,
standard techniques used in cosmology literature have been used.
This pragmatic approach represents the first step that one would reasonably move when embarking the problem and it
led to interesting results, as we will soon see.
Nevertheless, the modern approach of AQFT might provide a quite fertile ground for future works on the theoretical
aspects of the problem.\footnote{In this context, the interpretation of the mixing as an interaction (Section \ref{ssTI})
might be crucial: in absence of Hilbert spaces, the extension of concepts such as irreducible representations
and the definition of different vacua, as we did in Section \ref{sBVF},
might result arduous; nonetheless, from the perspective of interactive theories, these complications might be unnecessary.}

\subsectionItalic{One Free Dirac Spinor in Curved Spacetime}\label{ssOFDSICS}

In order to define a theory for two free Dirac fields with flavour mixing in curved spacetime,
we start by looking at a single Dirac field first.
If the action for such a field in flat spacetime is given by\footnote{This Chapter is mainly based on the works in \cite{Parker:1971pt,
Birrell:1982ix, mavrosarkar}; therefore it has been adopted their convention
for the metric: $\e_{\mu\nu}\equiv diag\{-1,1,1,1\}$ denotes the flat metric,
whereas the generic metric for curved spacetime is $g_{\mu\nu}=g_{\mu\nu}(x)$. For the same reasons, the analysis for the
FRW universe is worked out in the conformal framework, identified by the metric $g_{\mu\nu}=\Ce \e_{\mu\nu}$.
Finally, through all Section \ref{sSUTM1}, and only in this section,
\textit{Latin} indices in the Einstein summation convention are understood with respect to the flat metric ($\e_{ab}$ instead of $\e_{\mu\nu}$),
whereas \textit{Greek} indices are contracted by the means of the generic metric $g_{\mu\nu}$. For the rest of the work, just greek indices
will be used, with the meaning deducible from the context.}
\be
S_{\flat}=\int d^4x \left(\bar{\psi}(x)\gamma^a \partial_a \psi (x)+m\bar{\psi}(x)\psi (x)\right)+hc
\ee
with $\bar{\psi}(x)\equiv i \psi^\dagger (x) \gamma^0$, $\{\gamma^a,\gamma^b\}=2\e^{ab}$,
we can consider \cite{Birrell:1982ix,weinberg,Ford:1997hb} its generalization on a smooth ($C^\infty$), 4-dimensional,
globally hyperbolic\footnote{The
request of \textit{globally hyperbolicity} ensures that all events on the manifold are causally linked
to the event on a generic spacelike hypersurface.}, pseudo-Riemannian manifold\footnote{Pseudo-Riemannian manifolds are allowed
 to have metrics with negative eigenvalues. In particular we consider a manifold whose metric has signature
$(-1,+1,+1,+1)$.}, as
\be\label{actioncurved}
S=\int d^4x \sqrt{-g} \left( \bar{\psi}(x)\tgamma^\mu D_\mu \psi (x)+m\bar{\psi}(x)\psi (x) \right)+hc
\ee
in which
\begin{itemize}
	\item $d^4x \sqrt{-g}$ represents the volume element ($g=\det(g_{\mu\nu})$),
	\item $\bar{\psi}(x)$ is still defined as $\bar{\psi}(x)\equiv i \psi^\dagger (x)\gamma^0$,
	\item the partial derivative $\partial_a$ has been substituted by the covariant derivative $D_{\mu}=\partial_\mu +\Gamma_\mu$,
  	\item $\Gamma_{\mu}$ is the spin connection defined by
				$\Gamma_{\mu}=\frac{1}{2}\Sigma^{a b} V_{a}^{\;\;\nu}\left(V_{b\nu;\mu}\right)$,
	\item the matrices (\textsl{vierbein}) $V^{a}_{\;\;\mu}=V^{a}_{\;\;\mu}(x)$ are defined by the relation
				$g_{\mu\nu}(x)=V^{a}_{\;\;\mu}(x)V^{b}_{\;\;\nu}(x)\e_{ab}$,
	\item $\Sigma^{a b}$ is the generator of the Lorentz group associated with the spinorial representation under which $\psi$ transforms:
				$\Sigma^{a b}=\frac{1}{4}[\gamma^a,\gamma^b]$,
	\item $V_{b\nu;\mu}\equiv\partial_\mu V_{b\nu}-\Gamma^{\lambda}_{\nu\mu}V_{b\lambda}$,
	\item $\Gamma^{\lambda}_{\nu \mu}\equiv\frac{1}{2}g^{\lambda \kappa}\left(\partial_{\nu}g_{\kappa \mu}+\partial_{\mu}g_{\kappa \nu}
				-\partial_{\kappa}g_{\nu \mu}\right)$ is the Christofell symbol,
	\item the generalized gamma matrices\newnot{symbol:curvedgamma} $\tgamma^\mu=\tgamma^\mu(x)$ are defined by \be\{\tgamma^\mu,\tgamma^\nu\}=2g^{\mu\nu}.\ee
\end{itemize}
The action of the gravitational field on the matter field $\psi$ is encoded in the dependency of the action on the generic
metric $g_{\mu\nu}(x)$. Such a gravitational field is considered to be classic and external: $g_{\mu\nu}(x)$ is just a matrix
of c-numbers and their values are fixed \textit{a priori}, being not determined, as one would expect from General Relativity,
by an analogous of the Einstein equation. The latter condition implies that the background affects the evolution of
the matter field, but its curvature \textit{is not} determined by the matter field itself.
Choosing a foliation of the spacetime into spacelike hypersurfaces, labeled by the value of
a time parameter,
the quantization of the field $\psi$ is obtained by imposing on each spinor component
$\psi_a$ the rules
\begin{gather}\label{curvedrules}
\{\psi_a(\vx,t),\psi_b^\dagger(\vx',t)\}=\delta^3(\vx-\vx')\delta_{ab}\\
\{\psi_a(\vx,t),\psi_b(\vx',t)\}=\{\psi_a^\dagger(\vx,t),\psi_b^\dagger(\vx',t)\}=0
\end{gather}
with $\vx$ and $\vx'$ belonging to the same generic spacelike hypersurface $\Sigma$, labeled by $t$, and with
\be
\int_\Sigma \delta^3(\vx-\vx')d\Sigma=1.
\ee
being $d\Sigma$ the volume element of $\Sigma$.
From (\ref{actioncurved}) the equation of motion
\be\label{curvedeom}
\tgamma^\mu D_\mu \psi (x)+m\psi (x)=0
\ee
follows.

\subsectionItalic{Choice of the Vacuum in Curved Spacetime}\label{ssCOTVICS}

In analogy with the flat case, we now would like to decompose a generic solution of (\ref{curvedeom}) as
\be\label{curveddec}
\psi (x)=\sum_j \left(	a_j f_j+b^\dagger_j g_j	\right)
\ee
with $\{f_j,g_j\}$ a complete set of solutions (with a spinorial structure) for (\ref{curvedeom}),
orthonormal under the product
\be
(f_1,f_2)\equiv i\int \left((\partial_\mu f_1^T )f_2^*-f_1^T(\partial_\mu f^*_2)\right)d\Sigma^\mu
\ee
where $d\Sigma^\mu=d\Sigma n^\mu$, $d\Sigma$ being the volume element of a given spacelike hypersurface\footnote{It
is possible to show that such a product is independent on the specific choice of spacelike hypersurface.}
and $n^\mu$ the timelike unit vector
normal to this hypersurface, and $a_j^{(\dagger)}$ and $b_j^{(\dagger)}$ are annihilation (creation) operators.
From (\ref{curveddec}) and (\ref{curvedrules}), it is possible to prove that the operators $a^{(\dagger)}_j$ and
$b_j^{(\dagger)}$ would satisfy the usual CAR algebra.
Therefore it would be possible to build a Fock space for physical states, in analogy with the procedure explained
in Section \ref{ssQFL}.

However we now have to face an inherent ambiguity of the formalism: there is no obvious choice
of the set of solutions $\{f_j,g_j\}$.
In flat spacetime, there is a \textit{preferred} set of solutions, in connection with the Poincar\'e invariance of the theory,
namely the \textit{plane waves set}, multiplied by suitable constant spinors:
\be\label{proptowaves}
f_p\propto e^{ixp},\;g_p\propto e^{-ixp}.
\ee
Such a set allows us to build a vacuum state that is invariant under the action of the Poincar\'e group.
Physically, this choice is equivalent to say that \textit{all inertial} observers would not
detect particles in such a state.
This is a crucial point since a set of solutions different from (\ref{proptowaves}) would lead to different unphysical situations:
more specifically, any sets of modes define an \textit{inequivalent representation} of the algebra CAR, in the sense
specified in Section \ref{ssUIR}, and therefore lead to systems with different physical interpretations.
We can clarify this point, by noticing that, given the set in (\ref{proptowaves}),
we can choose another set of solutions $\{l_j,m_j\}$ and write
\be\label{psicldm}
\psi (x)=\sum_j \left(	c_j l_j+d^\dagger_j m_j	\right)
\ee
with $c_j^{(\dagger)}$ and $d_j^{(\dagger)}$ new ladder operators. Since the two set are both
complete, we can write one in terms of the other:
\be\label{fglm}\begin{split}
f_j=\sum_k (\alpha_{jk} l_k+\beta_{jk} m_k)\\
g_j=\sum_k (\gamma_{jk} l_k+\delta_{jk} m_k)
\end{split}\ee
with $\alpha_{\slot},\beta_{\slot},\gamma_{\slot},\delta_{\slot}\in \field{C}$ suitable coefficients.
By inserting (\ref{fglm}) into (\ref{curveddec}) and comparing it with (\ref{psicldm}), one can easily obtain
\be\begin{split}
c_j=\sum_k (a_k \alpha_{kj}+b^\dagger_k \gamma_{kj})\\
d_j=\sum_k (a^\dagger_k \beta^*_{kj}+b_k \delta^*_{kj}).
\end{split}\ee
These equations are the analogous of (\ref{bogoexample}): $(a^{(\dg)},b^{(\dg)})$ and $(c^{(\dg)},d^{(\dg)})$ are part of two inequivalent representations
of the CAR algebra, and the vacuum defined by $(a,b)$ is orthogonal to the vacuum annihilated by $(c,d)$.

In flat spacetime, the request that all inertial observers agree on the vacuum state allows us to pick
one specific set of solutions. In a generic spacetime, in which the Poincar\'e group no longer represents the group of isometries,
we are left without criteria for choosing the set of solutions.
As we said, in recent developments of the field, there is an effort for avoiding the problem via a formulation
of the theory in a completely algebraic way, without referring to a specific representation of the algebra
of the operators (and therefore a specific Hilbert space on which the fields act as operators).
Nevertheless, some specific backgrounds allow to circumvent the problem and enable us to use standard techniques of QFT.
This is the usual approach adopted in cosmology phenomenology, and followed in this work:
without the need of having a full theory defined on background of arbitrary curvature,
specific models can be considered in which the spacetime has enough symmetries to admit a treatment close to the spirit of
ordinary QFT.

Indeed, one would expect that ordinary QFT holds in regions that are approximately flat. In those regions
the particle interpretation of the Fock space is physically meaningful and well defined. We can then extend the formulation
of the theory from one (almost flat) region to another, by simply choosing those solutions of the equations of motion
over the whole spacetime  that are ``almost'' plane waves in the regions of flatness.

\begin{figure}[t]
\centering
\includegraphics[width=1\textwidth]{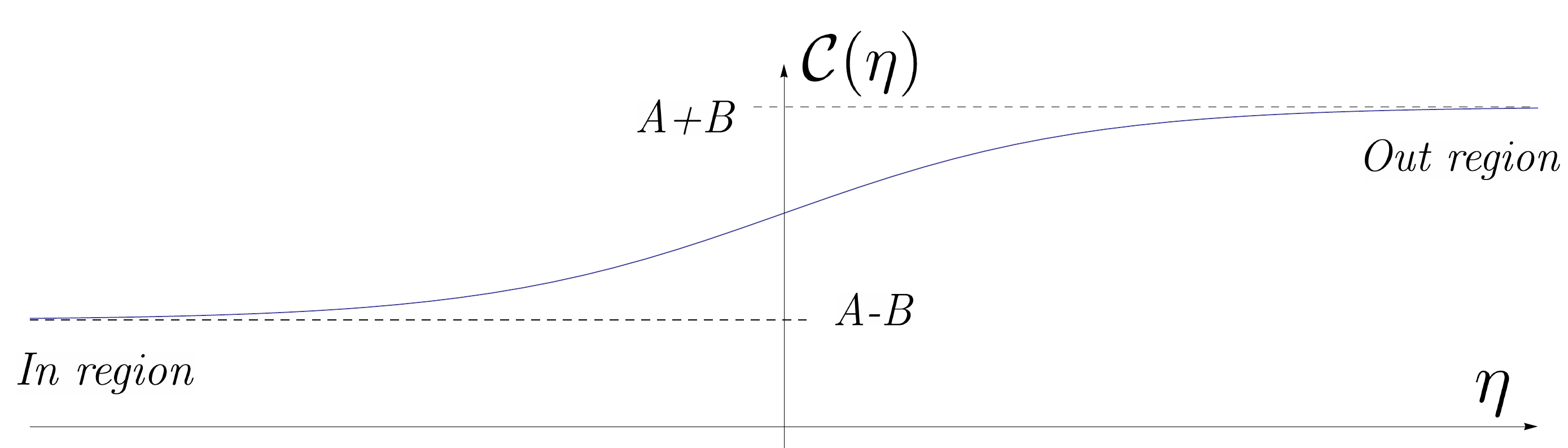}
\caption{\small{An example of conformal scale factor that smoothly interpolates between two asymptotically flat region is plotted:
$\C(\e)=A+B\tanh(x)$ with $A>B>0$.}}
\label{fig:conformal}
\end{figure}

To make this statement more precise, we consider an FRW universe, \textit{i.e.} an isotropic (no preferred directions) and
homogeneous (translational invariant) expanding and/or contracting
universe, whose line element, in conformal coordinates\footnote{Usually
a FRW universe is represented in the coordinate system such that $ds^2=dt^2-a^2(t)(dx^2+dy^2+dz^2)$.
Here conformal coordinates are used for conform to the notation used in \cite{mavrosarkar}  (\textit{cf} Section \ref{ssMCAI}).
One can notice that $t\rightarrow \pm\infty$ corresponds to $\e\rightarrow \pm\infty$,
therefore we are allowed to refer to the regions for which $\e\rightarrow \pm\infty$,
as the regions at \textit{early/late times}.},
is written as
\be
ds^2=\Ce (-d\e^2+dx^2+dy^2+dz^2)
\ee
with $\Ce$ a positive function of $\e$ (the \textit{conformal time}),
obeying to the constrain $\C(\e\rightarrow \pm \infty)=constant_\pm$:
such an universe is therefore asymptotically flat
for early (\textit{in} region) and late (\textit{out} region) times (see Figure \ref{fig:conformal}).
Being $g_{\mu\nu}=\Ce\;diag\{-1,1,1,1\}$, the equation of motion (\ref{curvedeom}) becomes
\be\label{frweom}
\left(\gamma^a \partial_a+\frac{3}{4}\frac{\Cp(\e)}{\C(\e)}\gamma^0+\sqrt{\C(\e)}m\right)\psi=0
\ee
(see Appendix \ref{aDIFRW}).
Furthermore, we formulate our field theory imposing boundary conditions on the field such as
$\psi(\e,\vx+\vec{n}L)=\psi(\e,\vx)$, $\vec{n}$ being a vector with integer Cartesian components.
This condition, that is equivalent to consider a spatially finite universe of volume $V=L^3$,
is relaxed at the end of the calculations, by taking the limit $L\rightarrow \infty$ of physically significant quantities.\footnote{The
importance of this initial assumption is remarked at the end of Appendix \ref{aC23fv}.}

Following \cite{Parker:1971pt}, we can consider the ansatz
\be\label{solutionparkerconformal}
\psi (\e,\vx)=\left(\frac{1}{L\sqrt{\Ce}}\right)^{\frac{3}{2}}\!\!\!\!\sum_{\mbox{\tiny{$\begin{array}{c} \vp  \\ a,b=\pm 1 \end{array}$}}}
\!\!\!\!a^{(a,b)}(\vp,\e)u^{(a,b)}(\vp,\e)e^{ia\left(\vp\cdot\vx-\int\w(p,\e) d\e\right)}
\ee
where
\begin{itemize}
	\item $L$ is the parameter that enters the periodic boundary condition,
	\item $\w(p,\e)\equiv\sqrt{p^2+m^2 \C(\e)}$\newnot{symbol:we},
	\item $u^{(a,b)}(\vp,\e)$ is a spinor defined by
		\be
		\left\{
		\begin{array}{l}
		(-i a\sqrt{p^2+m^2 } \gamma^0+i a \vec{\gamma}\cdot\vp+m\sqrt{\Ce})u^{(a,b)}(\vp,\e)=0\\
		u^{(a,b)\dagger}(\vp,\e)u^{(a',b')}(\vp,\e)=\delta_{a,a'}\delta_{b,b'}
		\end{array}
		\right.
		\ee
	\item $a^{(a,b)}(\vp,\e)$ are operators such that
				\be\{a^{(a,b)}(\vp,\e),a^{(a',b')\dagger}(\vq,\e)\}=\delta_{a,a'}\delta_{b,b'}\delta_{\vp,\vq}\ee.
\end{itemize}
The equation of motion (\ref{frweom}) imposes a specific evolution of the operators $a^{(a,b)}(\vp,\e)$:
\be\label{atoA}
a^{(a,b)}(\vp,\e)=\sum_{c=\pm 1}D^{a}_{c}(p,\e)A^{(c,abc)}(ac\vp)
\ee
with
\be\label{Ddefinition}
	D_{(a')}^{(a)}(p,\e)=
		\delta_{a'}^{a}+a'\int^{\e}_{-\infty}d\e'\frac{1}{4}\frac{\C'(\e')}{\sqrt{\C(\e')}}
			\frac{m p}{\w(p,\e')^2}\, e^{2 i a' \int\w(p,\e') d\e'}D_{(-a')}^{(a)}(p,\e')
\ee
and $a,a'=-1,1$. Combining (\ref{atoA}) and (\ref{Ddefinition}), it is easy to see that
the operator $A^{(a,b)}(\vp)$, that does not depend on time, represents
the initial value of the operator $a^{(a,b)}(\vp,\e)$: $A^{(a,b)}(\vp)=a^{(a,b)}(\vp,\e\rightarrow -\infty)$.
As such, it also respects the CAR algebra:
\be\label{carA}
\{A^{(a,b)}(\vp),A^{(a',b')\dagger}(\vq)\}=\delta_{a,a'}\delta_{b,b'}\delta_{\vp,\vq}
\ee
Although it is not known an exact solution of the integral equation (\ref{Ddefinition}),
once the metric (and therefore $\Ce$) has been fixed, equation (\ref{Ddefinition})
admits one and only one solution in $D_{(a')}^{(a)}(p,\e)$.
Such a decomposition of the field $\psi(\e,\vx)$ represents therefore a suitable candidate for building our quantum theory.

In order to define a Fock space,
we distinguish three different phases, in accordance with the behaviour of $\Ce$:

\begin{enumerate}
	\item in the first one, at very early times the metric is ``almost'' (asymptotically) flat $\Ce\sim constant_-$,
		QFT in flat spacetime provides a good approximation of the theory and indeed $\psi(\e,\vx)$ is decomposed in
		plane waves, being $\Cp(\e)\sim 0$ and therefore $D_{(a')}^{(a)}(p,\e)\sim\delta_{a'}^{a}$.
		The Fock space built by the means of the operators $A^{(a,b)}(\vp)$ can be correctly regarded as representing physical particle states,
		in the familiar way. In particular, the vacuum $\rmv$ defined by
		\be\label{in-vacuum}
		A^{(+1,b)}(\vp)\rmv=A^{(-1,b)\dagger}(\vp)\rmv=0
		\ee
		corresponds to the physical state that is empty for all inertial observers.
	\item In the second phase, an expansion/contraction of the universe occurs ($\Ce$ changes sensibly),
		ordinary QFT in flat spacetime does not apply; nevertheless, equation (\ref{frweom}) and (\ref{atoA})
		still determine the dynamics of our system. Although it is still possible to build a Fock space
		by using the operators $a^{(a,b)(\dagger)}(\vp,\e)$,
		the interpretation of vectors of this Fock space as physical particle states becomes meaningless.
	\item In the third phase, at late times, the metric is once more approximately flat, since $\Ce\sim constant_+$.
		Ordinary flat QFT applies, the plane-wave decomposition of the field $\psi(x)$ is again a good approximation (once more
		$D_{(a')}^{(a)}(p,\e)\sim\delta_{a'}^{a}$) and Fock space built out of the (asymptotic limit of the) operators
		$a^{(a,b)(\dagger)}(\vp,\e)$ is again an irriducible representation of the Poincar\'e group.
\end{enumerate}

However, the Fock space defined by $a^{(a,b)(\dagger)}(\vp,\e)$
at \textit{late} times might not correspond to the Fock space defined at \textit{early} times.
This happens if $a^{(a,b)}(\vp\rightarrow +\infty,\e)\neq A^{(a,b)}(\vp)$.
More specifically, the vacuum state defined at late times (that we can call ``out-vacuum'')
might not correspond to $\rmv$ defined in (\ref{in-vacuum}) (the ``in-vacuum'').
This remarkable mathematical result is usually interpreted as a \textit{particle creation} phenomenon, due to
the gravitational field influence on the system: an (inertial) observer in the \textit{out} region might actually detect particles
in the state that was defined as the vacuum in the \textit{in} region.

In analogy with Section \ref{ssTI}, we could regard at the theory as an \textit{interactive} theory,
in which an external field (the gravitational field) \textit{interacts} with (otherwise) free particles.
The vectors of the Fock space used to describe the physical states contains (completely)
free particles, but the interpretation of the vectors as physical particle states makes sense only in low interactive regimes,
\textit{i.e.} at early and late times.
It should be remarked that, unlike previous examples of Section \ref{sBVF}, now the theory is
\textit{imposing} us a change \textit{in time} of the Fock space:
the in- and the out-vacuum are not two unrelated objects but they
represent the \textit{dynamical evolution} of \textit{one} specific state: the \textit{physical} vacuum.
At early times this state is actually empty, and it is represented by $\rmv$.
But later on, after the influence of the gravitational field, it evolves in a more complex object,
labeled as \textit{out vacuum},
that can actually contain some particles.

The dynamical evolution of the \textit{physical} vacuum state (and consequently of all other particle states) due to the interaction
with the external gravitational field underlies most of the phenomenological applications of the formalism of QFT in curved spacetime
and plays also a crucial role in rest of our discussion.

\subsectionItalic{Choosing the Flavour Vacuum}\label{ssCTFV}

So far, we pointed out that one of the major problems in QFT in curved spacetime concerns the choice of
a suitable Hilbert space for the theory (or, equivalently, a vacuum state), when the curvature is kept arbitrary.
It is easy to understand the reasons of this problem, when considering the fact that the ordinary Hilbert space in QFT
in flat spacetime is chosen to be an irreducible representation of the Poincar\'e group: this choice ensures the
invariance of the vacuum state under Poincar\'e transformations, \textit{i.e.} all inertial observers agree that
the vacuum state so defined describes the physical state with \textit{no particles}.
In a generic background no symmetries are present to guide such a choice and the na\"ive formalism is affected by
an inherent ambiguity. More sophisticated tools have been considered in recent developments of the field, in the effort
of overcoming this specific problem. Nevertheless, in the majority of phenomenological applications, one can still
work in a pragmatic framework, in which the theory is defined on a subset of the entire spacetime manifold
that can be considered as flat, and then extend the theory to the whole manifold.
More specifically, we considered the metric $g_{\mu\nu}=\Ce\;diag\{-1,1,1,1\}$ with the condition
$\C(\e\rightarrow \pm \infty)=constant_\pm$ and we explained how
the Hilbert space in the \textit{in} and the \textit{out} region can make sense as the irreducible representation of the Poincar\'e group,
\textit{i.e.} representing \textit{physical} particles.

In order to introduce the BV formalism for flavour states, we shall rewrite such a procedure
in the language of Sections \ref{ssPSHSFS} and \ref{ssUIR}.
The Hilbert space describing our system is \hilbert{H},
the space of states with arbitrary (even infinite) number of of free Dirac particles of two types, distinguished
by two different masses, $m_1$ and $m_2$.

This is actually the same Hilbert space that one uses in describing a theory in \textit{flat} spacetime (as we did in Section \ref{ssPSHSFS}):
the curvature of the background is formally treated as an external classical field in a theory formulated on the usual Minkowskian manifold.
At early times ($\e\rightarrow-\infty$), the vacuum state is $\ket{0,0,0,...}\equiv \rmv$
and the Fock space considered is $\mathfrak{F_0}\subset\mathfrak{H}$.
When the system evolves, all information about the growth of the universe (\textit{i.e.} the action of the gravitational field
on our system) is encoded in the equation of motion (\ref{frweom}).
More specifically, as Parker pointed out \cite{Parker:1969au,Parker:1971pt},
at each (conformal) time $\e$, we can decompose the field $\psi(x)$ in such a way to be able to have
operators $a^{(a,b)}(\vp,\e)$ respecting the CAR algebra. These operators define a Fock space $\mathfrak{F}(t)$.
In general at different times one would expect that different Fock spaces are defined by $a^{(a,b)}(\vp,\e)$:
$\mathfrak{F}(t)\neq\mathfrak{F}(t')$ when $t\neq t'$.
Nonetheless, those spaces are all countable subsets of \hilbert{H}. In formulae, we have
\be
\forall t\in\field{R}\;\;\;\mathfrak{F}(t)\subset \mathfrak{H}
\ee
and
\be
\mathfrak{F}(-\infty)=\mathfrak{F_0}.
\ee
We can therefore say that \hilbert{H} is sufficient for describing our system at any times,
the influence of the gravitational field on the matter field being encoded in the evolution in time of $a^{(a,b)}(\vp,\e)$
(equation (\ref{atoA})) and therefore in the evolution in time of $\mathfrak{F}(t)$.

However, $\mathfrak{F}(t)$ does not admit a direct interpretation as space of \textit{physical} particle states,
unless considered at early and late times, when the metric is asymptotically flat.
\textit{Particle production}, as defined in the previous Section, occurs if
\be\label{parproFF}
\mathfrak{F}(-\infty)\neq \mathfrak{F}(+\infty)
\ee
since
\be
\ket{0(t\rightarrow+\infty)}\neq \ket{0(t\rightarrow-\infty)}= \rmv
\ee
with $\ket{0(t)}$ the ground state of $\mathfrak{F}(t)$.

From this perspective, it is easy to understand that the BV formalism can be implemented in this model
by simply requiring
\be\label{inflavourspace}
\mathfrak{F}(-\infty)=\mathfrak{F_f}
\ee
instead of
\be
\mathfrak{F}(-\infty)=\mathfrak{F_0},
\ee
with $\mathfrak{F_f}$ the Fock space for flavour states.

In particular, we are interested in the \textit{flavour vacuum} state and in its evolution.
Recalling once more that at early times the spacetime is asymptotically flat, we can require, without loss of generality,
that $\C(\e\rightarrow-\infty)\rightarrow 1$. We are therefore allowed to use the standard formalism in flat spacetime
and consider as \textit{flavour vacuum} at early times,
the state
\begin{multline}\label{earlyflavourvacuum}
\rfv\equiv\lim_{t_0\rightarrow-\infty}G_\theta^{\dagger}(t_0)\rmv=\\
\lim_{t_0\rightarrow-\infty}\exp \left[ -\theta\int d\vx \left(
\psi^{\dagger}_{\flat1}(x)\psi_{\flat2}(x)-\psi_{\flat2}^{\dagger}(x)\psi_{\flat1}(x)
\right)\right]\rmv
\end{multline}
in which we used the definition of $G_{\theta}(t)$ provided in \cite{Blasone:1995zc,Capolupo:2004av}
in terms of the fields $\psi_{\flat i}(x)$, the ordinary Dirac fields defined in flat ($\flat$) spacetime.
Since $\C(\e\rightarrow-\infty)\rightarrow 1$, the fields $\psi_\flat(x)$ correspond to the limit fields to whom
the fields $\psi(x)$ defined in (\ref{solutionparkerconformal}) approach at early times:
\be\label{psiflatpsi}
\psi_i(t\rightarrow-\infty,\vx)\approx \psi_{\flat i}(t\rightarrow-\infty,\vx).
\ee

Formula (\ref{earlyflavourvacuum}) defines the \textit{physical vacuum} state of our theory at early times,
in the sense of equation (\ref{inflavourspace}).
Once the state is defined at early times, the dynamics of the system is determined by the evolution in time
of the fields $\psi_{i}(x)$.

\subsectionItalic{MSW Gravitational Effect}\label{ssMGE}

If from one hand the action of the foam on the brane is encoded in our effective model by formulating the theory on a curved background,
on the other hand the interaction of the foam with the strings attached on the brane is modeled via the BV formalism and the MSW gravitational effect \cite{Barenboim:2004ev,Barenboim:2006xt,Mavromatos:2006yy}.

The standard Mikheyev-Smirnov-Wolfenstein (MSW) effect characterizes flavour oscillation of neutrinos in a medium \cite{MSW1,MSW2}\newnot{symbol:MSW}.
Such an effect can enhance flavour oscillation as it would occur in vacuum,
and might even being responsible for oscillations in cases in which they would be forbidden in vacuum
(for instance, if massless particles are considered).
The effect is especially relevant for the analysis of solar flavour neutrinos.
When a neutrino is traveling through a medium it can acquire an effective mass due to the interaction with the
microscopic structure of the medium.\footnote{A mechanism perhaps more familiar to the reader occurs with photons:
while traveling in a medium they have a slower average velocity due to the interaction with medium itself,
and therefore they can be regarded as free photons with a rest mass.} In the Sun this interaction is nearly exclusively present
for the first generation of leptons and quarks. Therefore electron neutrinos are singled out,
and are the only one receiving a relevant contribution for the effective mass.
Since flavour oscillations depend upon the squared mass difference of the neutrinos,
such an effective mass for one species of neutrino alters the oscillation with respect to the vacuum case.

A gravitational analogous of the above effect occurs in the stringy model of \cite{mavrosarkar}
(so called \textit{MSW gravitational effect}).
The crossing of D-particles through the brane
is seen on the brane itself as local topological defects flashing on and off instantly.
Such a microscopic structure, on a higher scale can be depicted as a medium in which particles living on the brane are
embedded. Particles experience the medium via gravitational interaction and we can incorporate in our model the
interplay between particles and medium via an extra contribution to their mass,
\textit{i.e.} an extra effective mass term in the Lagrangian. Assuming such a contribution small if compared with
the scale provided by $m_1\approx m_2$, we can implement the MSW effect in our model by substituting $m_i$ with $m_i+m_{eff}$,
without need of distinguishing between the contribution to $m_1$ from the one to $m_2$.

From a formal point of view, such a change of masses does not affect any calculations.
However, since it is the presence of D-particles that induces the effective mass $m_{eff}$,
we expect that the \textit{absence} of the effective mass $m_{eff}$ is a signal of absence of the D-particle foam.
This implies that we require that physically relevant quantities must vanish when $m_{eff}=0$.
In other words, we will subtract all contributions to the expectation value of the stress-energy tensor
that are independent on $m_{eff}$, since we would like to consider in our final expression only terms that are
related with the presence of the D-particle foam in the bulk space.

\section{FEATURES OF A FERMIONIC FLAVOUR VACUUM}\label{sFOAFFV}

\subsectionItalic{Definition of the model and relevant quantities}\label{ssDOTMARQ}

In the previous sections, we explained how to implement the BV formalism in a simple QFT model on curved background.
In summary, one starts with the action for two Dirac spinorial fields $\psi_1(x)$ and $\psi_2(x)$:
\begin{multline}\label{Spsi1psi2curved}
S=\int d^4x \sqrt{-g} \left( \bar{\psi}_1(x)\tgamma^\mu D_\mu \psi_1 (x)+m_1\bar{\psi}_1(x)\psi_1 (x) +\right.\\
\left.+\bar{\psi}_2(x)\tgamma^\mu D_\mu \psi_2 (x)+m_2\bar{\psi}_2(x)\psi_2 (x) \right)
\end{multline}
(compare formula (\ref{actioncurved}) for meaning of the symbols); then $g_{\mu\nu}(x)=\Ce \e_{\mu\nu}$,
with $\C(\e\rightarrow-\infty)=1$, is required in order to build, at early times, a Fock space that is an irreducible representation
of the Poincar\'e algebra. In absence of flavour mixing, this space would correctly describe physical states
at early times. The ladder operators $A^{(a,b)(\dagger)}(\vp)$ and the vacuum state $\rmv$, that define such a Fock space,
are also sufficient for describing physical states at any times, at least from a mathematical point of view:
an unambiguous physical interpretation is limited to (approximately)
flat regions of the spacetime, the concept of particle itself being meaningful only in sufficiently symmetric spacetimes.

In presence of flavour mixing, the action (\ref{Spsi1psi2curved}) would represent the linearized version of
\begin{multline}\label{SpsiApsiBcurved}
S=\int d^4x \sqrt{-g} \left( \bar{\psi}_A(x)\tgamma^\mu D_\mu \psi_A (x)+m_{A}\bar{\psi}_A(x)\psi_A (x) +\right.\\
+\bar{\psi}_B(x)\tgamma^\mu D_\mu \psi_B (x)+m_{B}\bar{\psi}_B(x)\psi_B (x)\\
\left. +m_{AB}\bar{\psi}_A(x)\psi_B (x)+m_{AB}\bar{\psi}_B(x)\psi_A (x)\right)
\end{multline}
where the flavoured fields $\psi_A(x)$ and $\psi_B(x)$ are linked to $\psi_1(x)$ and $\psi_2(x)$ via
\bea
\psi_A(x)&=&\psi_1(x)\ct+\psi_2(x)\st\\
\psi_B(x)&=&-\psi_1(x)\st+\psi_2(x)\ct
\eea
and
\bea
m_A&=&m_1 \cct+m_2 \sst\\
m_B&=&m_1 \sst+m_2 \cct\\
m_{AB}&=&(-m_1+m_2) \st \ct
\eea
Starting from the Fock space defined at early times for the theory without mixing,
one is able to define a new Fock space for flavour states, via the standard procedure described in \cite{Blasone:1995zc}
 and summarized in Section \ref{ssFSFS}.
The \textit{flavour vacuum} is therefore defined, at early times, in terms of $A^{(a,b)(\dagger)}(\vp)$ and $\rmv$
via formulae (\ref{earlyflavourvacuum}), (\ref{psiflatpsi}), (\ref{solutionparkerconformal}) and (\ref{atoA}):
\bea
\rfv&=&G^{\dagger}_{\theta}(-\infty)\rmv\nn
		&=&\exp\left[-\theta \int d\vx\left(\psi^\dagger_1(x)\psi_2(x)-\psi^\dagger_2(x)\psi_1(x)\right) \right]_{\e\rightarrow-\infty}\rmv
\eea
Furthermore, it is possible \cite{Blasone:1995zc,Capolupo:2004av} to reduce it to
\begin{multline}
\rfv =\prod_{\vk}\Big[1+\sin \theta \cos \theta (S_{-}(\vk)-S_{+}(\vk))+\frac{1}{2}\sin^2 \theta \cos^2 \theta ((S_{-}(\vk))^2+\\
				+(S_{+}(\vk))^2)-\sin^2\theta S_{+}(\vk)S_{-}(\vk)+\frac{1}{2}\sin^3\theta \cos \theta (S_{-}(\vk)(S_{+}(\vk))^2+\\
		 -S_{+}(\vk)(S_{-}(\vk))^2)+\frac{1}{4}\sin^4\theta(S_{+}(\vk))^2(S_{-}(\vk))^2\Big] \rmv
\end{multline}
with
\begin{multline}
S_{+}(\vp)\equiv\sum_{\mbox{\tiny{$\ba{c}a,b\\a',b'\ea$}}}\Big[\hat{A}^{(a,b)\dagger}_1(a\vp)\hat{A}^{(a',b')}_2(a'\vp)\times\\
\times\sqrt{\frac{m_1 m_2}{\sqrt{p^2+m_1^2}\sqrt{p^2+m_2^2}}}u_1^{(a,b)\dagger}(a\vp)u^{(a',b')}_2(a'\vp)\Big]
\end{multline}
\begin{multline}
S_-(\vp)\equiv\sum_{\mbox{\tiny{$\ba{c}a,b\\a',b'\ea$}}}\Big[\hat{A}^{(a,b)\dagger}_2(a\vp)\hat{A}^{(a',b')}_1(a'\vp)\times\\
\times\sqrt{\frac{m_1 m_2}{\sqrt{p^2+m_1^2}\sqrt{p^2+m_2^2}}}u_2^{(a,b)\dagger}(a\vp)u^{(a',b')}_1(a'\vp)\Big].
\end{multline}

The features of such a state are studied via the evolution in time of the quantity
\be\label{fvevsetcurvedst}
\lfv T_{\mu\nu}(x) \rfv.
\ee
In the Heisenberg picture, the stress-energy tensor operator $T_{\mu\nu}(x)$ encodes all the information about the
evolution in time of our system. For the theory described by (\ref{Spsi1psi2curved}), it is given by \cite{Birrell:1982ix,Weldon:2000fr}
\be
T_{\mu\nu}(x)=-g_{\mu\nu}(x)\lag +\frac{1}{2}\Big(\bar{\psi}_1(x)\tgamma_{\left(\mu\right.} D_{\left.\nu\right)} \psi_1 (x)
+\bar{\psi}_2(x)\tgamma_{\left(\mu\right.} D_{\left.\nu\right)} \psi_2 (x)+h.c.\Big)
\ee
and becomes (Appendix \ref{aDIFRW})
\bea\label{setfrwpsi1psi2}
T_{\mu\nu}(x)&=&\sum_{i=1,2}\bar{\psi}_i(x)
\left(\frac{\sqrt{\Ce}}{2} \gamma_{\left(\mu\right.}\partial_{\left.\nu\right)}
					  +\frac{\C'(\e)}{16 \sqrt{\Ce}} \gamma_{\left(\mu\right.}[\gamma_0,\gamma_{\left.\nu\right)}]
\right)\psi_i(x)+\nn
&&-\Ce \e_{\mu\nu}\lag+h.c.
\eea
when the metric $g_{\mu\nu}(x)=\Ce \e_{\mu\nu}$ is considered.

The action of this operator on the flavour vacuum is determined by its decomposition in terms of the operators $A^{(a,b)(\vp)}$
via (\ref{setfrwpsi1psi2}), (\ref{solutionparkerconformal}) and (\ref{atoA}).
Having both $\rfv$ and $T_{\mu\nu}(x)$ expressed in terms of $A^{(a,b)(\vp)}$
and $\rmv$, and knowing that
\be
\{A^{(a,b)}(\vp),A^{(a',b')\dagger}(\vq)\}=\delta_{a,a'}\delta_{b,b'}\delta_{\vp,\vq}
\ee
and
\be
A^{(+1,b)}(\vp)\rmv=A^{(-1,b)\dagger}(\vp)\rmv=0
\ee
(as stated in (\ref{carA}) and (\ref{in-vacuum})), the evaluation of $\lfv T_{\mu\nu}(x) \rfv$
reduces to a straightforward manipulation of algebraic quantities and simplification of spinorial functions in the momentum space.
Full details are provided in the dedicated Appendix \ref{aFVEVOTSET}.

In the continuous limit ($L\rightarrow\infty$), expression (\ref{fvevsetcurvedst}) simplifies to
\begin{multline}\label{set00befornormalordering}
\lfv T_{00}(x) \rfv=\lmv T_{00}(x)\rmv+\\
			+\sst \int_0^\infty dp V^2(p)\left(\mathcal{T}_{00}(\e,p,m_1)+\mathcal{T}_{00}(\e,p,m_2)\right)+\mathcal{O}(\ssst)
\end{multline}
and
\begin{multline}\label{setiibefornormalordering}
\lfv T_{jj}(x) \rfv=\lmv T_{jj}(x) \rmv+\\
			+\sst \int_0^\infty dp V^2(p)\left(\mathcal{T}_{jj}(\e,p,m_1)+\mathcal{T}_{jj}(\e,p,m_2)\right)+\mathcal{O}(\ssst)
\end{multline}
all other components vanishing, with
\be\label{innerset00befornormalordering}
\mathcal{T}_{00}(\e,p,m)\equiv \frac{8}{(2 \pi)^2}p^2\we\sqrt{\Ce}\left(1
																														 -|D^{(-1)}_{(1)}(p,\e)|^2-|D^{(1)}_{(-1)}(p,\e)|^2\right)
\ee
\be\label{innersetiibefornormalordering}
\mathcal{T}_{jj}(\e,p,m)\equiv \frac{1}{3}\frac{8}{(2 \pi)^2}\frac{p^4\sqrt{\Ce}}{\we}\left(1
																														 -|D^{(-1)}_{(1)}(p,\e)|^2-|D^{(1)}_{(-1)}(p,\e)|^2\right)
\ee
\be\label{Vsquarecurvedspacetime}
V^2(p)=\frac{\sqrt{p^2+m_1^2}\sqrt{p^2+m_2^2}-p^2-m_1 m_2}{\sqrt{p^2+m_1^2}\sqrt{p^2+m_2^2}}
\ee
\be
\we=\sqrt{p^2+m^2\Ce}
\ee
and the functions $D^{(a)}_{(b)}(p,\e)$ defined by (\ref{Ddefinition}). Once more, a general explicit solution
for the equation (\ref{Ddefinition}) that defines those functions is unknown. However, as we will see,
general considerations apply and enable us to analyze such results, even without full details for these functions.

Expressions (\ref{set00befornormalordering}) and (\ref{setiibefornormalordering})
correctly reproduce the energy and the pressure of the flavour vacuum, as stated
in \cite{Capolupo:2007hy,Blasone:2007jm,Capolupo:2006re,Blasone:2008rx}, for $\Ce=1$.
Moreover, when $\theta=0$ or $m_1=m_2$, the contribution provided by the flavour vacuum disappear,
and one is left only with $\lmv T_{\mu\nu}(x)\rmv$: for those values of the parameters $\theta$ and $m_i$,
the theory with mixing reduces to a theory \textit{without} mixing (\textit{i.e.} the action (\ref{SpsiApsiBcurved}) becomes
trivially identical to (\ref{Spsi1psi2curved})).
The function $V^2(p)$ defined in (\ref{Vsquarecurvedspacetime}) corresponds exactly to the distribution in the momentum space
of particles with masses $m_1$ and $m_2$, when the flavour vacuum is described as a gas of such particles (\textit{cf} Section \ref{ssVC}),
in \textit{flat spacetime}.

\subsectionItalic{Regularization and Normal Ordering}\label{ssRANO}

As one would expect, those expressions are formally divergent
and indeed they do not describe yet the energy and pressure of the flavour vacuum as prescribed by our model.
In first place, we need to remove the contribution to $\lfv T_{\mu\nu}(x) \rfv$ given by the vacuum $\rmv$,
by introducing the ordinary normal ordering
\be
\lfv: T_{\mu\nu}(x): \rfv\equiv \lfv T_{\mu\nu}(x) \rfv-\lmv T_{\mu\nu}(x) \rmv.
\ee
Since $\rmv$ is the vacuum state defined for $\mathfrak{F}(t\rightarrow-\infty)$,
we are requiring that the only contribution in energy or pressure \textit{at early times}
is provided by the flavour vacuum, the state $\rmv$ representing the truly empty state and therefore
carrying no energy or pressure. What is left from the normal ordering would therefore correspond to the
contribution to the energy and pressure provided by the flavour vacuum \textit{and} the particles produced by
the action of the external gravitational field (which at early times is suppose to be vanishing),
according to Section \ref{ssCOTVICS}.
Changing perspective, we can say that what is left is nothing but
the evolution in (conformal) time of the energy and pressure of the flavour vacuum, as the universe expands or contracts.

Then, recalling the D-particle foam model, we know that the order of magnitude of $\Ce$ is dictated by the interaction between the foam
and the brane representing our universe; more specifically, as motivated in \cite{mavrosarkar}, from general considerations
on the string model, the relation
\be\label{conformalscalefactorandsigma}
\Ce\approx1+\sigma^2 f(\e)
\ee
holds: $\sigma^2$ being a small parameter ($\sigma^2\ll 1$), that takes into account both the average distribution of particles
in the foam and the interaction with the brane\footnote{The parameter $\sigma^2$ describes
the statistical average over populations of D-particles of the variance of the stochastic Gaussian distribution of
the recoil velocities of D-particles, such a recoil being caused by a process of capture and subsequent emission
by stringy matter \cite{mavrosarkar}. In other words, the interaction of matter with D-particle defects at a given space-time point
implies a distortion of the neighboring metric of the form $(1 - u_j) \e_{0j}$
if the recoil is along the direction $j$.
Now, if one considers an average situation where an ensemble of defects appears in the foam,
and there are several directions of recoil in an isotropic manner on average, then one may arrive at (\ref{conformalscalefactorandsigma}),
with $\sigma^2$ proportional to the stochastic fluctuations of the recoil velocity averaged also over the ensemble of D-particles:
$\langle\langle u_i  u_j \rangle\rangle = \sigma^2 \delta_{ij}$, $\langle \langle u_i \rangle \rangle = 0$,
with $\langle\langle\slot\rangle\rangle$ denoting the statistical average.},
and $f(\e)$ being a function of the order of $1$. It follows that
\be\label{Dconformalscalefactorandsigma}
\C'(\e)\approx \sigma^2 f'(\e).
\ee
Focusing now on the term
\be
1-|D^{(-1)}_{(1)}(p,\e)|^2-|D^{(1)}_{(-1)}(p,\e)|^2
\ee
present in both (\ref{innerset00befornormalordering}) and (\ref{innersetiibefornormalordering}),
we can use the relation (\ref{Dconformalscalefactorandsigma}) to estimate  the functions $D^{(a)}_{(-a)}(p,\e)$ in terms of $\sigma^2$.
Recalling (\ref{Ddefinition}), we have that
\be
D^{(a)}_{(a')}=\delta^a_{a'}+a'\int d\e\sigma^2 g(p,\e)D^{(a)}_{(-a')}(p,\e)
\ee
with $g(p,\e)$ a specific function of the momentum and the conformal time, of the order of $\sigma^0$, that leads to
\be
D^{(a)}_{(-a)}=-a\;\sigma^2 \int d\e g(p,\e)\left(1+\mathcal{O}(\sigma^2)\right)=\sigma^2 G(p,\e)+\mathcal{O}(\sigma^4)
\ee
with $G(p,\e)\equiv -a\int d\e g(p,\e)$. We therefore have
\be
1-|D^{(-1)}_{(1)}(p,\e)|^2-|D^{(1)}_{(-1)}(p,\e)|^2=1+\mathcal{O}(\sigma^4).
\ee
Since in (\ref{innerset00befornormalordering}) and (\ref{innersetiibefornormalordering}) the first two leading orders in $\sigma$
are $\sigma^0$ and $\sigma^2$ (because of the explicit presence of $\we$ and $\Ce$), we can therefore neglect the
contribution given by $|D^{(a)}_{(-a)}(p,\e)|^2$:
\be\label{innerset00smallsigma}
\mathcal{T}_{00}(\e,p,m)\approx \frac{8}{(2 \pi)^2}p^2\we\sqrt{\Ce}
\ee
and
\be\label{innersetiismallsigma}
\mathcal{T}_{jj}(\e,p,m)\approx \frac{1}{3}\frac{8}{(2 \pi)^2}\frac{p^4\sqrt{\Ce}}{\we}
\ee
being valid for small values of $\sigma^2\approx\Ce-1$.

As anticipated in Section \ref{ssMGE}, the presence of the foam in the bulk space causes an extra effective mass term in the action, that
can be described by changing $m_i$ with $m_i+m_{eff}$
in (\ref{innerset00befornormalordering}) and (\ref{innersetiibefornormalordering}),
thanks to the MSW gravitational effect.\footnote{It
should be notice that the function $V^2(p)$ remains untouched by this procedure, since
its presence is only related with the existence of the  flavour vacuum,
dictated by the change in flavour of strings attached on the brane, and is not related with further effects concerning the foam and the
brane on which strings are attached.}
Vice versa, we expect that the absence of the MSW effect is a signal of the absence of the foam.
In order to take into account only the contribution to the physical vacuum induced by the presence of the foam,
we subtract to the physical relevant quantities any contributions that would not disappear when $m_{eff}=0$.
Since
\be
\mathcal{T}_{00}(\e,p,m)\approx \frac{8}{(2 \pi)^2}p^2\sqrt{\Ce}\left(\we + m_{eff}\frac{\Ce m }{\we}\right)
\ee
and
\be
\mathcal{T}_{jj}(\e,p,m)\approx \frac{1}{3}\frac{8}{(2 \pi)^2} p^4\sqrt{\Ce}\left(\frac{1}{\we}- m_{eff}\frac{\Ce m}{\we^3}\right)
\ee
it is clear that we are left with the terms
\be
\mathcal{\tilde{T}}_{00}(\e,p,m)\approx \frac{8}{(2 \pi)^2}p^2\C^{3/2}(\e)m_{eff} \frac{ m }{\we}
\ee
and
\be
\mathcal{\tilde{T}}_{jj}(\e,p,m)\approx -\frac{1}{3}\frac{8}{(2 \pi)^2} p^4\C^{3/2}(\e)m_{eff}\frac{ m }{\we^3}
\ee

Finally, we introduce a cutoff $K$ in the momentum space, imposing an upper bound to the integrals in
(\ref{set00befornormalordering}) and (\ref{setiibefornormalordering}).
Rather than being a regulator, in this context $K$
describes the scale of energy up to which the field theoretical model correctly describes the low energy limit
of the D-particle foam model. String theories are indeed expected to be finite at perturbative as well as non-perturbative
level, and the ultraviolet divergences of QFT models for their low energy limits are supposed to disappear thanks to
quantum-gravitational effects present at higher energies.

\subsectionItalic{Results}\label{ssR}

As a result of the further manipulation of the expressions (\ref{set00befornormalordering}), (\ref{setiibefornormalordering}),
(\ref{innerset00befornormalordering}) and (\ref{innersetiibefornormalordering}) above exposed, and recalling
that in the comoving frame the energy  per unit of volume of a relativistic fluid is given by $T_{00}(x)/\Ce$,
while $T_{jj}(x)/\Ce$ represents its pressure (\textit{cf} Section \ref{ssMCAI}),
we have that the \textit{energy density} of the flavour vacuum in our theory is described by
\bea\label{fvenergycurved}
\rho&=&\lfv:T_{00}:\rfv\nn
	&\approx&\sst \frac{8}{(2 \pi)^2}m_{eff}\int_0^K dp V^2(p)p^2\sqrt{\Ce}
					\left( \frac{ m_1}{\w_1(p,\e)}+\frac{ m_2}{\w_2(p,\e)}\right)
\eea
whereas its \textit{pressure} is given by
\begin{multline}\label{fvpressurecurved}
\pressure=\lfv:T_{jj}:\rfv\approx-\sst \frac{1}{3}\frac{8}{(2 \pi)^2}m_{eff}\times\\
\times\int_0^K dp V^2(p) p^4\sqrt{\Ce}\left( \frac{ m_1}{\w_1(p,\e)^3}+\frac{ m_2 }{\w_2(p,\e)^3}\right)
\end{multline}
with $\w_i(p,\e)\equiv\sqrt{p^2+m_i^2\Ce}$, $V^2(p)$ a function in the momentum space that encodes information about the
flavour vacuum state at early times and defined by (\ref{Vsquarecurvedspacetime}), $\Ce$ the conformal scale factor,
$\theta$ the angle that parameterize the rotational matrix linking flavour fields with fields with well defined mass,
$K$ a physical cutoff characterizing the scale of energies up to which our model can be consistently regarded
as the low energy limit of the underlying microscopic D-particle foam model, $m_{eff}$ an effective mass term that
accounts for the MSW gravitational effect induced by the D-particle foam on the matter fields.

\begin{figure}[ht]
\centering
\includegraphics[width=9cm]{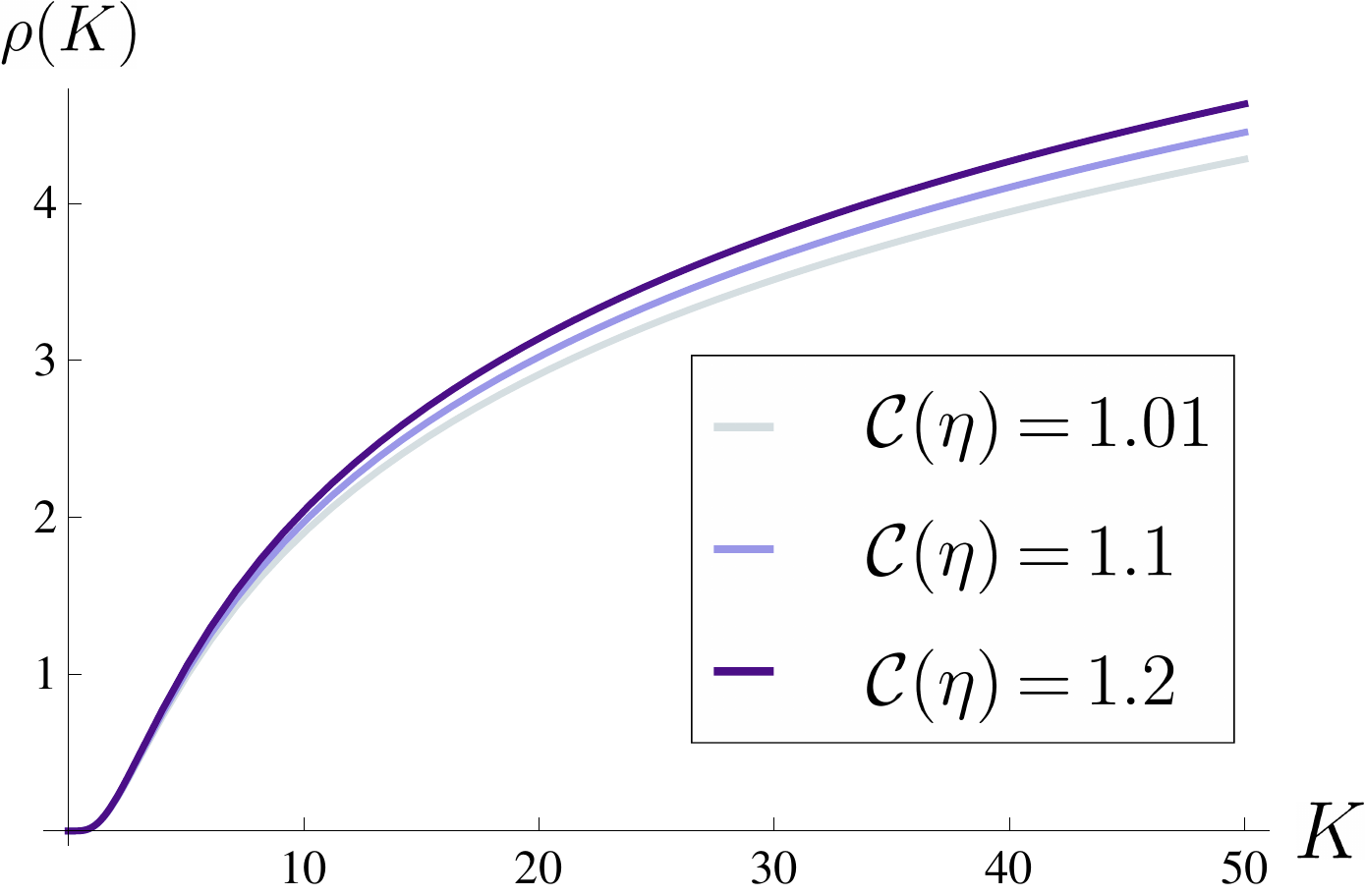}
\caption{\small{The energy density $\rho(K)$ as a function of the cut-off $K$ is plotted.
The parameters of (\ref{fvenergycurved}) take values:
$\sst8(2\pi)^{-2}m_{eff}=1$, $m_1=1$, $m_2=2$, in arbitrary units (sufficient for our purpose here,
which is only to show the dependence on $K$), whereas $\Ce=1.01,1.1,1.2$.}}
\label{fig:partII_e}
\end{figure}

\begin{figure}[ht]
\centering
\includegraphics[width=9cm]{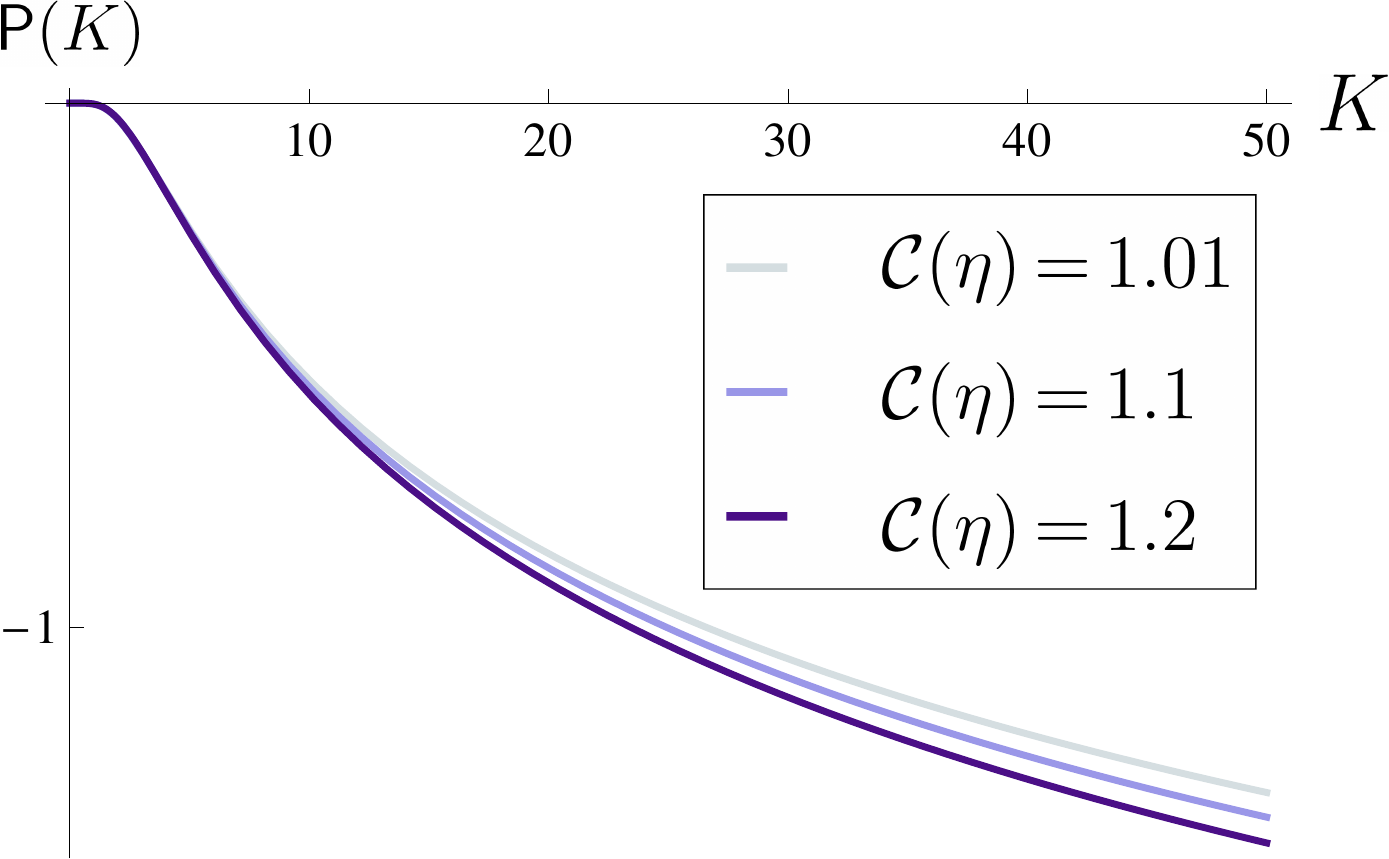}
\caption{\small{The pressure $\pressure(K)$ as a function of the cut-off $K$ is plotted.
As in Figure \ref{fig:partII_e}, the parameters of (\ref{fvpressurecurved}) take values:
$\sst8(2\pi)^{-2}m_{eff}=1$, $m_1=1$, $m_2=2$, in arbitrary units (sufficient for our purpose here,
which is only to show the dependence on $K$), whereas $\Ce=1.01,1.1,1.2$.}}
\label{fig:partII_p}
\end{figure}

Consistently with our expectations, those expressions depend only on the conformal time coordinate $\e$
and not also on the spatial coordinates $\vx$, although the operator $T_{\mu\nu}(x)$ depends explicitly on the four-vector $x^{\mu}$.
This result reflects both the required anisotropy and homogeneity of the space and
and the uniformity of the distribution in space of the fluid that fills the empty space,
that we called \textit{flavour vacuum}.
Since $m_{\slot}>0$, the energy density of the flavour vacuum is positive, whereas its pressure is negative.
As shown in Figure \ref{fig:partII_e}, the value of the energy density grows monotonically as the cutoff increases.
Such a value increases also when the value of $\Ce$ rises.
Similar considerations apply to the absolute value of the pressure: it increases either when $K$ or $\Ce$ increases,
as shown in Figure \ref{fig:partII_p}.

\begin{figure}[ht]
\centering
\includegraphics[width=9cm]{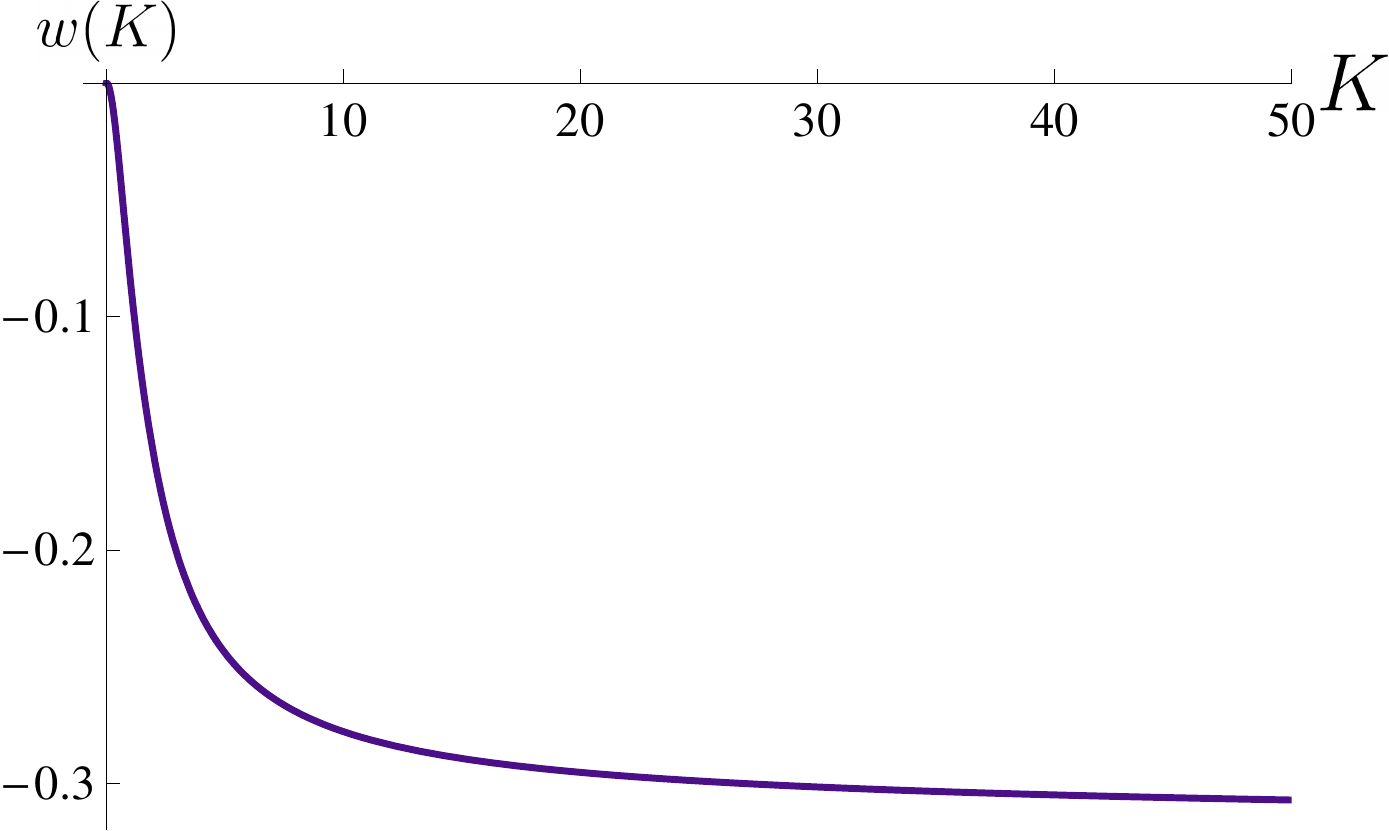}
\caption{\small{The equation of state $w(K)$ as a function of the cut-off $K$ is plotted, with
$m_1=1$, $m_2=2$, in arbitrary units, and $\Ce=1.01$.}}
\label{fig:partII_w}
\end{figure}

The equation of state $w$ is defined as the ratio between the pressure and the energy density.
Its value is therefore given by
\be\label{wcurvedspacetime}
w\equiv\frac{\pressure}{\rho}=-\frac{1}{3}\frac{\int_0^K dp V^2(p) p^4\left( \frac{ m_1}{\w_1(p,\e)^3}+
																																			 \frac{ m_2 }{\w_2(p,\e)^3}\right)}
									{\int_0^K dp V^2(p)p^2\left( \frac{ m_1}{\w_1(p,\e)}+
																											\frac{ m_2}{\w_2(p,\e)}\right)}.
\ee
The value of $w$ is confined in the interval $(-1/3,0)$, as one can see in Figure \ref{fig:partII_w}.

\begin{figure}[ht]
\centering
\includegraphics[width=9cm]{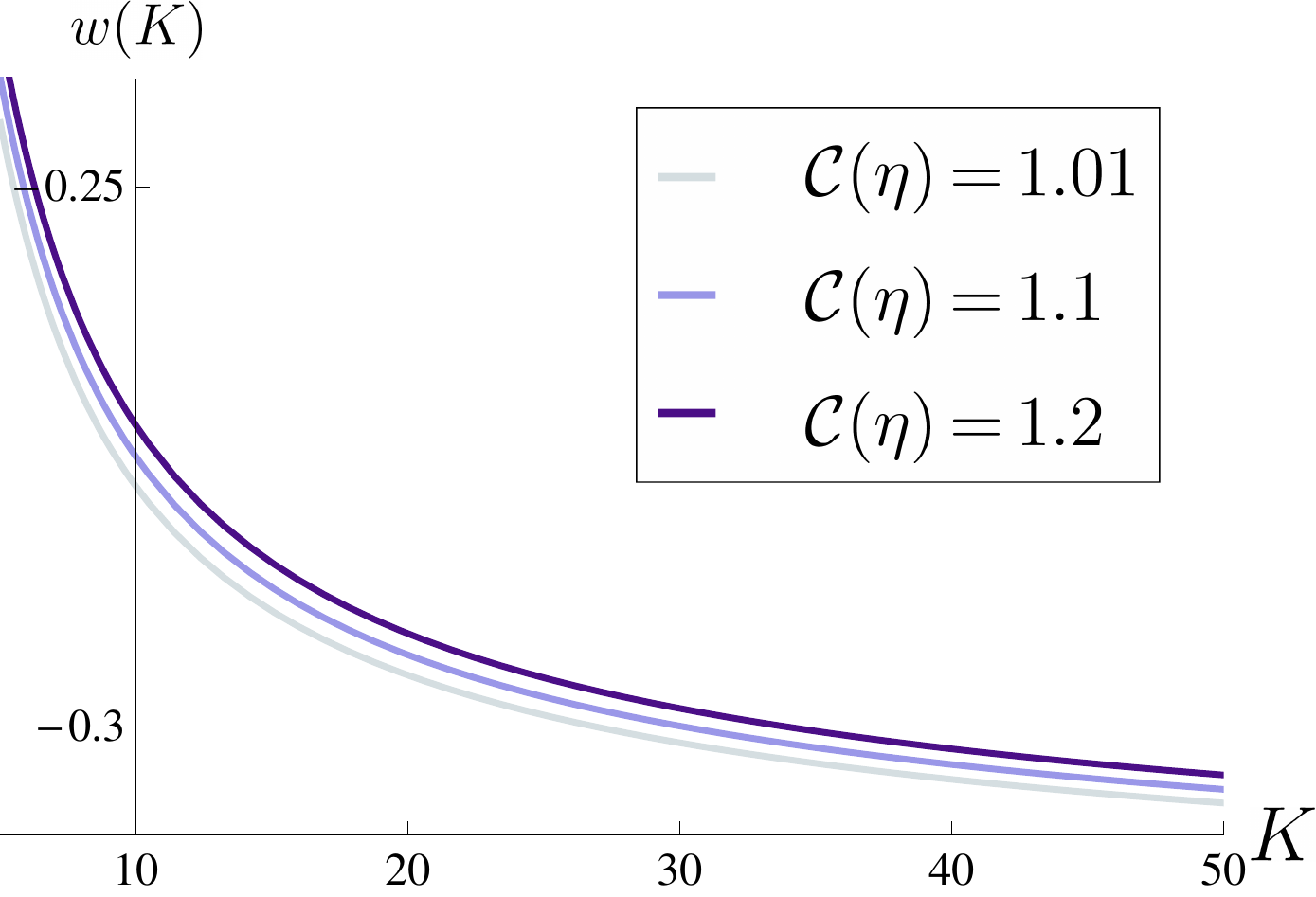}
\caption{\small{The equation of state $w(K)$ varies as $\Ce$ departs from 1. Here $w(K)$ is plotted
with $m_1=1$, $m_2=2$, in arbitrary units, and $\Ce=1.01,1.1,1.2$.}}
\label{fig:partII_w_eta}
\end{figure}

For small values of the
cutoff $K$, $w$ is approximately equal to zero. However, as $K$ increases, $w$ starts approaching
the value of $-1/3$, which behaves as an horizontal asymptote. Even slowly increasing the value of $\Ce$ starting from $\Ce=1$,
does not affect the confinement. As only result, when $\Ce$ grows, $w$ is just pushed more towards the horizontal axis,
\textit{i.e.} the higher the conformal scale factor, the higher must be the cutoff for $w$ to obtain a specific value.
(Figure \ref{fig:partII_w_eta}).

\section{DISCUSSION AND CONCLUSIONS}\label{sDAC}

\subsectionItalic{Phenomenology}\label{ssP}

A complete phenomenological analysis of the above results would require some knowledge of the parameters involved.
Unfortunately it is hard to make even an estimate of those parameters, since the brany model is far
from being fully understood. Moreover, the parameter $\sigma^2$ depends on the distribution of the D-particles
within the foam, that can be considered as a free parameter of the theory.\footnote{Certainly the parameter has to be
considered free within a \textit{finite} range of values: if the concentration of D-particle had been too high,
quantum-gravitational effects would become observable at macroscopic scales. However, such a range might
be wide enough to allow quite different physical scenarios.} It follows that $\sigma^2$ itself is a small but otherwise free
parameter in the effective theory.

However, if an estimate of the order of magnitude of the energy and pressure of the flavour vacuum state
are beyond our possibilities, we can at least observe that their signature is well defined, the energy density being
always positive and the pressure always negative.

Furthermore, the equation of state in (\ref{wcurvedspacetime}) does not depend on $m_{eff}$ and $\theta$, with $K$ and $\Ce$ the only
two parameters left. As already pointed out,
\be
-1/3< w < 0
\ee
independently of the choice of $K$ and $\Ce$. More specifically, the higher is the cutoff, the closer to $-1/3$
is the value of $w$.

Such a result has a clear and important phenomenological meaning. As explained in Section \ref{ssMCAI}, a fluid with negative
pressure, but characterized by an equation of state $w>-1/3$,
would cause a \textit{deceleration} of the expansion of the universe. We can therefore state that the flavour vacuum
provided by our simple model involving two free Dirac spinors \textit{is not} a suitable candidate for Dark Energy.

\subsectionItalic{Comparison With A Bosonic Case}\label{ssCWABC}

The study of this model has been motivated by an encouraging result obtained in a simpler model for bosons.
As anticipated in Section \ref{ssEFV}, in \cite{mavrosarkar} a free theory for two flavoured real scalars $\phi_{\slot}(x,\e)$
has been considered, characterized by a specific choice of the background:
a $1+1$ dimensional universe with a conformal scale factor
\be
\Ce=A+B \tanh (\rho \e)
\ee
with $A>B>0$, $\rho>0$ (\textit{cf} figure \ref{fig:conformal}).
Such a spacetime enabled the authors to find explicit solutions for the equations of motion,
to explicitly prove the orthogonality of the Fock space for flavour states to the ordinary Fock space on such a curved background,
to evaluate the effects of the expansion of the universe on flavour oscillations of one-particle states.

Furthermore, an equation of state was derived and found equal to
\be
w_{b}=-1
\ee
being (in our notation)
\bea\label{relevattermbosset}
\rho_{b}&=&\lfv:T_{00}^{b\;1+1}:\rfv=m_{eff}^2 \;\lfv ( \phi^2_1(x,\e)+ \phi^2_2(x,\e))\rfv=\nn
		&=&-\lfv:T_{xx}^{b\;1+1}:\rfv=-\pressure_{b}.
\eea

This term is the only one left from the full expression of the stress-energy tensor, whose diagonal terms are given by
\bea\label{setbos1+1}
T_{00}^{b\;1+1}(x,\e)&=&\sum_{i=1,2}\left((\partial_\e \phi_i(x,\e))^2+(\partial_x \phi_i(x,\e))^2+\Ce m^2_i \phi^2_i(x,\e)			 \right)\\
T_{xx}^{b\;1+1}(x,\e)&=&\sum_{i=1,2}\left((\partial_\e \phi_i(x,\e))^2+(\partial_x \phi_i(x,\e))^2-\Ce m^2_i \phi^2_i(x,\e)			\right)
\eea
after manipulations similar in spirit to the ones exposed in the previous sections.
In particular the only contribution to (\ref{relevattermbosset}) comes from
the last term
\be\label{relevantterm2}
\pm\sum_i \Ce m^2_i \phi^2_i(x,\e).
\ee
Quite notably, the expression in (\ref{relevattermbosset}) does not depend on the specific form of the conformal scale factor $\Ce$.
Moreover, when a $3+1$ background is considered, formulae (\ref{setbos1+1}) generalize to
\begin{multline}\label{setbos3+1}
T_{00}^{b\;3+1}(\vx,\e)=
	\sum_{i=1,2}\Big[(\partial_\e \phi_i(\vx,\e))^2+\sum_{l=1}^3(\partial_{x_l} \phi_i(\vx,\e))^2+\Ce m^2_i \phi^2_i(\vx,\e)\Big]
\end{multline}
\begin{multline}
T_{jj}^{b\;3+1}(\vx,\e)=
	\sum_{i=1,2}\Big[(\partial_\e \phi_i(\vx,\e))^2+\sum_{l=1}^3(\partial_{x_l} \phi_i(\vx,\e))^2
						+\\
	+2(\partial_{x_j})^2\phi_i(\vx,\e)-\Ce m^2_i \phi^2_i(\vx,\e)\Big]
\end{multline}
where the sums over spatial three-dimensional indices are explicitly denoted for clarity.
Since the kinetic terms, the ones involving derivatives of fields, are affected by the change of dimensionality,
but the equivalent of (\ref{relevantterm2}) remains the same, at least as a function of the fields,
one might infer from the $1+1$ dimensional result, that the relation $w_{b}=-1$ would hold also in the $3+1$ dimensions.

If this would actually be the case, we would have to face a discrepancy between the behaviour of the flavour vacuum
in the two different context, which differs just in virtue of the spinorial structure of the fields involved:
in the fermionic case the physical vacuum of the theory acts as a perfect fluid, characterized by an equation of state $-1/3<w_{f}<0$,
whereas in the bosonic case it would be characterized by $w_{b}=-1$.

The underlying microscopic theory is supersymmetric: fermions and bosons are treated equally.\footnote{More details about
supersymmetric theories will be provided in the forthcoming Chapter.} Although the two models here discussed
do not combine together to give a supersymmetric theory straightforwardly, nonetheless one would have expected similar results.

Such a difference in behaviour of the flavour vacuum might be caused by different factors: the spinorial structure (in contrast with
the underlying theory); the constraint of reality to the scalar field, a further reduction of the degrees of freedom of the bosonic theory,
in comparison with the Dirac spinors; the specific subtraction dictated by the microscopic model, that might affect different fields
in different ways.

The work presented in the next chapter is an attempt to better understand this problem: the features of the flavour vacuum
arising from a simple supersymmetric theory will be investigated.

\chapter{A SUPERSYMMETRIC FLAVOUR VACUUM}\label{cASFV}\label{cAFVIASM}

In the previous Chapter we examined the behaviour of the \textit{flavour vacuum}
in a specific context, dictated by a microscopic stringy-inspired theory (the \textit{D-particle foam model}).
Since it is still unclear how to fully formulate the underlying theory, and therefore to derive
its low energy limit (at scales on which we expect ordinary QFT to apply), we considered
a toy model consisting of two free Dirac spinors with mixing \textit{\`a la} BV in a FRW universe,
in the attempt of catching significant aspects of the microscopic theory.
In particular, the physical vacuum of the theory (a ground state carrying non zero energy)
was constructed by the means of the BV formalism, implemented on a specific class of curved spacetime backgrounds,
and further requests on the normal ordering for accounting of specific features of the D-particle foam model.

The resulting vacuum state presented a spatially uniform positive energy and an equally uniform
pressure, related with
the energy via the equation of state $-1/3<w<0$. Such a range is spanned when all different values
of a cutoff in the momentum space are taken into account.

An analogous model for a bosonic system was already known in literature \cite{mavrosarkar},
in which two real scalars in a $1+1$ dimensional expanding universe (obeying to a specific law for the expansion) were considered.
The flavour vacuum arising from that model was shown to obey an equation of state $w=-1$.
In Section \ref{ssCWABC} we argued that such a behaviour is preserved even when a generic $3+1$ FRW universe is considered.

Looking na\"ively at the two models apparently so similar, besides the spin of the fields considered,
one might ask why the ground state behaves so differently: is this discrepancy
due to an inappropriate comparison between two theories whose fields cannot be part of the same supersymmetric multiplet?
Is it caused by the way how the classical field representing the curved background couples with the other fields?
Or is the anomaly introduced by the particular prescriptions for the normal ordering adopted (induced by the MSW gravitational
effect, Section \ref{ssMGE}), and therefore limited to the specific context so far considered?

This Chapter is an attempt to enlighten these doubts:
we shall implement BV formalism on a simple (globally) supersymmetric model, without considering any gravitational effects,
such as a non-flat background or MSW gravitational effect.
Although the model will reflect just the supersymmetric invariance of the underlying microscopic model,
a more conventional set-up will enable us to compare our results with other works already existing in literature,
in which the \textit{flavour vacuum} has been considered as a phenomenon \textit{per se}
and was not motivated (at least directly) by any deeper theory.
The simplest supersymmetric model in which the flavour mixing \textit{\`a} la BV can be implemented,
is a free massive Wess-Zumino, that we shall realize with two species of a real scalar field,
two real pseudo-scalars and two Majorana spinors.
As we will see, such a set-up is sufficient to shed some light on our problem, the flavour vacuum
arising from the supersymmetric model being highly non supersymmetric.

After a brief overview on Supersymmetry in Section \ref{secSUSY},
Section \ref{sSUTM} will be devoted to introducing the model.
The features of the flavour vacuum will be presented in Section \ref{sFOTFVsusy}.
A discussion in Section \ref{sCsusy} about the phenomenology of the model will end the Chapter.

\section{SETTING UP THE MODEL}\label{sSUTM}

\subsectionItalic{Supersymmetry}\label{secSUSY}

Although the wave-packet duality in the early days of Quantum Mechanics seemed to have removed once for all
the matter-force dichotomy, that characterized all post-newtonian physics until then,
the successes of gauge theories, culminated in the Standard Model of particles,
forced physicists to reintroduce it in the new form of a bosonic/fermionic dichotomy,
as translated into the language of the new quantum formalism \cite{Sohnius:1985qm}:
according to our current understanding of particle physics, forces are mediated
by vector fields of spin one (the gauge potentials), while matter is represented by spin half fermionic fields
(quarks and leptons).

Supersymmetric theories remove once more this distinction between fields of different spin, treating
bosons and fermions in the same manner. Indeed, a theory is said to respect supersymmetry\newnot{symbol:SUSY} (SUSY)
if it is invariant under transformation of the bosonic fields into fermionic fields and vice versa.
Quite remarkably, this is not just an exotic transformation of the fields, invented \textit{ad hoc}:
SUSY is the only kind of symmetry that can extend the usual Poincar\'e symmetry of spacetime
(besides internal symmetries), in a realistic theory for non-trivial particle scattering
\cite{Coleman:1967ad,Haag:1974qh}.

Since their first appearance, SUSY theories have been found to display several good properties:
their are less affected by the infinities that plague non-SUSY theories (already mentioned in Sections \ref{ssCCP}),
via cancellations between bosonic and fermionic divergent terms \cite{Peskin:1995ev};
they offered a solution to the ``hierarchy problem'', \textit{i.e.} the high sensitivity
of the bare mass of the Higgs boson to physics at higher scales \cite{Aitchison:2005cf};
they provided naturally candidates for Dark Matter, before astronomers pointed out the necessity of non-baryonic
cold Dark Matter \cite{Kane:2010zz};
the natural generalization, that promotes SUSY to be a \textit{local} symmetry, is a theory invariant under
spacetime diffeomorphisms and therefore it includes gravity, as explained in \cite{deWit:2002vz}.
At the moment, a supersymmetric extension of the SM (the Minimal Supersymmetric Standard Model - \newnot{symbol:MSSM}MSSM)
is regarded as the most promising model for the physics \textit{beyond the SM}.
Unfortunately such a model is not complete, the major difficulty being due to the fact that SUSY
\textit{is not} a physical symmetry: if it were so,
one would have expected for each known particle of the
SM a correspondent ``superpartner'' particle, differing from the original just for the spin.

In order to do not lose all the desirable properties of a SUSY theory,
one would expect SUSY to be \textit{spontaneously} broken:
the Lagrangian being still invariant under supersymmetric transformations,
but the ground state being not supersymmetric. Thanks to a familiar mechanism  \cite{Peskin:1995ev},
perturbing the theory around a ground state that is not invariant under SUSY might lead to additional mass-like terms
for the superpartners, that pull them in a regime of energies not yet covered by experiments.

For a supersymmetric theory, the ground state is not supersymmetric (and therefore spontaneously breaking of SUSY is realized)
if and only if its energy is different than zero. Such a statement comes from the important relation that
connects the SUSY generators $Q$ and the Hamiltonian operator $H$, holding for any SUSY theory:
\be
 H\propto \sum_{all\;Q}\{Q^{\dagger},Q\},
\ee
which implies
\be
\bra{\Omega} H \ket{\Omega} \propto\sum_{all\;Q}\left( |Q^{\dagger}\ket{\Omega}|^2 +|Q\ket{\Omega}|^2\right)\geq0.
\ee
The breaking is achieved
by introducing a suitable (supersymmetric) potential in the Lagrangian that shifts conveniently the vacuum expectation value of $H$.
The search for such a potential for realistic theories so far has been unsuccessful, and at the moment the MSSM includes
terms that explicitly break SUSY, in the attempt of parameterizing the action of a yet-to-discover potential. In the MSSM
not all terms are accepted: to prevent the occurrence of quadratic divergencies and protect the theory from the hierarchy problem,
only terms with coupling parameters with positive mass dimension are allowed. This realizes the so called \textit{soft} SUSY breaking.
Even with this restriction, the MSSM adds to the SM 105 parameters that are not constrained by experiments,
leaving the model very little predictive.

Nevertheless, thanks also to the great importance that SUSY plays in String theories \cite{Polchinski:1998rr,Dine:2007zp},
a supersymmetric theory is regarded as the most natural way to extend the SM and
the search of superpartners of the particles already known is a prior target of current
experiments at LHC (Large Hadron Collider), the particle accelerator
just launched devoted to probe physics at TeV scale \cite{Aad:2011hh}.

\subsectionItalic{Flavoured Free Wess-Zumino Model}

As anticipated, we will implement the BV formalism on a supersymmetric model in flat spacetime.
A simple Lagrangian with flavour mixing and invariant under SUSY transformations is represented by\footnote{The Lagrangian
here presented can be derived from the most general supersymmetric renormalizable Lagrangian involving only chiral superfields
\cite{Wess:1992cp}. Nonetheless, we will avoid to use the superformalism.
Further references for the model considered can be found in \cite{Sohnius:1985qm} and
\cite{FigueroaO'Farrill:2001tr}.}
\begin{multline}\label{lagpsispfg}
\lag =\sum_{\iota=A,B}\Big[ i\bpsi (x)\diracpartial \psi_\iota(x) +\partial_\mu S_\iota(x)\partial^{\mu}S_\iota(x)
												+\partial_\mu P_\iota(x)\partial^{\mu}P_\iota(x)+\\
												+F^2_\iota(x)+G^2_\iota(x)\Big]+\\
	-2\sum_{\iota,\kappa=A,B}m_{\iota\kappa}\Big[ S_\iota(x)F_\kappa(x)+P_\iota(x)G_\kappa(x)+\frac{1}{2}\bpsi(x)_\iota\psi_\kappa(x)\Big]
\end{multline}
in which $\psi_\iota(x)$ is a Majorana spinorial field and  $\bpsi_\iota(x)=\psi^\dagger_\iota(x)\gamma^0$,
$S_\iota(x)$ and $F_\iota(x)$ scalar fields and  $P_\iota(x)$ and $G_\iota(x)$ pseudoscalar fields\footnote{In this context, we define a
Majorana spinor as a four-component spinorial object $\psi$ obeying to the condition
\be\label{majoranacondition1}
  C \gamma_0  \psi^*=\psi
\ee
with $C$ the charge-conjugation operator, following the notation of \cite{Itzykson:1980rh}. A Majorana spinorial field is used to describe
spin-\nicefrac{1}{2} particles that are also their own anti-particles.

Pseudoscalar fields are invariant under \textit{proper} Lorentz transformations, \textit{i.e.} not involving parity transformations.
Under the latter, they acquire a minus sigh. However, the pseudoscalar nature of $P_\iota(x)$,
the only pseudoscalar relevant to our discussion,  is not manifest in our discussion on the flavour vacuum.
It would have been, if an interactive theory was considered, through
the interactive term $g_{\iota \kappa \lambda}\bpsi_\iota P_\kappa \gamma_5 \psi_\lambda$
appearing both in the Lagrangian and in the stress-energy tensor,
in which the parity invariance is preserved thanks to $\gamma_5$ \cite{FigueroaO'Farrill:2001tr} (recalling that
$\gamma^5\equiv i\gamma^0\gamma^1\gamma^2\gamma^3$ changes sign if the orientation of the coordinate system changes).

$F_i(x)$ and $G_i(x)$, despite their scalar and pseudoscalar behaviour under Lorentz transformations, respectively,
have not the same dimensions of $S_i(x)$ and $P_i(x)$, in mass units. In fact, they are unphysical fields that, as we will see,
do not describe dynamical degrees of freedom, but are needed for the SUSY invariance of the model.},
with
\be
\e_{\mu\nu}=diag\{+1,-1,-1,-1\}.
\ee
This Lagrangian is invariant under the supersymmetric transformation
\begin{gather}
\delta S_\iota(x)=\bar{\epsilon}\psi_\iota(x)\;\;\;\;\delta P_\iota(x)=\bar{\epsilon}\gamma_5 \psi_\iota(x)\\
\delta F_\iota(x)=\bar{\epsilon}\diracpartial \psi_\iota(x)\;\;\;\;\delta G_\iota(x)=\bar{\epsilon}\gamma_5 \diracpartial \psi_\iota(x)\\
\delta \psi_\iota(x)=\diracpartial (S_\iota(x)+\gamma_5 P_\iota(x))\epsilon+(F_\iota(x)+\gamma_5 G_\iota(x))\epsilon\\
\end{gather}
where $\epsilon$ is an anticommuting Majorana spinor and $\bar{\epsilon}=\epsilon^\dagger\gamma^0$,
up to surface terms \cite{Sohnius:1985qm}.
In order to find the equations of motion, we first perform the usual rotation (\textit{cf} Section \ref{ssQFL})
\begin{gather}\label{fromflavourtomass1}
\varphi_A(x)=\varphi_1(x) \ct+\varphi_2(x)\st\\
\varphi_B(x)=-\varphi_1(x) \st+\varphi_2(x)\ct\\
\end{gather}
with $\varphi_{\slot}$ each one of the field present in (\ref{lagpsispfg}).
This allows us to rewrite (\ref{lagpsispfg}) as
\begin{multline}\label{lagpsispfgmass}
\lag = \sum_{i=1,2}\Big[ i\bpsi (x)\diracpartial \psi_i(x) +\partial_\mu S_i(x)\partial^{\mu}S_i(x)
												+\partial_\mu P_i(x)\partial^{\mu}P_i(x)+\\
	+F^2_i(x)+G^2_i(x)-2 m_{i}\left(	S_i(x)F_i(x)+P_i(x)G_i(x)	+\frac{1}{2}\bpsi(x)_i \psi_i(x)\right)\Big].
\end{multline}
We can now easily recognize two copies of the Wess-Zumino model, a first one labeled by the mass $m_1$,
and a second labeled by $m_2$ \cite{Wess:1973kz,Sohnius:1985qm}.
The equations of motion are given by
\bes\label{eompsispfg}
F_i(x)=&m_i S_i(x)\\
G_i(x)=&m_i P_i(x)\\
i\diracpartial \psi_i(x)=&m_i \psi_i(x)\\
\partial_\mu \partial^\mu S_i(x)=&-m_i F_i(x)\\
\partial_\mu \partial^\mu P_i(x)=&-m_i G_i(x)
\end{split}\ee
The fields $F_i(x)$ and $G_i(x)$ are called \textit{auxiliary fields}: since no kinetic terms (\textit{i.e.} involving
their derivatives) are present in the Lagrangian, their equations of motion do not describe an evolution in
space or time. Those equations are usually enforced on the Lagrangian in order to rewrite it just in terms of the other fields,
giving rise to the \textit{on-shell} Lagrangian \cite{Sohnius:1985qm}:
\begin{multline}\label{lag2}
\lag_{on-shell}=
	\sum_{i=1,2}\Big[\partial_\mu S_i(x)\partial^\mu S_i(x)-m_i^2 S_i^2(x)+\partial_\mu P_i(x)\partial^\mu P_i(x)+\\
	-m_i^2 P_i^2(x)+\bpsi_i(x)(i\diracpartial-m_i)\psi_i(x) \Big]
\end{multline}
For sake of simplicity, from now on we will work with this Lagrangian, in which no unphysical fields are present.
Such a choice will not affect our discussion, since
off-shell contributions (involving the auxiliary fields) are present when interactive terms are considered,
while here the whole theory trivially reduces to its full on-shell formulation
(both physical and unphysical fields satisfying the equations of motion (\ref{eompsispfg})).\footnote{For a discussion
about \textit{on-shell} and \textit{off-shell} formulation of a supersymmetric theory one may refer to \cite{Sohnius:1985qm}.}
The equations of motion for the other fields are now reduced (\textit{on-shell}) to
\bes
i\diracpartial \psi_i(x)=&m_i \psi_i(x)\\
\partial_\mu \partial^\mu S_i(x)=&-m_i^2 S_i(x)\\
\partial_\mu \partial^\mu P_i(x)=&-m_i^2 P_i(x)
\end{split}\ee
we can therefore decompose the fields in the usual plane waves expansion
\be\label{psimass}
\psi_i (x)=\sum_{r=1,2}\int{\frac{d^3k}{(2 \pi)^{3/2}} \left(  a^r_i(\vk) u^r_i (\vk)e^{-i\w_i(k)t}
						+ v_i^{r} (-\vk) a_i^{r\dagger}(-\vk)e^{i\w_i(k)t}\right)e^{i\vk \cdot \vx}}
\ee
\be\label{smass}
S_i(x)=\int{\frac{d^3k}{(2 \pi)^{3/2}}\frac{1}{\sqrt{2 \w_i(k)}} \left(  b_i(\vk)e^{-i\w_i(k)t}+ b_i^{\dagger}(-\vk) e^{i\w_i(k)t} \right) e^{i \vk \cdot \vx}  }
\ee
\be\label{pmass}
P_i(x)=\int{\frac{d^3k}{(2 \pi)^{3/2}}\frac{1}{\sqrt{2 \w_i(k)}} \left(  c_i(\vk)e^{-i\w_i(k)t}+ c_i^{\dagger}(-\vk) e^{i\w_i(k)t} \right) e^{i \vk \cdot \vx}  }
\ee
with $\w_i(k)=\sqrt{k^2+m_i^2}$, the spinor $u_i^r(\vk)$ being defined by
\be\left\{\begin{array}{l}
(\gamma^0 \w_i(k)+\vec{\gamma}\cdot \vk-m_i)u_i^r(\vk)=0\\
u^{r\dagger}_i(\vk)u^s_i(\vk)=\delta^{rs}
\end{array}\right.																	
\ee
and $v_i^r(\vk)$ determined by $\gamma_0 C u_i^r(\vk)^*=v_i^r(\vk)$, because of the Majorana condition (\ref{majoranacondition1}).
The quantum structure of the fields is encoded in the ladder operators, that satisfy the algebra
 $\{a^r_i(\vk),a^{s \dagger}_j(\vq)\}=\delta^{rs} [b_i(\vk),b^\dagger_j(\vq)]=\delta^{rs} [c_i(\vk),c^\dagger_j(\vq)]=\delta^{rs} \delta_{ij} \delta^3(\vk-\vq)$.

Since all other commutators vanish, $[a^{(\dagger)},b^{(\dagger)}]=[b^{(\dagger)},c^{(\dagger)}]=[c^{(\dagger)},a^{(\dagger)}]=0$, we can decompose the total Fock
 space into the tensorial product of six independent Fock spaces:
\be\label{focktotwz}
\mathfrak{F}_{tot}=\mathfrak{F}_{\psi_1} \otimes \mathfrak{F}_{\psi_2} \otimes \mathfrak{F}_{S_1} \otimes \mathfrak{F}_{S_2} \otimes \mathfrak{F}_{P_1} \otimes \mathfrak{F}_{P_2}
\ee
with $\mathfrak{F}_{\psi_i}$ the space generated by all polynomials in $a_i^{r\dagger}(\vk)$ acting on $\rmv_{\psi_i}$, and so on.
In particular, the ground state $\rmv$ is given by
\be
\rmv\equiv \rmv_{\psi_1} \otimes \rmv_{\psi_2} \otimes \rmv_{S_1} \otimes \rmv_{S_2} \otimes \rmv_{P_1} \otimes \rmv_{P_2},
\ee
with $\rmv_{\psi_i}$ being annihilated by $a_i^r(\vk)$, and so on.

The stress-energy tensor for the theory (\ref{lag2}), that we will need in order to study the properties of the flavour vacuum,
is written as
\begin{multline}\label{set2}
T_{\mu\nu}(x)=\sum_{i=1,2}\Big[ 2\partial_{\left(\mu\right.}S_i(x)\partial_{\left.\nu\right)}S_i(x)+2\partial_{\left(\mu\right.}P_i(x)\partial_{\left.\nu\right)}P_i(x)+
\\+i\bar{\psi}_i(x)\gamma_{\left(\mu\right.}\partial_{\left.\nu\right)}\psi_i(x)\Big]-\e_{\mu\nu}\lag_{on-shell}
\end{multline}
as shown in Sections \ref{bsetc} and \ref{fsetc}.

\subsectionItalic{The Flavour Fock Space}\label{ssTFFS}

In order to build a Fock space for the flavour states \textit{\`a la} BV, we define the operator $G_{\theta}$ as
satisfying the equations
\begin{equation}\begin{split}\label{fromflavourtomass2}
\varphi_A(x)=&G^{\dagger}_{\theta}(t) \varphi_1 (x)G_{\theta}(t)\\
\varphi_B(x)=&G^{\dagger}_{\theta}(t) \varphi_2 (x)G_{\theta}(t)
\end{split}\end{equation}
with $\varphi=\psi$, $S$, $P$ .
By comparing (\ref{fromflavourtomass1}) and (\ref{fromflavourtomass2}),
it is easy to show (Sections \ref{line}, \ref{appGb} and \ref{appGf}) that
\be
G_{\theta}(t)=e^{\theta \int d\vx \left(X_{12}(x)-X_{21}(x)\right)}
\ee
with
\be
X_{12}(x)\equiv \frac{1}{2}\psi^\dagger_1(x)\psi_2(x)+i \dot{S}_2(x) S_1(x)+i \dot{P}_2(x) P_1(x)
\ee
Flavour ladder operators are therefore defined by
\be
\left\{
\begin{array}{c}
a^r_A(\vk,t)\equiv G^{\dagger}_{\theta}(t) a^r_1(\vk)G_{\theta}(t)\\
a^r_B(\vk,t)\equiv G^{\dagger}_{\theta}(t) a^r_2(\vk)G_{\theta}(t)
\end{array}
\right.
\ee
\be
\left\{
\begin{array}{c}
b_A(\vk,t)\equiv G^{\dagger}_{\theta}(t) b_1(\vk) G_{\theta}(t)\\
b_B(\vk,t)\equiv G^{\dagger}_{\theta}(t) b_2(\vk) G_{\theta}(t)
\end{array}
\right.
\ee
\be
\left\{
\begin{array}{c}
c_A(\vk,t)\equiv G^{\dagger}_{\theta}(t) c_1(\vk) G_{\theta}(t)\\
c_B(\vk,t)\equiv G^{\dagger}_{\theta}(t) c_2(\vk) G_{\theta}(t)
\end{array}
\right.
\ee
satisfying the algebra
\be
\{a^r_\iota(\vk),a^{s \dagger}_\kappa(\vq)\}=\delta^{rs} [b_\iota(\vk),b^\dagger_\kappa(\vq)]=\delta^{rs} [c_\iota(\vk),c^\dagger_\kappa(\vq)]=\delta^{rs} \delta_{\iota\kappa} \delta^3(\vk-\vq)
\ee
with $\iota,\kappa=A,B$, all other commutators being zero.
It is possible to prove (Appendices \ref{flof} and \ref{flob}) that
\begin{equation}\label{fromflavourladdertomassiveladder1}\begin{split}
a^r_A(\vk,t)=&\ct a_1^r (\vk)-\st \sum_s \left( W^{rs}(\vk,t)a^s_2(\vk)+Y^{rs}(\vk,t)a^{s\dagger}_2(-\vk) \right)\\
a^r_B(\vk,t)=&\ct a_2^r(\vk)+\st \sum_s \left(  W^{sr*}(\vk,t) a_1^s(\vk)+Y^{sr}(-\vk,t)a_1^{s\dagger}(-\vk)  \right)
\end{split}\end{equation}
\begin{equation}\label{fromflavourladdertomassiveladder2}\begin{split}
b_A(\vk,t)=&\ct b_1(\vk) +\st \left( U^*(k,t) b_2(\vk)+V(k,t) b_2^\dagger (-\vk)\right)\\
b_B(\vk,t)=&\ct b_2(\vk) -\st \left( U^*(k,t) b_1(\vk)-V(k,t) b_1^\dagger (-\vk)\right)
\end{split}\end{equation}
\begin{equation}\label{fromflavourladdertomassiveladder3}\begin{split}
c_A(\vk,t)=&\ct c_1(\vk) +\st \left( U^*(k,t) c_2(\vk)+V(k,t) c_2^\dagger (-\vk)\right)\\
c_B(\vk,t)=&\ct c_2(\vk) -\st \left( U^*(k,t) c_1(\vk)-V(k,t) c_1^\dagger (-\vk)\right)
\end{split}\end{equation}
where
\begin{equation}\begin{split}
W^{rs}(\vk,t)\equiv& \frac{1}{2}\left( u^{r\dagger}_2(\vk)u_1^s(\vk)+v_2^{s\dagger}(\vk)v_1^r(\vk) \right)e^{i(\w_1-\w_2)t}\\
Y^{rs}(\vk,t)\equiv& \frac{1}{2}\left( u^{r\dagger}_2(\vk)v_1^s(-\vk)+u_1^{s\dagger}(-\vk)v_2^r(\vk) \right)e^{i(\w_1+\w_2)t}
\end{split}\end{equation}
\begin{equation}\begin{split}
U(k,t)\equiv& \frac{1}{2}\left(\sqrt{\frac{\w_1(k)}{\w_2(k)}}+\sqrt{\frac{\w_2(k)}{\w_1(k)}}\right)e^{-i(\w_1-\w_2)t}\\
V(k,t)\equiv& \frac{1}{2}\left(\sqrt{\frac{\w_1(k)}{\w_2(k)}}-\sqrt{\frac{\w_2(k)}{\w_1(k)}}\right)e^{i(\w_1+\w_2)t}
\end{split}\end{equation}
and $\w_i(k)\equiv\sqrt{\vk^2+m_i^2}$.
By solving those expressions for the operators $a_i$, $b_i$, $c_i$ ($i=1,2$) one gets
\begin{equation}\begin{split}\label{amassWT}
a_1^r(\vk)=&\ct a_A^r(\vk)+\st \sum_s \left(W^{rs}(\vk,t)a_B^s(\vk,t)+Y^{rs}(\vk,t) a_B^{s\dagger}(-\vk,t)\right)\\
a_2^r(\vk)=&\ct a_B^r(\vk)-\st \sum_s \left(W^{sr*}(\vk,t)a_A^s(\vk,t)+Y^{sr}(-\vk,t) a_A^{s\dagger}(-\vk,t)\right)
\end{split}\end{equation}
\begin{equation}\begin{split}\label{bmassWT}
b_1(\vk)=&\ct b_A(\vk,t)-\st\left(U^*(k,t)b_B(\vk,t)+V(k,t)b^\dagger_B(-\vk,t)\right)\\
b_2(\vk)=&\ct b_B(\vk,t)+\st\left(U(k,t)b_A(\vk,t)-V(k,t)b^\dagger_A(-\vk,t)\right)
\end{split}\end{equation}
\begin{equation}\begin{split}\label{cmassWT}
c_1(\vk)=&\ct c_A(\vk,t)-\st\left(U^*(k,t)c_B(\vk,t)+V(k,t)c^\dagger_B(-\vk,t)\right)\\
c_2(\vk)=&\ct c_B(\vk,t)+\st\left(U(k,t)c_A(\vk,t)-V(k,t)c^\dagger_A(-\vk,t)\right).
\end{split}\end{equation}
The flavour vacuum is defined as
\be
\rfv\equiv G^{\dagger}_{\theta}(t)\rmv
\ee
The Fock space for the flavour states is obtained by
considering all possible polynomials in the creation operators acting on the flavour vacuum state.

\subsectionItalic{Regularization of the Theory}

When the fermionic and bosonic cases are analyzed separately, a normal order with respect to the vacuum $\rmv$ is introduced
(Sections \ref{ssVC} and \ref{ssFVDE}).
In the context of Supersymmetry, such a normal ordering is not needed, being
\be
\lmv T_{\mu\nu}(x) \rmv=0.
\ee
However, the theory still needs a regulator, being $\lfv T_{\mu\nu}(x) \rfv$ formally divergent.
A cutoff in the momentum space $K$ is therefore considered.

\section{FEATURES OF A SUSY FLAVOUR VACUUM}\label{sFOTFVsusy}

\subsectionItalic{Energy Density, Pressure and Equation of State}

\begin{figure}[ht]
\centering
\includegraphics[width=9cm]{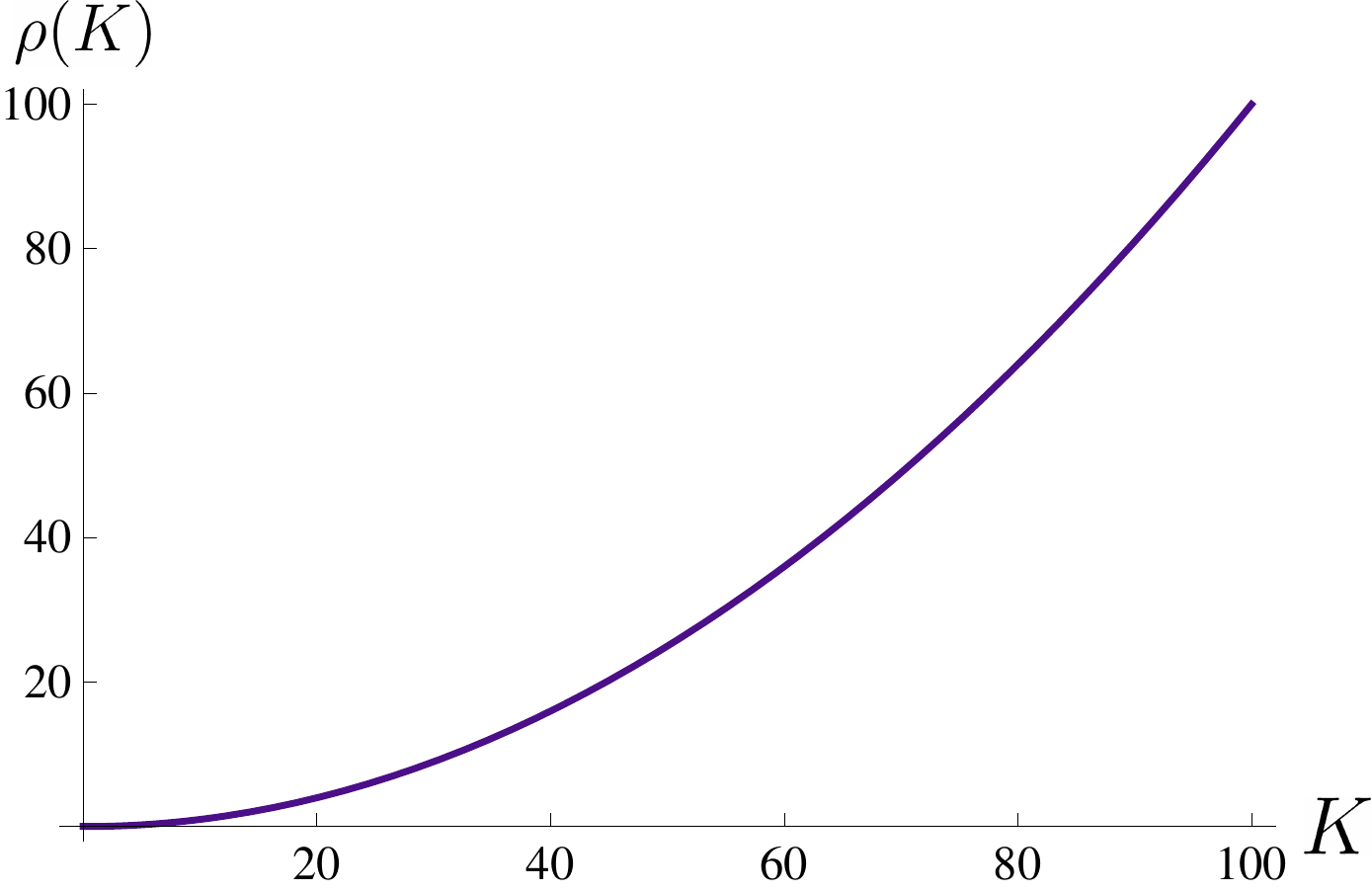}
\caption{\small{The energy density as a function of the cut-off is plotted, with values: $\st=2 \pi^{-2}$, $m_1=1$, $m_2=0.9$.}}
\label{fig:e1}
\end{figure}

\begin{figure}[ht]
\centering
\includegraphics[width=12cm]{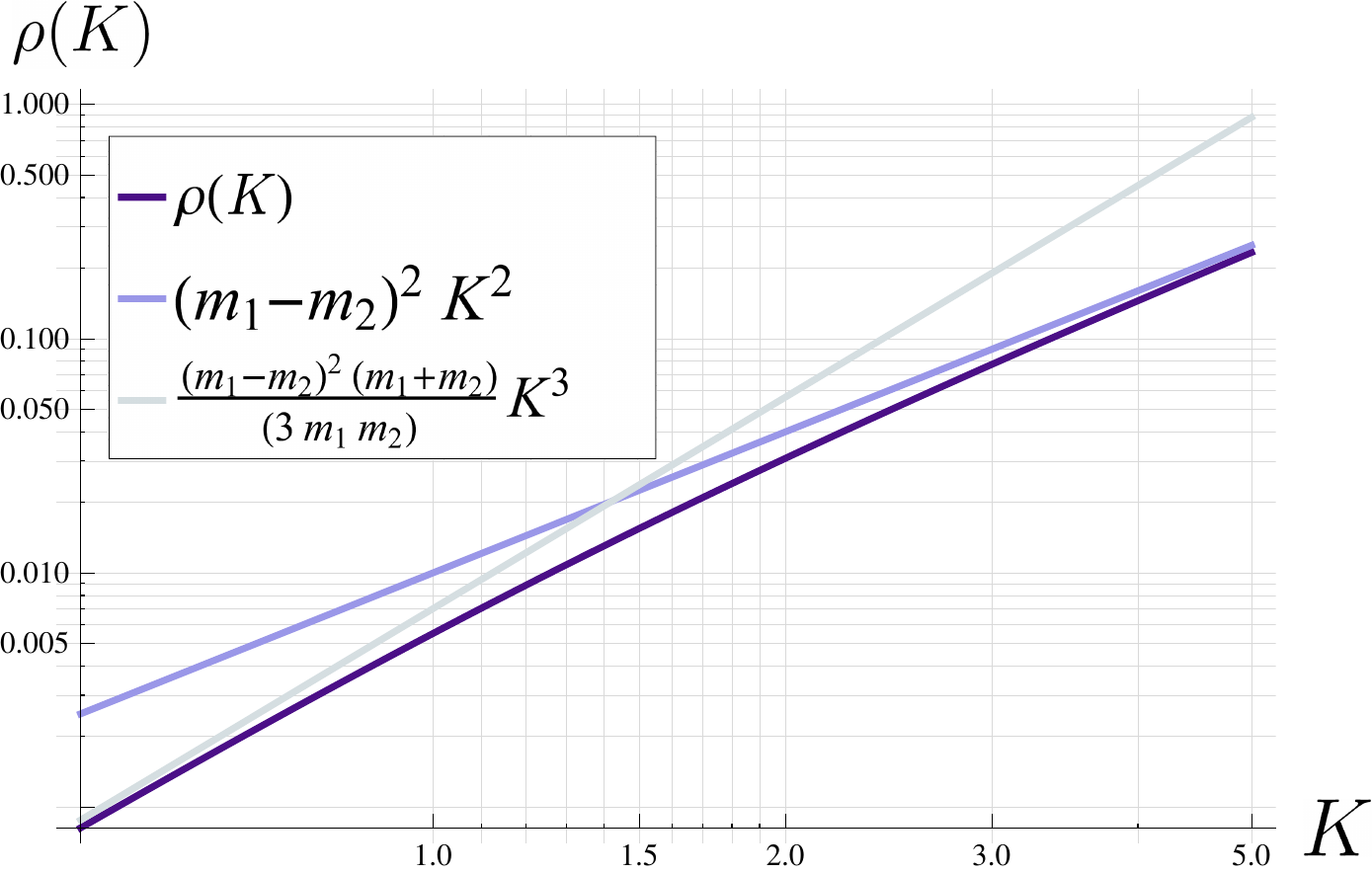}
\caption{\small{The energy density as a function of the cut-off, with values: $\st=2 \pi^{-2}$, $m_1=1$, $m_2=0.9$,
together with two approximating curves is plotted in a log-log graph.
For small $K$, the energy density approaches the zero as $\rho(K)\sim K^3$;
fore large $K$, we have that $\rho(K)\sim K^2$.
In this context, the scale of the theory (that distinguish the two different regimes)
is given $m_1 \approx m_2$, that in the plot is roughly $1$ (in arbitrary units).}}
\label{fig:e2}
\end{figure}

\begin{figure}[ht]
\centering
\includegraphics[width=7cm]{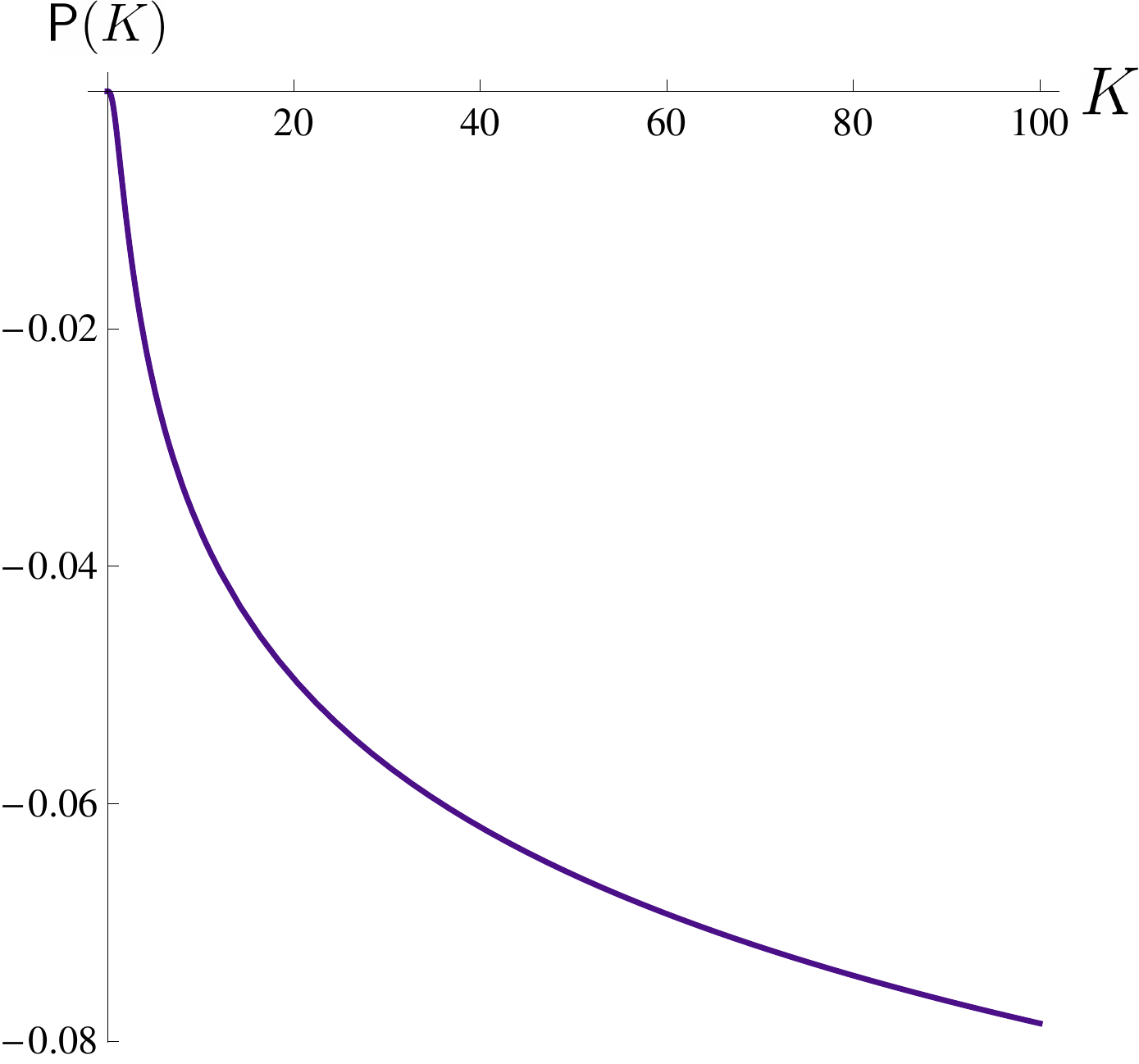} 
\caption{\small{The pressure as a function of the cut-off is plotted, with values: $\st=2 \pi^{-2}$, $m_1=1$, $m_2=0.9$.}}
\label{fig:p1}
\end{figure}

\begin{figure}[ht]
\centering
\includegraphics[width=10cm]{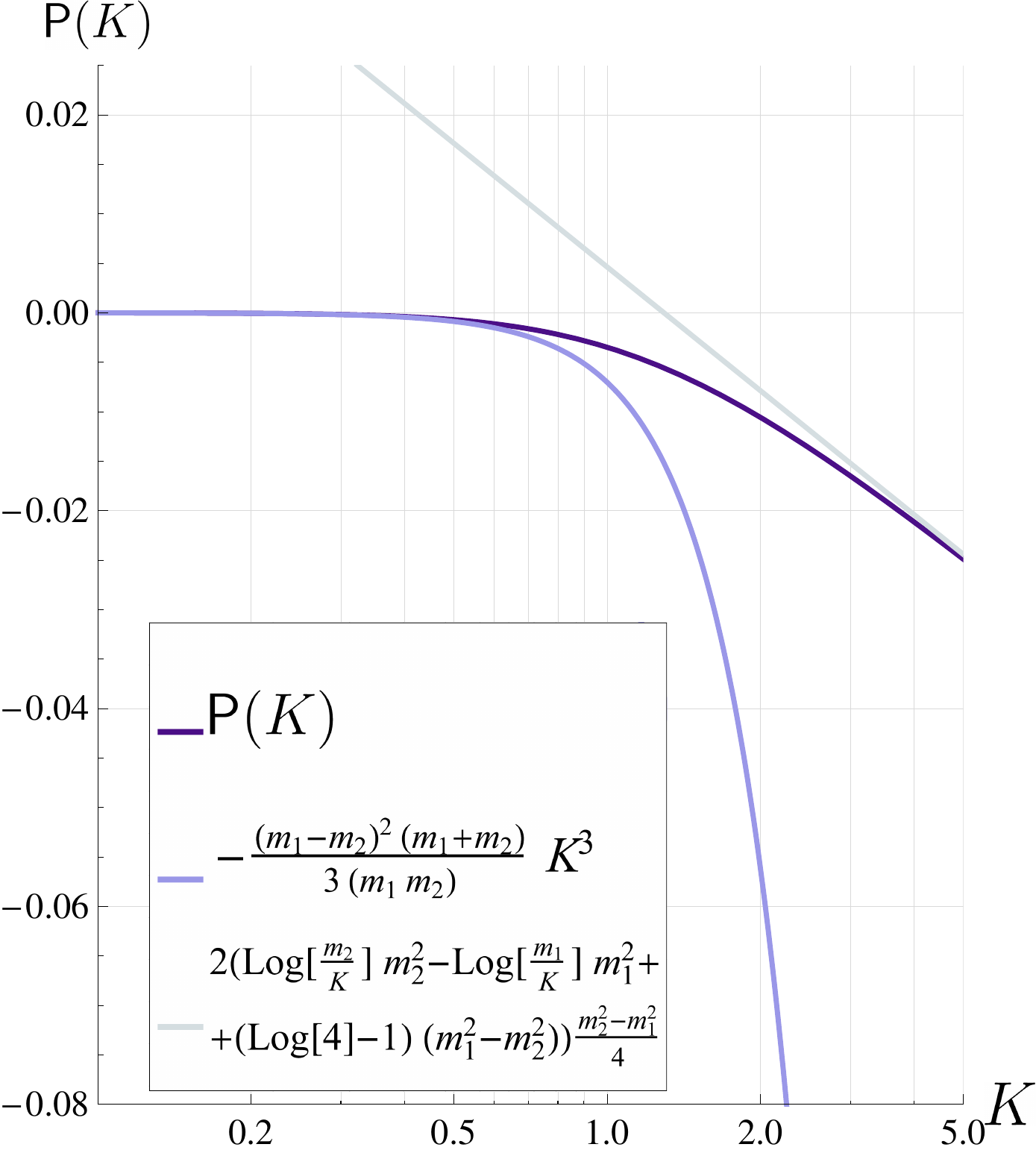}
\caption{\small{The pressure as a function of the cut-off, with values: $\st=2 \pi^{-2}$, $m_1=1$, $m_2=0.9$,
together with two approximating curves is plotted in a linear-log graph.
The behaviour of $\pressure(K)$ for small $K$ (compared with $m_2\approx m_1=1$)
is $\pressure(K)\sim K^3$, while large $K$, we have $\pressure(K)\sim \log (K)$.}}
\label{fig:p2}
\end{figure}

The energy density ($\rho$) and the pressure ($\pressure$) of the \textit{flavour vacuum} for the theory (\ref{lag}) are given by
\begin{equation}\begin{split}\label{e1}
\rho=&\lfv T_{00}(x)\rfv=\sst \frac{(m_1-m_2)^2}{2 \pi^2}\int_0^K dk\;k^2\left(\frac{1}{\w_1(k)}+\frac{1}{\w_2(k)}\right)=\\
 =&\sst \frac{(m_1-m_2)^2}{2 \pi^2}f(K)
\end{split}\end{equation}
\begin{equation}\begin{split}\label{p1}
\pressure=&\lfv T_{jj}(x)\rfv=-\sst \frac{m_1^2-m_2^2}{2 \pi^2}\int_0^K dk\; k^2 \left(\frac{1}{\w_2(k)}-\frac{1}{\w_1(k)}\right)\\
 =&\sst \frac{m_1^2-m_2^2}{2 \pi^2}g(K)
\end{split}\end{equation}
The derivation of these results is provided in Appendix \ref{aC4}.
The integrals in (\ref{e1}) and (\ref{p1}) can actually be solved, and function $f(K)$ and $g(K)$ are
\begin{multline}
f(K)=\frac{1}{2}\Bigg( K \w_1(K)+K \w_2(K)+\\
		-m_1^2 \log\left[\frac{K+\w_1(K)}{m_1}\right]-m_2^2 \log\left[\frac{K+\w_2(K)}{m_2}\right]  \Bigg)
\end{multline}
\begin{multline}
g(K)=\frac{1}{2}\Bigg( K \w_1(K)-K \w_2(K)+\\
-m_1^2 \log\left[\frac{K+\w_1(K)}{m_1}\right]+m_2^2 \log\left[\frac{K+\w_2(K)}{m_2}\right]  \Bigg)
\end{multline}
The behaviour of the energy density and the pressure as a function of the cut-off is shown in the graphics \ref{fig:e1},
\ref{fig:e2}, \ref{fig:p1} and \ref{fig:p2}.
As we were expecting, both energy and pressure are homogeneous in space (since $G_\theta$ does not carry any $\vx$ dependency).
However, quite notably, they are also \textit{time}-independent, although $G_\theta$ explicitly depends on time.

The equation of state $w$ is therefore
\be
w=\frac{(m_1+m_2)}{(m_1-m_2)}\frac{g(K)}{f(K)}
\ee
\begin{figure}[ht]
\centering
\includegraphics[width=9cm]{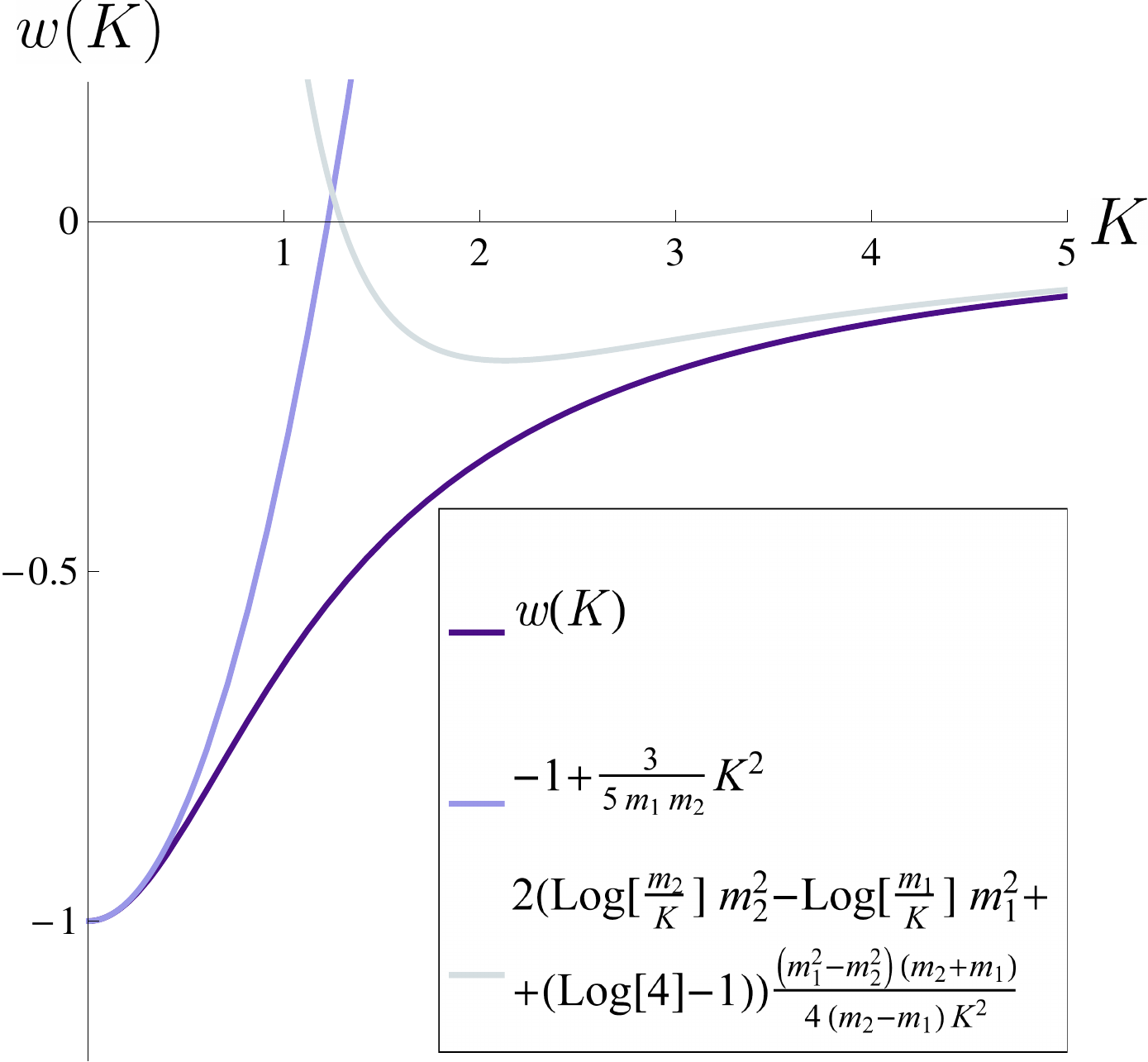} 
\caption{\small{The equation of state $w$ as a function of the cut-off is plotted, with values:
$m_1=1$, $m_2=0.9$.}}
\label{fig:w1}
\end{figure}
\begin{figure}[ht]
\centering
\includegraphics[width=9cm]{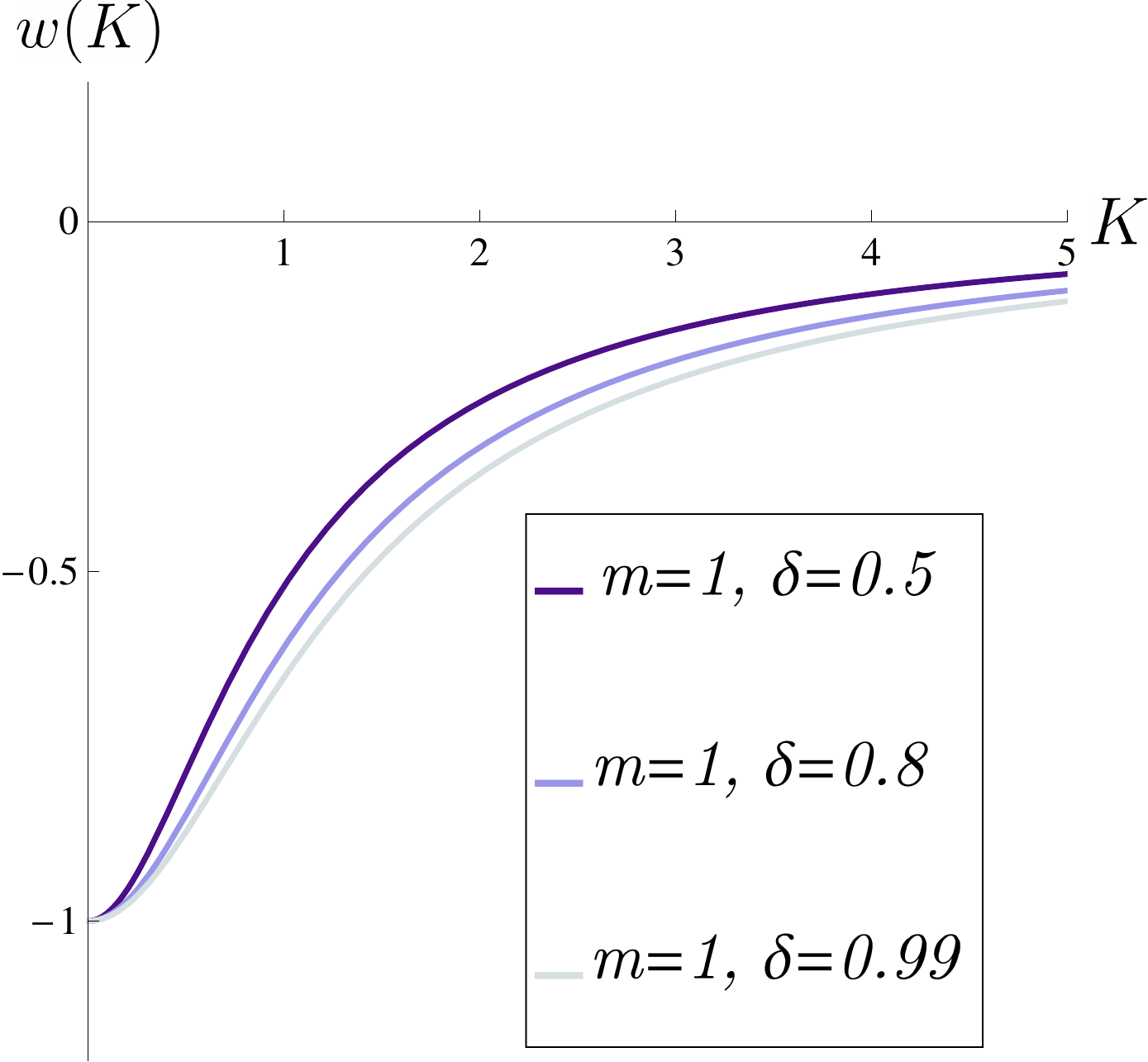}
\caption{\small{The equation of state $w$ as a function of the cut-off is plotted, for different values of these masses.
We have a set of three curves,
each of them corresponding to different values of $\delta$, which parametrises the difference of the masses
through $m_2\equiv m \delta$ and $m_1\equiv m$, as specified in the legend.}}
\label{fig:w2}
\end{figure}
As we can see in the graphic \ref{fig:w1}, for small values of the cut-off (compared with $m_1\approx m_2$),
the equation of state approaches $-1$. This is the region in which our \textit{flavour vacuum} might give
a contribution to th Dark Energy. Let us remind the reader that an acceleration of the expansion of the Universe is obtain when
$w<-1/3$ holds the fluid that fills it in. For grater values, the equation of state goes to zero as $\log(K)K^{-2}$.

\subsectionItalic{Disentangling The Two Contributions}

In order to understand these results and compare them with our previous work presented in Chapter \ref{cAFVICS},
it is useful to disentangle the contribution
to $\rho$ and $\pressure$ of the different fields: being the the Lagrangian (\ref{lag}) the sum of six free Lagrangians
(two scalars, two pseudo-scalars and two Majorana spinors),
the stress-energy tensor can be written as the sum of three different operators, each of them
involving just fields operators of one type (scalar, pseudo-scalar or spinor):
\be
T_{\mu\nu}(x)=T_{\mu\nu}^S(x)+T_{\mu\nu}^P(x)+T_{\mu\nu}^\psi(x).
\ee
Formally, quantities $\lfv T_{\mu\nu}^S(x) \rfv$, $\lfv T_{\mu\nu}^P(x) \rfv$ and $\lfv T_{\mu\nu}^\psi(x) \rfv$
are equal to the \textit{flavour vacuum} expectation value of the stress-energy tensor in a theory
with just two scalars ($\lfv T_{\mu\nu}^S(x) \rfv$), two pseudo-scalars ($\lfv T_{\mu\nu}^P(x) \rfv$)
and two Majorana spinors ($\lfv T_{\mu\nu}^\psi(x) \rfv$) respectively.
Recalling that
\be
\mathfrak{F}_{tot}=\left(\mathfrak{F}_{\psi_1} \otimes \mathfrak{F}_{\psi_2}\right) \otimes
					\left( \mathfrak{F}_{S_1} \otimes \mathfrak{F}_{S_2}\right) \otimes
					\left( \mathfrak{F}_{P_1} \otimes \mathfrak{F}_{P_2}\right)
\ee
we have
\be
\mathfrak{F}_{f\;tot}=\mathfrak{F}_{f\;\;\psi}\otimes \mathfrak{F}_{f\;\;S} \otimes \mathfrak{F}_{f\;\;P}
\ee
for flavour states, with $\mathfrak{F_{f\;\varphi}}$ the flavour Fock space corresponding to $\mathfrak{F_{\varphi_1}}\otimes\mathfrak{F_{\varphi_2}}$.
Therefore, we are allowed to look at single contributions of flavour vacua of each
$\mathfrak{F_{\varphi_1}}\otimes\mathfrak{F_{\varphi_2}}$, for $\varphi=\phi,P,S$.

For what concerns the energy density, we have
\begin{multline}\label{energydensityS}
\lfv \sum_{i=1,2}T_{00}^{S_i}(x)\rfv=\lfv \sum_{i=1,2}T_{00}^{P_i}(x)\rfv=\\
		=\int\!\!dk \frac{k^2}{2 \pi^2}\left(\w_1(k)+\w_2(k)\right)\left[1+ \sst \frac{\left(\w_1(k)-\w_2(k)\right)^2}{2\w_1(k) \w_2(k)}\right]
\end{multline}
\begin{multline}\label{energydensitypsi}
\lfv \sum_{i=1,2}T_{00}^{\psi_i}(x)\rfv=\int\!\!dk \frac{k^2}{\pi^2}\left(\w_1(k)+\w_2(k)\right)\times\\
\times\Bigg[-1+\sst\left( \frac{\w_1(k)\w_2(k)-m_1 m_2-k^2}{\w_1(k)\w_2(k)}\right)\Bigg]
\end{multline}
On the other hand, the contributions to the pressure are
\begin{multline}
\lfv \sum_{i=1,2}T_{jj}^{S_i}(x)\rfv=\lfv \sum_{i=1,2}T_{jj}^{P_i}(x)\rfv=\\
	=\int\!\!dk \frac{k^2}{2\pi^2}\left[\frac{k^2}{3}\left(\frac{1}{\w_1(k)}+\frac{1}{\w_2(k)}\right)
					 -\sst \frac{m_1^2-m_2^2}{2}\left(\frac{1}{\w_1(k)}-\frac{1}{\w_2(k)}\right)\right]
\end{multline}
\be
\lfv \sum_{i=1,2}T_{jj}^{\psi_i}(x)\rfv=-\int\!\!dk\frac{2k^4}{6 \pi^2}\left(\frac{1}{\w_1(k)}+\frac{1}{\w_2(k)}\right)
\ee
As already discussed, a so-defined theory needs a normal ordering, that is conventionally taken as
\be
\lfv :\mathcal O :\rfv\equiv \lfv \mathcal O \rfv-\lmv \mathcal O \rmv.
\ee

Now let us look at the single equations of states:
\be
w_b=\frac{\lfv: \sum_{i=1,2}T_{jj}^{S_i}(x):\rfv}{\lfv: \sum_{i=1,2}T_{00}^{S_i}(x):\rfv}
					=\frac{\lfv: \sum_{i=1,2}T_{jj}^{P_i}(x):\rfv}{\lfv: \sum_{i=1,2}T_{00}^{P_i}(x):\rfv}=-1
\ee
and
\be
w_f=\frac{\lfv: \sum_{i=1,2}T_{jj}^{\psi_i}(x):\rfv}{\lfv: \sum_{i=1,2}T_{00}^{\psi_i}(x):\rfv}=0
\ee
This means that the bosonic contribution to the vacuum condensate
by itself would lead to a pure cosmological constant-type equation of state.
This behaviour of the fluid is mitigated by the contribution of the fermionic fluid, that has a vanishing pressure.
When just low energies are taken into account, the bosonic contribution drives the equation of state
towards the $-1$, but as the cut-off increases, the fermionic contribution gets stronger and stronger,
leading $w$ to the value of zero. 

The difference of behaviour between the two contributions is not surprisingly, and, albeit not completely equal,
is quite similar to the ones that have been found in our previous analysis (Section \ref{sDAC}).
Nonetheless, here the fermionic contribution presents a very specific behaviour, for all values of the cutoff:
its pressure is always equal to zero, leading to an interesting phenomenological consequences, as we will soon see. 

\section{DISCUSSION AND CONCLUSIONS}\label{sCsusy}

In this Chapter we presented the behaviour of the \textit{flavour vacuum},
in a simple supersymmetric theory.
The model that was considered involves two free real scalars with mixing, two free real pseudo-scalars with mixing and
two free Majorana spinors with mixing.
The analysis confirmed the breaking of SUSY and Lorentz invariance induced by BV formalism,
via its non-trivial (flavour) vacuum state, for which $0\neq_f\langle T_{\mu\nu} \rangle_f \propto\!\!\!\!\!/ \;\e_{\mu\nu}$.

From the perspective of the previous Chapter, in which the flavour vacuum is regarded as the effective vacuum of
the D-particle foam model, this result clarifies the role of BV formalism, which breaks SUSY by itself.
This requires a deeper study of the stringy model, in order to understand how flavour mixing causes SUSY breaking
at a microscopic level.

On the other hand, being SUSY and BV formalism the only ingredients used in building our model,
the results here presented enlighten general features of BV formalism in a supersymmetric context,
that are independent of a possible underlying theory, and are worthy of a dedicated analysis.
In particular, we showed that:
\begin{enumerate}
	\item the energy of the \textit{flavour vacuum} is greater than zero, and in fact is divergent,
	\item the bosonic sector shows an equation of state $w=-1$, whereas for the fermionic sector we have $w=0$.
\end{enumerate}
The first statement is important from a theoretical point of view: it allows us to better understand the nature of the
flavour vacuum. The second result, on the other hand, has rich implications from a phenomenological perspective.

\subsectionItalic{SUSY Breaking?}

We shall now focus on the first point.
The model here considered is well known in literature \cite{Wess:1992cp}.
It is also known that in the usual representation of the supersymmetric algebra,
in terms of states describing particles with well defined masses,
there exists a state with zero energy (the lowest energy possible).
In the language of Section \ref{secSUSY}, we can say that in our model
\be
\exists \ket{\Omega};\;\;H\ket{\Omega}=0\;\Rightarrow\; Q\ket{\Omega}=0\;\Rightarrow\; \mbox{No spontaneous SUSY breaking.}
\ee
This fact implies that
\begin{itemize}
	\item SUSY is not spontaneously broken,
	\item the flavour vacuum is not the lowest energy state of the theory.
\end{itemize}
As explained in Chapter \ref{cB},
the coexistence of a zero-energy state and a flavour state with a formally divergent value of the energy
is possible in a Hilbert space bigger than the simple $\mathfrak{F}_{tot}$ described in (\ref{focktotwz}).
This argument
corroborates the interpretation of the flavour state as a
\textit{collection of an infinite number of particles}.
These particles are the usual particles with well defined mass, and the state with zero energy is nothing but
the state where none of such particles is present.
This also indicates that supersymmetry is not \textit{spontaneously} broken,
at least in the sense explained in Section \ref{secSUSY}.

Nonetheless, the flavour vacuum is the \textit{physical vacuum}:
since in laboratories we create/destroy/detect flavour particles,
what we can \textit{physically} consider as vacuum is a state with \textit{no flavour particles}.
In the theory, particle states with well defined mass are allowed and they are also used to describe
particle states with well defined flavour. Flavour states are actually states describing collections of
infinite particles with well defined mass. The ground state of the theory is defined as the state with
no particles of defined mass. However, in no experiment we would be able to build such a state,
since in our hardest attempt we would only get a state that contains no \textit{flavour} particles,
the state that in our theory is described by the \textit{flavour vacuum}.
Therefore the theoretical ground state is different from the \textit{physical} ground state, and
although supersymmetry is not spontaneously broken in the theory, it is \textit{effectively} broken
by the existence of a \textit{physical} vacuum carrying a non vanishing energy density.

Whether this \textit{effective} SUSY breaking (a novel mechanism, not yet explored in literature, to our knowledge)
leads to desirable features, such as differences in the mass
spectrum between fermions of bosons, is an issue that requires further investigations.

Interpreting the flavour vacuum as a condensate in which two different species of particles coexist,
namely a bosonic and a fermionic one, offers us also an intuitive explanation for the difference in pressure
of the two contributions. Indeed, one is tempted to interpret the zero total pressure of the fermionic fluid as a result 
of an extra contribution, as compared with the bosonic case, due to the Pauli exclusion principle, the so called \textit{degeneracy pressure}.
The latter is a positive contribution to the total pressure of a fermion fluid which is due to the extra force one has to exert
as a consequence of the fact that two identical fermions cannot occupy the same quantum state at the same time.
The force provided by this pressure sets a limit on the extent to which matter can be squeezed together without collapsing into,
\textit{e.g.} a neutron star or black hole. From a (semi-)classical perspective,
we can describe the mechanism as follow: when two fermions are ``squeezed'' too close together, the exclusion principle requires them
to have opposite spins in order to occupy the same energy level; to add another fermion to a given volume 
(as required by the formation of a condensate) requires raising the fermion's energy level, and this requirement for energy to compress
the material appears as a (positive) pressure. In our cosmological situation, it may be that this positive contribution to the pressure
cancels out the negative pressure of the fermionic fluid, leading to the dust-like behaviour we find here.

\subsectionItalic{The Dark Side of the Flavour Vacuum}

Coming to the second important result of the analysis presented in the present analysis,
namely the equation of state $w=-1$ for bosons and $w=0$ for fermions when considered separately,
it seems that our simple model suggests
that the physical vacuum is a combination of two fluids that behave differently:
both permeate the empty space uniformly and statically, but one has a cosmological constant behaviour ($w=-1$),
while the other behaves as dust ($w=0$, see Section \ref{ssDM}).
Can any trace of realism be recognized in this model?
Quite surprisingly, in our search for a mechanism to explain a cosmological problem, the accelerated rate of the expansion
of the universe, the solution offered by the model might shed some light on another cosmological conundrum:
the bosonic component of our supersymmetric physical vacuum
behaves indeed as a source of Dark Energy, while the fermionic component mimic the behaviour of Dark Matter.

The role of the flavour vacuum as source of Dark Energy has been extensively discussed in literature (Section \ref{ssFVDE}).
Here we present a new feature of the flavour vacuum: its possible contribution to Dark Matter.
A first objection that can be moved against this interpretation regards the uniformity of the distribution of the
flavour vacuum, being Dark Matter gathered in clusters around and inside galaxies.
However, one should consider that in our model we are actually considering
a vacuum state (as we said, a ``physical'' vacuum), with no other matter present in space:
we are actually modeling a simple ``empty'' universe.
If (other) matter is considered in our toy universe, this
matter would start interact with our physical vacuum gravitationally and might start forming clusters,
via the phenomenon called ``gravitational instability''.\footnote{Gravity tends to enhance irregularities, pulling
matter towards denser regions \cite{Liddle:2009zz}.}
It is known that such an effect, on the other hand, does not necessarily occurs for Dark Energy \cite{amendola},
that can persist in its state of spatial uniformity even in presence of clustered matter.
The evolution of our flavour vacuum, considering both its bosonic and fermionic component, in presence
of other matter and gravitational interaction represents necessarily an object for future studies.

Having solved this first apparent difficulty, we can now notice various reasons for flavour vacuum being
a possible candidate for Dark Matter:
\begin{itemize}
	\item it is ``dark'' (\textit{i.e.} it does not absorb	 or emit light, it is electromagnetically neutral);
	\item being the \textit{vacuum} state for flavour particles, we expect it to interact with other SM particles just
	via gravitational effects;\footnote{From a physical point of view, as already emphasized, flavoured neutrinos,
	rather then massive neutrinos, are actual \textit{observable} particles; in other words, in the QFT framework
	other particles from the SM are directly coupled to flavoured neutrinos, thanks to \textit{weak interactions},
	and the flavour vacuum is treated from those particles as a ``mere'' empty state.
	In a broader picture, when gravitational effects are considered, 
	the complex structure of the flavour vacuum becomes manifest and it acts gravitationally
	on other particles thanks to its non-vanishing energy,	at least at a classical level.}
	\item at least in the free case, it is cold (in the sense that $w=0$).\footnote{The possibility of extending
	this result to interactive models will be discussed in Section~\ref{sTIFV}.}
\end{itemize}

Clearly, the model is still in its embryonic stage and many consistency tests with astrophysical data
are in order, before start regarding the flavour vacuum as a plausible Dark Matter candidate (see for instance
Section 1.4 of \cite{BertoneBook} for a list of criteria that a proper candidate should fulfill).
However, the possibility that the same mechanism can lead to both Dark Energy and Dark Matter
is without any doubts an intriguing possibility, worth of further investigations.

\subsectionItalic{Testability}

Another interesting aspect of the model concerns the interplay between the two fluids.
Supersymmetry imposes that the energy density of the bosonic component is tied up
to the energy density of the fermionic component; in a more realistic theory, therefore,
one should be able to reproduce the current experimental value of the ratio between the Dark Energy density and
the Dark Matter energy density ($\sim 2.8$), in the optimist belief that the flavour vacuum is the only responsible for
both of them. The role of a curved background in the formulation of the theory might be crucial,
since the energy density of a dust-like fluid gets diluted by the expansion of the universe, whereas such an effect
does not occur for a cosmological-constant type (as seen in Section \ref{ssMCAI} for classic fields \cite{weinberg}),
and therefore the ratio between those two quantities changes dramatically with time.

Within a cutoff regularization framework, as the one here presented,
the two energy densities depend on such a cutoff.\footnote{One might object that 
the cut-off violates Lorentz invariance. However, we should recall the reader that, in the
context of cosmology, there is a preferred frame (which is at rest with respect to the Cosmic microwave background radiation)
so this regularization may not be as unjustified as it seems at first.}
Nonetheless, being the theory supersymmetric, the \textit{same} cutoff
applies to both quantities. However, the ratio between them can in general be cutoff dependent, as it actually is in the case
here presented. On one side, one might hope that in a more realistic theory (on a curved background, for instance)
the ratio might be cutoff independent, although at the moment there is no clear reason how this could happen.
On the other hand, one could consider such a situation as highly desirable:
if the ratio is fixed from the cutoff, the same value for the cutoff would also fix the value of the energy.
This implies that once the cutoff is decided on the basis of experimental data on the ratio,
the model gives a precise prediction for the absolute values of the energies,
that can be compared with their experimental values.

In order to illustrate these ideas we shall present a concrete example.
Let us assume that our supersymmetric model is effective up to the energy scale $K$ (that \textit{a priori} is not known).
In the standard Big-Bang  picture, this means that when the universe cools down to that energy,
the flavour vacuum becomes to be an effective description of the vacuum state of our theory.
We call $t_0$ the time corresponding with this transition and $a_0$ the corresponding scale factor.

In our toy universe, we assume that  at time $t_0$, in absence of any other source of energy or matter, the energy/matter content of our toy universe is only due to the flavour vacuum. Moreover, we assume that it can be described,
at a classic level (\textit{i.e.} on sufficiently large scales),  in terms of two fluids:
a first one, due to the bosonic component of the flavour vacuum and described by $\trho_b$ and $w=-1$,
and a second one, due to its fermionic component and described by
$\trho_f$ and $w=0$.
We will regard the bosonic component as the (only) source of Dark Energy and the fermionic as the (only) source of Dark Matter.
Both $\trho_b$ and $\trho_f$ are function of the following parameters:
neutrino masses, neutrino mixing angles and the cutoff $K$.
If we know (from observations) the neutrino masses and mixing angles, the cutoff is the only parameter left to determine.

As our toy universe expands, we assume that the two fluids obey Einstein equations and therefore they
scale as (see Section \ref{ssMCAI})
\be\begin{split}
\rho_f(t)\;\;a(t)^3=&\trho_f \;\;a_0^3\\
\rho_b (t)=&\trho_b.\\
\end{split}\ee
This means that \textit{today} their value is
\be\begin{split}\label{dedmnow}
\rho_f(t_{now})=&\trho_f \;\;a_0^3\\
\rho_b(t_{now})=&\trho_b.\\
\end{split}\ee
respectively, being $a(t_{now})=1$ by convention.
Those two quantities depends on the following parameters: neutrino masses, mixing angles, cutoff energy, scale
factor at $t_0$.
Provided with these expressions, we can then test our model in two ways.
\begin{enumerate}
	\item If observational data enable us to estimate neutrino masses, mixing angles, Dark Matter and Dark Energy,
	from (\ref{dedmnow}) we can derive the other parameters left: the cutoff energy and the scale factor $a_0$.
	Well equipped with all the parameters of the theory, we will then be able to
	check if the model is in reasonable agreement with other standard models.
	For example, if the scale factor $a_0$ fitting all data corresponds to a time \textit{in the future} (for $a_0>1$),
	the model has to be rejected, or corrected at least.
	\item On the other hand, theoretical reasons might suggest specific values for the cutoff (if for instance the flavour
	vacuum rises in the low energy limit of an underlying theory, as discussed earlier), and/or the scale factor $a_0$
	(since in the standard Big-Bang picture the energy of the universe is connected with its scale factor).
	In this case, we might be able to make \textit{a prediction} on the value of the Dark Matter and Dark Energy density,
	via formulae (\ref{dedmnow}), that we might compare with observational estimates. On this basis
	our model is therefore accepted or refused.
\end{enumerate}

The simple toy universe discussed in our example is not realistic enough to hold the comparison
with data already available (only two generations of neutrinos have been considered, SUSY is unbroken, no interactions are present...).
However, a preliminary test can be performed. We first choose the value of the masses to be $m_1=1\;eV$ and $m_2=1.5\;eV$,
so to preserve the hierarchy $m_1^2\sim m_2^2\sim \Delta m^2_{21}$;\footnote{It should be noticed that in our model bosons and fermions
have the same masses. A realistic mechanism of SUSY breaking might impose masses for sneutrinos (the bosonic partnes of neutrinos)
much higher than for neutrinos. In the absence of a trustworthy prediction, here we choose a sort of intermediate scale ($\sim 1\;eV$),
just for demonstrative purposes.} we then choose the energy scale of the cutoff to be $K=10^{30}\;eV$,
inspired by the ratio of Planck scale and the measured scale of neutrino physics (indeed of the order of $10^{30}$);
finally we choose the mixing angle to be such that $\sst =1/3$, since
direct observations on neutrino oscillations show that
two out of the three angles that describe the mixing matrix for real neutrinos are such that $\sin^2\theta_{12}\approx 1/4$
and  $\sin^2\theta_{23}\approx 1/2$ \cite{Bilenky:2010zza}.
These values lead to $\rho_b\approx 1\;eV^4$, that reasonably reflects the observed hierarchy between Dark Energy and neutrino physics
($\rho_{\Lambda}\sim (\Delta m^2)^2$), as discussed in Section \ref{ssCCP}.
Moreover, in order for Dark Matter density to be of the same order, the value of the scale factor must be $a_0\approx 10^{-20}$,
that sets the transition phase (when the flavour vacuum became effective)
well in the far past, when our toy universe was $10^{20}$ times smaller than ``now''.

The evolution of our flavour vacuum, considering both its bosonic and fermionic component, in presence of other matter and
gravitational interaction represents necessarily an object for future studies.

Despite its oversimplification, the example presented seems to be going in the right direction. One would expect that
a more realistic model would preserve the same good features here discussed.

To conclude, despite the speculative character of the ideas here presented,
we believe that the important results here discussed well motivate further studies of the flavour vacuum in more realistic theories.

\chapter{A NEW METHOD OF CALCULATION}\label{cANMOF}

\section{Introduction}

In this Chapter we shall present a method of calculating
$\lfv T_{\mu\nu} \rfv$, which applies to certain theories, including the supersymmetric model
studied in the previous Chapter.
To our knowledge, such an approach is not present in the
existing literature. However, it presents several interesting advantages, with respect to
other methods commonly used, that make it worthy of a detailed explanation. Above all,
this method reduces the length of calculations required to simplify $\lfv T_{\mu\nu} \rfv$ by standard approaches
of a significant amount.

Reproducibility is an essential part of the validation process of a result,
in experimental as well as in theoretical contexts.
It follows that simplicity and brevity are desirable attributes for all procedures (calculations included),
in order to decrease the chances of errors.

It is actually hard to underestimate the importance of the search for new methods of calculating complicated expressions.
Such a search quite often leads to a deeper understanding of relations between the objects involved:
faster methods of calculation usually rely on a cleaver usage of the symmetries of the problem
rather then the brute force that characterizes any first attempts.
Moreover, they might also shed light on new aspects of the theory that posed the problem in first place,
leading to further important developments.

As a first development and example of the potential of the method,
we shall attempt to generalize results concerning the equation of state
in free models to cases in which certain classes of interactions are present, in a completely non-perturbative way.
Whithout aiming at an exhaustive non-perturbative treatment of an interactive theory,
nonetheless, we shall indicate how important results could be achieved
under reasonable assumptions.

Standard approaches, borrowed from the relevant literature, have been used in the present work
to analyze the two models presented in Chapter \ref{cAFVICS} and \ref{cASFV}.
In Section \ref{sGI}, we shall explain the general idea underlying the novel approach,
in comparison with those.
In order to catch the important features of the new method,
in Section \ref{sWFVR}, we shall implement it on the supersymmetric model of Chapter \ref{cASFV}.
Although originally developed quite specifically for this model,
it can be easily generalized to other interesting cases.
In Section \ref{sAC} we shall clarify
its limits of applicability on other models.
In Section \ref{sA} we shall discuss its advantages and
relevance in our understanding of the BV formalism itself.
Finally, in Section \ref{sTIFV} we shall apply the method on certain self-interactive theories.

\section{General Idea}\label{sGI}

In BV formalism, the \textit{flavour vacuum} is defined via
\be\label{fvGmvpartiv}
\rfv\equiv G^{\dagger}_{\theta}\rmv=e^{iA}\rmv
\ee
with $A=A^\dagger$ a specific function of the fields\footnote{Explicit dependency on spacetime coordinates is omitted
here and forth.} with well defined mass (from now on \textit{massive fields}),
denoted here by $\varphi_i$,
and $\rmv$ the standard vacuum.
In the two previous Chapters, we studied the features of such a state in two different contexts.
In both cases we evaluated the expression
\be\label{setpartiv}
\lfv T_{\mu\nu} \rfv
\ee
with $T_{\mu\nu} $ denoting the stress-energy tensor.
Both the operator $T_{\mu\nu} $ and the state $\rfv$ depend on the specific model considered.
In order to evaluate expression (\ref{setpartiv}) two different approaches have been used for the two different models.
They represent the standard approaches adopted in the relevant literature \cite{Capolupo:2004av}.

According to the first one,
the flavour vacuum is written in terms of the standard vacuum $\rmv$ and the action
of the relevant ladder operators $a_i^{(\dagger)}(\vp)$ on it\footnote{Indices
denoting other degrees of freedom of the theory, such as spin or anti-particles,
are here omitted, for sake of clarity.}, thanks to its definition (\ref{fvGmvpartiv})
and the decomposition of $\varphi_i$ in terms of $a_i^{(\dagger)}(\vp)$.
Since it is possible to write also $T_{\mu\nu} $ in terms of the the massive fields, and therefore in terms of
$a_i^{(\dagger)}(\vp)$, the expression in (\ref{setpartiv}) reduces to an expression just involving the operators
$a_i^{(\dagger)}(\vp)$ and the state $\rmv$. Knowing the algebra of the operators and $a_i(\vp)\rmv=0$,
one is able to further reduce the algebraic structure of (\ref{setpartiv}), being left with a simple function of the
coordinates.\footnote{The expression ``simple'' function here denotes a map between c-numbers, \textit{i.e.}
does not involves operators and bra/ket explicitly.}
Such a procedure is graphically resumed in the following scheme:

$$
\xymatrix{
	\lfv\ar[d]		&				T_{\mu\nu} 	\ar[dd]				&					\rfv	\ar[d]					\\	
	\lmv G_\theta  \ar@/_1pc/[dddr]\ar[dr] &  &\ar[dl]\ar@/^1pc/[dddl]    G_\theta^{\dagger} \rmv \\
											&			\varphi_i 			\ar[d]	&								\\
							&				a_i^{(\dagger)}(\vp)	\ar[d]	&						\\
										&	\lmv  \int d\vp	\;	\mathcal{H}_{\mu\nu}(a_i^{(\dagger)}(\vp))						\rmv			 &						
}
$$
with $\mathcal{H}_{\mu\nu}(\slot)$ a specific function, whose details we can here omit.
The main disadvantage of this procedure is given by the exponential form of $G_{\theta}$,
that determines an infinite series of operators $a_i^{(\dagger)}$ acting on $\rmv$.
Manipulating such a series might be a rather difficult task \cite{Blasone:1995zc}.
This procedure was adopted in Chapter \ref{cAFVICS}, in which, further complications due to the presence of an external
classical field were present.

In the second approach, followed in Chapter \ref{cASFV},
instead of writing (\ref{setpartiv}) in terms of $\rmv$ and $a^{(\dagger)}_i(\vp)$, we first reduced $T_{\mu\nu} $
in terms of $a^{(\dagger)}_i(\vp)$, and then, with the help of the simple relations stated in (\ref{fromflavourladdertomassiveladder1}),
(\ref{fromflavourladdertomassiveladder2}) and (\ref{fromflavourladdertomassiveladder3}),
we were able to rewrite it in terms of the \textit{flavour ladder operators} $a^{(\dagger)}_\iota(\vp)$ only (with $\iota$
the index that runs over the possible flavours). Those operators not only respect the usual CAR/CCR algebra,
but they also satisfy $a_\iota(\vp)\rfv=0$.
Provided with these tools, we were able to further simplify the quantum algebraic expressions present in (\ref{setpartiv}),
being left, once more, with a simple function of the coordinates.
The corresponding scheme is given by
$$
\xymatrix{
	\lfv\ar@/_1pc/[ddddr]		&				T_{\mu\nu} 	\ar[d]				&					\rfv	\ar@/^1pc/[ddddl]					 \\
											&			\varphi_i 			\ar[d]		&								\\
											&				a_i^{(\dagger)}(\vp)	\ar[d]															 &								\\
											&			a_\iota^{(\dagger)}(\vp)\ar[d]																			 &								\\
								&		\lfv\int d\vp	\;	\mathcal{F}_{\mu\nu}(a_\iota^{(\dagger)}(\vp))					\rfv		&						 
}
$$
with $\mathcal{F}_{\mu\nu}(\slot)$ a suitable function.
This approach avoids the complications due to the infinite series, earlier mentioned.

In certain cases, a third approach might be adopted.
Recalling that the action of the operator $G_{\theta}$ on the field $\varphi_i$ is
\begin{equation}\begin{split}
G^{\dagger}_{\theta} \varphi_1  G_{\theta}=\varphi_1   \ct +\varphi_2  \st\\
G^{\dagger}_{\theta}\varphi_2  G_{\theta}=-\varphi_1   \st +\varphi_2  \ct
\end{split}\end{equation}
we may try to implement a similar transformation on the stress-energy tensor, as a function of the fields:
\be\label{GTGinv}
G_{\theta} T_{\mu\nu}  G^{\dagger}_{\theta}=\tilde{T}_{\mu\nu}
\ee
with $\tilde{T}_{\mu\nu} $ an expression that does not involve $G_{\theta}$ explicitly.
This is not a trivial task: the infinite series introduced by $G_{\theta}$ via (\ref{fvGmvpartiv})
appears on the left hand side of (\ref{GTGinv}), whereas on the
right hand side an expression in a closed
form is expected.
However, if our knowledge of the model enables us to find such a closed expression,
we can take advantage of this in the evaluation of (\ref{setpartiv}):
\be
\lfv T_{\mu\nu} \rfv=\lmv G_\theta T_{\mu\nu} G^{\dagger}_\theta \rmv= \lmv \tilde{T}_{\mu\nu} \rmv
\ee
the latter expression being a polynomial function\footnote{We are here considering simple theories in which only
polynomial functions of the fields and their derivatives are involved in the Lagrangian. Nonetheless, further generalizations
of the present discussion are rather straightforward.} of the massive fields $\varphi_i$.
Starting from there, we can decompose the fields in terms of $a^{(\dagger)}_i$,
hence we reduce the quantum algebra in the customary way.
Such a procedure is visualized in the scheme
$$
\xymatrix{
	&\lfv\ar[d]		&				T_{\mu\nu} 	\ar[d]			&					\rfv	\ar[d]&					\\
	&\lmv G_\theta\ar[ddl] \ar@{-->}[dr]	&  \varphi_i 	\ar@{-->}[d]&   G^{\dagger}_\theta \ar@{-->}[dl]	\rmv	\ar[ddr]&	\\
							& &	 		\varphi_\iota \ar[d] 					& &				\\
							\lmv\ar@/_/[ddrr]&&			\varphi_i 		\ar[d]									&&				 \rmv\ar@/^/[ddll]	\\
							  		&&			a_i^{(\dagger)}(\vp)	\ar[d]													&&				 \\
							 		&	&	\lmv  \int d\vp	\;	\mathcal{G}_{\mu\nu}(a_i^{(\dagger)}(\vp))						\rmv		 &&	
							\save "4,1"."4,5"*[F]\frm{}	\restore
}$$
Dashed lines denote the passage from the left hand side of (\ref{GTGinv}) to the right hand side,
for which no general procedure can be stated.
The boxed line denotes the expression (\ref{setpartiv}) when written just in terms of $\rmv$ and the massive fields $\varphi_i$.
As we will see, it will be useful to look at this intermediate step for a deeper understanding of the formalism.

In the last procedure no flavour fields or flavour ladder operators have been invoked at all.
However, the latter are needed to give a physical meaning to (\ref{setpartiv}):
without knowing what a \textit{flavour particle} is,
the expression ``flavour vacuum'' is meaningless.

\section{WZ-Flavour Vacuum Revisited}\label{sWFVR}

We shall now proceed to apply the last method to the supersymmetric model studied in Chapter \ref{cASFV}.

Recalling the discussion in Section \ref{sSUTM},
in the study of the free WZ model we can consider the bosonic and the fermionic component separately,
by evaluating relevant quantities in two separated contexts (a bosonic theory and a fermionic one) and eventually combining together the results.
Furthermore, the pseudoscalar and the scalar field are indistinguishable for our purposes, therefore
we are allowed to consider just the scalar field, keeping in mind to sum its contribution to the relevant quantities twice.

In the real scalar case, we have
\be
T^b_{00}(x)=\sum_i\left(\pi_i^2(x)+\left(\vec{\nabla}\phi_i(x)\right)^2+m_i^2\phi_i^2(x)\right).
\ee
with $\pi_i\equiv\dot{\phi}_i$, the conjugate momentum of $\phi_i$.
Since
\be
G_{\theta}(t)=e^{i\theta\int d\vx \left( \pi_2(x)\phi_1(x)-\pi_1(x)\phi_2(x) \right)}
\ee
from which
\begin{equation}\begin{split}
G_{\theta}(t) \phi_1(x)  G^{\dagger}_{\theta}(t)=G^{\dagger}_{-\theta}(t) \phi_1(x)  G_{-\theta}(t)=\phi_1(x)   \ct -\phi_2(x)  \st\\
G_{\theta}(t)\phi_2(x)  G^{\dagger}_{\theta}(t)=G^{\dagger}_{-\theta}(t)\phi_2(x)  G_{-\theta}(t)=\phi_1(x)   \st +\phi_2(x)  \ct
\end{split}\end{equation}
and
\begin{equation}\begin{split}
G_{\theta}(t) \pi_1(x)  G^{\dagger}_{\theta}(t)=G^{\dagger}_{-\theta}(t) \pi_1(x)  G_{-\theta}(t)=\pi_1(x)   \ct -\pi_2(x)  \st\\
G_{\theta}(t)\pi_2(x)  G^{\dagger}_{\theta}(t)=G^{\dagger}_{-\theta}(t)\pi_2(x)  G_{-\theta}(t)=\pi_1(x)   \st +\pi_2(x)  \ct
\end{split}\end{equation}
we can write
\be
 G_{\theta}(t)   \left(\sum_{i=1,2}\pi_i^2(x)\right)   G_{\theta}^\dagger (t) =   \left(\sum_{i=1,2}\pi_i^2(x)\right)  ,
\ee
\be
 G_{\theta}(t)  \left(\sum_{i=1,2}\left(\vec{\nabla}\phi_i(x)\right)^2\right)   G_{\theta}^\dagger (t) =
   \left(\sum_{i=1,2}\left(\vec{\nabla}\phi_i(x)\right)^2\right)
\ee
and
\begin{multline}
\lmv G_{\theta}(t) (m_1^2\phi_1^2(x)+m_2^2\phi_2^2(x)) G_{\theta}^\dagger (t) \rmv=\\
								\lmv  (m_1^2\phi_1^2(x)+m_2^2\phi_2^2(x))  \rmv+\sst(m_1^2-m_2^2)\lmv(\phi_2^2(x)-\phi_1^2(x))\rmv,
\end{multline}
as shown in Appendix \ref{aC5be}.
It follows that
\be
\lfv T_{00}(x) \rfv=\lmv T_{00}(x)\rmv +\sst(m_1^2-m_2^2)\lmv(\phi_2^2(x)-\phi_1^2(x))\rmv
\ee
and therefore
\be\label{energybospartiv}
\rho^b=\lfv: T_{00}(x) :\rfv=\sst(m_1^2-m_2^2)\lmv(\phi_2^2(x)-\phi_1^2(x))\rmv
\ee

Equivalently for $T_{jj}(x)$ we have:
\be
T^{b}_{jj}(x)= \sum_{i=1,2}\left(
      2(\partial_j \phi_i(x))^2+\pi_i^2(x)-\left(\vec{\nabla}\phi_i(x)\right)^2-m_i^2\phi_i^2(x)\right)
\ee
and consequently:
\begin{multline}
\lmv G_{\theta}(t)   \sum_{i=1,2}\left(2(\partial_j \phi_i(x))^2+\pi_i^2(x)-\left(\vec{\nabla}\phi_i(x)\right)^2\right)  G_{\theta}^\dagger (t) \rmv=\\
			=\lmv\sum_{i=1,2}\left(2(\partial_j \phi_i(x))^2+\pi_i^2(x)-\left(\vec{\nabla}\phi_i(x)\right)^2\right)\rmv
\end{multline}
\begin{multline}
\lmv G_{\theta}(t)   \sum_i(-m_i^2\phi_i^2(x))   G_{\theta}^\dagger (t) \rmv=\\
									=\lmv \sum_i(-m_i^2\phi_i^2(x)) \rmv-\sst(m_1^2-m_2^2)\lmv(\phi_2^2(x)-\phi_1^2(x))\rmv
\end{multline}
as shown in Appendix \ref{aC5bp}.
It follows that
\bea\label{pressurebospartiv}
\pressure^b&=&\lfv: T_{jj}(x) :\rfv=-\sst(m_1^2-m_2^2)\lmv(\phi_2^2(x)-\phi_1^2(x))\rmv=\nn
					&=&-\rho^b.
\eea
Once the fields in (\ref{energybospartiv}) and  (\ref{pressurebospartiv}) are decomposed in terms of the ladder operators and
the quantum algebra is simplified, expressions (\ref{energydensityS}) and (\ref{energydensitypsi}) are correctly reproduced.

In the fermionic case, a similar procedure leads to
\bea\label{energyferpartiv}
\rho^f&=&\lfv: T_{00}^{f}(x):\rfv=\nn
				&=&\sst\left(m_1-m_2\right)\lmv\left( \bpsi_2(x) \psi_2(x)-\bpsi_1(x) \psi_1(x)\right)\rmv
\eea
and
\be\label{pressureferpartiv}
\pressure^f=\lfv: T_{jj}^{f}(x):\rfv =0,
\ee
as proven in Appendices \ref{aC5fe} and \ref{aC5fp}.
By comparing (\ref{energyferpartiv}) and (\ref{energybospartiv}), the analogy between the fermionic and the bosonic condensate that earlier
was hidden in formulae (\ref{energydensityS}) and (\ref{energydensitypsi}) is now more evident.
Again, formula (\ref{energydensitypsi}) is correctly reproduced, once
the operatorial structure of the fields is simplified with respect to $\lmv \slot \rmv$.
The expression (\ref{energyferpartiv}) dispels any doubts concerning formula (\ref{energydensitypsi}) and its possible dependency on the
specific form of the gamma matrices and spinors used to achieve the results of Chapter \ref{cASFV},
being (\ref{energyferpartiv}) independent of such a choice \cite{Peskin:1995ev}.

Furthermore, Supersymmetry enables us to rewrite this result in terms of the bosonic fields only.
For the \textit{massive} vacuum $\rmv$ we know that
\be
\lmv T_{\mu\nu}(x)\rmv=0
\ee
which leads to (Appendix \ref{aC5fe}) 
\be
\lmv \bpsi_i(x) \psi_i(x) \rmv=-4m_i\lmv \phi^2_i(x)\rmv
\ee
and hence
\be
\rho^f=4\sst(m_1-m_2)\lmv\left(m_1\phi_1^2(x)-m_2\phi_2^2(x)\right)\rmv.
\ee

Combining the previous results into the full WZ model, we can write
\begin{multline}
\rho^{WZ}=2\sst(m_1-m_2)\Big[(m_1+m_2)\lmv(\phi_2^2(x)-\phi_1^2(x))\rmv+\\
						+\lmv\left( \bpsi_2(x) \psi_2(x)-\bpsi_1(x) \psi_1(x)\right)\rmv\Big]=\\
					 =2\sst(m_1-m_2)^2\lmv \left(\phi_1^2(x)+\phi_2^2(x)\right)\rmv
\end{multline}
\be
\pressure^{WZ}=-2\sst(m_1^2-m_2^2)\lmv(\phi_2^2(x)-\phi_1^2(x))\rmv.
\ee

\section{Applicability conditions}\label{sAC}

In order to understand limits and conditions of applicability of the method, we introduce the symbol\newnot{symbol:fnormal} $\fnormal{\slot}$,
defined as:
\be\label{fvevdef}
\fnormal{f(\varphi_1,\varphi_2)}\equiv f(\ct \varphi_1-\st \varphi_2,\st\varphi_1+\ct\varphi_2)- f(\varphi_1,\varphi_2)
\ee
with $f(\slot)$ an arbitrary function of the fields $\varphi_1$ and $\varphi_2$,
that might also involve their derivatives.
The linearity of $\fnormal{\slot}$ follows straightforwardly from its definition:
\be
\fnormal{\alpha f(\varphi_1,\varphi_2)+\beta g(\varphi_1,\varphi_2)}=\alpha \fnormal{f(\varphi_1,\varphi_2)}+\beta\fnormal{g(\varphi_1,\varphi_2)}
\ee
with $\alpha,\beta\in\field{C}$.
We can also notice that
\be
\fnormal{f(\varphi_1,\varphi_2)}=\st \left(-\varphi_1 f^{(0,1)}(\varphi_1,\varphi_2)+\varphi_2 f^{(1,0)}(\varphi_1,\varphi_2)  \right)+\mathcal{O}(\sst)
\ee
with $f^{(0,1)}(x,y)\equiv\partial_x f(x,y)$ and $f^{(1,0)}(x,y)\equiv \partial_y f(x,y)$.
It follows that
\be
\fnormal{\slot}=0\;\;\;\;\;\;\;\;if\;\;\;\;\theta=0.
\ee
Furthermore, it is possible to prove that for certain specific $f(\slot)$
\be\label{fvevequalnormfvev}
\lfv: f(\varphi_1,\varphi_2): \rfv=\fvev{f(\varphi_1,\varphi_2)}
\ee
holds.
For instance, this is true for polynomial functions:
\be\label{examplepoly}
f(\varphi_1,\varphi_2)=\sum_{n,m}c_{nm}\varphi_1^n\varphi_2^m
\ee
with $c_{nm}\in\field{C}$ and $n,m\in\field{N}$, as it is easy to show
\bea
\lfv f(\varphi_1,\varphi_2) \rfv&=&\lfv \sum_{n,m}c_{nm}\varphi_1^n\varphi_2^m \rfv\nn
													&=&\lmv G_{\theta}\sum_{n,m}c_{nm}\varphi_1^n\varphi_2^m G^\dagger_{\theta} \rmv=\nn
													&=&\lmv \sum_{n,m}c_{nm}(G_{\theta}\varphi_1G_{\theta}^\dagger)^n(G_{\theta}\varphi_2 G^\dagger_{\theta})^m \rmv=\nn
													&=&\lmv f(G_{\theta}\varphi_1G_{\theta}^\dagger,G_{\theta}\varphi_2G_{\theta}^\dagger)\rmv=\nn
													&=&\lmv f(\ct \varphi_1-\st \varphi_2,\st\varphi_1+\ct\varphi_2)\rmv=\nn
													&=&\fvev{f(\varphi_1,\varphi_2)}+\lmv f(\varphi_1,\varphi_2)\rmv,
\eea
or for expressions involving spatial derivatives, such as
\be\label{exampleder}
f(\varphi_1,\varphi_2)=\sum_{n,m} c_{nm}\partial_x \varphi_1^n\partial_x \varphi_2^m
\ee
when $\partial_x G_{\theta}=\partial_y G_{\theta}=\partial_z G_{\theta}=0$:
\bea
\lfv f(\varphi_1,\varphi_2) \rfv&=&\lmv G_{\theta}\sum_{n,m} c_{nm}\partial_x \varphi_1^n\partial_x \varphi_2^m G^\dagger_{\theta} \rmv=\\
							&=&\lmv \sum_{n,m} c_{nm}\partial_x (G_{\theta}\varphi_1G_{\theta}^\dagger)^n\partial_x (G_{\theta}\varphi_2G^\dagger_{\theta})^m  \rmv=\\
													&=&\lmv f(G_{\theta}\varphi_1G_{\theta}^\dagger,G_{\theta}\varphi_2G_{\theta}^\dagger)\\
													&=&\fvev{f(\varphi_1,\varphi_2)}+\lmv f(\varphi_1,\varphi_2)\rmv,
\eea

The method above exemplified simply reduces to write the stress-energy tensor in such a way that
\be\label{applicabilitycondition}
\lfv :T_{\mu\nu}: \rfv=\fvev{T_{\mu\nu}}
\ee
holds. The applicability of the method coincides therefore with the applicability of (\ref{applicabilitycondition}) to the
specific theory. It should be noticed that $T_{\mu\nu}$ can be written in several different ways,
using properties of the fields and equations of motion. Starting from simple known cases, such as (\ref{examplepoly}) and (\ref{exampleder}),
we might want to reduce a more complicated expression of $T_{\mu\nu}$ in terms of them.
For instance, when $f(\slot)$ contains $\partial_t\varphi_i$, (\ref{fvevequalnormfvev}) might not be true in general, since
$\partial_tG_{\theta}(t)\neq0$.
In the bosonic case, we showed explicitly that
\be
\lfv:\dot{\phi_i}\dot{\phi_i}:\rfv=\fvev{\dot{\phi_i}\dot{\phi_i}}
\ee
but the same proof cannot be implemented for the fermionic case as well.
However, a manipulation of the terms in the stress-energy tensor containing $\dot{\psi}$
leads to an expression involving only spatial derivatives and polynomials of $\psi$, for which
(\ref{fvevequalnormfvev}) holds.
More precisely, the fermionic energy is given by
\be
T_{00}=i\sum_{i=1,2}\bpsi_i\gamma_0\dot{\psi}_i
\ee
but since
\be
i\gamma_0\dot{\psi}_i=i\vec{\gamma}\cdot\vec{\nabla}\psi_i+m_i\psi_i
\ee
we can write
\be\label{examplefermionicenergy}
T_{00}=\sum_{i=1,2}\bpsi_i\left(i\vec{\gamma}\cdot\vec{\nabla}+m_i\right)\psi_i
\ee
for which (\ref{applicabilitycondition}) holds.

The notation $\fvev{\slot}$ is also useful to distinguish terms that effectively contributes to the final results.
For instance, starting from (\ref{examplefermionicenergy}):
\bea\label{relevantterms}
\lfv:T_{00}:\rfv&=&\fvev{\sum_{i=1,2}\bpsi_i\left(i\vec{\gamma}\cdot\vec{\nabla}+m_i\right)\psi_i}=\nn
								&=&\fvev{\sum_{i=1,2}i\bpsi_i\vec{\gamma}\cdot\vec{\nabla}\bpsi_i}+\fvev{\sum_i m_i\bpsi_i\psi_i}=\nn
								&=&\fvev{\sum_i m_i\bpsi_i\psi_i}
\eea
being
\be
\fvev{\sum_{i=1,2}\left(i\vec{\gamma}\cdot\vec{\nabla}\psi_i\right)}=0.
\ee
By applying definition (\ref{fvevdef}), the term $\fvev{\sum_i m_i\bpsi_i\psi_i}$ reduces to the second expression
of (\ref{energyferpartiv}) correctly.

\section{Advantages}\label{sA}

Besides providing us with nice formulae in terms of the empty vacuum $\rmv$ and the massive fields $\varphi_i$,
the method of calculation just exposed presents several practical advantages.

We already mentioned that once the energy of the flavour vacuum for two different theories (real scalar and Majorana)
is written in terms of the fields, similarities between them emerge quite naturally (compare formulae (\ref{energybospartiv})
and (\ref{energyferpartiv}), with respect to (\ref{energydensityS}) and (\ref{energydensitypsi})).
Moreover, formula (\ref{energyferpartiv}) is manifestly independent of the a specific representation of the gamma
matrices and the explicit form of the spinors involved in calculations of (\ref{energydensityS}).

However, the most important advantage of the method relies, perhaps, on the total number of calculations required.
As an illustrative and quite informative example, we can compare it with the second method exposed in the Section \ref{sGI}, in
evaluating the energy of the flavour vacuum in the scalar theory.
By simply following the summary scheme of the second method, using the relevant formulae ((\ref{set2}), (\ref{psimass}),
(\ref{smass}), (\ref{pmass}), (\ref{fromflavourladdertomassiveladder1}),
(\ref{fromflavourladdertomassiveladder2}) and (\ref{fromflavourladdertomassiveladder3})),
we end up with an expression in the form
\be\label{questoquiii}
\lfv \int d\vp \mathcal{T}_{00}(a_\iota^{(\dagger)}(\vp))\rfv.
\ee
More specifically, $\mathcal{T}_{00}$ is bilinear in the ladder operators, therefore it can be written as
\be
\lfv \int d\vp d\vq \sum_{\iota\kappa} c_{\iota \kappa}(\vp,\vq)a_\iota^{(\dagger)}(\vp)a_\kappa^{(\dagger)}(\vq)\rfv.
\ee
This expression counts $2\times3\times2\times2\times3=72$ addenda\footnote{When $T_{\mu\nu}$ is written in terms of the fields, we
have 6 addenda, each of them counting 2 fields (formula (\ref{set2})); when the fields are written in terms of the ladder operators,
each field doubles the addenda (formulae (\ref{psimass}), (\ref{smass}) and (\ref{pmass})); when the flavour ladder fields operators are considered, we triple the addenda,
being each massive ladder operator function of three addenda involving flavour operators
(formula (\ref{fromflavourladdertomassiveladder1}),
(\ref{fromflavourladdertomassiveladder2}) and (\ref{fromflavourladdertomassiveladder3})).
We are here neglecting ``on the way'' simplifications, due to symmetries between different terms,
that we suppose to occur equally proportioned in both procedures.}
in the form $\lfv \slot a_\iota^{(\dagger)}a_\kappa^{(\dagger)} \rfv$.
On the other hand, the corresponding expression (\ref{energyferpartiv}), just counts $2\times2\times2=8$ terms.
The effort of finding such a simpler expression is largely compensated by the much lower number of terms
to evaluate and simplify in the last place. Indeed those terms can be quite complicated, especially
in the fermionic case, in which they involve spinors and matrices: their evaluation usually represents the majority of the
total work. Moreover, in more realistic models with three flavours, the number of those terms
increases significantly and their explicit evaluation starts becoming prohibitive.

As already pointed out, the method does not require an explicit decomposition of the flavour fields in terms
of flavour ladder operators. Such a decomposition has been object of a debate in literature,
raised by the authors of \cite{Fujii:1998xa}.
Although the problem was exhaustively discussed in \cite{Blasone:1999jb}
and \cite{Fujii:2001zv}, not all the community was convinced by the arguments presented \cite{Giunti:2003dg}.
Without entering into the details of the dispute, here we would like to suggest that a different point of view on the formalism,
such as the one offered by formulae (\ref{energyferpartiv}) and (\ref{energybospartiv}),
where an observable quantity concerning a flavour state has been calculated
without the explicit use of the controversial decomposition, might help in a deeper understanding of the problem
and the formalism itself.

The procedure exemplified in the previous section can be easily implemented for other fields:
one might want to consider Dirac or two component Weyl spinors as well as complex scalar fields, getting to analogous results.
For mere speculative reasons, applications to vector fields or even more complex objects might be thought:
the method involves a manipulation of the stress-energy tensor, with the use of equation of motion of the field and
its (anti-)commutation rules, regardless of the tensorial or spinorial structure of the field itself.

Furthermore, by looking at (\ref{relevantterms}), we can distinguish among all the terms of the stress-energy tensors the ones
that really contribute to the final result. This might be helpful in constructing interactive toy models,
to test the properties of the flavour vacuum under different assumptions, as we shall see in the forthcoming section.

\section{Towards Interactive Flavour Vacua}\label{sTIFV}

\subsubsection{Self-interactive Bosons}

A neat example of the applications above discussed is offered by a $\lambda \phi^4$ model.
The theory
\be\label{phiforthlinear}
\lag =\sum_{i=1,2}\left(\partial_\mu \phi_i \partial^\mu \phi_i-m_i^2 \phi_i^2-\lambda \phi_i^4\right)
\ee
can be regarded as derived from a model with flavour mixing:
\be\label{flavourphiforth}
\lag = \partial_\mu \phi_A \partial^\mu \phi_A+\partial_\mu \phi_B \partial^\mu \phi_B-m_A^2 \phi_A^2-m_B^2 \phi_B^2-m_{AB}^2\phi_A\phi_B
			-\!\!\!\!\!\!\!\!\sum_{\iota,\kappa,\lambda,\rho=A,B}\!\!\!\!\!\!\!\!
									g_{\iota\kappa\lambda\rho} \phi_\iota \phi_\kappa \phi_\lambda \phi_\rho
\ee
with the usual rotation
\be\label{usualrotation}\begin{split}
\phi_A=\ct \phi_1-\st \phi_2\\
\phi_B=\st \phi_1+\ct \phi_2
\end{split}\ee
and a specific choice of the coupling constants $g_{\slot}$.
Since the expression of $G_{\theta}$ in terms of the fields can be deduced from
\be\begin{split}
G_{\theta}^{\dagger}\phi_1G_{\theta}=\ct \phi_1-\st \phi_2\\
G_{\theta}^{\dagger}\phi_2G_{\theta}=\st \phi_1+\ct \phi_2
\end{split}\ee
just using commutation relations between fields and conjugate momenta,
that are not modified by the form of the Lagrangian \cite{Peskin:1995ev},
expression
\be
G_{\theta}=e^{i\theta\int d\vx (\dot{\phi}_2 \phi_1-\dot{\phi}_1 \phi_2)}
\ee
that was found valid in the free case, holds also in the interactive one.

If we \textit{assume} that the flavour vacuum is defined as
\be
\rfv\equiv G^{\dagger}_{\theta}\rmv
\ee
na\"ively generalizing the free case, with $\rmv$ the ground state of the theory described by (\ref{phiforthlinear}),
we can easily see that
\bea
\lfv T_{\mu\nu} \rfv&=&\lfv \sum_i\left( 2\partial_\mu \phi_i \partial_\nu \phi_i-\e_{\mu\nu}\lag \right)\rfv=\nn
										&=&\lfv\left(T^{free}_{\mu\nu}-\e_{\mu\nu}\sum_i\lambda_i\phi_i^4 \right)\rfv=\nn
										&=&\lmv T_{\mu\nu} \rmv+\e_{\mu\nu}\left(\sst (m_1^2-m_2^2)\lmv \phi_2^2-\phi_1^2\rmv\right)+\nn
										& &+\e_{\mu\nu}\lambda \lmv\left( -\phi_1^4-\phi_2^4+(\phi_2 \cos (\theta)-\phi_1 \sin (\theta ))^4+\right.\nn
										& &+\left.(\phi_1 \cos (\theta)+\phi_2 \sin (\theta ))^4  \right)\rmv
\eea
in which we used
\be\label{freesetbosonicfvev}
\lfv T^{free}_{\mu\nu}\rfv=\lmv T^{free}_{\mu\nu} \rmv+\e_{\mu\nu}\left(\sst (m_1^2-m_2^2)\lmv \phi_2^2-\phi_1^2\rmv\right)
\ee
that is possible to recover by following the same exact steps of the free case (Section \ref{sWFVR}).\footnote{A
remark on the notation is in order: in this Section $T^{free}_{\mu\nu}$ denotes the functional expression of the
stress-energy tensor in terms of the fields in the free theory. However, one should not forget that the field
itself, of which is not a \textit{free} field, as well as $\rmv$ here does not represent the \textit{free} vacuum.}
We can therefore state that the equation of state is given by
\be\label{wphiforth}
w=\frac{\lfv : T_{jj}:\rfv}{\lfv : T_{00}:\rfv}=\frac{-\fvev{\sum_{i=1,2}(m_i^2\phi_i^2+\lambda\phi_i^4)}}{\fvev{\sum_{i=1,2}(m_i^2\phi_i^2+\lambda\phi_i^4)}}=-1
\ee

Quite notably, this result generalizes the analogous result for the free theory,
in a completely \textit{non-perturbative} way:
equation (\ref{wphiforth}) is independent of the explicit form of the fields, that we might be able to recover
just in a perturbative treatment of the model.

In fact, it is possible the further generalize the above result for \textit{any} interactive theory for two scalar fields with flavour mixing.
If we consider
\be
\lag = \partial_\mu \phi_A \partial^\mu \phi_A+\partial_\mu \phi_B \partial^\mu \phi_B-m_A^2 \phi_A^2-m_B^2 \phi_B^2-m_{AB}^2\phi_A\phi_B
			+\lag_{int}(\phi_A,\phi_B)
\ee
with $\lag(\phi_A,\phi_B)$ any polynomial function of $\phi_A$ and $\phi_B$, we can write
\bea
\lfv :T_{\mu\nu}: \rfv&=&\lfv :T^{free}_{\mu\nu}: \rfv-\e_{\mu\nu}\;\lfv :\lag_{int}: \rfv=\nn
										&=&\e_{\mu\nu}\lmv \fnormal{\sum_{i=1,2}m_i^2 \phi^2_i}\rmv -\e_{\mu\nu}\;\lfv: \lag_{int}: \rfv=\nn
										&=&\e_{\mu\nu}\lmv \fnormal{\sum_{i=1,2}m_i^2 \phi^2_i}\rmv -\e_{\mu\nu}\lmv \fnormal{\lag_{int}}\rmv=\nn
										&=&\e_{\mu\nu}\lmv \fnormal{\sum_{i=1,2}m_i^2 \phi^2_i-\lag_{int}}\rmv
\eea
in which we used (\ref{freesetbosonicfvev}),
(\ref{fvevequalnormfvev}), (\ref{examplepoly}) and (\ref{usualrotation}),
leading to the equation of state
\be
w=-1.
\ee

\subsubsection{Self-interactive fermions}

Analogously, we can generalize the result presented in Section \ref{sWFVR} for fermionic fields (namely, $w=0$)
for a certain class of self-interactive theories.
We start by considering a theory written in terms of the massive fields $\psi_1$ and $\psi_2$:
\be\label{lag}
\lag=\sum_i \bpsi_i (i\diracpartial-m_i)\psi_i+\lag_{int}
\ee
with $\lag_{int}$ a suitable polynomial function of $\psi_i$ and $\bpsi_i$.
Again, we regard (\ref{lag}) as the \textit{diagonalized} Lagrangian:
in case of flavour mixing, $\psi_1$ and $\psi_2$ come from a rotation of the flavoured fields $\psi_A$ and $\psi_B$.
The equations of motion
\be\label{interactiveeom}
(i\diracpartial-m_i)\psi_i=\gamma^0 \left[L_{int},\psi_i\right]
\ee
can be derived from Heisenberg equation
\be
\dot{\psi}_i=i\left[H,\psi_i\right]
\ee
with $H=H_{free}-L_{int}$ the Hamiltonian operator and $L_{int}\equiv \int d^4 x\lag_{int}$.

Recalling that the stress-energy tensor is written as
\be
T_{\mu\nu}=\sum_i i\bpsi_i \gamma_{\left(\mu\right.}\partial_{\left.\nu\right)}\psi_i-\e_{\mu\nu}\lag,
\ee
its $00$ component is given by
\be
T_{00}=\sum_i\left(i\bpsi_i\gamma_0 \partial_0\psi_i\right)-\e_{00}\lag
\ee
which on-shell can be written as
\be
T_{00}=\sum_i \bpsi_i\left(i\vec{\gamma}\cdot\vec{\partial}+m_i\right)\psi_i+\lag_{int}
\ee
via (\ref{interactiveeom}).

Combining our previous discussion on the bosonic case and results of Section \ref{sAC},
we can write
\be
\lfv : T_{00} : \rfv= \fvev{T_{00}} =\fvev{\sum_i m_i \bpsi_i \psi_i+\lag_{int}}
\ee
in which we used
\be	
\fnormal{\bpsi_i \vec{\gamma}\cdot\vec{\partial} \psi_i}=0.
\ee

Analogously, the $jj\neq 00$ component of the stress energy tensor is given by
\be
T_{jj}=\sum_i\left(\bpsi_i\gamma_j \partial_j\psi_i\right)-\e_{jj}\lag
\ee
that on-shell can be written as
\be
T_{jj}=\sum_i\left(\bpsi_i\gamma_j \partial_j\psi_i+\psi^\dagger_i\left[L_{int},\psi_i\right]\right)+\lag_{int}
\ee
leading to
\be
\lfv : T_{jj} : \rfv= \fvev{T_{jj}} =\fvev{\sum_i \bpsi_i \left[L_{int},\psi_i \right]+\lag_{int}}
\ee

We can now distinguish two cases:
\begin{enumerate}
	\item the interactive term of the Lagrangian $\lag_{int}$ (and consequently $L_{int}$) is invariant under the transformation
				\begin{equation}\label{su2transformation}\begin{split}
				\psi_1\rightarrow&\ct \psi_1-\st \psi_2\\
				\psi_2\rightarrow&\st \psi_1+\ct \psi_2
				\end{split}\end{equation}
	\item $\lag_{int}$ is not invariant under (\ref{su2transformation}).
\end{enumerate}

If $\lag_{int}$ \textit{is} invariant, we can then write
\be\label{conditionselfinteraction}
\fnormal{\lag_{int}}=\fnormal{L_{int}}=\fnormal{\sum_i\bpsi_i\left[L_{int},\psi_i\right]}=0
\ee
leading to 
\be
\lfv : T_{jj} : \rfv=0
\ee
and hence
\be
w=0.
\ee
Therefore, we can argue that the pressureless character of the fermionic condensate is preserved if one considers
self-interactions that satisfy condition (\ref{conditionselfinteraction}).

If $\lag_{int}$ \textit{is not} invariant,
we cannot push our analysis further and we are unable to decide whether the pressure is zero or not,
provided just with the tools here presented.
It should be emphasized that other cancellation mechanisms might occur, leading to a full generalization of $w=0$
for \textit{all} self-interactive cases, just like in the bosonic case.
However, these mechanisms are not reproduced within our method.

\subsubsection{Final Remarks}

To conclude the discussion of these examples, a few remarks are in order.
Throughout our analysis we assumed that the flavour vacuum was \textit{defined} by
\be\label{interactivefv}
\rfv\equiv G^{\dagger}_{\theta}\rmv
\ee
with $G_{\theta}$ the operator mapping flavour fields into massive fields, and vice versa,
and $\rmv$ being the massive vacuum state of the interactive theory.
The derivation of our results was purely formal and did not require any other knowledges of the theory.
Nonetheless, although it might look reasonable, the assumption (\ref{interactivefv}) remains a mere guess
in absence of a complete (either perturbative or non-perturbative) interactive theory.

An interactive theory is a rather different object than a free one, from a non-perturbative level.
In Section \ref{ssTI}, we already mentioned that the usual Fock space \hilbert{F_0}
is not sufficient for fully describing the theory.
More generally we can say that in the framework of Second Quantization few
progresses on a coherent definition of the theory have been made so far \cite{Streater:1989vi}, and
the explicit construction of physical states in interactive theories
still represents an open issue (\textit{cf} Section 3.1 of \cite{Borne:2002} and references therein).

Moreover, the familiar Perturbation Theory scheme,
in the formulation of Lehmann, Symanzik and Zimmermann \cite{Lehmann:1954rq},
is thought specifically for scattering processes and it might be unfit for describing
the \textit{flavour vacuum}. Since it relies on the assumption that particles
are free at early and late times,
all relevant quantities (scattering probabilities)
are expressed in terms of time ordered products of field acting on the vacuum of the free theory $\rmv$,
which is suppose to coincide with the true vacuum of theory at early and late times.
However, the features of the flavour vacuum are not expressed in these terms, \textit{i.e.}
as probabilities of having certain states at late times, given some initial conditions.

It follows that implementing BV formalism on interactive theories is not a trivial task and requires
very much care.
Such a generalization is not among the aims of the present work.
However, the purpose of this Section was to indicate a possible path for further developments of the formalism,
taking advantage of the method of calculation discussed so far.
Although an interactive theory might suffer from serious problems
when it comes to construct particle states, as above mentioned, we believe that
certain quantities, such as the equation of state of the flavour vacuum,
might not require an explicit expression of such a state.
Our analysis is valid under the assumption that (\ref{interactivefv}) holds,
irrespectively of a detailed knowledge of $\rmv$ or any particle states in
the interactive theory.
Therefore, we might expect to be able to get some features of the phenomenology of the flavour
vacuum, even though the underlying theory is not understood in full detail.
However, a dedicated analysis is in order to fully justify the use of (\ref{interactivefv}).

\chapter{CONCLUSIONS}\label{cC}
The aim of this work was to develop the phenomenology of BV formalism at cosmological scales.
The non-perturbative nature of the formalism requires the use of a subset of \hilbert{H}, the Hilbert space of 
physical states, different than the usual Fock space for massive particles \hilbert{F_0} commonly used in Particle Physics.
As a consequence, the state that represents the physical vacuum is not the usual vacuum state, characterized by zero energy,
but the so called \textit{flavour vacuum}, whose structure is highly non-trivial and depends on the model considered.

This thesis was dedicated to study the features of the \textit{flavour vacuum} in different contexts,
with special attention on possible connections with cosmological problems.
In particular we have been interested in embedding the formalism in a model developed in the context of Braneworld/String Theory,
the D-particle Foam model. 
Generalizing the analysis of \cite{mavrosarkar}, we investigated the features of the flavour vacuum
in a more realistic theory, which included two Dirac fermions with flavour mixing \textit{\`a la} BV in a FRW universe
and the MSW gravitational effect, according to the requests of the underlying microscopical brany model.
In order to identify the correct Hilbert space for physical states,
we required the metric to be flat at early times.
In addition, two approximations has been considered: a small mixing angle and an adiabatic expansion of the universe.
As a result, the equation of state for the flavour vacuum was found to satisfy
\be
-1/3<w<0
\ee
the exact value being fixed by the choice of a specific background metric and a physical cutoff for momenta,
both quantities in principle unambiguously determined by the underlying microscopical model, but in fact unknown.
Such a result, which by itself would lead to a decelerating universe,
was inconsistent with the hypothesis of flavour vacuum as a source of Dark Energy.
Furthermore, it was in contrast with the analogous result for the bosonic model examined in \cite{mavrosarkar},
for which $w=-1$.

Motivated by this discrepancy, a supersymmetric theory (free WZ model) has been studied.
As a first step towards a more complete treatment, a flat universe has been considered.
No approximations have been performed and the equation of state for the flavour vacuum was found to be
\be
-1<w<0
\ee
the precise value depending on the choice of the cutoff.
Two components with distinct features characterize this supersymmetric flavour vacuum,
one arising from the bosonic sector of the model, whereas the second from the fermionic one.
By disentangling these two components,
we found the seeds of the discrepancy occurring in the models on curved spacetime:
the bosonic sector contributes to the total condensate via a component characterized by
\be
w=-1
\ee
whereas for the fermionic component
\be
w=0
\ee
holds.

Since no extra ingredients dictated by the stringy framework, besides Supersymmetry, were present,
these results are quite general and can be regarded as a genuine development of the BV formalism.
A hint for a physical interpretation of the difference in behaviour between the bosonic and the fermionic
component might come from the concept of \textit{degeneracy pressure}: thanks to Pauli exclusion principle,
it is harder to compress a fluid of fermions rather than bosons.
These results seem to indicate a novel mechanism for SUSY breaking, for which the physical vacuum differs from
the theoretical vacuum state, characterized by zero energy, even in presence of a SUSY-invariant Lagrangian.
However, this point remains an open problem and requires further investigations.
What is already clear is that the flavour vacuum is not a Lorentz invariant. This feature is actually
not a surprise, being inherent in $\lfv :T_{\mu\nu}: \rfv\propto\!\!\!\!\!/ \e_{\mu\nu}$.
Furthermore, it should be noticed that the expression that we found for $\lfv T_{\mu\nu}\rfv$ is not
time-dependent, as one might have expected from the fact that $G_{\theta}$ is a time-dependent operator,
holding $[H,G_\theta]\neq 0$ for theories with flavour mixing ($H$ being the total Hamiltonian).\footnote{It
is easy to remove any doubts about this point by considering the following simple example:
if we define our Hamiltonian as
\be
H=\sum_k \w(k) a^\dagger_k a_k
\ee
with $a^{(\dg)}_{\slot}$ respecting the usual CCR algebra, and we consider the time-dependent operator
\be
\tilde{G}_\theta=1+\st \sin t \;a_p
\ee
we then have that $\tilde{G}_\theta \rmv =\rmv$, which is still Lorentz invariant, even though 
\be
[G_\theta,H]=\w(p)a_p\neq 0.
\ee
Time-depended operators applied on the vacuum state do not in general define time-dependent states.}

We have pointed out that the phenomenology coming from gravitational effects is rich and interesting,
the fermionic component of the supersymmetric flavour vacuum acting as Dark Matter,
whereas the bosonic one providing a source for Dark Energy.
This interpretation is actually quite appealing: the flavour vacuum satisfies basic requests for Dark Energy
and Dark Matter candidates; moreover, despite the vast majority of approaches, the model here considered
presents a very limited number of parameters, which in principle can all be determined by observations.
Even better, we explained how in more realistic pictures there might even be the possibility of making a prediction for one of them
(for instance in a renormalized theory, where no cutoff is present).
Finally, despite the huge differences in orders of magnitude of the parameters involved
(ranging from the Plank scale to the mass of the neutrinos, from the density of Dark Matter today and its density at very early times),
a preliminary analysis with ``semi-realistic'' parameters showed
that very different scales can fit in our simple model quite naturally.

A new method of calculation has also been presented. Thanks to it, we were able not only to derive
the above-mentioned results for a supersymmetric theory in a few lines, but we also started developing a treatment for interactive theories
in a completely non-perturbative way.

The possibility here discussed that a source for both Dark Matter and Dark Energy might arise from flavour physics,
whether it derives from new physics beyond the Standard Model or from non-perturbative aspects of QFT, is quite attractive.
However, before any sorts of claim the models here presented need to be understood and developed much further.
In particular, we list a few developments that we consider to be in order:
\begin{itemize}
	\item More realistic theories need to be constructed and analyzed: three flavours, SM/MSSM interactions, a realistic profile for the
				expansion of the universe are the necessary ingredients required.\footnote{A major obstacle in this sense relies on the
				current problem of particle physicists of embedding massive neutrinos in the SM. No to mention difficulties to define
				chiral fermions as neutrinos, even massless, in supersymmetric contexts \cite{Sohnius:1985qm}.}
	\item Renormalization techniques should be applied to the problem. One might want to implement BV formalism in the
				framework of the path integral formulation of QFT, in order to use tools from the usual perturbation theory.\footnote{The
				problem of defining asymptotic states might be avoided by
				assuming that free states are boundary conditions for the interactive theory at \textit{finite} times $t_{-}$
				and $t_{+}$ and then taking the limit $t_{\pm}\rightarrow \pm\infty$. Despite flavour states,
				it is reasonable to expect the features of the flavour vacuum to be well defined in that limit.}
				However, other tools might be more effective (see for instance \cite{calzetta2008nonequilibrium}). In particular,
				techniques developed for QFT in curved spacetime might be more appropriate \cite{Ford:1997hb}.\footnote{The
				equivalence of the problem from a formal point of view is rather straightforward: 
				curved spacetime techniques have been developed specifically for evaluating the expectation value of the stress-energy tensor
				with respect to a vacuum that is not part of \hilbert{F_0}.}
				Zeta function regularizations developed for
				the Casimir effect \cite{Bytsenko:2003tu} might also be a suitable choice, for the closeness of the formalisms.
	\item How can one probe the features of the flavour vacuum, besides gravitational effects?
				The average number of particles that are present in such a state is always zero, therefore one might think that
				scattering processes would not be helpful. However, the possible role of the flavour vacuum
				in scattering processes needs a dedicated analysis, perhaps involving non-perturbative tools specifically developed
				in order to not lose the non-perturbative character of BV formalism.
	\item In order to corroborate the interpretation of the fermionic component of the condensate as a source of Dark Matter
				further investigations are in order. Above all, the hypothesis that gravitational instability leads the condensate
				to cluster in ways that are in agreement with observational data on Dark Matter densities must be verified.
				Moreover, does a different distribution of the flavour vacuum energy density affect neutrino oscillation rates?
				And if so, would that be observable?
	\item A deeper understanding of the SUSY breaking is in order: is there \textit{any} SUSY breaking, in the usual sense?
				What are the observable consequences of considering the flavour vacuum as the physical ground state for SUSY?
				Does it cause any mechanism for generating a split of the spectrum between fermions and bosons?
	\item Quarks are not contributing to the flavour vacuum in the context of the D-particle Foam Model, because
				only neutral strings can interact with D-particles of the foam. However, if we relax this condition,
				what is their contribution to the flavour vacuum?\footnote{Quark-flavour-vacuum would not contribute
				to DM, presumably: because of the electromagnetic interaction of quarks, the corresponding flavour vacuum
				might not be ``dark''.}
	\item Looking at the flavour vacuum as it has been defined so far, can we recognize any clear signature of physics beyond Standard Model
				or even at the Planck scale? The relationship between BV formalism and the D-particle Foam Model needs be explored, clarified
				and supported with more arguments.
\end{itemize}
Despite the many questions that remain unanswered, the models presented in this work suggest an interesting possibility
for a deeper understanding of fundamental problems in cosmology. The promising results here discussed certainly motivate
further developments of the approach.


\appendix
\makeatletter
\def\thickhrulefill{\leavevmode \leaders \hrule height 1ex \hfill \kern \z@}
\def\@makechapterhead#1{%
  \vspace*{10\p@}%
  {\parindent \z@ 
        \raggedright
        \reset@font\huge\bfseries
          \Huge #1
        \par\nobreak
    \vskip 100\p@
  }}
\makeatother
\chapter{APPENDICES}
\pagestyle{myheadings}
\markright{\textit{APPENDICES}}
\section*{INTRODUCTION}\addcontentsline{toc}{section}{\textit{Introduction}}
The following appendices include most relevant calculations needed to achieve the results presented in the previous chapters.
In particular, Appendix \ref{aC23} completes Chapters \ref{cB} and \ref{cAFVICS},
Appendices \ref{aC4} and \ref{aC5} refer to Chapters \ref{cAFVIASM} and \ref{cANMOF},
respectively.

As already mentioned, two slightly different notations
have been used through the thesis (\textit{cf} section \textbf{Notation}, on page \pageref{section:notation}).
It follows that in Appendix \ref{aC23} (which completes the analysis of the model in curved spacetime) 
the notation of Chapter \ref{cAFVICS} and works in \cite{mavrosarkar,Birrell:1982ix,Parker:1971pt,Blasone:1995zc} is adopted,
whereas in Appendices \ref{aC4} and \ref{aC5} (which refers to the supersymmetric model)
the notation of Chapters \ref{cAFVIASM} and \ref{cANMOF}, and works in
\cite{Sohnius:1985qm,FigueroaO'Farrill:2001tr,Capolupo:2004av,Blasone:2003hh} is used.

A further remark. All appendices are quite rich of details, perhaps even more than one would require from a standard
appendix.\footnote{Although they
might lack of \textit{other} relevant details, according to someone else's taste!}
A few textbook calculations have also been reproduced. Moreover, in Appendix \ref{aC4} some basic results of
BV formalism for real fields (scalar and Majorana spinor)
have been derived from scratch: the operator $G_{\theta}(t)$, which maps mass eigenstates
into flavour eigenstates, in terms of field operators,
and the flavour ladder operators\newnot{symbol:FLO} (FLO), as simple expressions in terms of massive ladder operators\newnot{symbol:MLO} (MLO),
have been calculated.

These results were already present in literature, even though the full derivation was missing (as one would rightly expect from
works that are not reviews or pedagogical texts on the subject).
However, we decided to include them in this appendix in order to make Appendices \ref{aC4} and \ref{aC5}
a self-consistent piece of work, for future benefits of someone which aims to enter into the field.
\newpage
\section{APPENDIX TO CHAPTERS \ref{cB} AND~\ref{cAFVICS}}\label{aC23}
\markright{\textit{\thesection \quad APPENDIX TO CHAPTERS \ref{cB} AND~\ref{cAFVICS}}}
\subsectionItalic{Outline}

In this appendix we shall consider two fermionic fields (Dirac spinors) in a FRW-universe with flavour mixing.

In Section \ref{aC23dfifewu} we shall define the theory for one single Dirac field in curved spacetime. 
A more specific metric will be then considered (FRW metric in conformal coordinates) and relevant expressions (stress-energy tensor,
equations of motion, etc.) will be derived for this background.
Details about the decomposition of the field in terms of ladder operators will be provided in Section \ref{aC23fd}.
Flavour mixing and the \textit{flavour vacuum} state will be introduced in Section \ref{aC23fv}.
Everything will be expressed in terms of the massive Fock space \hilbert{F_0} \textit{at early times}
and in \textit{finite volume}.

Provided with all relevant tools, we shall then proceed with calculating the \textit{flavour} vacuum expectation value of the stress energy tensor
$_f\langle T_{\mu\nu}\rangle_f$, in Section \ref{aFVEVOTSET}.
Since both this operator and the flavour vacuum will be given in terms of massive ladder operators \textit{at early times}, we will
first simplify the operatorial structure of $_f\langle T_{\mu\nu}\rangle_f$ and then we will
reduce the remaining spinorial structure.

In order to give a physical interpretation of the evaluated quantities, we will finally relate the energy density and the pressure
of a perfect classical fluid with the stress-energy tensor, in conformal coordinates (Section \ref{sectioneom}).
\subsectionItalic{Notation}\label{aC23n}
\paragraph{Metric} Minkowski metric is chosen to be
\be
\e_{\mu\nu}=diag\{-1,+1,+1,+1\}
\ee
\paragraph{Tensorial indices} In the first part of Section \ref{aC23dfifewu}
Minkowksi metric and a metric for generic curved backgrounds will be present at the same time.
In order to distinguish whether a tensor is written in flat spacetime or in curved we shall adopt to different
notation for tensorial indices: \textit{Latin} indices ({\scriptsize$a,b,c,\dotso$}) will be used for tensors in \textit{flat} spacetime,
whereas \textit{Greek} indices ({\scriptsize$\lambda,\mu,\mu,\dotso$}) will be used for tensors on curved backgrounds.
Therefore, Minkowski metric will be denoted by $\e_{ab}$, whereas a generic metric will be $g_{\mu\nu}$.
In the second part of the Section and all other Sections
(in which just Minkowski metric will appear) only Greek indices will be used (\textit{e.g.} $\e_{\mu\nu}$).
\paragraph{Einstein summation} Tensors follow Einstein summation convention. In the first part of Section \ref{aC23dfifewu}
the type of index determine the summation convention: for \textit{Latin} indices we have
$A_a B^a=A^a B^b \e_{ab}$, while for Greek indices $A_\mu B^\mu=A^\mu B^\nu g_{\mu\nu}$ holds.
In the second part of Section \ref{aC23dfifewu} all other Sections of this appendix it is understood that $A_\mu B^\mu=A^\mu B^\nu \e_{\mu\nu}$.
Summations over spinorial or other sorts of index are explicitly denoted.
\paragraph{Spinors} In decomposing the free Dirac field $\psi$ in flat spacetime, the spinor associated with the annihilation operator
is usually denoted by $u^r(\vp)$, whereas the one associated with the creation operator is denoted by $v^r(\vp)$,
according to standard literature on QFT. Here, it is more convenient to use only one symbol, with an index: $u^{(d,r)}(\vp)$,
being understood that $u^{(1,r)}(\vp)$ corresponds to $u^r(\vp)$, whereas $u^{(2,r)}(\vp)$ corresponds to $v^r(\vp)$
(see Section \ref{aC23fd} for the precise definition).
\paragraph{Operators} In Section \ref{aFVEVOTSET},
because of long expressions involving \textit{q}- and \textit{c}-numbers
at the same time, we shall distinguish the former from the latter by a \textit{hat}:
the expression $\hat{\slot}$ will then univocally identify operators (\textit{q}-numbers).
\paragraph{Gamma matrices} The following representation for Gamma matrices (in flat spacetime) has been chosen:
\be
\gamma^{0}=\begin{pmatrix}-i \mathbb{I}&0\\0&i \mathbb{I}\end{pmatrix}
\;\;\;
\gamma^i=\begin{pmatrix}0&-i\hat{\sigma}_i\\i\hat{\sigma}_i&0\end{pmatrix}
\ee
with
\be
\hat{\sigma}_1=\begin{pmatrix} 0 & 1 \\ 1 & 0 \end{pmatrix}
\;\;\;
\hat{\sigma}_2=\begin{pmatrix} 0 & -i \\ i & 0 \end{pmatrix}
\;\;\;
\hat{\sigma}_3=\begin{pmatrix} 1 & 0 \\ 0 & -1 \end{pmatrix}
\;\;\;
\mathbb{I}=\begin{pmatrix}1&0\\0&1\end{pmatrix}.
\ee

\subsectionItalic{Dirac Fields in FRW Universe}\label{aC23dfifewu}
\subsubsection{First Part: Dirac field in curved space-time}\label{fermioncurved}

 Spinor field theories are generalized in curved spacetime via \textit{Veirbeins formalism} \cite{Birrell:1982ix,weinberg}.
According to the prescriptions of the method, we generalize the action for the flat spacetime
\be
S_{\flat}=\int{d^4 x\left[\frac{1}{2}\left(\bar{\psi}\gamma^a \partial_a \psi-\partial_a \bar{\psi}\gamma^a \psi \right)+m\bar{\psi} \psi\right]}
\ee
where
\begin{itemize}
	\item $\bar{\psi}\equiv \psi^{\dagger}\gamma^4$
	\item $\gamma^4 \equiv i \gamma^{0}$
	\item $\{\gamma^a,\gamma^b\}=2\e^{ab}$
\end{itemize}
with
\be\label{actionvierbein}
S
=\int{d^4 x \det V \left[\frac{1}{2}\left(\bar{\psi}\gamma^a \mathcal{D}_a \psi-\mathcal{D}_a \bar{\psi}\gamma^a \psi \right)+m\bar{\psi} \psi\right]}
\ee
where
\begin{itemize}
	\item the matrices (\textsl{vierbein}) $V^{a}_{\;\;\mu}(x)$ are defined by the relation
				$g_{\mu\nu}=V^{a}_{\;\;\mu}V^{b}_{\;\;\nu}\e_{ab}$,
	\item $\mathcal{D}_a=V_a^{\;\;\mu}D_{\mu}=V_{a}^{\;\;\mu}(\partial_{\mu}+\Gamma_{\mu})$,
	\item $\Gamma_{\mu}$ is the spin connection defined by\newnot{symbol:;cov}
				$\Gamma_{\mu}=\frac{1}{2}\Sigma^{a b} V_{a}^{\;\;\nu}\left(V_{b\nu;\mu}\right)$,
	\item $\Sigma^{a b}$ is the generator of the Lorentz group associated with the spinorial representation under which $\psi$ transforms:
				$\Sigma^{a b}=\frac{1}{4}[\gamma^a,\gamma^b]$,
	\item $V_{b\nu;\mu}\equiv\partial_\mu V_{b\nu}-\Gamma^{\lambda}_{\nu\mu}V_{b\lambda}$,
	\item $V_{b\nu}=\g V_{b}^{\;\;\mu}$,
	\item $\Gamma^{\lambda}_{\nu \mu}\equiv\frac{1}{2}g^{\lambda \kappa}\left(\partial_{\nu}g_{\kappa \mu}+\partial_{\mu}g_{\kappa \nu}
				-\partial_{\kappa}g_{\nu \mu}\right)$ is the Christofell symbol.
\end{itemize}
We can rewrite this action as
\begin{multline}\label{actioncovariant}
S
=\int{d^4 x \det V \left[\frac{1}{2}\left(\bar{\psi}\gamma^a \mathcal{D}_a \psi-\mathcal{D}_a \bar{\psi}\gamma^a \psi \right)+m\bar{\psi} \psi\right]}=\\
=\int{d^4 x \det V \left[\frac{1}{2}\left(\bar{\psi}\gamma^a V_a^{\;\;\mu}D_{\mu} \psi-V_a^{\;\;\mu}D_{\mu} \bar{\psi}\gamma^a \psi \right)+m\bar{\psi} \psi\right]}=\\
=\int{d^4 x \sqrt{-g} \left[\frac{1}{2}\left(\bar{\psi}\gamma^{\mu} D_{\mu} \psi-D_{\mu} \bar{\psi}\gamma^{\mu} \psi \right)+m\bar{\psi} \psi\right]}
\end{multline}
since $\det(g)=\det(VV\e)=-(\det(V))^2$ and by defining the $\gamma$-matrices with greek index as $\gamma^a V_{a}^{\;\;\mu}\equiv \gamma^{\mu}$.
From this position follows that
\be
\{\gamma^{\mu},\gamma^{\nu}\}=\{\gamma^a V_{a}^{\;\;\mu},\gamma^b V_{b}^{\;\;\nu}\}=2 V_{a}^{\;\;\mu} V_{b}^{\;\;\nu} \e^{ab}=2g^{\mu\nu}.
\ee

The Lagrangian can be therefore written as
\be\label{eccolalagrangiana}
\lag=\frac{1}{2}\bar{\psi}\left( \gamma_\mu D_\mu m \right) \psi
\ee
that is obtained combining $S=\int d\vx \sqrt{-det(g)}\; \lag$ and (\ref{actioncovariant}).
Equations of motion read
\be\label{eccolequazionedelmoto}
\gamma^\mu D_\mu \psi+m\psi=0
\ee
whereas the stress-energy tensor is given by \cite{Birrell:1982ix}
\be\label{setcurved2bis}
T_{\mu\nu}=-g_{\mu\nu} \lag+\frac{1}{2}\left(   \bar{\psi}\gamma_\mu D_\nu  \psi-  D_\nu \bar{\psi}\gamma_\mu \psi   \right).
\ee
If we take into account equations of motion (\ref{eccolequazionedelmoto}) and (\ref{eccolalagrangiana}),
we can simplify (\ref{setcurved2bis}) to
\be\label{setcurved2}
T_{\mu\nu}=\frac{1}{2}\left(   \bar{\psi}\gamma_\mu D_\nu  \psi-  D_\nu \bar{\psi}\gamma_\mu \psi   \right)
\ee
that is valid \textit{on-shell}\footnote{We should recall the reader that in all our calculations fields are always considered 
\textit{on-shell}, \textit{i.e.} as solution of equations of motion.}.

\subsubsection{Second Part: Dirac field in a FRW Universe}\label{aDIFRW}

We want now to calculate some of the expressions defined in the previous section in case of a FRW universe.
In particular, we will use the following metric, in conformal time $g_{\mu\nu}=\Ce \e_{\mu\nu}$.
\subsection*{$\eqbox{\Gamma^{\lambda}_{\nu\mu}}$}

\bea
\Gamma^{\lambda}_{\nu\mu}&=&\frac{1}{2}\frac{1}{\Ce}\e^{\lambda \kappa}\left[\partial_\nu (\Ce \e_{\kappa \mu})+\partial_\mu (\Ce \e_{\kappa \nu})
														-\partial_\kappa (\Ce \e_{\nu \mu})   \right]=\nn
												 &=&\frac{1}{2}\frac{\Cpe}{\Ce}	\e^{\lambda \kappa}\left[\delta^0_\nu \e_{\kappa\mu}+\delta^0_\mu \e_{\kappa\nu}
												    -\delta^0_\kappa \e_{\nu\mu}\right]=\nn
												 &=&\frac{1}{2}\frac{\Cpe}{\Ce} \left[ \delta^0_\nu \delta^\lambda_\mu+\delta^0_\mu \delta^\lambda_\nu
												 		-\e^{\lambda 0}\e_{\nu\mu}\right]
\eea
using $g_{\mu\nu}=\Ce \e_{\mu\nu}$, $g^{\mu\nu}=\Ce^{-1} \e^{\mu\nu}$ and $\e^{\alpha \beta}\e_{\beta \gamma}=\delta^\alpha_\gamma$.
\bea
\Gamma^0_{00}&=&\frac{1}{2}\frac{\Cpe}{\Ce}(2-1)\;\;=\;\;\frac{1}{2}\frac{\Cpe}{\Ce}\nn
\Gamma^0_{j0}&=&\frac{1}{2}\frac{\Cpe}{\Ce}(0+0-0)\;\;=\;\;0\;\;\;\Gamma^0_{0i}\;\;=\;\;0\nn
\Gamma^0_{jj}&=&\frac{1}{2}\frac{\Cpe}{\Ce}(+1)\;\;=\;\;\frac{1}{2}\frac{\Cpe}{\Ce}\nn
\Gamma^i_{j0}&=&\Gamma^i_{0i}\;\;=\;\;0\;\;\;\Gamma^i_{jj}\;\;=\;\;0
\eea
(no sum over $j$, and $j\neq0$).

\subsection*{$\eqbox{V^a_{\;\;\nu}}$}
\be
\left\{\ba{rcl}g_{\mu\nu}&=&\Ce \e_{\mu\nu}=\Ce \delta^a_\mu \delta^b_\nu \e_{ab}\\
							 g_{\mu\nu}&=&V^a_{\;\;\mu} V^b_{\;\;\nu}
			 \ea\right. \Rightarrow
V^a_{\;\;\mu}=\delta^a_{\mu}\sqrt{\Ce}
\ee
\be
V^a_{\;\;\mu} V_b^{\;\;\mu}=\delta^a_b \Rightarrow V_b^{\;\;\mu}=\frac{1}{\sqrt{\Ce}}\delta_b^\mu
\ee

\subsection*{$\eqbox{V_{b\nu;\mu}}$}
\bea
V_{b\nu;\mu}&=&(\partial_\mu V_{b\nu}-\Gamma^{\lambda}_{\nu\mu}V_{b\lambda})=
        \partial_\mu(g_{\nu\rho}V_b^\rho)-\Gamma^{\lambda}_{\nu\mu}g_{\lambda \rho}V_b^\rho=\nn
      &=&\partial_\mu \left(\frac{\C}{\sqrt{\C}}\e_{\nu\rho}\delta_b^\rho\right)\
         -\frac{1}{2}\frac{\Cp}{\C}( \delta^0_\nu \delta^\lambda_\mu+\delta^0_\mu \delta^\lambda_\nu
         -\e^{\lambda 0}\e_{\nu\mu})\sqrt{\C}\e_{\lambda\rho}\delta^\rho_b=\nn
      &=&\frac{\Cp}{2\sqrt{\C}}\left(\delta^0_\mu \e_{\nu\rho}\delta^\rho_b-\delta^0_\nu \delta^\lambda_\mu \e_{\lambda \rho}\delta^\rho_b
        -\delta^0_\mu \delta^\lambda_nu \e_{\lambda \rho}\delta^\rho_b+\e^{\lambda 0}\e_{\nu\mu}\e_{\lambda\rho}\delta_b^\rho\right)=\nn
      &=&\frac{\Cp}{2\sqrt{\C}}\left(\delta^0_\mu \e_{\nu b}-\delta^0_\nu \e_{\mu b}-\delta^0_\mu\e_{\nu b}+\delta^0_b \e_{\nu\mu}\right)
\eea
\subsection*{$\eqbox{\Gamma_\mu}$}
\bea
\Gamma_\mu&=&\frac{1}{8}[\gamma^a,\gamma^b]\frac{1}{\sqrt{\C}}\delta^\nu_a \frac{\Cp}{2 \sqrt{\C}}(-\delta^0_\nu \e_{\mu b}+\delta^0_b \e_{\nu\mu})=\nn
					&=&\frac{1}{16}\frac{\Cp}{\C}[\gamma^\nu,\gamma^b](-\delta^0_\nu\e_{\mu b}+\delta^0_b \e_{\nu\mu})=\nn
					&=&\frac{1}{16}\frac{\Cp}{\C}(-[\gamma^0,\gamma_\mu]+[\gamma_\mu,\gamma^0])=\nn
					&=&\frac{1}{8}\frac{\Cp}{\C}(-[\gamma^0,\gamma_\mu])=\frac{1}{8}\frac{\Cp}{\C}[\gamma_0,\gamma_\mu]
\eea
with $\gamma^0=-\gamma_0$.

\subsection*{$\eqbox{\gamma^\mu D_\mu \psi+m\psi=0}$}

$$
\gamma^a V_a^{\mu}D_\mu \psi+m\psi=0 \Rightarrow \gamma^a \frac{1}{\sqrt{\C}}\delta_a^\mu \left(\partial_\mu
+\frac{1}{8}\frac{\Cp}{\C}[\gamma_0,\gamma_\mu]\right)\psi+m\psi=0
$$
$$
\Rightarrow\left(\gamma^a\partial_a+\frac{1}{8}\frac{\Cp}{\C}\gamma^a[\gamma_0,\gamma_a]+\sqrt{\C}m\right)\psi=0\Rightarrow
$$
\be\label{eom}
\Rightarrow\left(\gamma^a \partial_a+\frac{3}{4}\frac{\Cp}{\C}\gamma^0+\sqrt{\C}m\right)\psi=0
\ee
since
\be
\gamma^a[\gamma_0,\gamma_a]+\gamma^a\gamma_0\gamma_a-\gamma^a\gamma_a\gamma_0=-6\gamma_0=6\gamma^0
\ee
being
\be
\gamma^a\gamma_a=4\mathbb{I}\;\;\;\mbox{and}\;\;\;\gamma^a\gamma_0\gamma_a=-2\gamma_0
\ee


\subsection*{$\eqbox{\lag}$}

\bea
\lag&=&\frac{1}{2}\left(\bar{\psi}\gamma^a V_a^\rho D_\rho\psi+m\bpsi \psi\right)+h.c.=\nn
		&=&\frac{1}{2}\left(\bpsi \gamma^a\frac{1}{\sqrt{\C}}\delta_a^\rho (\partial_\rho+\frac{1}{8}\frac{\Cp}{\C}[\gamma_0,\gamma_\rho])\psi+\C m \bpsi \psi\right)+h.c.=\nn		
		&=&\frac{1}{2\sqrt{\C}}\left(\bpsi\gamma^a \partial_a \psi+\frac{1}{8}\frac{\Cp}{\C} \bpsi\gamma^a [\gamma_0,\gamma_a]\psi+\C m\bpsi\psi\right)+h.c.
\eea

\subsection*{$\eqbox{T_{\mu\nu}}$}

\bea\label{setclassic}
T_{\mu\nu}&=&\frac{1}{2}\left(\bpsi \gamma_aV^a_{\left(\mu\right.}D_{\left.\nu\right)}\psi\right)+h.c.=\nn
					&=&\frac{1}{2}\left[\bpsi \gamma_a \sqrt{\C} \delta^a_{\left(\mu\right.}\left(\partial_{\left.\nu\right)}
					 +\frac{1}{8}\frac{\Cp}{\C}[\gamma_0,\gamma_{\left.\nu\right)}]\right)\psi\right]+h.c.=\nn
					&=&\frac{1}{2}\sqrt{\C}\left( \bpsi \gamma_a \delta^a_{\left(\mu\right.}\partial_{\left.\nu\right)}\psi
					  +\frac{1}{8}\frac{\Cp}{\C}\bpsi \gamma_a \delta^a_{\left(\mu\right.}[\gamma_0,\gamma_{\left.\nu\right)}]\psi\right)+h.c.=\nn
					&=&\frac{1}{2}\sqrt{\C}\left( \bpsi \gamma_{\left(\mu\right.}\partial_{\left.\nu\right)}\psi
					  +\frac{1}{8}\frac{\Cp}{\C}\bpsi \gamma_{\left(\mu\right.}[\gamma_0,\gamma_{\left.\nu\right)}]\psi\right)+h.c.
\eea

\subsectionItalic{Field Decomposition}\label{aC23fd}
\subsubsection{Parker's Ansatz}
A generic solution of equation (\ref{eom}) can be written, following \cite{Parker:1971pt}, as
\begin{multline}\label{psi}
\psi(\eta,\vec{x})=\left(\frac{1}{L \sqrt{\C(\eta)}}\right)^{\frac{3}{2}} \!\!\!\!\!\!\!\!
\sum_{\mbox{\tiny{$\begin{array}{c} \vp  \\ a,b,c=\pm 1 \end{array}$}}}  \!\!\!\!\!\!\!\!
A^{(a,b)}(\vp)\sqrt{\frac{m}{\w(p,\eta)}}D^{(a)}_{(c)}(p,\e)\times\\
\times u^{(c,abc)}(ac\vp,\e) e^{ia\vp\cdot\vx-ic\int \w d\e}
\end{multline}
where:
\begin{itemize}
	\item $L$ is the parameter of our boundary condition: $\psi(\e,\vx+\vec{n}L)=\psi(\e,\vx)$, where $\vec{n}$ is a vector with integer Cartesian 								
				components;
	\item $A^{(a,b)}(\vp)$ are operators defined by:
				\be\{A^{(a,b)}(\vp),A^{(a',b')\dagger}(\vq)\}=\delta_{a,a'}\delta_{b,b'}\delta_{\vp,\vq};\ee
	\item $\w(p,\e)\equiv\sqrt{p^2+\C(\e)m^2}$;
	\item $u^{(a,b)}(\vp,\e)\equiv u^{(a,b)}(\vp/\sqrt{\C(\e)})$, with 
				\be(-a\sqrt{p^2+m^2 } \gamma^4+i a \vec{\gamma}\cdot\vp+m)u^{(a,b)}(\vp)=0\ee
				and 
				\be u^{(a,b)\dagger}(\vp)u^{(a',b')}(\vp)=\frac{\sqrt{p^2+m^2 }}{m}\delta_{a,a'}\delta_{b,b'};\ee
	\item functions $D_{(a')}^{(a)}(p,\e)$ are defined by: 
				\be\label{D}
				D_{(a')}^{(a)}(p,\e)=\delta_{a'}^{a}+a'\int^{\e}_{\e_0}d\e'\frac{1}{4}\frac{\C'(\e)}{\sqrt{\C(\e)}}\frac{m p}{\w^2}e^{2 i a' \int\w 										
				d\e'}D_{(-a')}^{(a)}(p,\e')
				\ee
				with $a,a'=-1,1$, that can be deduced by requiring the ansatz (\ref{psi}) to obey equation of motion (\ref{eom});
				a useful relation that holds between them is \cite{Parker:1971pt}
				\be
				\sum_b D^{(b)}_{(a)}(p,\e)D^{(b)*}_{(a')}(p,\e)=\delta_{a,a'}
				\ee
				that explicitly it is written as
				\bea
				1&=&|D^{(1)}_{(1)}|^2+|D^{(-1)}_{(1)}|^2\\ \label{d1}
				1&=&|D^{(-1)}_{(-1)}|^2+|D^{(1)}_{(-1)}|^2\\ \label{d2}
				0&=&D^{(1)}_{(1)}D^{(1)*}_{(-1)}+D^{(-1)}_{(1)}D^{(-1)*}_{(-1)}\\ \label{d3}
				0&=&D^{(1)}_{(-1)}D^{(1)*}_{(1)}+D^{(-1)}_{(-1)}D^{(-1)*}_{(1)}.\label{d4}
				\eea
				
\end{itemize}

\subsubsection{Spinors}\label{spinorsection}

Spinors $u^{(a,d)}(\vp)$ (with $a,d=\pm1$) are defined by the following conditions

\be\label{sconditions}
\left\{
\begin{array}{l}
(\gamma^a \partial_a +m)e^{i a \vp \cdot \vx-i a \w t} u^{(a,d)}(\vp)=0\\
\sigma_p u^{(a,d)}(\vp)=d u^{(a,d)}(\vp)\\
u^{(a,d)}(\vp)^{\dagger}u^{(a,d)}(\vp)=\sqrt{p^2+m^2}/m
\end{array}\right.
\ee

where 
\be
\gamma^{0}=\begin{pmatrix}-i \mathbb{I}&0\\0&i \mathbb{I}\end{pmatrix}
\;\;\;
\gamma^i=\begin{pmatrix}0&-i\hat{\sigma}_i\\i\hat{\sigma}_i&0\end{pmatrix}
\ee
\be
\sigma_p=\vec{\sigma}\cdot\vp/p=(\sigma^1 p_x+\sigma^2 p_y+\sigma^3 p_z)/p
\ee
with
\be
\sigma^i=\begin{pmatrix} \hat{\sigma}_i & 0 \\ 0 & \hat{\sigma}_i \end{pmatrix}
\ee

\be
\w=\sqrt{p^2+m^2}
\ee
\be
p=\sqrt{p_x^2+p_y^2+p_z^2}
\ee
$\mathbb{I}$ being the $2\times2$ unit matrix and the $\hat{\sigma}_i$ the $2\times2$ Pauli matrices:
\be
\hat{\sigma}_1=\begin{pmatrix} 0 & 1 \\ 1 & 0 \end{pmatrix}
\;\;\;
\hat{\sigma}_2=\begin{pmatrix} 0 & -i \\ i & 0 \end{pmatrix}
\;\;\;
\hat{\sigma}_3=\begin{pmatrix} 1 & 0 \\ 0 & -1 \end{pmatrix}.
\ee

Starting from the first equation in (\ref{sconditions}) we have
\begin{gather}
\left(\gamma^{0} \partial_t+\vec{\gamma}\cdot\vec{\nabla}+m\right)e^{i a \vp \cdot \vx-i a \w t}\uad=0\\
\Rightarrow\;\;\;\;\left(-i a \w \gamma^{0}+i a \vec{\gamma}\cdot\vec{p}+m\right)\uad=0\\
\Rightarrow\;\;\;\;\left(a \w-a \gamma^{0} \vec{\gamma}\cdot\vec{p}+i m\right)\uad=0.
\end{gather}
Since
\be
\gamma^{0} \gamma^i=-\begin{pmatrix}0&\hat{\sigma}_i\\ \hat{\sigma}_i&0\end{pmatrix}
\ee
and writing
\be
\uad=\begin{pmatrix} \varphi \\ \chi \end{pmatrix}
\ee
we have
\be
\left\{\begin{array}{rcl}
a \w \varphi&=&m \varphi+a \vec{\hat{\sigma}}\cdot \vp \chi\\
a \w \chi&=&a \vec{\hat{\sigma}}\cdot \vp \varphi-m \chi
\end{array}\right.
\Rightarrow
\left\{\begin{array}{rcl}
(a \w-m) \varphi&=&a \vec{\hat{\sigma}}\cdot \vp \chi\\
(a \w+m) \chi&=&a \vec{\hat{\sigma}}\cdot \vp \varphi
\end{array}\right.
\ee
By imposing now
\be
\chi=x\varphi
\ee
we have
\be
\left\{\begin{array}{rcl}
\frac{1}{x}(\w-a m) \varphi&=&\vec{\hat{\sigma}}\cdot \vp \varphi\\
x(\w+a m) \varphi&=&\vec{\hat{\sigma}}\cdot \vp \varphi
\end{array}\right.
\Rightarrow
x=\pm\sqrt{\frac{\w-a m}{\w+a m}}.
\ee
The sign of the factor $x$ is decided by the second condition in (\ref{sconditions}):
\be
\sigma_p u^{(a,d)}(\vp)=d u^{(a,d)}(\vp)\Rightarrow \vec{\hat{\sigma}}\cdot \vp \varphi=d p \varphi
\ee
and knowing that eigenvectors of $\vec{\hat{\sigma}}\cdot \vp$ are proportional to
\be
\begin{pmatrix}1\\ \frac{p_x +i p_y}{p_z + d p}\end{pmatrix}
\ee
we have
\begin{gather}
\varphi=\kappa \begin{pmatrix}1\\\frac{p_x +i p_y}{p_z + d p}\end{pmatrix}\Rightarrow 
\pm(\w+a m)\sqrt{\frac{\w-a m}{\w+a m}}\kappa =d p \kappa \\
\Rightarrow x=d\sqrt{\frac{\w-a m}{\w+a m}}
\end{gather}
since $\sqrt{(\w+a m)(\w-a m)}=p$.
Therefore we can write
\be
\uad=\kappa \begin{pmatrix}1\\ \frac{p_x +i p_y}{p_z + d p}\\  d\sqrt{\frac{\w-a m}{\w+a m}}  \\  
 d\sqrt{\frac{\w-a m}{\w+a m}}   \frac{p_x +i p_y}{p_z + d p} \end{pmatrix}.
\ee
Imposing finally the last condition, we have
\be
\uadd\uad=\frac{\w}{m}\Rightarrow \kappa=\pm \sqrt{\frac{(\w+a m)(p+d p_z)}{4 m p}}
\ee
and choosing the positive solution, we finally arrive to
\be
\uad=\sqrt{\frac{(\w+a m)(p+d p_z)}{4 m p}}
\begin{pmatrix}
v_d \\
d\sqrt{\frac{\w-a m}{\w+a m}}\;\;v_d 
\end{pmatrix}
\ee
with 
\be
v_d\equiv \begin{pmatrix}1 \\ \frac{p_x +i p_y}{p_z + d p}\end{pmatrix}.
\ee

\subsectionItalic{Flavour Mixing}\label{sectionmixing}\label{aC23fv}
In order to introduce flavour mixing, we consider the Lagrangian for two non-interactive Dirac fields with different masses:
\be\label{l1l2}
\mathcal{L}=\sum_{i=1,2}
\frac{1}{2}\left(\bpsi_i\gamma^{\mu} D_{\mu} \psi_i+m_i\bpsi_i \psi_i\right)+h.c.
\ee
This Lagrangian can be regarded as the diagonalized version of
\be\label{lAlB}
\mathcal{L}=\sum_{\iota,\kappa=A,B}
\frac{1}{2}\left(\bpsi_\iota\gamma^{\mu} D_{\mu} \psi_\iota+m_{\iota\kappa}\bpsi_\iota \psi_\kappa\right)+h.c.
\ee
with $m_{\iota\kappa}=m_{\kappa\iota}$, where
\bea
\psi_A(x)&=&\psi_1(x) \ct+\psi_2(x)\st\nn
\psi_B(x)&=&-\psi_1(x) \st+\psi_2(x)\ct
\eea
and
\bea
m_{AA}&=&m_1 \cct+m_2 \sst\nn
m_{BB}&=&m_1 \sst+m_2 \cct\nn
m_{AB}&=& (m_2-m_1) \st \ct.
\eea
The stress-energy tensor for our theory now reads
\be\label{set1plusset2}
T_{\mu\nu}=T_{\mu\nu}^{\psi_1}+T_{\mu\nu}^{\psi_2}
\ee
with
\bea\label{set1set2}
T_{\mu\nu}^{\psi_1}&=&\frac{1}{2}\left( \bar{\psi}_1\gamma_\mu D_\nu  \psi_1-  D_\nu \bar{\psi}_1\gamma_\mu \psi_1   \right)\nn
T_{\mu\nu}^{\psi_2}&=&\frac{1}{2}\left( \bar{\psi}_2\gamma_\mu D_\nu  \psi_2-  D_\nu \bar{\psi}_2\gamma_\mu \psi_2   \right).
\eea
Furthermore, we restrict our analysis to FRW metrics, obeying the following requests:\footnote{The two requests are not independent:
the second one implicitly includes the first one. However, we treat them separately because of their different purposes.}
\bea
\lim_{\e\rightarrow-\infty}\C'(\e)&=&0\nn
\lim_{\e\rightarrow-\infty}\C(\e)&=&1,
\eea
being $\Ce$ the conformal scale factor and $\e$ the conformal time.
As explained in  \ref{cAFVICS}, these two requests enable us to implement BV formalism on curved spacetime,
by choosing the flavour vacuum as our physical vacuum state at \textit{early times}.
The former, $\C'(-\infty)=0$, allows us to correctly identify the Fock space
that describes particle states at early times via Parker's ansatz. The latter,
$\C(-\infty)=1$, enables us to use standard BV formalism at $\e\rightarrow-\infty$
with no corrections deriving from a possible coordinates rescaling induced by $\C(-\infty)$.

In the assumption of \textit{finite volume} of our universe, we can write both the
flavour vacuum and the field operators in terms of vectors of \hilbert{F_0},
the Fock space that at early times describes (a finite number of) massive particles.
It should be emphasized that this choice is arbitrary: we could have expressed everything in terms
of \hilbert{F_f} (the Fock space for flavour states at early times) or whatever $\mathfrak{F}\in\mathfrak{H}$,
with no changes in final results. The physics of the problem is determined by the choice of the vacuum state, not
the choice of the \textit{basis} in which the vacuum state is expressed.

\hilbert{F_0} represents a convenient choice firstly because our theory in curved spacetime is naturally
formulated in \hilbert{F_0}: the field is decomposed in ladder operators that creates/annihilates 
single particles in \hilbert{F_0} via Parker's ansatz (\ref{psi}) (thanks to the condition $\C'(-\infty)=0$).
Secondarily, also \cite{Blasone:1995zc,Capolupo:2004av} provide us with as expression of the flavour vacuum in terms
of states of \hilbert{F_0} (thanks to the other condition $\C(-\infty)=1$), viz.,
\begin{multline}\label{fv}
\rfv =\prod_{\vk}\Big[1+\sin \theta \cos \theta \big(S_{-}(\vk)-S_{+}(\vk)\big)+\frac{1}{2}\sin^2 \theta \cos^2 \theta S_{-}^2(\vk)+\\
				+S_{+}^2(\vk)-\sin^2\theta S_{+}(\vk)S_{-}(\vk)+\frac{1}{2}\sin^3\theta \cos \theta \big(S_{-}(\vk)(S_{+}(\vk)\big)^2+\\
		 -S_{+}(\vk)S_{-}^2(\vk)+\frac{1}{4}\sin^4\theta S_{+}^2(\vk)S^2_{-}(\vk)\Big] \rmv
\end{multline}
with
\begin{multline}\label{Splus}
S_{+}(\vp)\equiv\sum_{\mbox{\tiny{$\ba{c}a,b\\a',b'\ea$}}}\Big[\hat{A}^{(a,b)\dagger}_1(a\vp)\hat{A}^{(a',b')}_2(a'\vp)\times\\
\times\sqrt{\frac{m_1 m_2}{\sqrt{p^2+m_1^2}\sqrt{p^2+m_2^2}}}u_1^{(a,b)\dagger}(a\vp)u^{(a',b')}_2(a'\vp)\Big]
\end{multline}
\begin{multline}\label{Sminus}
S_-(\vp)\equiv\sum_{\mbox{\tiny{$\ba{c}a,b\\a',b'\ea$}}}\Big[\hat{A}^{(a,b)\dagger}_2(a\vp)\hat{A}^{(a',b')}_1(a'\vp)\times\\
\times\sqrt{\frac{m_1 m_2}{\sqrt{p^2+m_1^2}\sqrt{p^2+m_2^2}}}u_2^{(a,b)\dagger}(a\vp)u^{(a',b')}_1(a'\vp)\Big].
\end{multline}

A final remark.\\
The \textit{finite volume} condition plays a fundamental role here: in the \textit{infinite} volume limit, \hilbert{F_0} would
not have been a suitable choice, because of its orthogonality with \hilbert{F_f}.
More specifically, in the following we will consider an expression similar to
\be\label{Gseries}
G_{\theta}=e^{\theta A}=1+\theta A+\theta^2 A^2/2+\mathcal{O}(\theta^3)
\ee
with a suitable operator $A$ (compare formulae (\ref{Gseries1}), (\ref{Gseries2}) and (\ref{Gseries3})),
which will lead to
\be
\lfv 0\rangle=\lmv G_{\theta}\rmv=1+\theta \langle A \rangle+\theta^2 \langle A^2/2 \rangle+\dotso
\ee
On the other hand, in the infinite volume limit the orthogonality of the two Fock spaces implies that
\be
\lfv 0 \rangle=0
\ee
and therefore the expansion (\ref{Gseries}) would not be allowed.
However, the finite volume condition implies that the total number of particles is a finite itself (\textit{cf} Section \ref{ssVC})
and therefore flavour states belonging to \hilbert{F_f} \textit{can be expressed} in terms of \hilbert{F_0}.
In order to better clarify this point, we shall consider a simpler analogous.
If we consider the function $e^{-x}$ of a real variable $x$, we have that
\be\label{expseries}
e^{-x}=1-x+x^2/2-\dotso
\ee
whereas
\be
\lim_{x\rightarrow \infty}e^{-x}=0.
\ee
The limit of the series (\ref{expseries}) for $x\rightarrow\infty$ is not well defined, since all terms diverge.
Analogously, the expression $\langle A \rangle$, as well as $\langle A^2 \rangle$ etc. in (\ref{Gseries}),
is proportional to the volume \cite{Ji:2002tx,Blasone:1995zc,Blasone:2001du} (that in terms of the parameter $L$ introduced earlier (formula (\ref{psi})) is written as $L^3$).
Therefore, the expansion (\ref{Gseries}) will make sense \textit{only} when $L$ is \textit{finite}.
However, the equation of state $w$ will turn out to be $L$ independent. We will then be allowed to
consider our final results valid also in the \textit{infinite volume} limit.


\subsectionItalic{Flavour-vev of the Stress-Energy Tensor}\label{aFVEVOTSET}
\subsubsection{Quantum algebraic structure}

As explained above, the flavour vacuum is defined by
\be\label{justmentioned}
\rfv\equiv \hat{G}_{-\theta}\rmv
\ee
with a $\hat{G}_{\theta}$ a specific operator determined by the specific theory considered. It will be useful to recall that
\be
\hat{G}_{-\theta}=\hat{G}_{\theta}^{\dagger}=\hat{G}_{\theta}^{-1}.
\ee
In our specific case, formula (\ref{justmentioned}) reduces to (\ref{fv}).
Let us first notice that, although $\hat{G}_{-\theta}$ is defined
as a series of terms containing both creation ad annihilation operators,
the expression $\hat{G}_{-\theta}\rmv$ can be written as a linear combination of terms
that just contain creation operators acting on the vacuum state. Therefore, we can write
\be\label{Gseries1}
\rfv\equiv \hat{G}_{-\theta}\rmv=\hat{g}\rmv
\ee
where the operator $\hat{g}$ is just made of creation operators. Moreover $\hat{g}$ is a series of terms that contain an \textsl{even} number
of creation operators. This information will help us in our next calculation.
At the moment the explicit form of $\hat{g}$ is not needed and it will be specified later on. 
We shall now concentrate on the momentum decomposition of both the operator $\hat{g}$ and $\hat{T}_{\mu\nu}$:
\be
\left\{
\begin{array}{l}
\hat{T}_{\mu\nu}\equiv \sum_{\vp,\vq} \hat{t}_{\vp,\vq}\\
\hat{g}\equiv \prod_{\vk} \hat{g}_{\vk,-\vk}
\end{array}
\right.
\ee
The terms with different momenta commute, since anticommuting creation/annihilation operators $\hat{a}_{\vp}$, $\hat{b}_{\vp}$, etc.
appear in couples both in $\hat{t}_{\vp,\vq}$ and $\hat{g}_{\vk,-\vk}$ (being $\hat{T}_{\mu\nu}$ bilinear in $\hat{\bpsi}$ and $\hat{\psi}$,
\textit{cf} (\ref{setcurved2})); therefore we can write:
\begin{multline}
\lfv \hat{T}_{\mu\nu} \rfv=\lmv \hat{g}^{\dagger} \hat{T}_{\mu\nu} \hat{g} \rmv=\\
=\lmv\left(\prod_{\vk}\hat{g}_{\vk,-\vk}\right)^{\dagger} \left(\sum_{\vp,\vq}\hat{t}_{\vp,\vq}\right) \left(\prod_{\vk}\hat{g}_{\vk,-\vk}\right)\rmv=\\
=\sum_{\vp,\vq}\lmv\left(\prod_{\vk}\hat{g}_{\vk,-\vk}\right)^{\dagger} \hat{t}_{\vp,\vq}\left(\prod_{\vk}\hat{g}_{\vk,-\vk}\right)\rmv=\nonumber
\end{multline}
\begin{multline}
=\sum_{\vp,\vq}\lmv\left(\prod_{\vk}\hat{g}_{\vk,-\vk}\right)^{\dagger}\left(\prod_{\mbox{\tiny{$\ba{c} \vk\neq \\ \{\vp,-\vp,\\ \vq,-\vq\}\ea$}}}\hat{g}_{\vk,-\vk}\right)
\hat{t}_{\vp,\vq}\left(\prod_{{\mbox{\tiny{$\vk=\{\vp,-\vp,\vq,-\vq\}$}}}}\hat{g}_{\vk,-\vk}\right)\rmv=\\
=\sum_{\vp,\vq}\lmv\left(\prod_{\mbox{\tiny{$\ba{c} \vk= \\ \{\vp,-\vp,\\ \vq,-\vq\}\ea$}}}\hat{g}_{\vk,-\vk}\right)^{\dagger}
\left(\prod_{\mbox{\tiny{$\ba{c} \vk\neq \\ \{\vp,-\vp,\\ \vq,-\vq\}\ea$}}}\hat{g}_{\vk,-\vk}\right)^{\dagger}\times\\
\times\left(\prod_{\mbox{\tiny{$\ba{c} \vk\neq \\ \{\vp,-\vp,\\ \vq,-\vq\}\ea$}}}\hat{g}_{\vk,-\vk}\right)
\hat{t}_{\vp,\vq}\left(\prod_{\mbox{\tiny{$\ba{c} \vk= \\ \{\vp,-\vp,\\ \vq,-\vq\}\ea$}}}\hat{g}_{\vk,-\vk}\right)\rmv
\end{multline}
Since $\hat{g}_{\vk,-\vk}^{\dagger}\hat{g}_{\vk,-\vk}=1$, 
we have
$$
=\sum_{\vp,\vq}\lmv
\left(\prod_{\mbox{\tiny{$\ba{c} \vk= \\ \{\vp,-\vp,\\ \vq,-\vq\}\ea$}}}\hat{g}_{\vk,-\vk}\right)^{\dagger}
\hat{t}_{\vp,\vq}\left(\prod_{\mbox{\tiny{$\ba{c} \vk= \\ \{\vp,-\vp,\\ \vq,-\vq\}\ea$}}}\hat{g}_{\vk,-\vk}\right)\rmv=
$$
$$
=\sum_{\vp=\vq}\lmv
\left(\prod_{\mbox{\tiny{$\vk=\{\vp,-\vp\}$}}}\hat{g}_{\vk,-\vk}\right)^{\dagger}
(\hat{t}_{\vp,\vp}+\hat{t}_{-\vp,-\vp}+\hat{t}_{-\vp,\vp}+\hat{t}_{\vp,-\vp})\left(\prod_{\mbox{\tiny{$\vk=\{\vp,-\vp\}$}}}\hat{g}_{\vk,-\vk}\right)\rmv+
$$
\be\label{secondsum}
+\sum_{\vp\neq\vq}\lmv
\left(\prod_{\mbox{\tiny{$\ba{c} \vk= \\ \{\vp,-\vp,\\ \vq,-\vq\}\ea$}}}\hat{g}_{\vk,-\vk}\right)^{\dagger}
\hat{t}_{\vp,\vq}\left(\prod_{\mbox{\tiny{$\ba{c} \vk= \\ \{\vp,-\vp,\\ \vq,-\vq\}\ea$}}}\hat{g}_{\vk,-\vk}\right)\rmv
\ee
Let us now prove that the second series vanishes identically. Again using the bilinearity of $\hat{T}_{\mu\nu}$ in $\hat{\bpsi}$ and $\hat{\psi}$,
we can write
\be\label{tminuscolo}
\hat{t}_{\vp,\vq}=\sum_i \hat{m}_{\vp}^i\hat{n}_{\vq}^i
\ee
where both $\hat{m}_{\vp}^i$ and $\hat{n}_{\vq}^i$ contain just one creation/annihilation operator. From this it follows that
$$
\sum_{\vp\neq\vq}\lmv
\left(\prod_{\mbox{\tiny{$\ba{c} \vk= \\ \{\vp,-\vp,\\ \vq,-\vq\}\ea$}}}\hat{g}_{\vk,-\vk}\right)^{\dagger}
\hat{t}_{\vp,\vq}\left(\prod_{\mbox{\tiny{$\ba{c} \vk= \\ \{\vp,-\vp,\\ \vq,-\vq\}\ea$}}}\hat{g}_{\vk,-\vk}\right)\rmv=
$$
$$
=\sum_{\vp\neq\vq}\lmv
\left(\hat{g}_{\vp,-\vp}\hat{g}_{-\vp,\vp}\hat{g}_{\vq,-\vq}\hat{g}_{-\vq,\vq}\right)^{\dagger}
\hat{t}_{\vp,\vq}\left(\hat{g}_{\vp,-\vp}\hat{g}_{-\vp,\vp}\hat{g}_{\vq,-\vq}\hat{g}_{-\vq,\vq}\right)\rmv=
$$
$$
\sum_{\vp\neq\vq}\sum_i \lmv
\left(\hat{g}_{\vp,-\vp}\hat{g}_{-\vp,\vp}\hat{g}_{\vq,-\vq}\hat{g}_{-\vq,\vq}\right)^{\dagger}
(\hat{m}_{\vp}^i\hat{n}_{\vq}^i)\left(\hat{g}_{\vp,-\vp}\hat{g}_{-\vp,\vp}\hat{g}_{\vq,-\vq}\hat{g}_{-\vq,\vq}\right)\rmv=
$$
$$
=\sum_{\vp\neq\vq}\sum_i \lmv
\left(\hat{g}_{\vp,-\vp}\hat{g}_{-\vp,\vp}\right)^{\dagger}
\hat{m}_{\vp}^i
\left(\hat{g}_{\vp,-\vp}\hat{g}_{-\vp,\vp}\right)
\left(\hat{g}_{\vq,-\vq}\hat{g}_{-\vq,\vq}\right)^{\dagger}
\hat{n}_{\vq}^i
\left(\hat{g}_{\vq,-\vq}\hat{g}_{-\vq,\vq}\right)\rmv=	
$$
\be
=\sum_{\vp\neq\vq}\sum_i 
\lmv
\left(\hat{g}_{-\vp,\vp}^{\dagger}\hat{g}_{\vp,-\vp}^{\dagger}
\hat{m}_{\vp}^i
\hat{g}_{\vp,-\vp}\hat{g}_{-\vp,\vp}\right)
\left(g^{\dagger}_{-\vq,\vq}\hat{g}_{\vq,-\vq}^{\dagger}
\hat{n}_{\vq}^i
\hat{g}_{\vq,-\vq}\hat{g}_{-\vq,\vq}\right)
\rmv
\ee
We already said that $\hat{g}_{\vp,-\vp}$ is a linear combination of terms that contain an even number of creation operators,
whereas $\hat{m}^{i}_{\vp}$ and $\hat{n}^{i}_{\vp}$ contain either one creation or an annihilation operator.

Let us first consider:
\be
\lmv \left(\hat{g}_{-\vp,\vp}^{\dagger}\;\hat{g}_{\vp,-\vp}^{\dagger}\;
\hat{m}_{\vp}^i\;\hat{g}_{\vp,-\vp}\;\hat{g}_{-\vp,\vp}\right)
\rmv.
\ee
This object is  a linear combination of terms that contain $4n$ annihilation operators (with $n\in\mathbb N$),
due to $\hat{g}_{-\vp,\vp}^{\dagger}\hat{g}_{\vp,-\vp}^{\dagger}$,
one operator that can be either annihilation or creation, due to $\hat{m}_{\vp}^i$,
and $4m$ creation operators ($m\in\mathbb N$), due to $\hat{g}_{\vp,-\vp}\hat{g}_{-\vp,\vp}$. In both cases, $(4n+1)+4m$ or $4n+(1+4m)$,
it vanishes: we can regard at it as the product of a ket defined by
the creation annihilation operator acting on the right on the vacuum ket, and
a bra defined by the annihilation operators acting on the left on the
vacuum bra; but, since the number of operators is different, they are defining
two different state of the same basis that are orthogonal between each other, therefore the product is zero.
This can be also proven by using the specific anticommutation relations for the operators.
Moreover, this results does not change if we insert a new operator that anticommutes with
all the others.
And specifically, this is the case of
$$
\lmv
\left(\hat{g}_{-\vp,\vp}^{\dagger}\hat{g}_{\vp,-\vp}^{\dagger}
\hat{m}_{\vp}^i
\hat{g}_{\vp,-\vp}\hat{g}_{-\vp,\vp}\right)
\left(\hat{g}^{\dagger}_{-\vq,\vq}\hat{g}_{\vq,-\vq}^{\dagger}
\hat{n}_{\vq}^i
\hat{g}_{\vq,-\vq}\hat{g}_{-\vq,\vq}\right)
\rmv
$$
since all new creation/annihilation operators we added (with index $\vq$) anticommute
with the old ones (with index $\vp$). Therefore we can assert that the second series in (\ref{secondsum}) vanishes identically.

Looking now at the first series:
$$
\sum_{\vp}\lmv
\left(\prod_{\mbox{\tiny{$\vk=\{\vp,-\vp\}$}}}\hat{g}_{\vk,-\vk}\right)^{\dagger}
(\hat{t}_{\vp,\vp}+\hat{t}_{-\vp,-\vp}+\hat{t}_{-\vp,\vp}+\hat{t}_{\vp,-\vp})\left(\prod_{\mbox{\tiny{$\vk=\{\vp,-\vp\}$}}}\hat{g}_{\vk,-\vk}\right)\rmv=
$$
\be
=\sum_{\vp}
\lmv
\left(\hat{g}_{\vp,-\vp}\hat{g}_{-\vp,\vp}\right)^{\dagger}
(\hat{t}_{\vp,\vp}+\hat{t}_{-\vp,-\vp}+\hat{t}_{-\vp,\vp}+\hat{t}_{\vp,-\vp})
\left(\hat{g}_{\vp,-\vp}\hat{g}_{-\vp,\vp}\right)
\rmv
\ee
applying the above reasoning to operators with indices $\vp$ and $-\vp$ we can argue that terms with
an odd number of operators with a specific index will vanish; the only surviving terms will be
\be\label{chisopravvive}
\sum_{\vp}
\lmv
\left(\hat{g}_{\vp,-\vp}\hat{g}_{-\vp,\vp}\right)^{\dagger}
(\hat{t}_{\vp,\vp}+\hat{t}_{-\vp,-\vp})
\left(\hat{g}_{\vp,-\vp}\hat{g}_{-\vp,\vp}\right)
\rmv.
\ee

Before writing down the explicit form of $\hat{g}_{\vp,\vq}$,
we introduce an approximation.
Considering the parameter of the mixing $\sin \theta$ to be small, we can take in account
just the leading orders in the expansion of the the vev of $\hat{T}_{\mu\nu}$ in $\sin \theta$.
So far we have seen that
\be\label{operatorel1}
\lfv \hat{T}_{\mu\nu}\rfv=
\sum_{\vp}
\lmv
\left(\hat{g}_{\vp,-\vp}g_{-\vp,\vp}\right)^{\dagger}
(\hat{t}_{\vp,\vp}+\hat{t}_{-\vp,-\vp})
\left(\hat{g}_{\vp,-\vp}\hat{g}_{-\vp,\vp}\right)
\rmv
\ee
that we can write as
\be\label{operatorel2}
\lfv \hat{T}_{\mu\nu}\rfv=
\sum_{\vp}
\lmv
\hat{l}^{\dagger}
\hat{O}
\hat{l}
\rmv
\ee
with 
\be\label{Gseries2}
\hat{l}\equiv \hat{g}_{\vp,-\vp}\hat{g}_{-\vp,\vp}\;\;\;\;\;
\hat{O}\equiv \hat{t}_{\vp,\vp}+\hat{t}_{-\vp,-\vp}.
\ee
Furthermore, comparing (\ref{operatorel1}), (\ref{operatorel2}), (\ref{Gseries1}) and (\ref{fv}),
we understand that the dependency on $\sin \theta$ of our expression
is hidden in the operator $\hat{l}$. If we write it as
\be\label{Gseries3}
\hat{l}=1+\sin \theta \hat{l}_{(1)}+\sin^2 \theta \hat{l}_{(2)}+\mathcal{O}(\sin^3 \theta)
\ee
we then have 
\begin{multline}
\lmv \hat{l}^{\dagger}\hat{O}\hat{l}\rmv=\lmv \hat{l}^{\dagger}\hat{l}\hat{O}\rmv+\lmv \hat{l}^{\dagger}[\hat{O},\hat{l}]\rmv=\\
\approx \lmv \hat{O}\rmv+\lmv (1+\sin \theta \hat{l}_{(1)}+\sin^2 \theta \hat{l}_{(2)})[\hat{O},1+\sin \theta \hat{l}_{(1)}+\sin^2 \theta \hat{l}_{(2)}]\rmv=\\
=\lmv \hat{O}\rmv+\lmv (1+\sin \theta \hat{l}_{(1)}+\sin^2 \theta \hat{l}_{(2)})(\sin \theta[\hat{O},\hat{l}_{(1)}]+\sin^2 [\hat{O},\theta \hat{l}_{(2)}])\rmv=\\
\approx \lmv \hat{O}\rmv+\sin\theta \lmv [\hat{O},\hat{l}_{(1)}]\rmv+\\
+\sin^2\theta\left(\lmv [\hat{O},\hat{l}_{(2)}]\rmv+\lmv \hat{l}_{(1)}^{\dagger}[\hat{O},\hat{l}_{(1)}]\rmv\right)
\end{multline}
We now need the explicit expression for $\hat{l}_{(1)}$ and $\hat{l}_{(2)}$, being $\hat{l}_{(0)}=1$.

Let us first study the contribution of the first order in $\sin \theta$.
In order to simplify the notation, we shall adopt the follwing notation
\begin{equation}\label{mn2}
\eqbox{
\begin{IEEEeqnarraybox}[][c]{rCl}
&\mbox{{\small\textit{Upper-Case Symbols}}}&\\
\hat{A}^r\equiv \hat{A}^{(1,r)}_1(\vp)&&\hat{B}^r\equiv \hat{A}^{(-1,r)\dagger}_1(\vp)\\
\hat{C}^r\equiv \hat{A}^{(2,r)}_1(\vp)&&\hat{D}^r\equiv \hat{A}^{(-1,r)\dagger}_2(\vp)\\
M^{rs}\equiv &\sqrt{\frac{m_1 m_2}{\sqrt{k^2+m^2_1}\sqrt{k^2+m^2_2}}}\;\;\;&u_1^{(1,r)\dagger}(\vk)u_2^{(-1,s)}(-\vk)\\
N^{rs}\equiv &\sqrt{\frac{m_1 m_2}{\sqrt{k^2+m^2_1}\sqrt{k^2+m^2_2}}}\;\;\;&u_2^{(1,r)\dagger}(\vk)u_1^{(-1,s)}(-\vk)
\end{IEEEeqnarraybox}
}
\end{equation}
(left hand side symbols being defined in Section \ref{aC23fd}) and
\begin{equation}
\eqbox{
\begin{IEEEeqnarraybox}[][c]{rCl}
&\mbox{{\small\textit{Lower-Case Symbols}}}&\\
\hat{a}^r\equiv \hat{A}^r\Big|_{\vp\rightarrow-\vp}&&\hat{b}^r\equiv \hat{B}^r\Big|_{\vp\rightarrow-\vp}\\
\hat{c}^r\equiv \hat{C}^r\Big|_{\vp\rightarrow-\vp}&&\hat{d}^r\equiv \hat{D}^r\Big|_{\vp\rightarrow-\vp}\\
m^{rs}\equiv M^{rs}\Big|_{\vp\rightarrow-\vp}&&n^{rs}\equiv N^{rs}\Big|_{\vp\rightarrow-\vp}
\end{IEEEeqnarraybox}
}
\end{equation}
Comparing (\ref{fv}), (\ref{Splus}), (\ref{Sminus}), and (\ref{Gseries1}), we have
\be
\hat{g}_{\vp,-\vp}\approx 1+\st \sum_{r,s}\left(M^{rs}\hat{C}^{r \dagger}\hat{b}^{s \dagger} -N^{rs}\hat{A}^{r \dagger}\hat{d}^{s \dagger}\right)
\ee
and
\begin{multline}
\hat{g}_{\vp,-\vp}\hat{g}_{-\vp,\vp}\approx\Bigg[1+\st \sum_{r,s}\left(M^{rs}\hat{C}^{r \dagger}\hat{b}^{s \dagger}
														-N^{rs}\hat{A}^{r \dagger}\hat{d}^{s \dagger}\right)\Bigg]\times\\
\times\Bigg[ 1 +\st \sum_{r,s}\left(m^{rs}\hat{c}^{r \dagger}\hat{B}^{s \dagger}
							-n^{rs}\hat{a}^{r \dagger}\hat{D}^{s \dagger}\right)\Bigg]\approx\\
\approx1+\st \sum_{r,s}\left(M^{rs}\hat{C}^{r \dagger}\hat{b}^{s \dagger}-N^{rs}\hat{A}^{r \dagger}\hat{d}^{s \dagger}
+m^{rs}\hat{c}^{r \dagger}\hat{B}^{s \dagger}-n^{rs}\hat{a}^{r \dagger}\hat{D}^{s \dagger}\right)
\end{multline}
This allow us to write
\be
l_{(1)}=\sum_{r,s}\left(M^{rs}\hat{C}^{r \dagger}\hat{b}^{s \dagger}-N^{rs}\hat{A}^{r \dagger}\hat{d}^{s \dagger}
+m^{rs}\hat{c}^{r \dagger}\hat{B}^{s \dagger}-n^{rs}\hat{a}^{r \dagger}_{-\vk}\hat{D}^{s \dagger}\right)
\ee
We now turn our attention to the operatorial structure of $\hat{T}_{\mu \nu}$. 
Since this operator is bilinear in the two fields
$\hat{\psi}_1(x)$ and $\hat{\psi}_2(x)$ (cf (\ref{setcurved2})), we can write:
\begin{multline}
\hat{t}_{\vp,\vp}+\hat{t}_{-\vp,-\vp}=\sum_{r,s}\left(
\mfA_{r,s}\hat{A}^{\dagger}_r \hat{A}_s+
\mfB_{r,s}\hat{A}^{\dagger}_r \hat{B}^{\dagger}_s+
\mfC_{r,s}\hat{B}_r \hat{A}_s+
\mfD_{r,s}\hat{B}_r \hat{B}^{\dagger}_s+\right.\\
+\left.\mfa_{r,s}\hat{a}^{\dagger}_r \hat{a}_s+
\mfb_{r,s}\hat{a}^{\dagger}_r \hat{b}^{\dagger}_s+
\mfc_{r,s}\hat{b}_r \hat{a}_s+
\mfd_{r,s}\hat{b}_r \hat{b}^{\dagger}_s\right)+(m_1 \rightleftarrows m_2)
\end{multline}
with some specific function $\mfA_{r,s}$, $\mfB_{r,s}$, $\mfC_{r,s}$, and $\mfD_{r,s}$
(for which the \textit{upper/lower-case} convention also applies: $\mfa_{r,s}\equiv\mfA_{r,s}\Big|_{\vp\rightarrow-\vp}$ etc.),
and $(m_1 \rightleftarrows m_2)$ denoting the preceding expression with $m_1$ and $m_2$ swapped.
Moreover, looking at (\ref{chisopravvive}), we see that in $t_{\vp,\vp}+t_{-\vp,-\vp}$ only terms that have one creation operator
and one annihilation operator survive, as already explained.
Hence we have
$$
[\hat{t}_{\vp,\vp}+\hat{t}_{-\vp,-\vp},\hat{l}_{(1)}]=\left[\sum_{r,s}\left(
\mfA_{r,s}\hat{A}^{\dagger}_r \hat{A}_s+
\mfB_{r,s}\hat{A}^{\dagger}_r \hat{B}^{\dagger}_s+
\mfC_{r,s}\hat{B}_r \hat{A}_s+
\mfD_{r,s}\hat{B}_r \hat{B}^{\dagger}_s+\right.\right.
$$
$$
\left.+\mfa_{r,s}\hat{a}^{\dagger}_r \hat{a}_s+
\mfb_{r,s}\hat{a}^{\dagger}_r \hat{b}^{\dagger}_s+
\mfc_{r,s}\hat{b}_r \hat{a}_s+
\mfd_{r,s}\hat{b}_r \hat{b}^{\dagger}_s\right)+(m_1 \rightleftarrows m_2),
$$
$$
\left.\sum_{t,v}\left( M_{tv}\hat{C}_t^{\dagger}b_v^{\dagger}-N_{tv}\hat{A}^{\dagger}_t\hat{d}^{\dagger}_v
+m_{tv}\hat{c}^{\dagger}_t \hat{B}^{\dagger}_v-n_{tv}a^{\dagger}_t \hat{D}_v^{\dagger} \right)\right]=
$$
$$
=\sum_{\mbox{\tiny{$r,s,t,v$}}}\left[\left(\mfA_{rs}\hat{A}^{\dagger}_{r}\hat{A}_{s},-N_{tv}\hat{A}^{\dagger}_{t}\hat{d}^{\dagger}_v\right)+
\left(\mfC_{rs}\hat{B}_r \hat{A}_s,-N_{tv}\hat{A}^{\dagger}_t \hat{d}^{\dagger}_v+m_{tv}\hat{c}^{\dagger}_t \hat{B}^{\dagger}_v\right)+\right.
$$
$$
+\left(\Kq \hat{B}_r \hat{B}^{\dagger}_s,m_{tv} \hat{c}^{\dagger}_t \hat{B}^{\dagger}_v\right)
+\left(\ku \hat{a}^{\dagger}_r \hat{a}_s,-n_{tv}\hat{a}^{\dagger}_t \hat{D}^{\dagger}_v\right)+
$$
$$
\left.+\left(\kt \hat{b}_r \hat{a}_s,M_{tv} \hat{C}^{\dagger}_t \hat{b}^{\dagger}_v -n_{tv} \hat{a}^{\dagger}_t \hat{D}^{\dagger}_v\right)+
\left(\kq \hat{b}_r \hat{b}^{\dagger}_s,M_{tv}\hat{C}^{\dagger}_t \hat{b}^{\dagger}_v\right)\right]=
$$
$$
=\sum_{\mbox{\tiny{$r,s,t,v$}}}\left(-\Ku N_{tv} \hat{A}^{\dagger}_r \hat{d}^{\dagger}_v \delta_{st}
-\Kt N_{tv} \hat{B}_r \hat{d}^{\dagger}_v \delta_{rv}
-\Kt m_{tv} \hat{c}_t \hat{A}_s \delta_{rv}+\right.
$$
$$
-\Kq m_{tv} \hat{c}^{\dagger}_t \hat{B}^{\dagger}_s \delta_{rv}-\ku n_{tv} \hat{a}^{\dagger}_r \hat{D}^{\dagger}_v \delta_{st}
-\kt M_{tv} \hat{C}^{\dagger}_t \hat{a}_s \delta_{rv}
$$
$$
\left.-\kt n_{tv} \hat{b}_r \hat{D}^{\dagger}_v \delta_{st}-\kq M_{tv} \hat{C}^{\dagger}_t \hat{b}^{\dagger}_s \delta_{rv}\right)
$$
$$
\Rightarrow \lmv [\hat{t}_{\vp,\vp}+\hat{t}_{-\vp,-\vp},\hat{l}_{(1)}] \rmv=0
$$
being the same reasoning valid also for terms $(m_1\rightleftarrows m_2)$ , as we shall prove at the end of this section.
The contribution of the first order in $\sin \theta$ vanishes.
Let us now look at the second order.
We have to consider the terms
$$
\lmv [\hat{O},\hat{l}_{(2)}]\rmv+\lmv \hat{l}_{(1)}^{\dagger}[\hat{O},\hat{l}_{(1)}]\rmv
$$
It is easy to see that the first term is identically zero:
first we can notice that it is composed of two pieces
$$
\lmv \hat{O}\hat{l}_{(2)}\rmv-\lmv \hat{l}_{(2)}\hat{O}\rmv
$$
and the second vanishes because the $\hat{l}_{(2)}$ operator, being composed of $\hat{g}_{\vp,-\vp}$ terms and therefore just of creation operators,
vanishes when it act on the vacuum bra $\rmv$; the other term includes $\hat{O}$, that is composed of couples of creation/annihilation operators,
and $\hat{l}_{(2)}$, that is composed of terms with four creation operators. Since only terms with an equal number of creation and annihilation operators
survive
$$
\lmv \hat{O}\hat{l}_{(2)}\rmv=0.
$$
This reasoning applies in general to all terms in the form
$$
\lmv [\hat{O},\hat{l}_{(i)}]\rmv.
$$
We have then to evaluate $\lmv \hat{l}_{(1)}^{\dagger}[\hat{O},\hat{l}_{(1)}]\rmv$. The operator $\hat{O}$
is the sum of two terms, one coming from the contribution of $\hat{T}_{\mu \nu}^{\psi_1}$, and the other one
coming from $\hat{T}_{\mu \nu}^{\psi_2}$; we will consider just the first one,
since an equivalent reasoning holds also for the other one, as we will soon see.
\be
\lmv \hat{l}_{(1)}^{\dagger}[\hat{O},\hat{l}_{(1)}]\rmv \equiv
\lmv \hat{l}_{(1)}^{\dagger}[\hat{O}_{m_1},\hat{l}_{(1)}]\rmv+\lmv \hat{l}_{(1)}^{\dagger}[\hat{O}_{m_2},\hat{l}_{(1)}]\rmv
\ee
\begin{multline}
\lmv \hat{l}_{(1)}^{\dagger}[\hat{O}_{m_1},\hat{l}_{(1)}]\rmv=\lmv  \sum_{l,m}\Big[M^*_{tv}\hat{b}_v \hat{C}_t-
N^*_{lm}\hat{d}_v \hat{A}_{t}+m^*_{lm}\hat{D}_v \hat{a}_t \Big]\times\\
\times \sum_{\mbox{\tiny{$r,s,t,v$}}} \Big[\left(-\Ku N_ {tr}\delta_{st}\right)\hat{A}^{\dagger}_r \hat{d}^{\dagger}_v+
\left(-\Kt N_{tv} \delta_{st}\right) \hat{B}_r \hat{d}^{\dagger}_v+\\
+\left(-\Kt m_ {tv}\delta_{rv}\right)\hat{c}_t \hat{A}_s+\left(-\Kq m_{tv} \delta_{rv}\right) \hat{c}^{\dagger}_t \hat{B}^{\dagger}_s+\\
+\left(-\ku n_ {tv}\delta_{st}\right)\hat{a}^{\dagger}_r \hat{D}^{\dagger}_v+\left(-\kt M_{tv} \delta_{rv}\right) \hat{C}^{\dagger}_t \hat{a}_s+\\
+\left(-\kt n_ {tv}\delta_{st}\right)\hat{b}_r \hat{D}^{\dagger}_v+
\left(-\kq M_{tv} \delta_{rv}\right) \hat{C}^{\dagger}_t \hat{b}^{\dagger}_s\Big)\Big]\rmv=\nonumber
\end{multline}
\begin{multline}
=\sum_{\mbox{\tiny{l,m,r,s,t,v}}}
\Big[M^*_{lm}M_{tv}(-\kq)\delta_{rv}\delta_{lt}\delta_{ms}-
N^*_{lm}N_{ts}(-\mfA_{rv})\delta_{vt}\delta_{lr}\delta_{ms}+\\
+m^*_{lm}m_{tv}(-\Kq)\delta_{rv}\delta_{lt}\delta_{ms}-n^*_{lm}n_{ts}(-\mfa_{rv})\delta_{vt}\delta_{lr}\delta_{ms}\Big)=\nonumber
\end{multline}
\begin{multline}
=\sum_{r,s,t}
\Big[M^*_{ts}M_{tr}(-\kq)-N^*_{rs}N_{ts}(-\mfA_{rt})+\\
+m^*_{ts}m_{tr}(-\Kq)-n^*_{rs}n_{ts}(-\mfa_{rt})\Big].
\end{multline}

We can now concentrate on $(m_1\rightleftarrows m_2)$ terms. Since (recalling (\ref{set1plusset2}) and (\ref{set1set2}))
\bea
\lfv \hat{T}_{\mu \nu}\rfv&=&\lfv \hat{T}_{\mu \nu}^{\psi_1}\rfv+\lfv \hat{T}_{\mu \nu}^{\psi_2}\rfv=\nn
&=&\lfv \hat{T}_{\mu \nu}^{\psi_1} \rfv+\lfv C[\hat{T}_{\mu \nu}^{\psi_1}] \rfv
\eea
where the formal operator $C[\slot]$ exchanges the masses $m_1$ and $m_2$ of its argument.
Holding
\be
C[\hat{A}\hat{B}]=C[\hat{A}]C[\hat{B}]
\ee
and
\be
C[G_{\theta}]=G^{\dagger}_{\theta}=G_{-\theta}
\ee
we can write
\be
\lfv \hat{T}_{\mu \nu}^{\psi_2} \rfv=\lfv C[\hat{T}_{\mu \nu}^{\psi_1}] \rfv=\lmv G_{\theta} C[\hat{T}_{\mu \nu}^{\psi_1}] G_{-\theta}\rmv=
\ee
\be
=\lmv C[G_{-\theta} \hat{T}_{\mu \nu}^{\psi_1} G_{\theta}] \rmv=C[\lmv G_{-\theta} \hat{T}_{\mu \nu}^{\psi_1} G_{\theta}\rmv]
\ee
Since
\be
\lfv \hat{T}_{\mu \nu}^{\psi_1} \rfv=f_0 (m_1,m_2)+\sst f_2 (m_1,m_2)+\mathcal{O}(\ssst)
\ee
we have that
$$
\lfv \hat{T}_{\mu \nu}^{\psi_2} \rfv=f_0 (m_2,m_1)+\sin^2(-\theta) f_2 (m_2,m_1)+\mathcal{O}(\ssst)=
$$
$$
=f_0 (m_2,m_1)+\sin^2(\theta) f_2 (m_2,m_1)+\mathcal{O}(\ssst)
$$
and consequently
\be
\lfv \hat{T}_{\mu \nu}  \rfv=\lfv \hat{T}_{\mu \nu}^{\psi_1} \rfv+(m_1\rightleftarrows m_2)
\ee
at least up to the second order in $\st$.


\subsubsection{Spinorial structure}

Resuming results of the previous part, we found that
\begin{multline}\label{dacalcolare}
\lfv \hat{T}_{\mu\nu}\rfv=\\
=\lmv \hat{T}_{\mu\nu}\rmv+\sst \Bigg[
\sum_{rst,\vp}\left(N^*_{rs}N_{ts}\mfA_{rt}(\vp,\vp)+n^*_{rs}n_{ts}\mfa_{rt}(\vp,\vp)+\right.\\
\left.-m^*_{ts}m_{tr}\Kq(\vp,\vp)-M^*_{ts}M_{tr}\kq(\vp,\vp)\right)+(m_1\rightleftarrows m_2)\Bigg]+\\
+\mathcal{O}(\ssst).
\end{multline}
where the functions denoted by Gothic letters are defined by:
\begin{multline}\label{gothic}
\hat{T}_{\mu\nu} =\sum_{\mbox{\tiny{$\ba{c}\vp,\vq\\r,s\ea$}}}
\left(\Ku (\vp,\vq)\hat{A}_r^{\dagger}\hat{A}_s+\Kd (\vp,\vq)\hat{A}_r^{\dagger}\hat{B}_s^{\dagger}+\right.\\
\left.+\Kt (\vp,\vq)\hat{B}_r\hat{A}_s+\Kq (\vp,\vq)\hat{B}_r\hat{B}_s^{\dagger}\right)+(m_1\rightleftarrows m_2)
\end{multline}
The first term, indipendent on $\st$, is the usual contribution to the Stress-energy tensor,
that the normal order defined in absence of mixing cuts away.

Formula (\ref{dacalcolare}) can be simplified, by taking into account relations between spinors occuring in $M_{rs}$ and $N_{rs}$.
More specifically, starting from with (\ref{mn2}), we can write
\begin{gather}
M_{rs}=m_{rs}=V(k,m_1,m_2)\delta_{r,-s} \\
N_{rs}=n_{rs}=V(k,m_2,m_1)\delta_{r,-s}
\end{gather}
with
\begin{multline} \label{mn3}
V(k,m_i,m_j)\equiv\\
\equiv\frac{(\sqrt{k^2+m_i^2}-m_i)(\sqrt{k^2+m_j^2}+m_j)-k^2}{2\sqrt{\sqrt{k^2+m_i^2}\sqrt{k^2+m_j^2}}\sqrt{\sqrt{k^2+m_i^2}-m_i^2}\sqrt{\sqrt{k^2+m_j^2}+m_j^2}}
\end{multline}
From this it follows that
\bea\label{mn4}
M^*_{ts}M_{tr}&=&\delta_{r,s}V^2(k,m_1,m_2)\equiv \delta_{r,s}V^2_M (k)\nn
N^*_{ts}N_{tr}&=&\delta_{r,s}V^2(k,m_2,m_1)\equiv \delta_{r,s}V^2_N (k)
\eea
Therefore, we can write:
\begin{multline}
\lfv \hat{T}_{\mu\nu}\rfv=\lmv \hat{T}_{\mu\nu}\rmv+\sst \sum_{r,\vp}\Big[\VN \left(\mfA_{rr}(\vp,\vp)+\mfa_{rr}(\vp,\vp)\right)+\\
-\VM \left(\mfD_{rr}(\vp,\vp)+\mfd_{rr}(\vp,\vp)\right)\Big]+(m_1\rightleftarrows m_2)+\mathcal{O}(\ssst).
\end{multline}

Now, let us focus on the functions denoted with Gothic letters. 
Recalling that
\begin{equation}\begin{split}
\ku(\vp,\vq)\equiv\Ku(-\vp,-\vq)&\kd(\vp,\vq)\equiv\Kd(-\vp,-\vq)\\
\kt(\vp,\vq)\equiv\Kt(-\vp,-\vq)&\kq(\vp,\vq)\equiv\Kq(-\vp,-\vq)
\end{split}\end{equation}
we now want to work out explicit expressions for $\Ku(\vp,\vq)$, $\Kd(\vp,\vq)$, $\Kt(\vp,\vq)$, $\Kq(\vp,\vq)$.
Following their definition in (\ref{gothic}), they are related with
the quantum operatorial structure of $\hat{T}_{\mu\nu} $. 

It should be noticed that tensorial indices have been omitted, regarding Gothic functions.
However, to each different value of $(\mu,\nu)$ will correspond a different expression of $\hat{T}_{\mu\nu}$
and therefore a different expression for our Gothic functions.

Moreover,
we have that $\Ku$, $\Kd$, $\Kt$, and $\Kq$ derive from $\hat{T}^{\psi_1}_{\mu\nu}$ only.
As a result, it is suffient to consider just the stress-energy tensor for one single field $\hat{T}^{\psi}_{\mu\nu}$,
with no mass labels ($m$ instead of $m_1$ or $m_2$). Gothic function will then be recovered by considering $m=m_1$.

Recalling (\ref{psi}), we write
\be
\hat{\psi}(\e,\vx)=\sum_{r,\vp}\left(\Psi^{+}_r(\e,\vx,\vp)\hat{A}^{(1,r)}(\vp)+\Psi^{-}_r(\e,\vx,\vp)\hat{A}^{(-1,r)}(\vp)\right)
\ee
that in the notation here adopted becomes
\be\label{psipm}
\hat{\psi}(\e,\vx)=\sum_{r,\vp}\left(\Psi^{+}_r(\e,\vx,\vp)\hat{A}_{r}(\vp)+\Psi^{-}_r(\e,\vx,\vp)\hat{B}^{\dagger}_{r}(\vp)\right)
\ee
with
\begin{multline}
\Psi^{+}_r(\e,\vx,\vp)=\sum_{a}\left(\frac{1}{L \sqrt{\C}}\right)^{\frac{3}{2}}\sqrt{\frac{m_1}{\w_1(\eta)}}\times\\
\times D^{(1)}_{(a)}(p,\e)u^{(a,r a)}(a\vp,\e)e^{i\vp\cdot\vx-ia\int \w_1 d\e}
\end{multline}
\begin{multline}
\Psi^{-}_r(\e,\vx,\vp)=\sum_{a}\left(\frac{1}{L \sqrt{\C}}\right)^{\frac{3}{2}}\sqrt{\frac{m_1}{\w_1(\eta)}}\times\\
\times D^{(-1)}_{(a)}(p,\e)u^{(a,-r a)}(-a\vp,\e)e^{-i\vp\cdot\vx-ia\int \w_1 d\e}
\end{multline}
We consider now the stress-energy tensor for one singe field $\hat{T}^{\psi}_{\mu\nu}$.
Such an operator is bilinear in the fields $\hat{\psi}^{\dagger}$ and $\hat{\psi}$, therefore
we can write
\be\label{tcal}
\hat{T}_{\mu \nu}^{\psi}=\hat{\psi}^{\dagger}\mathcal{T}_{\mu\nu}\hat{\psi}
\ee
with $\mathcal{T}_{\mu\nu}$ the classical (non-quantum) operator acting on $\hat{\psi}$ (from (\ref{setclassic})):
\be\label{settcal}
\mathcal{T}_{\mu\nu}\equiv \frac{\sqrt{\C(\e)}}{2}
\left(\gamma^4 \gamma_{\left(\mu\right.}\partial_{\left.\nu\right)}
+\frac{1}{8}\frac{\C'(\e)}{\C(\e)}\gamma^4\gamma_{\left(\mu\right.}[\gamma_{0},\gamma_{\left.\nu\right)}]\right)+h.c.
\ee

Comparing (\ref{tcal}), (\ref{gothic}) and (\ref{psipm})
we can easily see that
\be\label{gothic2}
\ba{rcl}
\Ku(\vp,\vq)&=&\Psi^{+\dagger}_r(\e,\vx,\vp)\mathcal{T}_{\mu\nu}\Psi^{+}_s(\e,\vx,\vq)|_{m\rightarrow m_1}\\
\Kd(\vp,\vq)&=&\Psi^{+\dagger}_r(\e,\vx,\vp)\mathcal{T}_{\mu\nu}\Psi^{-}_s(\e,\vx,\vq)|_{m\rightarrow m_1}\\
\Kt(\vp,\vq)&=&\Psi^{-\dagger}_r(\e,\vx,\vp)\mathcal{T}_{\mu\nu}\Psi^{+}_s(\e,\vx,\vq)|_{m\rightarrow m_1}\\
\Kq(\vp,\vq)&=&\Psi^{-\dagger}_r(\e,\vx,\vp)\mathcal{T}_{\mu\nu}\Psi^{-}_s(\e,\vx,\vq)|_{m\rightarrow m_1}
\ea
\ee
The explicit evaluation of the $\Ku$ and $\Kq$ for different values of $\mu$ and $\nu$ can
be performed with the help of a computer program.
Moreover, it will be useful to know that
\begin{multline}\label{parpsi}
(L\sqrt{\Ce})^{\frac{3}{2}}\partial_\e \hat{\psi}(\e,\vx)=\!\!\!\sum_{\mbox{\tiny{$\ba{c}\vp\\a,d,a'\ea$}}}\!\!\!\!\!\!\hat{A}^{(a,d)}D^{(a)}_{(a')}(p,\e)e^{ia\vp\cdot\vx-ia'\int{\we d\e}}\times\\
\times\Bigg[-\frac{(2\we^2 +m^2 \Ce)\Cpe}{4 \we^2 \Ce}+a'\frac{mp \Cpe}{4\sqrt{\Ce}\we^2}e^{2ia'\intw}+\\
-ia'\we+a'\frac{\Cpe}{4\sqrt{\Ce}}\frac{m}{\we}\gamma^4\Bigg]u^{(a',a a' d)}(a a' \vp,\e).
\end{multline}
Such a formula will be explicitly derived at the end of this Section.

The explicit form of the Gothic functions are below summarized.


\subsubsection*{$\eqbox{\mu=0\;\nu=0}$}

From (\ref{settcal}) it follows that
\be
\mathcal{T}_{00}=\frac{\sqrt{\C(\e)}}{2}\left(\gamma^4 \gamma_{0} \partial_\e\right)+h.c.
\ee
Helping us with a computer and using (\ref{parpsi}) and (\ref{d1}) it is easy to show that 
\bea
\sum_{d}\mfA_{dd}(\vp,\vp)&=&\frac{\we}{L^3 \Ce}\left(|D^{(1)}_{(1)}|^2-|D^{(1)}_{(-1)}|^2\right)+\nn
						   &&-i\frac{\Cpe}{2L^3\Ce^2}\left(|D^{(1)}_{(-1)}|^2+|D^{(1)}_{(1)}|^2\right)+h.c.\Big|_{m\rightarrow m_1}=\nn
	&=&\frac{2\we}{L^3 \Ce}\left(|D^{(1)}_{(1)}|^2-|D^{(1)}_{(-1)}|^2\right)\Big|_{m\rightarrow m_1}=\nn
	&=&\frac{2\we}{L^3 \Ce}\left(1-|D^{(-1)}_{(1)}|^2-|D^{(1)}_{(-1)}|^2\right)\Big|_{m\rightarrow m_1}
\eea
Since $\mfa_{dd}(\vp,\vp)=\mfA_{dd}(-\vp,-\vp)$ we have that
\be
\sum_{d}\left[\mfA_{dd}(\vp,\vp)+\mfa_{dd}(\vp,\vp)\right]=\frac{4\we}{L^3 \Ce}\left(1-|D^{(-1)}_{(1)}|^2-|D^{(1)}_{(-1)}|^2\right)\Big|_{m\rightarrow m_1}
\ee
all quantities involved depending just on $p=|\vp|$.
In a similar way it is possible to derive
\be
\sum_{d}\left[\mfD_{dd}(\vp,\vp)+\mfd_{dd}(\vp,\vp)\right]=\frac{4\we}{L^3 \Ce}\left(1-|D^{(-1)}_{(1)}|^2-|D^{(1)}_{(-1)}|^2\right)\Big|_{m\rightarrow m_1}
\ee
and therefore
$$
\sum_{d}\left[\VN\left(\mfA_{dd}(\vp,\vp)+\mfa_{dd}(\vp,\vp)\right)-\VM \left(\mfD_{dd}(\vp,\vp)+\mfd_{dd}(\vp,\vp)\right)\right]=
$$
\be\label{t00discrete}
=\left[\frac{4\we}{L^3 \Ce}\left(1-|D^{(-1)}_{(1)}|^2-|D^{(1)}_{(-1)}|^2\right)\right]\left(\VN+\VM\right)\Big|_{m\rightarrow m_1}
\ee
in which we used
\be
\left(\VM+\VN\right)=\left(\VN+\VM\right)\Big|_{m\rightarrow m_1}
\ee
recalling their definition (\ref{mn4}).

\subsubsection*{$\eqbox{\mu=0\;\nu=i}$}
Being
\be
\mathcal{T}_{0i}=\frac{\sqrt{\C(\e)}}{4}\left(\gamma^4 \gamma_{0} \partial_i+\gamma^4 \gamma_{i} \partial_{0}
+\frac{1}{4}\frac{\Cpe}{\Ce}\gamma^4\gamma_0[\gamma_0,\gamma_i]\right)+h.c.
\ee
a computer program can help us to get to
\be
\sum_{d}\left[\mfA_{dd}(\vp,\vp)+\mfa_{dd}(\vp,\vp)\right]=0
\ee
and
\be
\sum_{d}\left[-\mfD_{dd}(\vp,\vp)-\mfd_{dd}(\vp,\vp)\right]=0.
\ee


\subsubsection*{$\eqbox{\mu=i\;\nu=j}$}
Since
\begin{multline}
\mathcal{T}_{ij}=\frac{\sqrt{\C(\e)}}{4}\left(\gamma^4 \gamma_{i} \partial_j+\gamma^4 \gamma_{j} \partial_{i}
+\frac{1}{8}\frac{\Cpe}{\Ce}\gamma^4\gamma_i[\gamma_0,\gamma_j]+\right.\\
\left.+\frac{1}{8}\frac{\Cpe}{\Ce}\gamma^4\gamma_j[\gamma_0,\gamma_i]\right)+h.c.
\end{multline}
we have
$$
\sum_{d}\left[\VN\left(\mfA_{dd}(\vp,\vp)+\mfa_{dd}(\vp,\vp)\right)-\VM\left(\mfD_{dd}(\vp,\vp)+\mfd_{dd}(\vp,\vp)\right)\right]=
$$
\be\label{tijdiscrete}
=\left[\frac{4 p_i p_j}{L^3 \we \Ce}\left(1-|D^{(-1)}_{(1)}|^2-|D^{(1)}_{(-1)}|^2\right)\right]\left(\VN+\VM\right)\Big|_{m\rightarrow m_1}
\ee
in which (\ref{d1}), (\ref{d2}) and (\ref{d3}) have been used.

\subsubsection*{$\eqbox{\mu=j\;\nu=j}$}
Since
\be
\mathcal{T}_{jj}=\frac{\sqrt{\C(\e)}}{2}\left(\gamma^4 \gamma_{j} \partial_j
+\frac{1}{8}\frac{\Cpe}{\Ce}\gamma^4\gamma_j[\gamma_0,\gamma_j]\right)+h.c.
\ee
we have
$$
\sum_{d}\left[\VN\left(\mfA_{dd}(\vp,\vp)+\mfa_{dd}(\vp,\vp)\right)-\VM\left(\mfD_{dd}(\vp,\vp)+\mfd_{dd}(\vp,\vp)\right)\right]=
$$
\be\label{tiidiscrete}
=\left[\frac{4 p_j^2}{L^3 \we \Ce}\left(1-|D^{(-1)}_{(1)}|^2-|D^{(1)}_{(-1)}|^2\right)\right]\left(\VN+\VM\right) \Big|_{m\rightarrow m_1}
\ee
again using (\ref{d1}), (\ref{d2}) and (\ref{d3}).


\subsubsection{Continous limit}

We shall we take the continuos limit of expressions (\ref{t00discrete}),
(\ref{tijdiscrete}) and (\ref{tiidiscrete}) (being the contribution of $\lfv \hat{T}_{0i} \rfv$ identically equal to zero).
When the parameter $L$ occurring in the boundary conditions is taken $L\rightarrow +\infty$, sums over momenta become integrals
\be
\sum_{\vp}\rightarrow \left(\frac{L \sqrt{\Ce}}{2 \pi}\right)^{3}\int{d\vp}=\left(\frac{L \sqrt{\Ce}}{2 \pi}\right)^{3}\int_0^{\infty}p^2dp\int d\Omega_p
\ee
Moreover, since a cutoff of momenta $K$ is required by the theory, we set
\be
\sum_{\vp}\rightarrow \left(\frac{L \sqrt{\Ce}}{2 \pi}\right)^{3}\int_0^{K}p^2dp\int d\Omega_p
\ee
Expression (\ref{t00discrete}) (time-time component) therefore becomes
\begin{multline}
\sum_{\vp}\frac{4\we}{L^3 \Ce}\left[\left(1-|D^{(-1)}_{(1)}|^2-|D^{(1)}_{(-1)}|^2\right)V^2(p)\right]\rightarrow\\
\rightarrow\frac{4}{(2 \pi)^3} \int^K\!\!\! d\vp \;\we\sqrt{\Ce}\Big[\left(1-|D^{(-1)}_{(1)}|^2-|D^{(1)}_{(-1)}|^2\right)V^2(p)\Big]=\\
=\frac{4}{(2 \pi)^3} \int_{0}^{2 \pi}\!\!\!\!\!\!d\phi \int_{0}^{\pi}\!\!\!d \theta \st \int^K dp\;p^2  \we\sqrt{\Ce}\times\\
\times\left[\left(1-|D^{(-1)}_{(1)}|^2-|D^{(1)}_{(-1)}|^2\right)V^2(p)\right]=\\
=\frac{8}{(2 \pi)^2} \int^K\!\!dp p^2 V^2(p) \we\sqrt{\Ce}\Bigg(\!1-|D^{(-1)}_{(1)}|^2-|D^{(1)}_{(-1)}|^2\!\Bigg)
\end{multline}
in which
\be
V^2(p)\equiv \VN+\VM=\frac{\sqrt{p^2+m_1^2}\sqrt{p^2+m_2^2}-p^2-m_1 m_2}{\sqrt{p^2+m_1^2}\sqrt{p^2+m_2^2}}
\ee
is understood.
Then, from (\ref{tijdiscrete}) (off-diagonal spatial-spatial components) we have
\begin{multline}
\sum_{\vp}\frac{4 p_i p_j}{L^3 \we \Ce}V^2(p)\left[\left(1-|D^{(-1)}_{(1)}|^2-|D^{(1)}_{(-1)}|^2\right)\right]\rightarrow\\
\rightarrow\frac{4}{(2 \pi)^3} \int^K \!\!\!d\vp \;\frac{p_i p_j\sqrt{\Ce}}{\we}V^2(p)\left(1-|D^{(-1)}_{(1)}|^2-|D^{(1)}_{(-1)}|^2\right)=0
\end{multline}
since
\be
\int{d \vp \;p_i p_j f(p)}=\left\{\ba{rl}\int{d\vp\;p_x p_y f(p)}\propto& \int^{2 \pi}_{0}{\cos \phi \sin \phi}=0\\
																			\int{d\vp\;p_x p_z f(p)}\propto& \int^{2 \pi}_{0}{\cos \phi }=0\\
																			\int{d\vp\;p_y p_z f(p)}\propto& \int^{2 \pi}_{0}{\sin \phi}=0
														\ea\right.
\ee
with 
\be
\left\{
\ba{rcl}p_x&=&p \st \cos \phi \\
 				p_y&=&p \st \sin \phi \\
 				p_z&=&p \ct
\ea\right. 				
\;\;\;\mbox{and}\;\;\;d \vp=dpd\theta d\phi\;p^2\st
\ee
Finally, from (\ref{tiidiscrete}) (diagonal spatial-spatial components)
$$
\sum_{\vp}\frac{4 p_i^2}{L^3 \we \Ce}V^2(p)\left(1-|D^{(-1)}_{(1)}|^2-|D^{(1)}_{(-1)}|^2\right)\rightarrow
$$
$$
\rightarrow\frac{4}{(2 \pi)^3} \int^K  \frac{p_i^2\sqrt{\Ce}}{\we}V^2(p)\left(1-|D^{(-1)}_{(1)}|^2-|D^{(1)}_{(-1)}|^2\right)=
$$
\be
=\frac{1}{3}\frac{8}{(2 \pi)^2} \int^K \frac{p^4\sqrt{\Ce}}{\we}V^2(p)\left(1-|D^{(-1)}_{(1)}|^2-|D^{(1)}_{(-1)}|^2\right)
\ee
being
\begin{multline}
\int{d \vp \;p_i^2 f(p)}=\\
				=\left\{\ba{rl}\!\!\int{\!d\vp\;p_x^2 f(p)}=&\!\!\int\! dp\;p^4 f(p)\int^{\pi}_0 d\theta\;\ssst \int^{2 \pi}_0 d\phi \cos^2 \phi\\
									  \!\!\int{\!d\vp\;p_y^2 f(p)}=&\!\!\int\! dp\;p^4 f(p)\int^{\pi}_0 d\theta\;\ssst \int^{2 \pi}_0 d\phi \sin^2 \phi\\
									  \!\!\int{\!d\vp\;p_z^2 f(p)}=&\!\!\int\! dp\;p^4 f(p)\int^{\pi}_0 d\theta\;\st \ccct \int^{2 \pi}_0 d\phi \phi
							 \ea\!\!\right\}=\\
						=\frac{4 \pi }{3} \int dp\;p^4 f(p).
\end{multline}
Formulae (\ref{innerset00befornormalordering}) and (\ref{innersetiibefornormalordering})
presented in Chapter \ref{cAFVICS} are therefore recovered.

\subsubsection{A useful formula}

We shall now derive formula (\ref{parpsi}).
Starting from (\ref{psi}), we first notice that
\be
\partial_\e D_{(a')}^{(a)}(p,\e)=a'\frac{1}{4}\frac{\C'(\e)}{\sqrt{\C(\e)}}\frac{m p}{\w^2}e^{2 i a' \intw}D_{(-a')}^{(a)}(p,\e)
\ee
which derives from (\ref{D}).
Next, we prove that the derivative of the spinor can be written as:
\be\label{spinorderivative}
\partial_{\e}\uade=a\frac{\C'(\e)}{4 \sqrt{\C(\e)}}\frac{m}{\sqrt{p^2+m^2\C(\e)}}\gamma^4 \uade
\ee
provided that $\uade\equiv u^{(a,d)}=(\vp/\sqrt{\C(\e)})$.
First, we notice that:
\bea
\uade&=&\sqrt{\frac{\left(\sqrt{\frac{p^2}{\C}+m^2}+a m\right)\left(\frac{p}{\sqrt{\C}}+d \frac{p_z}{\sqrt{\C}}\right)}{4 m \frac{p}{\sqrt{\C}}}}
\begin{pmatrix}v_d(\e)\\d\frac{\frac{p}{\sqrt{\C}}}{\sqrt{\frac{p^2}{\C}+a m}}\;v_d(\e)\end{pmatrix}=\nn
&=&\sqrt{\frac{\left(\sqrt{p^2 + m^2\sqrt{\C}}\right)\left(p+d p_z\right)}{4 m p}}
\begin{pmatrix}v_d\\ \frac{d p}{\sqrt{p^2+m^2 \sqrt{\C}}+a m \sqrt{\C}}\;v_d\end{pmatrix}=\nn
&=&\sqrt{\frac{g_a(\e)(p+d p_z)}{4 m p}}\begin{pmatrix}v_d\\\frac{dp}{g_a(\e)}\;v_d\end{pmatrix}\nn
&=&
\sqrt{\frac{(p+d p_z)}{4 m p}}\begin{pmatrix}\sqrt{g_a(\e)}v_d\\\frac{dp}{\sqrt{g_a(\e)}}\;v_d\end{pmatrix}
\eea
and
\bea
\partial_{\e}g_a(\e)&=&\frac{m^2 \C'(\e)}{2\sqrt{p^2+m^2 \C(\e)}}+a m \frac{\C'(\e)}{2\sqrt{\C(\e)}}\nn
										&=&\frac{\C'}{2}\left(\frac{m^2}{\we}+a \frac{m}{\sqrt{\C}}\right)=
												\frac{\C'}{2}\frac{m}{\we}\left(\frac{m\sqrt{\C}+a \we}{\sqrt{\C}}\right)\nn
										&=&\frac{\C'}{2\sqrt{\C}}\frac{m}{\we}a(\we+a m \sqrt{\C})=\nn
										&=&\frac{\C'}{2\sqrt{\C}}\frac{m}{\we}a g_a(\e)
\eea
with $\we \equiv \sqrt{p^2+m^2\C(\e)}$, and using the fact that $a=1/a$ since $a=\pm1$ and being $v_d(\e)$ in fact independent of $\e$
(therefore denoted by $v_d$).
Hence we can write:
\bea\label{partialu}
\partial_{\e}\uade&=&\sqrt{\frac{p+d p_z}{4 m p}}\begin{pmatrix}\frac{g_a'(\e)}{2\sqrt{g_a(\e)}}\;v_d\\
																								-\frac{d p g_a'(\e)}{2g_a^{3/2}(\e)}\;v_d\end{pmatrix}=\nn
									&=&a\frac{\C'}{4\sqrt{\C}}\frac{m}{\we}\sqrt{\frac{p+dp_z}{4mp}}\begin{pmatrix}\sqrt{g_a(\e)}\;v_d\\
																								-\frac{d p}{\sqrt{g_a(\e)}}\;v_d\end{pmatrix}=\nn
									&=&a\frac{\C'}{4\sqrt{\C}}\frac{m}{\we}\sqrt{\frac{p+dp_z}{4mp}}
											\begin{pmatrix}\mathbb{I}&0\\0&-\mathbb{I}\end{pmatrix}\begin{pmatrix}\sqrt{g_a(\e)}\;v_d\\
																								\frac{d p}{\sqrt{g_a(\e)}}\;\;v_d\end{pmatrix}=\nn
									&=&a\frac{\C'}{4\sqrt{\C}}\frac{m}{\we}\gamma^4 \uade
\eea
Collecting all together, we have
\begin{multline}
\partial_\e \psi(\eta,\vec{x})
	=\sum A^{(a,b)}(\vp)\Bigg[\left(\frac{1}{L \sqrt{\C(\eta)}}\right)^{\frac{3}{2}} \sqrt{\frac{m}{\w(p,\eta)}}	e^{ia\vp\cdot\vx-ic\int \w d\e}\times\\
			\times\left(\partial_\e (D^{(a)}_{(c)}(p,\e)) u^{(c,abc)}(ac\vp,\e)+ D^{(a)}_{(c)}(p,\e) \partial_\e (u^{(c,abc)}(ac\vp,\e))\right)+\\
	+\partial_\e\left(	\left(\frac{1}{L \sqrt{\C(\eta)}}\right)^{\frac{3}{2}}\sqrt{\frac{m}{\w(p,\eta)}}	e^{ia\vp\cdot\vx-ic\int \w d\e}\right)\times\\
			\times D^{(a)}_{(c)}(p,\e) u^{(c,abc)}(ac\vp,\e)\Bigg]=\nn
\end{multline}
\begin{multline}
=\sum A^{(a,b)}(\vp)\Bigg[\left(\frac{1}{L \sqrt{\C(\eta)}}\right)^{\frac{3}{2}} \sqrt{\frac{m}{\w(p,\eta)}}	e^{ia\vp\cdot\vx-ic\int \w d\e}\times\\
\Bigg(\left(-i\Ce \w(p,\e)-\frac{3\C'(\e)}{4\Ce}-\frac{m^2\C'(\e)}{4\w^2(p,\e)}\right)D_{(-c)}^{(a)}(p,\e) u^{(c,abc)}(ac\vp,\e)+\\
+c\frac{1}{4}\frac{\C'(\e)}{\sqrt{\C(\e)}}\frac{m p}{\w^2(p,\e)}e^{2 i c \intw}D_{(-c)}^{(a)}(p,\e) u^{(c,abc)}(ac\vp,\e)+\\
+c\frac{\C'(\e)}{4 \sqrt{\C(\e)}}\frac{m}{\w(p,\e)}D_{(-c)}^{(a)}(p,\e)\gamma^4 u^{(c,abc)}(ac\vp,\e)\Bigg)\Bigg]=\nn
\end{multline}
\begin{multline}
=\sum A^{(a,b)}(\vp)\Bigg[\left(\frac{1}{L \sqrt{\C(\eta)}}\right)^{\frac{3}{2}} \sqrt{\frac{m}{\w(p,\eta)}}	e^{ia\vp\cdot\vx-ic\int \w d\e}\times\\
\Bigg(
-\frac{(2\w^2(p,\e)+m^2\C(\e)\C'(\e)}{4\w^2(p,\e)\Ce}+\frac{c\;m\;p\;\C'(\e)}{4\sqrt{\C(\e)}\w^2(p,\e)} e^{2ic\!\!\int\!\!\we d\e}+\\
-ic\we+\frac{\C'(e)}{4\sqrt{\Ce}}\frac{c\;m}{\we}\gamma^4\Bigg)u^{(c,abc)}(ac\vp,\e)\Bigg]
\end{multline}
\textit{Q.E.D.}
\subsectionItalic{Stress-Energy Tensor and Equation of State}\label{wandset}\label{aCS}\label{sectioneom}
In the frame in which $\g=diag\{-1,a^2(t),a^2(t),a^2(t)\}$, the Stress-energy tensor a perfect fluid is described by the equation
\be
T_{\mu\nu}=\pressure \g+U_{\mu}U_{\nu}(\rho+\pressure)
\ee
$\pressure$ is the pressure, $\rho$ the energy density, $U^0=1$ and $U^i=0$, that means:
\be\label{setrhop}
T_{\mu\nu}=\begin{pmatrix}\rho&0&0&0\\
													0&a^2(t)\pressure&0&0\\
													0&0&a^2(t)\pressure&0\\
													0&0&0&a^2(t)\pressure
					\end{pmatrix}
\ee
In our case we evaluated $\lfv T_{\mu\nu}\rfv$ in the frame in which $g_{\mu\nu}=\Ce \e_{\mu\nu}$;
let us see how our Stress-energy tensor changes when expressed in this frame:
\be
T'_{\mu\nu}\rightarrow T_{\mu\nu}=\frac{\partial x^\alpha}{\partial x^{'\mu}}\frac{\partial x^\beta}{\partial x^{'\nu}}T'_{\alpha \beta}
\ee
with $\{\e,\vx\}\rightarrow \{t,\vx\}$ and $\sqrt{\C(\e(t))}=a(t)$. We are here denoting with $T'_{\mu\nu}$ the stress-energy tensor in the frame
with the metric $\g=\Ce \e_{\mu\nu}$, whereas $T_{\mu\nu}$ is the stress-energy tensor in the frame with the metric 
$$\g=diag\{-1,a^2(t),a^2(t),a^2(t)\}$$.
Therefore
\bea
T_{00}&=&\frac{1}{a^2(t)}T'_{00}(\e(t))\\
T_{\mu\neq0\;\nu\neq0}&=&T_{\mu\neq0\;\nu\neq0}(\e(t)).
\eea
Comparing this result with (\ref{setrhop}) we have
\bea
\rho&=&T_{00}=\frac{1}{a^2(t)}T'_{00}(\e(t))=\frac{1}{\C(\e(t))}T'_{00}(\e(t))\\
p&=&\frac{1}{a^2(t)}T_{jj}=\frac{1}{\C(\e(t))}T'_{jj}(\e(t))
\eea
that can be expressed as
\be\label{setconf}
T'_{\mu\nu}=\C(\e(t))\begin{pmatrix}\rho&0&0&0\\
													0&\pressure&0&0\\
													0&0&\pressure&0\\
													0&0&0&\pressure
					\end{pmatrix}
\ee

The equation of state is stated as
\be\label{w1}
w=\frac{\pressure}{\rho}
\ee
that in virtue of (\ref{setconf}) in our case becomes
\be\label{w2}
w=\frac{T'_{jj}}{T'_{00}}.
\ee
It will be useful for our calculations to remark that for any rescaling of the metric $\Ce\e_{\mu\nu}\rightarrow \kappa\;\Ce\e_{\mu\nu}$ (with $\kappa$
independent on the coordinates) does not affect the expression for the equation of state:
\be
T''_{\mu\nu}\rightarrow T'_{\mu\nu}=\frac{\partial x^{'\alpha}}{\partial x^{''\mu}}\frac{\partial x^{'\beta}}{\partial x^{''\nu}}T''_{\alpha \beta}=
\kappa T'_{\mu\nu}\;\;\Rightarrow\;\; T''_{\mu\nu}=\frac{1}{k}T'_{\mu\nu}
\ee
\be
\Rightarrow w=\frac{\pressure}{\rho}=\frac{T''_{jj}}{T''_{00}}
\ee

\newpage
\section{APPENDIX TO CHAPTER \ref{cAFVIASM}}\label{aC4}
\markright{\textit{\thesection \quad APPENDIX TO CHAPTER \ref{cAFVIASM}}}
\subsectionItalic{Line of reasoning}\label{line}

As already explained in Section \ref{sGI}, in order to evaluate the quantity
\be\label{set1c}
\lfv T_{\mu\nu}(x)\rfv
\ee
we first simplify the operatorial structure and then the remaining functions over the momenta.
In more details:
the stress-energy tensor $T_{\mu\nu}(x)$ is written in terms of the fields $\psi_i(x)$, $S_i(x)$ and $P_i(x)$ as
\begin{multline}
T_{\mu\nu}(x)=\sum_{i=1,2}\Big( 2\partial_{\left(\mu\right.}S_i(x)\partial_{\left.\nu\right)}S_i(x)+2\partial_{\left(\mu\right.}P_i(x)\partial_{\left.\nu\right)}P_i(x)+\\
+i\bar{\psi}_i(x)\gamma_{\left(\mu\right.}\partial_{\left.\nu\right)}\psi_i(x)\Big)-\e_{\mu\nu}\lag
\end{multline}
with
\begin{multline}\label{lagc}
\lag =\sum_{i=1,2}\Big(\bpsi_j(x)(i\diracpartial-m_j)\psi_j(x)
				+\partial_\mu S_j(x)\partial^\mu S_j(x)-m_j^2S_j^2(x)+\\
				+\partial_\mu P_j(x)\partial^\mu P_j(x)-m_j^2P_j^2(x)\Big)
\end{multline}
Those fields are decomposed in terms of the ladder \textit{mass} operators:
\be\label{psimassc}
\psi_i (x)=\sum_{r=1,2}\int{\frac{d^3k}{(2 \pi)^{3/2}} \left[  a^r_i(\vk) u^r_i (\vk)e^{-i\w_i(k)t} 
						+ v_i^{r} (-\vk) a_i^{r\dagger}(-\vk)e^{i\w_i(k)t}\right]e^{i\vk \cdot \vx}}
\ee
\be\label{smassc}
S_i(x)=\int{\frac{d^3k}{(2 \pi)^{3/2}}\frac{1}{\sqrt{2 \w_i(k)}} \left[  b_i(\vk)e^{-i\w_i(k)t}+ b_i^{\dagger}(-\vk) e^{i\w_i(k)t} \right] e^{i \vk \cdot \vx}  }
\ee
\be\label{pmassc}
P_i(x)=\int{\frac{d^3k}{(2 \pi)^{3/2}}\frac{1}{\sqrt{2 \w_i(k)}} \left[  c_i(\vk)e^{-i\w_i(k)t}+ c_i^{\dagger}(-\vk) e^{i\w_i(k)t} \right] e^{i \vk \cdot \vx}  }
\ee
therefore we can write
\begin{multline}\label{msetc}
T_{\mu\nu}(x)= \sum_{i=1,2} \sum_{r,s} \int d\vp d\vq \Big[ L_{\mu\nu}^{rs}(\vp,\vq,x,m_i) a_i^{r\dagger}(\vp)a_i^s(\vq)+\\
+M_{\mu\nu}^{rs}(\vp,\vq,x,m_i) a_i^{r\dagger}(\vp)a_i^{s\dagger}(\vq)+N_{\mu\nu}^{rs}(\vp,\vq,x,m_i) a_i^{r}(\vp)a_i^{s}(\vq)+\\
+K_{\mu\nu}^{rs}(\vp,\vq,x,m_i) a_i^r(\vp) a_i^{s\dagger}(\vq)  \Big]             
\end{multline}
with $r,s=-1,0,1,2$, defining $a_i^{0}(\vk)\equiv b_i(\vk)$ and $a_i^{-1}(\vk)\equiv c_i(\vk)$, with $L$, $M$, $N$, $K$
suitable functions of the momenta.
When we consider (\ref{set1c}) we are taking the expectation value of this operator with respect the \textit{flavour vacuum} state:
when the stress-energy tensor is written as in (\ref{msetc}), 
the bra $\lfv$ and the ket $\rfv$ act on the \textit{mass} ladder operators (from now on \textit{\textbf{MLO}})
\begin{multline}\label{mset1c}
\lfv T_{\mu\nu}(x)\rfv= \sum_{i=1,2} \sum_{r,s} \int d\vp d\vq \Big[ L_{\mu\nu}^{rs}(\vp,\vq,x,m_i) \lfv a_i^{r\dagger}(\vp)a_i^s(\vq)\rfv+\\
+M_{\mu\nu}^{rs}(\vp,\vq,x,m_i) \lfv a_i^{r\dagger}(\vp)a_i^{s\dagger}(\vq)\rfv+N_{\mu\nu}^{rs}(\vp,\vq,x,m_i) \lfv a_i^{r}(\vp)a_i^{s}(\vq)\rfv+\\
				+K_{\mu\nu}^{rs}(\vp,\vq,x,m_i) \lfv a_i^r(\vp) a_i^{s\dagger}(\vq) \rfv \Big].             
\end{multline}
In order to simplify these expectation values of couples of \textit{mass} operators, we 
rewrite those operators in terms of the \textit{flavour} ladder operators (\textit{\textbf{FLO}}) and then we use the familiar algebra
\be
\{a^{r}_{\iota}(\vq),a^{s\dagger}_{\kappa}(\vp)\}=\delta_{rs}[b_\iota(\vq),b^\dagger_\kappa(\vp)]=
\delta_{rs}[c_\iota(\vq),c^\dagger_\kappa(\vp)]=\delta_{rs}\delta_{\iota \kappa}\delta^3(\vq-\vp)
\ee
(all others being zero) and
\be
a^r_\iota(\vk)\rfv=b_\iota(\vk)\rfv=c_\iota(\vk)\rfv=0.
\ee
More precisely, starting from the definition of the FLO in terms of the MLO:
\be
\left\{
\begin{array}{c}
a^r_A(\vk)=G^{\dagger}_{\theta}(t) a^r_1(\vk)G_{\theta}(t)\\
a^r_B(\vk)=G^{\dagger}_{\theta}(t) a^r_2(\vk)G_{\theta}(t)
\end{array}
\right.
\ee
\be
\left\{
\begin{array}{c}
b_A(\vk)=G^{\dagger}_{\theta}(t) b_1(\vk) G_{\theta}(t)\\
b_B(\vk)=G^{\dagger}_{\theta}(t) b_2(\vk) G_{\theta}(t)
\end{array}
\right.
\ee
\be
\left\{
\begin{array}{c}
c_A(\vk)=G^{\dagger}_{\theta}(t) c_1(\vk) G_{\theta}(t)\\
c_B(\vk)=G^{\dagger}_{\theta}(t) c_2(\vk) G_{\theta}(t)
\end{array}
\right.
\ee
with
\be
G_{\theta}(t)=e^{\theta \int d\vx \left(X_{12}(x)-X_{21}(x)\right)}
\ee
and
\be
X_{12}(x)\equiv \frac{1}{2}\psi^\dagger_1(x)\psi_2(x)+i \dot{S}_2(x) S_1(x)+i \dot{P}_2(x) P_1(x)
\ee
one is able to derive simpler relations
\begin{equation}\begin{split}
a^r_A(\vk)=&\ct a_1^r (\vk)-\st \sum_s \left( W^{rs}(\vk,t)a^s_2(\vk)+Y^{rs}(\vk,t)a^{s\dagger}_2(-\vk) \right)\\
a^r_B(\vk)=&\ct a_2^r(\vk)+\st \sum_s \left(  W^{sr*}(\vk,t) a_1^s(\vk)+Y^{sr}(-\vk,t)a_1^{s\dagger}(-\vk)  \right)
\end{split}\end{equation}
\begin{equation}\begin{split}
b_A(\vk)=&\ct b_1(\vk) +\st \left( U^*(k,t) b_2(\vk)+V(k,t) b_2^\dagger (-\vk)\right)\\
b_B(\vk)=&\ct b_2(\vk) -\st \left( U^*(k,t) b_1(\vk)-V(k,t) b_1^\dagger (-\vk)\right)
\end{split}\end{equation}
\begin{equation}\begin{split}
c_A(\vk)=&\ct c_1(\vk) +\st \left( U^*(k,t) c_2(\vk)+V(k,t) c_2^\dagger (-\vk)\right)\\
c_B(\vk)=&\ct c_2(\vk) -\st \left( U^*(k,t) c_1(\vk)-V(k,t) c_1^\dagger (-\vk)\right)
\end{split}\end{equation}
where
\begin{equation}\begin{split}
W^{rs}(\vk,t)\equiv& \frac{1}{2}\left( u^{r\dagger}_2(\vk)u_1^s(\vk)+v_2^{s\dagger}(\vk)v_1^r(\vk) \right)e^{i(\w_1-\w_2)t}\\
Y^{rs}(\vk,t)\equiv& \frac{1}{2}\left( u^{r\dagger}_2(\vk)v_1^s(-\vk)+u_1^{s\dagger}(-\vk)v_2^r(\vk) \right)e^{i(\w_1+\w_2)t}
\end{split}\end{equation}
\begin{equation}\begin{split}
U(k,t)\equiv& \frac{1}{2}\left(\sqrt{\frac{\w_1}{\w_2}}+\sqrt{\frac{\w_2}{\w_1}}\right)e^{-i(\w_1-\w_2)t}\\
V(k,t)\equiv& \frac{1}{2}\left(\sqrt{\frac{\w_1}{\w_2}}-\sqrt{\frac{\w_2}{\w_1}}\right)e^{i(\w_1+\w_2)t}
\end{split}\end{equation}
By solving those expressions for the MLO one gets
\begin{equation}\begin{split}\label{amass}
a_1^r(\vk)=&\ct a_A^r(\vk)+\st \sum_s \left(W^{rs}(\vk,t)a_B^s(\vk)+Y^{rs}(\vk,t) a_B^{s\dagger}(-\vk)\right)\\
a_2^r(\vk)=&\ct a_B^r(\vk)-\st \sum_s \left(W^{sr*}(\vk,t)a_A^s(\vk)+Y^{sr}(-\vk,t) a_A^{s\dagger}(-\vk)\right)
\end{split}\end{equation}
\begin{equation}\begin{split}\label{bmass}
b_1(\vk)=&\ct b_A(\vk)-\st\left(U^*(k,t)b_B(\vk)+V(k,t)a^\dagger_B(-\vk)\right)\\
b_2(\vk)=&\ct b_B(\vk)+\st\left(U(k,t)b_A(\vk)-V(k,t)a^\dagger_A(-\vk)\right)
\end{split}\end{equation}
\begin{equation}\begin{split}\label{cmass}
c_1(\vk)=&\ct c_A(\vk)-\st\left(U^*(k,t)c_B(\vk)+V(k,t)a^\dagger_B(-\vk)\right)\\
c_2(\vk)=&\ct c_B(\vk)+\st\left(U(k,t)c_A(\vk)-V(k,t)a^\dagger_A(-\vk)\right)
\end{split}\end{equation}
Formulae (\ref{amass}), (\ref{bmass}) and (\ref{cmass}) enable us to evaluate the expressions $\lfv MLO \rfv$ in (\ref{mset1c}).

Being left with functions of the momenta, the last step
is to simplify and evaluate the integrals over $\vp$ and $\vq$.

\subsectionItalic{Outline}\label{outlinec}

In the rest of this Section we will derive all the formulae stated in the previous section (\ref{line}) and
we will get to the results exposed in Section \ref{sFOTFVsusy}.

The Lagrangian for the free WZ model that we have considered (\ref{lagc}) is the sum of free Lagrangians for six
different fields: two scalars, two pseudo-scalars and two Majorana spinors. Being free, the fields
are independent and we can consider the whole theory as a simple ``superposition'' of six different free theories,
each of them describing a single field.
More precisely, the fields do not interact with each other, therefore the Hilbert space that describes our theory
is actually a tensorial product of six different Hilbert spaces, each of them describing a different field.

Such a structure is reflected in the Fock space for flavours, with one single difference: 
we are now able to split the total Hilbert space into three different and independent Hilbert spaces,
instead of six, since fields with same behaviour under Lorentz transformations but of different flavours mix up together.

As a consequence of this, we are allowed to analyze the three theories \textit{separately}:
the expectation value for the stress-energy tensor of the full theory is just the sum of the three contribution
coming from those theories.
Moreover, for our purposes, pseudo-scalars and scalars are indistinguishable (thinks would be different
in the interactive case), therefore we can just consider a theory for two scalars and a theory for two Majorana spinors
with mixing.

The rest of the present Section is indeed divided into two parts: one for the bosonic theory (Section \ref{asusyBS}),
the other for the fermionic (Section \ref{asusyFS}).
In each of them we will evaluate the flavour vacuum expectation value of the stress-energy tensor,
and at the end we will sum up the different contributions to get to the WZ flavour vacuum.
The derivation of such quantities in the bosonic and the fermionic theories are conceptually equivalent,
and technically different just because of the spinorial structure of the fermionic fields.
A computer program (\textit{Mathematica}) has been used for simplifying most of the expressions in which spinors and matrices
(but not operators) were involved, adopting the explicit form stated in \cite{Itzykson:1980rh}, and reported in Section \ref{prel}.
Nonetheless, final results are independent on the specif form used for the spinors, as we will see in Appendix \ref{aC5}.

For sake of simplicity we will use a similar notation: for instance, we will use $a_i^{(\dagger)}(\vk)$ both for the 
bosonic MLO and $a_i^{r(\dagger)}(\vk)$ for the fermionic ones, instead of the notation introduced Section \ref{ssTFFS}.
This will not lead to any confusion since the theories will be treated completely separately and
similar symbols will never appear in the same expression at the same time.
This will also allow the reader to consider the two single Parts separately for further references.

After some preliminary calculations and convention exposed in Section \ref{prel},
we will retrace the path exposed in the previous Section. More specifically, we shall:
\begin{enumerate}
	\item start with deriving the stress energy tensor from the Lagrangian of the theory;
	\item define the $G(t)$ operator, that maps FLO into MLO and vice versa;
	\item use such an operator to get to simple relations between FLO and MLO;
	\item solve relation with respect to the MLO;
	\item use them to evaluate expressions such as $\lfv MLO \rfv$;
	\item evaluate the flavour-vev of the stress energy tensor, collecting all together.
\end{enumerate}

\subsectionItalic{Preliminaries}\label{prel}
\subsubsection{Conventions}

Metric:
\be\label{metricWZ}
\e_{\mu\nu}=diag\{+1,-1,-1,-1\}
\ee

We will often use the following expression for the delta function:
\be
\int \frac{d^3x}{(2\pi)^3}e^{-i(\vp-\vk)\vx}=\delta^3(\vp-\vk)
\ee

For sake of clarity of notation, we will often omit the explicit coordinate or momentum dependency of operators or functions.
Unless is otherwise stated, through all the section it is understood that
\begin{equation}\label{convamp}\begin{split}
\psi^{\slot  }_{\slot  } 	\rightarrow&	 \psi^{\slot  }_{\slot  } (x)\\
\phi^{\slot  }_{\slot  }	\rightarrow&	 \phi^{\slot  }_{\slot  } (x)\\
\pi^{\slot  }_{\slot  }	\rightarrow&	 \pi^{\slot  }_{\slot  } (x)\\
G^{\slot  }	\rightarrow&	G^{\slot  }_\theta(t)\\
X^{\slot  }	\rightarrow&	X^{\slot  }_\theta(t)\\
U,V	\rightarrow& U(\vk,t),V(\vk,t)\\
w_{\slot  } \rightarrow& w_{\slot  } (\vk)\\
a_{\slot  }	\rightarrow&	a_{\slot  } (\vk)\\
W^{\slot  }(\vk),Y^{\slot  }(\vk)	\rightarrow&	W^{\slot  }(\vk,t),Y^{\slot  }(\vk,t)
\end{split}\end{equation}
the $ \slot $ denoting any relevant index or set of indices.
We will also adopt the notation\newnot{symbol:aplusminus}
\be
a^\mp_{\slot} \equiv a^\dagger_{\slot} (-\vk).
\ee

The notation\newnot{symbol:overscore} $A\stackrel{\mbox{\tiny{(n)}}}{=}B$ means that $A=B$ follows from equation ($n$).

For the momentum space coordinates we will use the symbols $\vk$, $\vp$, $\vq$. We will also use the convention
$k=\sqrt{k_1^2+k_2^2+k_3^3}$, $p=\sqrt{p_1^2+p_2^2+p_3^3}$ and $q=\sqrt{q_1^2+q_2^2+q_3^3}$.
An exception will occur: in writing $x k$ we will mean the tensorial notation $x k=x^\mu k_\mu$ and not $x \sqrt{k_1^2+k_2^2+k_3^3}$.

Some of the calculations have been worked out by means of the computer program \textit{Mathematica}.
This is denoted by the symbol\newnot{symbol:overscoreM} $A\stackrel{\mbox{\tiny{M}}}{\slot }B$, as in
$A\viam B$ or $A\stackrel{\mbox{\tiny{M}}}{\Rightarrow}B$.

The symbol\newnot{symbol:dagger} $\slot^\dagger$ acts on operators, giving back their hermitian conjugate, on vectors, giving back the conjugate transpose, and
on c-numbers, giving back their conjugate. The symbol\newnot{symbol:cc} $\slot^*$ acts on c-numbers, giving back their conjugate, and on vectors,
giving back their conjugate \textit{but not} their transpose.

Several stress-energy tensors for different theories have been used. Here we summarize which theory they refer to
\begin{equation*}\begin{split}
T^{b}_{\mu\nu}	\rightarrow&	 \mbox{two free real scalars with mixing}\\
T^{f}_{\mu\nu}	\rightarrow&	 \mbox{two free Majorana spinors with mixing}\\
T^{scalar}_{\mu\nu}	\rightarrow&	 \mbox{one free real scalar}\\
T^{majorana}_{\mu\nu}	\rightarrow&	 \mbox{one free Majorana spinor}\\
T^{WZ}_{\mu\nu}	\rightarrow&	 \mbox{free WZ with mixing}
\end{split}\end{equation*}
holding
\begin{equation}\label{prebos}\begin{split}
T^b_{\mu\nu}=&T^{scalar}_{\mu\nu}\Big|_{m\rightarrow m_1}+T^{scalar}_{\mu\nu}\Big|_{m\rightarrow m_2}\\
T^f_{\mu\nu}=&T^{majorana}_{\mu\nu}\Big|_{m\rightarrow m_1}+T^{majorana}_{\mu\nu}\Big|_{m\rightarrow m_2}\\
T^{WZ}_{\mu\nu}=&2 T^b_{\mu\nu}+T^f_{\mu\nu}.
\end{split}\end{equation}

Finally, let us evaluate some integrals that we will often use.
Using the spherical coordinates system
\be
\left\{
\ba{rcl}p_x=&p \st \cos \phi \\
 				p_y=&p \st \sin \phi \\
 				p_z=&p \ct
\ea\right. 				
\;\;\;\mbox{and}\;\;\;d \vp=dpd\theta d\phi\;p^2\st
\ee
we have that
\be\label{intpipj}
\int{d \vp \;p_i p_j f(p)}=\left\{\ba{rl}\int{d\vp\;p_x p_y f(p)}\propto& \int^{2 \pi}_{0}{\cos \phi \sin \phi}=0\\
																			\int{d\vp\;p_x p_z f(p)}\propto& \int^{2 \pi}_{0}{\cos \phi }=0\\
																			\int{d\vp\;p_y p_z f(p)}\propto& \int^{2 \pi}_{0}{\sin \phi}=0
														\ea\right.
\ee
\begin{multline}\label{intpi2f}
\int{d \vp \;p_i^2 f(p)}=\\
				=\left\{\ba{rl}\!\!\int{\!d\vp\;p_x^2 f(p)}=&\!\!\int\! dp\;p^4 f(p)\int^{\pi}_0 d\theta\;\ssst \int^{2 \pi}_0 d\phi \cos^2 \phi\\
									  \!\!\int{\!d\vp\;p_y^2 f(p)}=&\!\!\int\! dp\;p^4 f(p)\int^{\pi}_0 d\theta\;\ssst \int^{2 \pi}_0 d\phi \sin^2 \phi\\
									  \!\!\int{\!d\vp\;p_z^2 f(p)}=&\!\!\int\! dp\;p^4 f(p)\int^{\pi}_0 d\theta\;\st \ccct \int^{2 \pi}_0 d\phi \phi
							 \ea\!\!\right\}=\\
						=\frac{4 \pi }{3} \int dp\;p^4 f(p).
\end{multline}
and
\be\label{intf}
\int d\vp\;f(p)=4\pi\int dp\;p^2f(p).
\ee

\subsubsection{Spinorial Representation}
Gamma matrices \cite{Itzykson:1980rh}:
\be
\gamma^0=\begin{pmatrix}0&0&0&-i\\0&0&i&0\\0&-i&0&0\\i&0&0&0\end{pmatrix}\;\;\;\;\;\;
\gamma^1=\begin{pmatrix}i&0&0&0\\0&-i&0&0\\0&0&i&0\\0&0&0&-i\end{pmatrix}
\ee
\be
\gamma^2=\begin{pmatrix}0&0&0&i\\0&0&-i&0\\0&-i&0&0\\i&0&0&0\end{pmatrix}\;\;\;\;\;\;
\gamma^3=\begin{pmatrix}0&-i&0&0\\-i&0&0&0\\0&0&0&-i\\0&0&-i&0\end{pmatrix}
\ee
Spinors:
\be
u^1(\vp)=\begin{pmatrix}\frac{(i m+p_x)\sqrt{\frac{-p_y+\w(p)}{\w(p)}}}{\sqrt{2}(p_y-\w(p))}\\
												\frac{p_z\sqrt{\frac{-p_y+\w(p)}{\w(p)}}}{\sqrt{2}(-p_y+\w(p))}\\
												0\\
												\frac{1}{\sqrt{2}}\sqrt{\frac{-p_y+\w(p)}{\w(p)}}
					\end{pmatrix}
\;\;\;\;\;\;u^2(\vp)=\begin{pmatrix}\frac{p_z\sqrt{\frac{-p_y+\w(p)}{\w(p)}}}{\sqrt{2}(p_y-\w(p))}\\
												\frac{i(m+ip_x)\sqrt{\frac{-p_y+\w(p)}{\w(p)}}}{\sqrt{2}+\w(p)}\\
												\frac{1}{\sqrt{2}}\sqrt{\frac{-p_y+\w(p)}{\w(p)}}\\
												0
					\end{pmatrix}
\ee
and
\be
v^r=M u^{r*}
\ee
with $\w(p)=\sqrt{\vp^2+m^2}$ and
\be
M=\begin{pmatrix}0&0&0&-1\\0&0&1&0\\0&-1&0&0\\1&0&0&0\end{pmatrix}.
\ee

\subsubsection{Operatorial Identities}
In the following we will make extensive use of the Baker-Campbell-Hausdorff formula
\be\label{BCH}
e^Y X e^{-Y}=X+[Y,X]+\frac{1}{2}[Y,[Y,X]]+\frac{1}{3!}[Y,[Y,[Y,X]]]+...
\ee
that is valid for two generic linear operators $X$ and $Y$.
If we define the notation\newnot{symbol:Xtothenth}
\be
X^n[Y]\equiv \underbrace{[X,[X,[X,...}_{n},Y\underbrace{]]...]}_{n}
\ee
we can recast (\ref{BCH}) as
\be\label{BCH2}
e^X Y e^{-X}=\sum_{n=0}^\infty \frac{X^n[Y]}{n!}.
\ee
We can notice that
\be
X^0[Y]=Y\;\;\;\;\;\;X^1[Y]=[X,Y]\;\;\;\;\;\;...
\ee
\be
[X,X^n[Y]]=X^{n+1}[Y]\;\;\;\;\;\;X^{n+1}[Y]=X[X^n[Y]]
\ee
and if $z\in\field{C}$ we have
\be\label{propXY}
X^n[zY]=zX^n[Y]\;\;\;\;\;\;\left(zX\right)^n[Y]=z^nX^n[Y].
\ee
In this work we will consider specific operators $X$ that will depend on a parameter $\theta$, therefore
in the rest of the current section we will write $X_\theta$, representing such a dependency.
Moreover, in all cases considered $X_\theta$ will act on operators $Y$ such that
\be\label{xythetaz}
X_\theta[Y]=(i\theta)Z
\ee
with $Z$ some linear operator \textit{independent} of $\theta$,
and
\be\label{xythetay}
X^2_\theta[Y]=(i\theta)^2 Y.
\ee
Provided with (\ref{xythetaz}) and (\ref{xythetay}), it is possible to prove by induction that
\be\label{x2ky}
X_\theta^{2k}[Y]=(i\theta)^{2k}Y
\ee
with $k\in \field{N}_0$. Indeed we have
\begin{equation}\begin{split}
X_\theta^{2(k+1)}[Y]=&X_\theta^{2k}[X_\theta^2[Y]]\via{xythetay}X_\theta^{2k}[(i\theta)^2 Y]=\\
								\via{propXY}&(i\theta)^2 X_\theta^{2k}[ Y]\via{x2ky}(i\theta)^2(i\theta)^{2k}Y=\\
								=&(i\theta)^{2(k+1)}Y
\end{split}\end{equation}
and
\be
X_\theta^{2\cdot0}[Y]=X_\theta^{0}[Y]=Y=(i\theta)^0 Y.
\ee
This result allows us to write
\begin{equation}\begin{split}
e^Y X e^{-Y}=&\sum_{n=0}^\infty \frac{X_\theta ^n[Y]}{n!}\\
		=&\sum_{k=0}^\infty\left(\frac{X_\theta ^{2k}[Y]}{(2k)!}+\frac{X_\theta ^{2k+1}[Y]}{(2k+1)!}\right)=\\												
		=&\sum_{k=0}^\infty\left(\frac{X_\theta ^{2k}[Y]}{(2k)!}+\frac{X_\theta [X_\theta ^{2k}[Y]]}{(2k+1)!}\right)=\\
		\via{x2ky}&\sum_{k=0}^\infty\left(\frac{(i\theta )^{2k}Y}{(2k)!}+\frac{X_\theta [(i\theta )^{2k}Y]}{(2k+1)!}\right)=\\			
		=&\sum_{k=0}^\infty\left(\frac{(i\theta )^{2k}Y}{(2k)!}+\frac{(i\theta )^{2k}X_\theta [Y]}{(2k+1)!}\right)=\\
		\via{xythetaz}&\sum_{k=0}^\infty\left(\frac{(i\theta )^{2k}}{(2k)!}Y+\frac{(i\theta )^{2k+1}}{(2k+1)!}Z\right)=\\
		=&\ct Y+\st Z
\end{split}\end{equation}
In summary
\be\label{BCHsc}
\left\{
\begin{array}{c}
X_\theta[Y]=(i\theta)Z\\
X^2_\theta[Y]=(i\theta)^2 Y
\end{array}
\right.
\Rightarrow e^Y X e^{-Y}=\ct Y+\st Z.
\ee

\subsectionItalic{Bosonic Sector}\label{asusyBS}
\subsubsection{Bosonic Stress Energy Tensor}\label{setb}

Using the definition of the stress-energy tensor provided in \cite{Peskin:1995ev}
\be
T^\mu_\nu(x)\equiv \frac{\partial \lag }{\partial(\partial_\mu \phi)}\partial_\nu \phi-\lag \delta^\mu_\nu
\ee
and using
\be
\lag^{scalar} =\e^{\rho\sigma}\partial_\rho \phi \partial_\sigma \phi-m^2\phi^2
\ee
we have
\be
\frac{\partial \lag^{scalar} }{\partial(\partial_\mu \phi)}=2 \e^{\mu\rho}\partial_\rho \phi=2\partial^\mu \phi
\ee
in which we used $\e_{\mu\nu}=\e_{\nu\mu}$, and therefore
\be\label{bsetbset}
T^{scalar}_{\mu\nu}=2\partial_\mu \phi \partial_\nu \phi-\e_{\mu\nu}\left(\partial^\rho \phi \partial_\rho \phi-m^2\phi^2\right).
\ee\label{bsetc}
\subsubsection{Bosonic $G_\theta(t)$}\label{Gtb}\label{appGb}

Flavour fields $\phi$ are defined as
\begin{equation}\begin{split}\label{phiAB}
\pa (x) =&\po (x) \ct+\pt (x)\st\\
\pb (x)=&-\po (x)\st+\pt (x)\ct
\end{split}\end{equation}
We want to proven that
\begin{equation}\begin{split}\label{paG}
\pa (x)=& G^{\dagger}_\theta (t)\po (x)G_\theta (t)\\
\pb (x)=& G^{\dagger}_\theta (t)\pt (x)G_\theta (t)
\end{split}\end{equation}
with
\be\label{G1}
G_\theta (t)\equiv e^{i\theta \int d^3 x \left(\pi_2 (x)\phi_1 (x)-\pi_1(x)\phi_2(x) \right)}
\ee
with $\pi_j(x)$ the conjugate momentum of $\phi_j(x)$, following the algebra
\be\label{ccrphipi}
[\phi_i(x),\pi_j(y)]_{x_0=y_0}=i\delta_{ij}\delta^3(\vec{x}-\vec{y})\;\;\;\;\;\;\;\;\;i,j=1,2.
\ee

\cdel

In order to apply formula (\ref{BCHsc}) we denote
\be\label{Xboson}
X\equiv i\theta\int d^3x \left(\pi_1 \phi_2-\pi_2 \phi_1 \right)
\ee
and then we have that (\ref{G1}) is written as
\be
G=e^{-X}
\ee
and therefore (\ref{paG}) are written as
\begin{equation}\begin{split}\label{paeX}
\pa =& e^X \po e^{-X}\\
\pb =& e^X \pt e^{-X}
\end{split}\end{equation}
We now evaluate
\begin{multline}\label{Xpo}
[X,\po]=i\theta \int d^3 x \left( [\pi_1,\po]\pt \right)=\\=-i\theta \int d^3 x \left( [\po,\pi_1]\pt \right)=-i\theta i \pt=\theta \pt
\end{multline}
\begin{multline}\label{Xpt}
[X,\pt]=i\theta \int d^3 x \left(-[\pi_2,\pt]\po \right)=\\=i\theta \int d^3 x \left( [\pt,\pi_2]\po \right)=i\theta i \po=-\theta \po
\end{multline}
Looking first at 
\be
\pa = e^X \po e^{-X}
\ee
we have that
\be
X[\po]\via{Xpo}\theta \pt
\ee
and
\be
X^2[\po]\via{Xpo}\theta X[\pt]\via{Xpt}(i\theta)^2\po
\ee
therefore, by applying (\ref{BCHsc}), we obtain
\be
\pa = e^X \po e^{-X}=\po \ct+\pt \st
\ee
On the other hand
\be
X[\pt]\via{Xpt} -\theta \po
\ee
and
\be
X^2[\pt]\via{Xpt}-\theta X[\po]=(i\theta)^2 \pt.
\ee
Therefore
\be
\pb = e^X \pt e^{-X}=\pt \ct-\po \st
\ee

\subsubsection{Bosonic FLO}\label{flob}

Flavour ladder operators (FLO) are defined in terms of the Mass ladder operators (MLO) by
\begin{equation}\begin{split}\label{abosonAB}
a_A(\vk) =&G^{\dagger}_\theta (t)a_1(\vk)G_\theta (t)\\
a_B(\vk)=&G^{\dagger}_\theta (t)a_2(\vk) G_\theta (t)
\end{split}\end{equation}
with $G(t)$ defined in (\ref{G1}). By the means of (\ref{BCHsc}), a simpler relation between
FLO and MLO can be deduced.
To be able to apply (\ref{BCHsc}), we need to evaluate $X[a_i]$ and $X^2[a_i]$.
In order to do that, we first start by deducing some useful relations:
\begin{equation}\begin{split}\label{aph}
		[a_i(\vk),\phi_j(x)]=&\int\frac{d^3p}{(2\pi)^{3/2}}\frac{1}{\sqrt{2\w_i(k)}}[a_i(\vk),a^\dagger_j(-\vp)]e^{i\w_j(p)t+i\vp\cdot\vx}=\\
				=&\int\frac{d^3p}{(2\pi)^{3/2}}\frac{1}{\sqrt{2\w_i(k)}}\delta_{ij}\delta^3(\vk-\vp)e^{i\w_j(p)+i\vp\cdot\vx}=\\
				=&\frac{1}{(2\pi)^{3/2}}\frac{1}{\sqrt{2\w_i(k)}}e^{i\w_j(k)-i\vk\cdot\vx}\delta_{ij}
\end{split}\end{equation}
\be\label{ap}
				[a_i(\vk),\pi_j(x)]=\frac{i}{(2 \pi)^{3/2}}\sqrt{\frac{\w_i(k)}{2}}e^{i\w_i(k)-i\vk\cdot\vx}\delta_{ij}
\ee
\begin{multline}\label{inteph}
				\int \frac{d^3x}{(2\pi)^{3/2}}e^{-i\vk\cdot\vx}\phi_i(x)=\\
	=\!\int\!\! \frac{d^3x}{(2\pi)^{3/2}}e^{-i\vk\cdot\vx}\frac{d^3p}{(2\pi)^{3/2}}\frac{1}{\sqrt{2\w_i(p)}}
						\left(a_i(\vp)e^{-i\w_i(p)t}+a_i^\dagger(-\vp)e^{i\w_i(p)t}\right)e^{i\vp\cdot\vx}=\\
		=\!\int\!\!\frac{d^3p}{\sqrt{2\w_i(p)}}\left(a_i(\vp)e^{-i\w_i(p)t}+a_i^\dagger(-\vp)e^{i\w_i(p)t}\right)
																	\int \frac{d^3x}{(2\pi)^3}e^{-i(\vk-\vp)\vx}=\\
		=\!\int\!\!\frac{d^3p}{\sqrt{2\w_i(p)}}\left(a_i(\vp)e^{-i\w_i(p)t}+a_i^\dagger(-\vp)e^{i\w_i(p)t}\right)\delta^3(\vk-\vp)=\\
		=\frac{1}{\sqrt{2\w_i(k)}}\left(a_i(\vk)e^{-i\w_i(k)t}+a_i^\dagger(-\vk)e^{i\w_i(k)t}\right)
\end{multline}
\be\label{intep}
				\int
				\frac{d^3x}{(2\pi)^{3/2}}e^{-i\vk\cdot\vx}\pi_i(x)=-i\sqrt{\frac{w_i(k)}{2}}
				\left(a_i(\vk)e^{-i\w_i(k)t}-a_i^\dagger(-\vk)e^{i\w_i(k)t}\right)
\ee
in which we used
\be\label{scalardec}
\phi_i(x)=\int{\frac{d^3k}{(2 \pi)^{3/2}}\frac{1}{\sqrt{2 \w_i(k)}} \left[  b_i(\vk)e^{-i\w_i(k)t}+ b_i^{\dagger}(-\vk) e^{i\w_i(k)t} \right] e^{i \vk \cdot \vx}  }
\ee
and $\pi_i=\dot{\phi}_i$.

We are now ready for $X[a_1(\vk)]$:
\begin{multline}\label{Xa1}
X[a_1(\vk)]=[X,a_1(\vk)]=\\
		=i\theta \int d^3x\left([\pi_1(x),a_1(\vk)]\phi_2(x)-\pi_2(x)[\phi_1(x),a_1(\vk)]\right)\doublevia{ap}{aph}\\
	  =i\theta \int \frac{d^3x}{(2\pi)^{3/2}}\left(-i\sqrt{\frac{\w_1(k)}{2}}\phi_2(x)
	  		+\pi_2(x)\frac{1}{\sqrt{2\w_1(k)}}\right)e^{i\w_1(k)t-i\vk\cdot\vx}=\\
	  =i\theta\left(-i\sqrt{\frac{\w_1(k)}{2}}e^{i\w_1(k)t}\int\frac{d^3x}{(2\pi)^{3/2}}\phi_2(x)e^{-i\vk\cdot\vx}+\right.\\
	  			\left.+\frac{1}{\sqrt{2\w_i(k)}}e^{i\w_i(k)t}\int\frac{d^3x}{(2\pi)^{3/2}}\pi_2(x)e^{-i\vk\cdot\vx}\right)\doublevia{inteph}{intep}\\
	  =i\theta\left[  -i\sqrt{\frac{\w_1(k)}{2}}e^{i\w_1(k)t}\frac{1}{\sqrt{2\w_2(k)}}\left(a_2(\vk)e^{-i\w_2(k)t}+
	  					a^\dagger_2(-\vk)e^{i\w_2(k)t}\right)+\right.\\
	 \left.+\frac{1}{\sqrt{2\w_1(k)}}e^{i\w_1(k)t}(-i)\sqrt{\frac{\w_2(k)}{2}}\left(a_2(\vk)e^{-i\w_2(k)t}
	 					-a_2^\dagger(-\vk)e^{i\w_2(k)t}\right)\right]=\\	
	 	=\frac{\theta}{2}\left(\sqrt{\frac{\w_1}{\w_2}}e^{i\w_-t}a_2(\vk)+\sqrt{\frac{\w_1}{\w_2}}e^{i\w_+t}a_2^\dagger(-\vk)+\right.\\
	 	\left.+\sqrt{\frac{\w_2}{\w_1}}e^{i\w_-t}a_2(\vk)-\sqrt{\frac{\w_2}{\w_1}}e^{i\w_+t}a_2^\dagger(-\vk)\right)=\\
	 	=\frac{\theta}{2}\left[\left(\sqrt{\frac{\w_1}{\w_2}}+\sqrt{\frac{\w_2}{\w_1}}\right)e^{i\w_-t}a_2(\vk)+
	 			\left(\sqrt{\frac{\w_1}{\w_2}}-\sqrt{\frac{\w_2}{\w_1}}\right)e^{i\w_+t}a_2^\dagger(-\vk)\right]
\end{multline}
with $\w_\pm\equiv\w_1\pm\w_2$ and $\w_{\slot} = \w_{\slot} (k)$.

We can now easily evaluate $X[a_2(\vk)]$ by considering
\be\label{ottoA}
\otto{X}=-X\Rightarrow
\ee
with $\otto{\slot}$ a symbol that denotes the interchange of the indices 1 and 2.
In virtue of (\ref{ottoA}) we have
\begin{multline}\label{Xa2}
X[a_2(\vk)]=[X,a_2(\vk)]=\otto{[-X,a_1(\vk)]}=-\otto{[X,a_1(\vk)]}=\\
				=-\frac{\theta}{2}\left(\left(\sqrt{\frac{\w_1}{\w_2}}+\sqrt{\frac{\w_2}{\w_1}}\right)e^{-i\w_-t}a_1(\vk)-
	       \left[\sqrt{\frac{\w_1}{\w_2}}-\sqrt{\frac{\w_2}{\w_1}}\right)e^{i\w_+t}a_1^\dagger(-\vk)\right]
\end{multline}
The next step consists in evaluating $X^2[a_i(\vk)]$. Looking carefully at (\ref{Xa1}) and (\ref{Xa2}),
we can see that the knowledge of $X[a^{\dagger}_i(\vk)]$ is required.
Once we notice that
\begin{multline}
X^\dagger=-i\theta \int d^3x \left(\left(\pi_1\phi_2\right)^\dagger-\left(\pi_2\phi_1\right)^\dagger\right)=\\
=-i\theta\int d^3x\left(\pi_1\phi_2-\pi_2\phi_1\right)=-X
\end{multline}
being $\pi^\dagger=\pi$ and $\phi^\dagger=\phi$, it is easy to derive
\begin{multline}
[X,a_i^\dagger]=[-X^\dagger,a_i^\dagger]=-\left(X^\dagger a_i^\dagger-a^\dagger_i X^\dagger\right)=\\
=-\left(\left(a_iX\right)^\dagger-\left(X a_i\right)^\dagger\right)=\left(Xa_i-a_iX\right)^\dagger=\left([X,a_i]\right)^\dagger
\end{multline}
We are now ready to compute $X^2[a_i(\vk)]$.
Calling 
\be\label{UVdef}\begin{array}{c}
U(k,t)\equiv \frac{1}{2}\left(\sqrt{\frac{\w_1(k)}{\w_2(k)}}+\sqrt{\frac{\w_2(k)}{\w_1(k)}}\right)e^{-i\w_- (k)t}\\
V(k,t)\equiv \frac{1}{2}\left(\sqrt{\frac{\w_1(k)}{\w_2(k)}}-\sqrt{\frac{\w_2(k)}{\w_1(k)}}\right)e^{i\w_+ (k)t}\end{array}
\ee
with
$\w_\pm(k)=\w_1(k)\pm\w_2(k)$, we can write
\begin{equation}\begin{split}\label{XXa1}
X^2[a_1]=&[X,[X,a_1]]=\theta\left(U^*[X,a_2]+V[X,a_2^\mp]\right)=\\
		=&\theta \left(U^* (-\theta)\left(Ua_1-Va_1^\mp\right)+V(-\theta)\left(U^*a_1^\mp-V^*a_1\right)\right)=\\
		=&-\theta^2\left(|U|^2a_1-U^*Va_1^\mp+VU^*a_1^\mp-|V|^2a_1\right)=\\
		=&-\theta^2\left(\left(|U|^2-|V|^2\right)a_1+\left(VU^*-U^*V\right)a_1^\mp\right)=\\
		=&-\theta^2 a_1
\end{split}\end{equation}
being
\begin{multline}\label{UV1}
|U|^2-|V|^2=\frac{1}{4}\left(\left(\frac{\w_1+\w_2}{\sqrt{\w_1 \w_2}}\right)^2-\left(\frac{\w_1-\w_2}{\sqrt{\w_1 \w_2}}\right)^2\right)=\\
					=\frac{1}{4}\left(\frac{\w_1^2+\w_2^2+2\w_1\w_2}{\w_1\w_2}-\frac{\w_1^2+\w_2^2-\w_1\w_2}{\w_1\w_2}\right)=\\
					=\frac{1}{4}\left(\frac{2\w_1\w_2+2\w_1\w_2}{\w_1\w_2}\right)=\\
					=1.
\end{multline}
And, thanks to (\ref{ottoA}), we have
\be\label{XXa2}
[X,[X,a_2]]=\otto{[-X,[-X,a_1]]}=\otto{[X,[X,a_1]]}=-\theta^2 a_2
\ee
that can be explicitely verified
\begin{equation}\begin{split}
X^2[a_2]=&[X,[X,a_2]]=-\theta\left(U[X,a_1]-V[X,a_1^\mp]\right)=\\
		=&-\theta \left(U \theta\left(U^*a_2+Va_2^\mp\right)-V\theta\left(Ua_2^\mp+V^*a_2\right)\right)=\\
		=&-\theta^2\left(\left(|U|^2-|V|^2\right)a_2+\left(VU-UV\right)a_2^\mp\right)=\\
		=&-\theta^2 a_2.
\end{split}\end{equation}

Provided with explicit expressions for $X[a_i]$, in formulae (\ref{Xa1}) and (\ref{Xa2}), and $X^2[a_i]$,
in formulae (\ref{XXa1}) and (\ref{XXa2}), we can now apply (\ref{BCHsc}):
\begin{equation}\begin{split}
a_A(\vk)=&G^{-1}_\theta(t)a_1(\vk)G_\theta(t)=e^X a_1(\vk) e^{-X}=\\
	 =&\ct a_1(\vk)+\st (U^*(k,t) a_2(\vk)+V(k,t) a_2^\dagger(-\vk))
\end{split}\end{equation}
and
\begin{equation}\begin{split}
a_B(\vk)=&G^{-1}_\theta(t)a_2(\vk)G_\theta(t=e^Xa_2(\vk)e^{-X}=\\
	 =&\ct a_2(\vk)-\st (U(k,t) a_1(\vk)-V(k,t) a_1^\dagger(-\vk))
\end{split}\end{equation}
with (\ref{UVdef}).
\subsubsection{Bosonic MLO}\label{mlob}
In the previous section we found a simple relations between FLO and MLO.
We want now to explicit such expressions with respect to MLO.
Adopting the convention (\ref{convamp}) and (\ref{convamp}), we have
\be
\left\{
\begin{array}{rcl}
a_A=&\ct a_1+\st \left(U^*a_2+Va_2^\mp \right)\\
a_B=&\ct a_2-\st\left( Ua_1-Va_1^\mp \right)
\end{array}
\right.\Rightarrow
\ee
\begin{equation}\begin{split}\label{a12aAB}
\stackrel{M}{\Rightarrow}\left\{
\begin{array}{rcl}
a_1=&\left(\ct a_A-\st\left( V a_B^\mp + U^* a_B\right)\right)/(*)\\
a_2=&\left(a_B \ct+\st\left( U a_A-V A_A^\mp \right)\right)/(*)
\end{array}
\right.
\end{split}\end{equation}
with
\begin{multline}
(*)=\cct+|U|^2 \sst-|V|^2\sst=\\
=\cct+\sst (|U|^2-|V|^2)=\cct+\sst=1
\end{multline}
being (\ref{UV1}).

\subsubsection{Bosonic Quantum Algebra}\label{qab}

Provided with relations (\ref{a12aAB}),
we are ready now to evaluate the flavour-vacuum expectation value of couples of MLO,
as they appear in (\ref{mset1c}).
Omitting the explicit momentum and time dependency of $U(\vk,t)$ and $V(\vk,t)$,
and recalling that
\be
\lfv 0  \rangle_f=\lmv G\theta(t)G^{-1}_\theta(t)\rmv=\lmv 0  \rangle=1
\ee
we can write
\begin{multline}\label{number1}
\lfv a_1^\dagger(\vp)a_1(\vq)\rfv =\lfv\left(-\st V^* a_B(-\vp)\right)\left(-\st V a^\dagger_B(-\vq)\right)\rfv=\\
							=\sst |V|^2 \delta^3(\vp-\vq) \lfv 0  \rangle_f=\sst |V|^2 \delta^3(\vp-\vq)\;\;\;\;
\end{multline}
\be\label{number2}
\lfv a_2^\dagger(\vp)a_2(\vq)\rfv =\sst |V|^2 \delta^3(\vp-\vq)
\ee
\be
\lfv a_1^\dagger(\vp)a_1^\dagger(\vq)\rfv =\sst U V^* \delta^3(\vp+\vq)
\ee
\be
\lfv a_2^\dagger(\vp)a_2^\dagger(\vq)\rfv =-\sst U^* V^* \delta^3(\vp+\vq)
\ee
\be
\lfv a_1(\vp)a_1(\vq)\rfv =\sst U^* V \delta^3(\vp+\vq)
\ee
\be
\lfv a_2(\vp)a_2(\vq)\rfv =-\sst U V \delta^3(\vp+\vq)
\ee
\be
\lfv a_1(\vp)a_1^\dagger(\vq)\rfv =\left(\cct+\sst |U|^2\right) \delta^3(\vp-\vq)
\ee
\be
\lfv a_2(\vp)a_2^\dagger(\vq)\rfv =\left(\cct+\sst |U|^2\right) \delta^3(\vp-\vq)
\ee
From (\ref{number1}) and (\ref{number2}), it is also possible to deduce the number of massive particles
that are present in the flavour vacuum state per unit volume. For finite volumes we have
\begin{multline}
\frac{1}{L^3}\lfv N\rfv=\frac{1}{L^3}\sum_{\vk}{\lfv\left(a^\dagger_1(\vk)a_1(\vk)+a^\dagger_2(\vk)a_2(\vk)\right)\rfv}=\\
=\sst\frac{1}{L^3}\sum_{\vk}2|V|^2
\end{multline}
where $N$ denotes the particle number operator and $L^3$ the total volume. In the infite volume limit,
for which 
\be
\sum_{\vk}\rightarrow\frac{L^3}{(2\pi)^3}\int d\vk
\ee
holds, the average of massive particles in the flavour vacuum state becomes
\begin{multline}
\lfv \frac{N}{L^3}\rfv\rightarrow \frac{1}{(2\pi)^3}\int d\vk 2 |V|^2=\\
=\frac{\sst}{(2\pi)^3}\int d\vk \frac{1}{2} \frac{(\w_1-\w_2)^2}{\w_1 \w_2}=\frac{\sst}{(2\pi)^2}\int dk \frac{(\w_1-\w_2)^2}{\w_1 \w_2}.
\end{multline}\label{bqac}
\subsubsection{Bosonic Flavour Vacuum}\label{fvevb}

Looking back to formula (\ref{mset1c}), since by now we know how $\lfv \slot \rfv$ are simplified,
we only need explicit expressions for the functions $L_{\mu\nu}(\vp,\vq,x,m_i)$, $M_{\mu\nu}(\vp,\vq,x,m_i)$,
$N_{\mu\nu}(\vp,\vq,x,m_i)$ and $K_{\mu\nu}(\vp,\vq,x,m_i)$, before finally computing the whole formula.
As it is clear from (\ref{mset1c}), we can just consider a theory with one field of mass $m$,
evaluate $L$, $M$, $N$ and $K$ as functions of $m$ and then consider $m\rightarrow m_i$.

We start by recalling the expression of the stress-energy tensor
\be
T^{scalar}_{\mu\nu}(x)=2\partial_{\left(\mu\right.}S(x)\partial_{\left.\nu\right)}S(x)-\e_{\mu\nu}\lag^{scalar}
\ee
\be
\lag^{scalar}=\partial_\lambda S(x)\partial^\lambda S(x)-m^2S^2(x).
\ee
Regarding the energy we have
\begin{multline}
T^{scalar}_{00}(x)=\\
=2\partial_{0}S(x)\partial_{0}S(x)-\e_{00}\left(\partial_0S(x)\partial_0S(x)-\vec{\nabla} S(x)\cdot\vec{\nabla} S(x)-m^2S^2(x) \right)=\\
				=\left(\partial_t S(x)\right)^2+\sum_{j=1}^3 \left(\partial_j S(x)\right)^2+m^2 S^2(x)
\end{multline}
and being (\ref{scalardec}) (in which we omit the mass index $i$)
\begin{multline}
T^{scalar}_{00}(x)=\int \frac{d^3p d^3q}{(2 \pi)^3}\frac{1}{2\sqrt{\w(p)\w(q)}}\times\\
\times\Big[
	\left(-i\w(p)e^{-ipx}a(\vp)+i\w(p)e^{ipx}a^\dagger(\vp)\right)\left(-i\w(q)e^{-iqx}a(\vq)+i\w(q)e^{iqx}a^\dagger(\vq)\right)+\\
				+\left(i\vp e^{-ipx}a(\vp)-i\vp e^{ipx}a^\dagger(\vp)\right)\left(i\vq e^{-iqx}a(\vq)-i\vq e^{iqx}a^\dagger(\vq)\right)+\\
				+\left(m e^{-ipx}a(\vp)+m e^{ipx}a^\dagger(\vp)\right)\left(m e^{-iqx}a(\vq)+m e^{iqx}a^\dagger(\vq)\right)\Big]=\\
				=\int \frac{d^3p d^3q}{(2 \pi)^3}\frac{1}{2\sqrt{\w(p)\w(q)}}
				\Big[\left(-\w(p)\w(q)-\vp\cdot\vq+m^2\right)e^{-ipx}e^{-iqx}a(\vp)a(\vq)+\\
				+\left(\w(p)\w(q)+\vp\cdot\vq+m^2\right)e^{-ipx}e^{iqx}a(\vp)a^\dagger(\vq)+\\
				+\left(\w(p)\w(q)+\vp\cdot\vq+m^2\right)e^{ipx}e^{-iqx}a^\dagger(\vp)a(\vq)+\\
				+\left(-\w(p)\w(q)-\vp\cdot\vq+m^2\right)e^{ipx}e^{iqx}a^\dagger(\vp)a^\dagger(\vq)\Big]
\end{multline}
with the convention $px\equiv \w(p)t-\vp\cdot\vx$.
If we want to write $T_{00}(x)$ as
\begin{multline}
T^{scalar}_{00}(x)=\int d^3p\;d^3q\;\Big[N_{00}(\vp,\vq,x,m)a(\vp)a(\vq)+\\
		+K_{00}(\vp,\vq,x,m) a(\vp)a^\dagger(\vq)+L_{00}(\vp,\vq,x,m) a^\dagger(\vp)a(\vq)+\\
		+M_{00}(\vp,\vq,x,m) a^\dagger(\vp)a^\dagger(\vq)\Big]
\end{multline}
one can easily recognize
\begin{equation}\begin{split}
N_{00}(\vp,\vq,x,m)\equiv& \frac{\left(-\w(p)\w(q)-\vp\cdot\vq+m^2\right)e^{-ipx}e^{iqx}}{(2 \pi)^3 2\sqrt{\w(p)\w(q)}}\\
K_{00}(\vp,\vq,x,m)\equiv& \frac{\left(\w(p)\w(q)+\vp\cdot\vq+m^2\right)e^{-ipx}e^{iqx}}{(2 \pi)^3 2\sqrt{\w(p)\w(q)}}\\
L_{00}(\vp,\vq,x,m)\equiv& \frac{\left(\w(p)\w(q)+\vp\cdot\vq+m^2\right)e^{ipx}e^{-iqx}}{(2 \pi)^3 2\sqrt{\w(p)\w(q)}}\\
M_{00}(\vp,\vq,x,m)\equiv& \frac{\left(-\w(p)\w(q)-\vp\cdot\vq+m^2\right)e^{ipx}e^{iqx}}{(2 \pi)^3 2\sqrt{\w(p)\w(q)}}
\end{split}\end{equation}
Clearly
\begin{equation}\begin{split}
L_{00}(\vp,\vp,x,m)=&\frac{\w^2(p)+\vp^2+m^2}{(2 \pi)^32\w(p)}=\frac{2\w^2(p)}{(2 \pi)^32\w(p)}=\frac{\w(p)}{(2 \pi)^3}\\
M_{00}(\vp,-\vp,x,m)=&\frac{-\w^2(p)+\vp^2+m^2}{(2 \pi)^32\w(p)}e^{2ipx}=0\\
N_{00}(\vp,-\vp,x,m)=&\frac{-\w^2(p)+\vp^2+m^2}{(2 \pi)^32\w(p)}e^{-2ipx}=0\\
K_{00}(\vp,\vp,x,m)=&\frac{\w(p)}{(2 \pi)^3}=L_{00}(\vp,\vp,x,m)
\end{split}\end{equation}

Similarly for the pressure
\be
T^{scalar}_{jj}(x)=2(\partial_jS(x))^2
	+\left(\left(\partial_t S(x)\right)^2-\left(\vec{\nabla}S(x)\right)^2-m^2S^2(x)\right)
\ee
being
\be
\partial_j e^{ipx}=\frac{\partial}{\partial x^j}e^{i(p_0x^0+p_jx^j)}=ip_je^{ipx}
\ee
with $P^{\mu}=(p^0,\vp)$ and hence $\vp=\{p^j\}=\{-p_j\}$, we have
\begin{multline}
T^{scalar}_{jj}(x)=\int \frac{d^3p d^3q}{(2 \pi)^3}\frac{1}{2\sqrt{\w(p)\w(q)}}\times\\
	\times\Big[2 \left(-i\;p_je^{-ipx}a(\vp)+i\;p_je^{ipx}a^\dagger(\vp)\right)\left(-i\;q_je^{-iqx}a(\vq)+i\;q_je^{iqx}a^\dagger(\vq)\right)+\\
	+\left(-i\w(p)e^{-ipx}a(\vp)+i\w(p)e^{ipx}a^\dagger(\vp)\right)\left(-i\w(q)e^{-iqx}a(\vq)+i\w(q)e^{iqx}a^\dagger(\vq)\right)+\\
	-\left(i\vp e^{-ipx}a(\vp)-i\vp e^{ipx}a^\dagger(\vp)\right)\left(i\vq e^{-iqx}a(\vq)-i\vq e^{iqx}a^\dagger(\vq)\right)+\\
	-\left(m e^{-ipx}a(\vp)+m e^{ipx}a^\dagger(\vp)\right)\left(m e^{-iqx}a(\vq)+m e^{iqx}a^\dagger(\vq)\right)\Big]=
\end{multline}
\begin{multline}
	=\int \frac{d^3p d^3q}{(2 \pi)^3}\frac{1}{2\sqrt{\w(p)\w(q)}}\times\\
	\times\Big[a(\vp)a(\vq)\left(2ip_jiq_j+(-i\w(p))(-i\w(q))-(i\vp)(i\vq)-m^2\right)e^{-ipx}e^{-iqx}+\\
	+a^\dagger(\vp)a(\vq)\left(-2ip_jiq_j+i\w(p)(-i\w(q))+(i\vp)(i\vq)-m^2\right)e^{ipx}e^{-iqx}+\\
	+a(\vp)a^\dagger(\vq)\left(-2ip_jiq_j-i\w(p)(i\w(q))-i\vp(-i\vq)-m^2\right)e^{-ipx}e^{iqx}+\\
	+a^\dagger(\vp)a^\dagger(\vq)\left(2ip_j iq_j+i\w(p)(i\w(q))-(-i\vp)(-i\vq)-m^2\right)e^{ipx}e^{iqx}\Big]=\nonumber
\end{multline}
\begin{multline}
	=\int \frac{d^3p d^3q}{(2 \pi)^3}\frac{1}{2\sqrt{\w(p)\w(q)}}\times\\
	\times\Big[a(\vp)a(\vq)\left(-2p_jq_j-\w(p)\w(q)+\vp\cdot\vq-m^2\right)e^{-ipx}e^{-iqx}+\\
	+a^\dagger(\vp)a(\vq)\left(2 p_jq_j+\w(p)\w(q)-\vp\cdot \vq-m^2\right)e^{ipx}e^{-iqx}+\\
	+a(\vp)a^\dagger(\vq)\left(2 p_j q_j+\w(p)\w(q)-\vp\cdot \vq-m^2\right)e^{-ipx}e^{iqx}+\\
	+a^\dagger(\vp)a^\dagger(\vq)\left(-2p_j q_j-\w(p)\w(q)+\vp\cdot\vq-m^2\right)e^{ipx}e^{iqx}\Big]\nonumber
\end{multline}
\begin{equation}\begin{split}
N_{jj}(\vp,\vq,x,m)\equiv& \frac{\left(-2p_j q_j-\w(p)\w(q)+\vp\cdot\vq-m^2\right)}{(2 \pi)^3 2\sqrt{\w(p)\w(q)}}e^{-ipx-iqx}\\
K_{jj}(\vp,\vq,x,m)\equiv& \frac{\left(2 p_j q_j+\w(p)\w(q)-\vp\cdot \vq-m^2\right)}{(2 \pi)^3 2\sqrt{\w(p)\w(q)}}e^{-ipx+iqx}\\
L_{jj}(\vp,\vq,x,m)\equiv& \frac{\left(2 p_j q_j+\w(p)\w(q)-\vp\cdot \vq-m^2\right)}{(2 \pi)^3 2\sqrt{\w(p)\w(q)}}e^{ipx-iqx}\\
M_{jj}(\vp,\vq,x,m)\equiv& \frac{\left(-2p_j q_j-\w(p)\w(q)+\vp\cdot\vq-m^2\right)}{(2 \pi)^3 2\sqrt{\w(p)\w(q)}}e^{ipx+iqx}
\end{split}\end{equation}


\cdel


We are now ready to evaluate (\ref{mset1c}) for our real scalar theory.
We have for the energy
\begin{multline}\label{enebosV}
\lfv T^{b}_{00}\rfv=\!\!\sum_{i=1,2}\int d\vp d\vq \Big[L_{00}(\vp,\vq,x,m_i)\lfv a_i^\dagger(\vp)a_i(\vq)\rfv+\\
			+M_{00}(\vp,\vq,x,m_i)\lfv a_i^\dagger(\vp)a_i^\dagger(\vq)\rfv+N_{00}(\vp,\vq,x,m_i)\lfv a_i(\vp)a_i(\vq)\rfv+\\
			+K_{00}(\vp,\vq,x,m_i)\lfv a_i(\vp)a_i^{\dagger}(\vq)\rfv\Big]=\\
								=\int \frac{d\vp }{(2\pi)^3}\Big[\w_1(p)\left(\sst |V|^2+\cct+\sst |U|^2\right)+\\
										+\w_2(p)\left(\sst |V|^2+\cct+\sst |U|^2\right)\Big]=\\
								\via{UV1}\int \frac{d\vp}{(2\pi)^3}(\w_1(p)+\w_2(p))\left(\sst |V|^2+\cct+\sst+\sst |V|^2\right)=\\
								=\int \frac{d\vp}{(2\pi)^3}(\w_1(p)+\w_2(p))\left(1+2\sst |V|^2\right)
\end{multline}
in which we used
\be
M_{00}(\vp,-\vp,x,m_i)=N_{00}(\vp,-\vp,x,m_i)=0
\ee
\be
L_{00}(\vp,\vp,x,m_i)=\frac{\w_i(p)}{(2\pi)^3}=K_{00}(\vp,\vp,x,m_i)
\ee
and previous relations on the quantum algebra.
Recalling the explicit expression for $V$ from formula (\ref{UVdef}), we can finally write
\begin{multline}\label{enebos1}
\lfv T^{b}_{00}\rfv=\\
	=\int \frac{d\vp}{(2\pi)^3}(\w_1(p)+\w_2(p))\left[1
									+2\sst\frac{1}{4}\left(\sqrt{\frac{\w_1(p)}{\w_2(p)}}-\sqrt{\frac{\w_2(p)}{\w_1(p)}}\right)^2\right]=\\
									=\int \frac{d\vp}{(2\pi)^3}(\w_1(p)+\w_2(p))\left[1
									+\sst \frac{1}{2}\left( \frac{(\w_1(p)-\w_2(p))^2}{\w_1(p)\w_2(p)} \right)\right]
\end{multline}
that can be also written in the form, that we will later use,
\begin{multline}\label{phienergyc}
\lfv T^{b}_{00}\rfv=\int \frac{d\vp}{(2\pi)^3}(\w_1(p)+\w_2(p))+\\
									+\sst\int \frac{d\vp}{2^4\pi^3}(\w_1(p)+\w_2(p))(\w_1(p)-\w_2(p))\left( \frac{\w_1(p)-\w_2(p)}{\w_1(p)\w_2(p)} \right)=\\
									=\int \frac{d\vp}{(2\pi)^3}\Big[(\w_1(p)+\w_2(p))
									+\sst \frac{(m_1^2-m_2^2)}{2}\left( \frac{1}{\w_2(p)}- \frac{1}{\w_1(p)}\right)\Big]
\end{multline}
in which we used $\w_1(p)^2-\w_2(p)^2=m_1^2-\vp^2-m_2^2+\vp^2=m_1^2-m_2^2$.

We define now $\U$ and $\V$ 
\be\label{curlUcurlVdef}
\U\equiv \frac{1}{2}\left(\sqrt{\frac{\w_1(k)}{\w_2(k)}}+\sqrt{\frac{\w_2(k)}{\w_1(k)}}\right)\;\;\;\;\;
\V\equiv \frac{1}{2}\left(\sqrt{\frac{\w_1(k)}{\w_2(k)}}-\sqrt{\frac{\w_2(k)}{\w_1(k)}}\right)
\ee
and therefore
\be\label{curlUcurlVUV}
U= \U e^{-i\w_- (k)t}\;\;\;\;\;\;\;\;\;\;
V= \V e^{i\w_+ (k)t}.
\ee
For the pressure we have
\begin{multline}
\lfv T_{jj}^{b}(x)\rfv=\int\frac{d^3pd^3q}{(2\pi)^3}\frac{1}{2}\times\\
		\times\Bigg[\left( \lfv a_1(\vp)a_1(\vq)\rfv e^{-i(p+q)x}+\lfv a^\dagger_1(\vp)a_1^\dagger(\vq)\rfv e^{i(p+q)x} \right)\times\\
	\times\left( \frac{-2p_jq_j-\w_1(p)\w_1(q)+\vp\cdot\vq-m_1^2}{\sqrt{\w_1(p)\w_1(q)}}\right)+\\
	+\left( \lfv a_1^\dagger(\vp)a_1(\vq)\rfv e^{i(p-q)x}+\lfv a_1(\vp)a_1^\dagger(\vq)\rfv e^{-i(p-q)x} \right)	\times\\
	\times	\left( \frac{+2p_jq_j+\w_1(p)\w_1(q)-\vp\cdot\vq-m_1^2}{\sqrt{\w_1(p)\w_1(q)}}\right)+\\
	+\left( \lfv a_2(\vp)a_2(\vq)\rfv e^{-i(p+q)x}+\lfv a^\dagger_2(\vp)a_2^\dagger(\vq)\rfv e^{i(p+q)x} \right)\times\\
	\times	\left( \frac{-2p_jq_j-\w_2(p)\w_2(q)+\vp\cdot\vq-m_2^2}{\sqrt{\w_2(p)\w_2(q)}}\right)+\\
	+\left( \lfv a_2^\dagger(\vp)a_2(\vq)\rfv e^{i(p-q)x}+\lfv a_2(\vp)a_2^\dagger(\vq)\rfv e^{-i(p-q)x} \right)\times\\
	\times	\left( \frac{+2p_jq_j+\w_2(p)\w_2(q)-\vp\cdot\vq-m_2^2}{\sqrt{\w_2(p)\w_2(q)}}\right)\Bigg]=\nonumber
\end{multline}
\begin{multline}
	=\int\frac{d^3pd^3q}{(2\pi)^3}\frac{1}{2}\times\\
	\times\Bigg[2\sst\;\; \U \V \delta^3(\vp+\vq)\left(\frac{-2p_jq_j-\w_1(p)\w_1(q)+\vp\cdot\vq-m_1^2}{\sqrt{\w_1(p)\w_1(q)}}\right)+\\
	+(1+2 \sst \V^2)\delta^3(\vp-\vq)\left( \frac{+2p_jq_j+\w_1(p)\w_1(q)-\vp\cdot\vq-m_1^2}{\sqrt{\w_1(p)\w_1(q)}}\right)+\\
	+(-2\sst \U\V)\delta^3(\vp+\vq)\left( \frac{-2p_jq_j-\w_2(p)\w_2(q)+\vp\cdot\vq-m_2^2}{\sqrt{\w_2(p)\w_2(q)}}\right)+\\
	+(1+2 \sst \V^2)\delta^3(\vp-\vq)\left( \frac{+2p_jq_j+\w_2(p)\w_2(q)-\vp\cdot\vq-m_2^2}{\sqrt{\w_2(p)\w_2(q)}}\right)\Bigg]=\nonumber
\end{multline}
\begin{multline}
	=\int\frac{d^3p}{(2\pi)^3}\frac{1}{2}\times\\
	\times\Bigg[ 2\sst \U \V \left(\frac{2p_j^2-\w_1^2-\vp^2-m_1^2}{\w_1}-\frac{2p_j^2-\w_2^2-\vp^2-m_2^2}{\w_2}\right)+\\
	+(1+2\sst \V^2)\left(\frac{2p_j^2+\w_1^2-\vp^2-m_1^2}{\w_1}-\frac{2p_j^2+\w_2^2-\vp^2-m_2^2}{\w_2}\right)\Bigg]=\\
	=\int\frac{d^3p}{(2\pi)^3}\Bigg[2\sst \U \V\left(\frac{p_j^2-\w_1^2}{\w_1}-\frac{p_j^2-\w_2^2}{\w_2}\right)+\\
				+(1+2\sst \V^2)p_j^2\left(\frac{1}{\w_1}+\frac{1}{\w_2}\right)\Bigg]=\\
	=\int\frac{d^3p}{(2\pi)^3}\Bigg[p^2_j\left(\frac{1}{\w_1}+\frac{1}{\w_2}\right)+\\
					+2\sst \left(\U \V \left(\frac{p_j^2-\w_1^2}{\w_1}-\frac{p_j^2-\w_2^2}{\w_2}\right)
					+\V^2 p^2_j\left(\frac{1}{\w_1}+\frac{1}{\w_2}\right) \right)\Bigg]
\end{multline}
then, using 
\be\label{smallusmallv1}
\U\V=\frac{1}{2}\frac{(\w_1+\w_2)(\w_1-\w_2)}{\w_1\w_2}\;\;\;\;\V^2=\frac{1}{4}\frac{(\w_1-\w_2)^2}{\w_1\w_2}
\ee
\begin{multline}\label{smallusmallv2}
\frac{p_j^2-\w_1^2}{\w_1}-\frac{p_j^2-\w_2^2}{\w_2}=\frac{p_j^2\w_2-\w_1^2\w_2-p_j^2\w_1+\w_2^2\w_1}{\w_1\w_2}=\\
=\frac{p_j^2(\w_2-\w_1)+\w_1\w_2(\w_2-\w_1)}{\w_1\w_2}=-\frac{(\w_1-\w_2)(p_j^2+\w_1\w_2)}{\w_1\w_2}
\end{multline}
and
\be\label{simplw}
(\w_1-\w_2)(\w_1+\w_2)=\w_1^2-\w_2^2=m_1^2+\vp^2-m_2^2-\vp^2=m_1^2-m_2^2
\ee
we have
\begin{multline}
\lfv T_{jj}^{b}(x)\rfv\doublevia{smallusmallv1}{smallusmallv1}\int\frac{d^3p}{(2\pi)^3}\Bigg[p^2_j\left(\frac{1}{\w_1}+\frac{1}{\w_2}\right)+\\
				+\frac{\sst}{2}\left(\frac{(\w_1-\w_2)}{(\w_1\w_2)^2}(\w_1+\w_2)(\w_1-\w_2)(-p_j^2-\w_1\w_2)+\right.\\
				\left.+\frac{(\w_1-\w_2)^2}{\w_1\w_2}\frac{(\w_1+\w_2)}{\w_1\w_2}p_j^2\right)\Bigg]=\nonumber
\end{multline}
\begin{multline}
	=\int\frac{d^3p}{(2\pi)^3}\Bigg[p^2_j\left(\frac{1}{\w_1}+\frac{1}{\w_2}\right)+\\
				+\frac{\sst}{2}\frac{(\w_1-\w_2)^2}{(\w_1\w_2)^2}(\w_1+\w_2)(-p_j^2-\w_1\w_2+p_j^2)\Bigg]=\nonumber
\end{multline}
\begin{equation*}
	=\int\frac{d^3p}{(2\pi)^3}\Bigg[p^2_j\left(\frac{1}{\w_1}+\frac{1}{\w_2}\right)
				-\frac{\sst}{2}\frac{(\w_1-\w_2)^2(\w_1+\w_2)}{\w_1\w_2}\Bigg]=\end{equation*}
\begin{equation*}\via{intpi2f}\int dp\;p^2\frac{4\pi}{4\pi2\pi^2}\Bigg[\frac{p^2}{3}\left(\frac{1}{\w_1}+\frac{1}{\w_2}\right)
				-\frac{\sst}{2}\frac{(\w_1-\w_2)^2(\w_1+\w_2)}{\w_1\w_2} \Bigg]=\end{equation*}
\begin{equation*}	=\int dp\Bigg[\frac{p^4}{6\pi^2}\left( \frac{1}{\w_1}+\frac{1}{\w_2}  \right)
		-\frac{\sst}{(2\pi)^2}p^2\frac{(\w_1-\w_2)^2(\w_1+\w_2)}{\w_1\w_2}\Bigg]=\end{equation*}
\be	\via{simplw}\int dp \Bigg[\frac{p^4}{6\pi^2}\left(\frac{1}{\w_1}+\frac{1}{\w_2}\right)
	-\sst\frac{(m_1^2-m_2^2)}{(2\pi)^2}p^2\frac{(\w_1-\w_2)}{\w_1\w_2}\Bigg].
\ee

\label{bfvevc}
\subsectionItalic{Fermionic Sector}\label{asusyFS}
\subsubsection{Fermionic Stress Energy Tensor}\label{setf}
Starting from \cite{Peskin:1995ev}
\be
T^\mu_\nu(x)\equiv \frac{\partial \lag }{\partial(\partial_\mu \psi)}\partial_\nu \psi-\lag \delta^\mu_\nu
\ee
and using
\be
\lag^{majorana} =\bpsi\left(i\gamma^\rho \partial_\rho -m\right)\psi
\ee
we have
\be
\frac{\partial \lag^{majorana} }{\partial(\partial_\mu \phi)}=i\bpsi \gamma^\mu
\ee
and therefore
\be\label{fsetfset}
T^{scalar}_{\mu\nu}= i\bpsi \gamma_{\left(\mu\right.} \partial_{\left.\nu\right)} \psi-\e_{\mu\nu}\lag^{majorana}
\ee
after symmetrization over indices $\mu$ and $\nu$.\label{fsetc}
\subsubsection{Fermionic $G_\theta(t)$}\label{Gtf}\label{appGf}
In this section we want to prove that
\begin{equation}\begin{split}\label{psiapsibpsi1psi2}
G^{\dagger}_{\theta}(t) \psi_1 (x)G_{\theta}(t)=&\ct \psi_1+\st \psi_2 \\
G^{\dagger}_{\theta}(t) \psi_2 (x)G_{\theta}(t)=&\ct \psi_2-\st \psi_1
\end{split}\end{equation}
with
\be\label{Gferm}
G_{\theta}(t)=e^{-X}
\ee
and
\be
X=-\theta \int d\vx \frac{1}{2}\left(\psi^\dagger_1(x)\psi_2(x)-\psi^\dagger_2(x)\psi_1(x)\right)
\ee

\cdel

We start by recalling the field decomposition
\be
\psi_\alpha (x)=\sum_{r=1,2}\int{\frac{d^3k}{(2 \pi)^{3/2}} \left[   u^r_\alpha (\vk)a^r(\vk)e^{-i\w(k)t} 
						+ v_\alpha^{r} (-\vk) a^{r\dagger}(-\vk)e^{i\w(k)t}\right]e^{i\vk \cdot \vx}}
\ee
\be
\psi^\dagger_\beta (y)=\sum_{s=1,2}\int{\frac{d^3k}{(2 \pi)^{3/2}} \left[   u^{s\dagger}_\beta (\vk)a^{s\dagger}(\vk)e^{i\w(k)t} 
						+ v_\beta^{s\dagger} (-\vk) a^{s}(-\vk)e^{-i\w(k)t}\right]e^{-i\vk \cdot \vx}}
\ee
and the Majorana condition
\be\label{majocon}
\psi=M\psi^*
\ee
(as we will see, we will not need a specific form of the matrix $M$) from which
\be\label{uno}
v^s_\beta=M_{\beta \gamma} u^{s*}_\gamma\;\;\;\;\mbox{and}\;\;\;\;u^s_\beta=M_{\beta \gamma} v^{s*}_\gamma
\ee

Before applying formula (\ref{BCHsc}), we notice that
\begin{multline}\label{psipsid}
\{\psi_\alpha (x),\psi^\dagger_\beta (y)\}_{x_0=y_0}=\sum_{rs}\int \frac{d^3p\;d^3q}{(2 \pi)^3}\times\\
		\times\Big[u_\alpha^r(\vp)u^{s\dagger}_\beta (\vq) \{a^r(\vp),a^{s\dagger}(\vq)\}e^{-i(\w(p)-\w(q))t}e^{i\vp \cdot \vx-i\vq \cdot \vec{y}}+\\
		+v_\alpha^r(-\vp)v_\beta^{s\dagger}(-\vq)\{a^{r\dagger}(-\vp),a^s(-\vq)\}e^{i(\w(p)-\w(q))t}e^{i\vp\cdot\vx-i\vq\cdot\vec{y}}\Big]=\\
		=\sum_{r}\int \frac{d^3p}{(2 \pi)^3}\left[
				u^r_\alpha (\vp)u^{r\dagger}_\beta(\vp)e^{i\vp(\vx-\vec{y})}+v^r_\alpha(-\vp)v^{r\dagger}_\beta(-\vp)e^{i\vp (\vx-\vec{y})}\right]=\\
		=\int \frac{d^3p}{(2 \pi)^3}\sum_{r}\left[
				u^r_\alpha (\vp)u^{r\dagger}_\beta(\vp)e^{i\vp(\vx-\vec{y})}+v^r_\alpha(-\vp)v^{r\dagger}_\beta(-\vp)e^{i\vp (\vx-\vec{y})}\right]=\\
		=\delta_{\alpha \beta} \int \frac{d^3p}{(2 \pi)^3} e^{i\vp (\vx-\vec{y})}=\delta_{\alpha\beta}\delta^3(\vx-\vec{y})
\end{multline}
in which we used
\be
\sum_{r}\left[
				u^r_\alpha (\vp)u^{r\dagger}_\beta(\vp)+v^r_\alpha(-\vp)v^{r\dagger}_\beta(-\vp)\right]=\delta_{\alpha \beta}
\ee
and
\begin{multline}\label{psipsi}
\{\psi_\alpha (x),\psi_\beta (y)\}_{x_0=y_0}=\sum_{rs}\int \frac{d^3p\;d^3q}{(2 \pi)^3}\times\\
	\times\left[u_\alpha^r(\vp)v_\beta^s(-\vq)\{ a^r(\vp),a^{s\dagger}(-\vq)\}e^{-i\w(p)t+i\vp \cdot \vx}e^{i \w(q) t+i\vq \cdot \vec{y}}+\right.\\
	\left.+v_\alpha^r(-\vp)u_\beta^s(\vq)\{ a^{r\dagger}(-\vp),a^{s}(\vq)\}e^{i\w(p)t+i\vp \cdot \vx}e^{-i \w(q) t+i\vq \cdot \vec{y}}\right]=\\
	=\sum_r \int \frac{d^3p}{(2 \pi)^3}\left(u^r_\alpha (\vp)v^r_\beta (\vp)e^{i\vp (\vx-\vec{y})}
					+v_\alpha^r (-\vp)u^r_\beta (-\vp)e^{i\vp (\vx-\vec{y})}\right)=\\
	=\int \frac{d^3p}{(2 \pi)^3}e^{i\vp (\vx-\vec{y})}\sum_r \left( u^r_\alpha (\vp)v^r_\beta (\vp)+  v_\alpha^r (-\vp)u^r_\beta (-\vp)\right)=\\
	\via{uno}\int \frac{d^3 p}{(2 \pi)^3}e^{i\vp (\vx-\vec{y})}\sum_r \left(u^r_\alpha (\vp)M_{\beta \gamma}u^{s*}_\gamma (\vp)+
								v_\alpha^r(-\vp)M_{\beta \gamma}v^{s*}_\gamma (-\vp) \right)=\\
	=\int \frac{d^3 p}{(2 \pi)^3}e^{i\vp (\vx-\vec{y})}M_{\beta \gamma}\sum_r \left(u^r_\alpha (\vp)u^{s*}_\gamma (\vp)+
								v_\alpha^r(-\vp)v^{s*}_\gamma (-\vp) \right)=\\
	\via{due}{=}\int \frac{d^3 p}{(2 \pi)^3}e^{i\vp (\vx-\vec{y})}M_{\beta \alpha}=\delta^3(\vx-\vec{y})M_{\beta \alpha}
\end{multline}
\be\label{majosym}
\{ \psi_\alpha(x),\psi_\beta (y)\}_{x_0=y_0}=\{\psi_\beta (y),\psi_\alpha(x)\}_{x_0=y_0}\Rightarrow M_{\beta \alpha}=M_{\alpha \beta}
\ee
using (\ref{uno})
and
\be\label{due}
\sum_r \left(u^r_\alpha (\vp)u^{s*}_\gamma (\vp)+v_\alpha^r(-\vp)v^{s*}_\gamma (-\vp) \right)=\delta_{\alpha \gamma}
\ee
We are now ready to evaluate $X[\psi_i(x)]=[X,\psi_i(x)]$. Starting from $X[\psi_1(x)]$, we have
\begin{multline}\label{Xpsi1}
[X,\psi_{1\alpha}(x)]=\\
	=-\frac{\theta}{2}\int d^3y\left([\psi^{\dagger}_{1\beta}(y)\psi_{2\beta}(y),\psi_{1\alpha}(x)]-
															[\psi_{2\beta}^\dagger (y)\psi_{1\beta}(y),\psi_{1\alpha}(x)]\right)=\\
		\via{tre}-\frac{\theta}{2}\int d^3y\left( -\{\psi_{1\beta}^\dagger(y),\psi_{1\alpha}(x)\}\psi_{2\beta}(y)-
																	\psi^\dagger_{2\beta}(y)\{\psi_{1\beta}(y),\psi_{1\alpha}(x)\}\right)=\\
												=\frac{\theta}{2}\int d^3y\left(\{\psi_{1\beta}^\dagger(y),\psi_{1\alpha}(x)\}\psi_{2\beta}(y)+
																	\psi^\dagger_{2\beta}(y)\{\psi_{1\beta}(y),\psi_{1\alpha}(x)\}\right)=\\
												=\frac{\theta}{2}\int d^3y\left(\delta_{\alpha \beta}\delta^3(\vx-\vec{y})\psi_{2\beta}(y)+
																													\psi_{2\beta}^\dagger (y) M_{\beta\alpha}\delta^3(\vx-\vec{y})\right)=\\
	=\frac{\theta}{2}\left( \psi_{2\alpha}(x)+\psi_{2\beta}^\dagger M_{\beta \alpha}\right)\doublevia{psipsi}{quattro}\theta \psi_{2\alpha}(x)
\end{multline}
with $y_0=x_0=t$
using
\begin{equation}\begin{split}\label{tre}
[\psi_1^\dagger \psi_2,\psi_1]=&\psi^\dagger_1 \psi_2 \psi_1-\psi_1\psi_1^\dagger \psi_2=-\{\psi^\dagger_1,\psi_1\}\psi_2\\
\left[ \psi_2^\dagger(x) \psi_1(x),\psi_1(x') \right] =&
		\psi^\dagger_2(x) \psi_1(x) \psi_1(x')-\psi_1(x')\psi_2^\dagger(x) \psi_1(x)=\\
		=&\psi_2^\dagger(x) \{\psi_1(x),\psi_1(x')\}
\end{split}\end{equation}
and
\be\label{quattro}
\psi^\dagger_\beta M_{\beta \alpha}= M_{\alpha \beta} \psi^*_\beta=\psi_\alpha
\ee
that follows from (\ref{due}) and from (\ref{majosym}).
In order to evaluate $X[\psi_2(x)]$, we just have to notice that
\begin{equation}\begin{split}
\otto{X}=&\otto{-\theta \int d\vx \frac{1}{2}\left(\psi^\dagger_1(x)\psi_2(x)-\psi^\dagger_2(x)\psi_1(x)\right)}=\\
				=&-\theta \int d\vx \frac{1}{2}\left(\psi^\dagger_2(x)\psi_1(x)-\psi^\dagger_1(x)\psi_2(x)\right)=-X
\end{split}\end{equation}
and therefore
\be\label{Xpsi2}
[X,\psi_2]=\left[-\otto{X},\otto{\psi_1}\right]=-\otto{\left[X,\psi_1\right]}=-\theta \otto{\psi_2}=-\theta \psi_1
\ee
Provided with (\ref{Xpsi1}) and (\ref{Xpsi2}) and consequently of
\begin{equation}\begin{split}
X^2[\psi_1(x)]=&X[\theta \psi_2(x)]=(i\theta)^2\psi_1\\
X^2[\psi_2(x)]=&X[-\theta \psi_1(x)]=(i\theta)^2\psi_2
\end{split}\end{equation}
we can finally refer to (\ref{BCHsc}), from which formula (\ref{psiapsibpsi1psi2}) follows straightforwards.
\subsubsection{Fermionic FLO}\label{flof}

Starting from the definition of FLO
\begin{equation}\begin{split}\label{flomlof}
a_A^r(\vk) =&G^{\dagger}_\theta (t)a_1^r(\vk)G_\theta (t)\\
a_B^r(\vk)=&G^{\dagger}_\theta (t)a_2^r(\vk) G_\theta (t)
\end{split}\end{equation}
with $G(t)$ defined in (\ref{Gferm}),
we want to see how (\ref{BCHsc}) applies to this case.

In order to to so, we first deduce some formulae required in the calculation.

\cdel

Since
\be
\{\psi_i,a^r_j\}_{i\neq j}=0
\ee
we have, omitting the $r$ index for $a$,
\begin{equation}\begin{split}\label{formulauno}
\left[\psi_1^\dagger\psi_2,a_1\right]=&\psi_1^\dagger\psi_2a_1-a_1\psi_1^\dagger\psi_2=-\psi^\dagger_1a_1\psi_2-a_1\psi_1^\dagger\psi_2=
	-\{\psi_1^\dagger,a_1\}\psi_1\\
\left[\psi_2^\dagger\psi_1,a_1\right]=&\psi_2^\dagger\psi_1a_1-a_1\psi_2^\dagger\psi_1=\psi^\dagger_2\{\psi_1,a_1\}\\
\left[\psi^\dagger_1\psi_2,a_2\right]=&\psi^\dagger_1\psi_2a_2-a_2\psi^\dagger_1\psi_2=\psi^\dagger_1\{\psi_2,a_2\}\\
\left[\psi_2^\dagger\psi_1,a_2\right]=&\psi_2^\dagger\psi_1a_2-a_2\psi_2^\dagger\psi=-\{\psi_2^\dagger,a_2\}\psi_1
\end{split}\end{equation}
Then
\begin{equation}\begin{split}\label{formuladue}
\{\psi_i(x),a_i^s(\vp)\}
		=&\sum_{r=1,2}\int\frac{d^3k}{(2\pi)^{3/2}}v_i^r(-\vk)\{a_i^{r\dagger}(-\vk),a_i^s(\vp)\}e^{i\w_i(k)t+i\vk\cdot\vx}\\
		=&\sum_{r=1,2}\int \frac{d^3k}{(2\pi)^{3/2}}v_i^r(-\vk) \delta^{rs}\delta^3(\vp+\vk)e^{i\w_i(k)t+i\vk\cdot\vx}\\
		=&\frac{1}{(2 \pi)^{3/2}}v^s_i(\vp)e^{i\w_i(p)t-i\vp\cdot\vx}\\
\{\psi_i^\dagger(x),a_i^s(\vp)\}
		=&\sum_{r=1,2}\int \frac{d^3k}{(2 \pi)^{3/2}}u^{r\dagger}_i(\vk)\{a_i^{r\dagger}(\vk),a_i^s(\vp)\}e^{i\w_i(k)t-i\vk\cdot\vx}\\
		=&\frac{1}{(2 \pi)^{3/2}}u^{s\dagger}_i(\vp)e^{i\w_i(p)t-i\vp\cdot\vx}\\
\{\psi_i(x),a_i^{s\dagger}(-\vp)\}
		=&\sum_{r=1,2}\int \frac{d^3k}{(2 \pi)^{3/2}}u^{r}_i(\vk)\{a_i^{r}(\vk),a_i^{s\dagger}(-\vp)\}e^{-i\w_i(k)t+i\vk\cdot\vx}\\
		=&\frac{1}{(2 \pi)^{3/2}}u^{s}_i(-\vp)e^{-i\w_i(p)t-i\vp\cdot\vx}\\
\{\psi_i^\dagger(x),a_i^{s\dagger}(-\vp)\}
		=&\sum_{r=1,2}\int \frac{d^3k}{(2 \pi)^{3/2}}v^{r\dagger}_i(-\vk)\{a_i^{r}(-\vk),a_i^{s\dagger}(-\vp)\}e^{-i\w_i(k)t-i\vk\cdot\vx}\\
		=&\frac{1}{(2 \pi)^{3/2}}v^{s\dagger}_i(-\vp)e^{-i\w_i(p)t-i\vp\cdot\vx}
\end{split}\end{equation}
\begin{multline}\label{formulatre1}
\int \frac{d^3x}{(2 \pi)^{3/2}}e^{-\vp \cdot \vx}\psi_i(x)=\sum_{r=1,2}\int d^3k \frac{d^3x}{(2\pi)^3}e^{-i(\vp-\vk)\vx}\times\\
		\times\left(u^r_i(\vk)a_i^r(\vk)e^{-i\w_i(k)t}+v_i^r(-\vk)a^{r\dagger}_i(-\vk)e^{i\w_i(k)t}\right)=\\
		=\sum_{r=1,2}\int d^3k \delta^3(\vp-\vk)\left(u^r_i(\vk)a_i^r(\vk)e^{-i\w_i(k)t}+v_i^r(-\vk)a^{r\dagger}_i(-\vk)e^{i\w_i(k)t}\right)=\\
									=\sum_{r=1,2}\left(u^r_i(\vp)a_i^r(\vp)e^{-i\w_i(p)t}+v_i^r(-\vp)a^{r\dagger}_i(-\vp)e^{i\w_i(p)t}\right)
\end{multline}
\begin{multline}\label{formulatre2}									
\int \frac{d^3x}{(2 \pi)^{3/2}}e^{-\vp \cdot \vx}\psi^\dagger_i(x)=\sum_{r=1,2}\int d^3k \frac{d^3x}{(2\pi)^3}e^{-i(\vp+\vk)\vx}\times\\
				\times\left(u^{r\dagger}_i(\vk)a_i^{r\dagger}(\vk)e^{i\w_i(k)t}+v_i^{r\dagger}(-\vk)a^{r}_i(-\vk)e^{-i\w_i(k)t}\right)=\\
									=\sum_{r=1,2}\left(u^{r\dagger}_i(-\vp)a_i^{r\dagger}(-\vp)e^{i\w_i(p)t}+v_i^{r\dagger}(\vp)a^{r}_i(\vp)e^{-i\w_i(p)t}\right)
\end{multline}
\bea\label{formulaquattro}
\otto{X}&=&-\frac{\theta}{2}\int d^3x\left( \psi_2^\dagger \psi_1-\psi^\dagger_1\psi_2  \right)=\frac{\theta}{2}\int d^3x\left( \psi_1^\dagger \psi_2-\psi^\dagger_2\psi_1  \right)\nn
		&=&-X
\eea

\begin{multline}\label{formulacinque}
X^\dagger=-\frac{\theta}{2}\int d^3x\left( \left(\psi_1^\dagger \psi_2\right)^\dagger-\left(\psi^\dagger_2\psi_1 \right)^\dagger\right)=\\
				=-\frac{\theta}{2}\int d^3x\left( \psi_2^\dagger \psi_1-\psi^\dagger_1\psi_2  \right)
			=\frac{\theta}{2}\int d^3x\left( \psi_1^\dagger \psi_2-\psi^\dagger_2\psi_1  \right)=-X
\end{multline}

\cdel

We are now ready to evaluate $X[a^s_1(x)]=[X,a^s_1(x)]$:
\begin{multline}\label{Xaf1}
[X,a_1^s(\vp)]=-\frac{\theta}{2}\int d^3x \left(\left[\psi_1^\dagger\psi_2,a_1^s(\vp)\right]-\left[\psi_2^\dagger\psi_1,a^s_1(\vp)\right]\right)=\\
						\via{formulauno}-\frac{\theta}{2}\int d^3x\left(-\{\psi_1^\dagger,a_1^s(\vp)\}\psi_2-\psi^\dagger_2\{\psi_1,a_1^s(\vp)\}\right)=\\
						=\frac{\theta}{2}\int d^3x \left(\{\psi_1^\dagger,a^s_1(\vp)\}\psi_2+\psi_2^\dagger\{\psi_1,a_1^s(\vp)\}\right)=\\
						\via{formuladue}\frac{\theta}{2}\int d^3x \left(
     																\frac{1}{(2 \pi)^{3/2}} u^{s\dagger}_1(\vp) e^{i\w_1(p)t-i\vp\cdot\vx}\psi_1+
     																\psi_2^\dagger\frac{1}{(2 \pi)^{3/2}}v^s_1(\vp)e^{i\w_1(p)t-i\vp\cdot\vx}\right)=\\
     				=\frac{\theta}{2}\left( 
     							u^{s\dagger}_1(\vp)e^{i\w_1(p)t}\int \frac{d^3x}{(2\pi)^{3/2}}e^{-i\vp\cdot\vx}\psi_2(x)+
     							\int\frac{d^3x}{(2\pi)^{3/2}}e^{-i\vp\cdot\vx}\psi_2^\dagger(x) v^s_1(\vp)e^{i\w_1(p)t}
     				 		 \right)=\\
     				\doublevia{formulatre1}{formulatre2}\frac{\theta}{2}\left(
     					u^{s\dagger}_1(\vp)e^{i\w_1(p)t}\sum_{r=1,2}\left(u_2^r(\vp)a_2^r(\vp)e^{-i\w_2(p)t}+
     					v_2^r(-\vp)a_2^{r\dagger}(-\vp)e^{i\w_2(p)t}\right)+\right.\\
     				+\left.\sum_{r=1,2}\left(u^{r\dagger}_2(-\vp)a^{r\dagger}_2(-\vp)e^{i\w_2(p)t}+
     					v^{r\dagger}_2(\vp)a^r_2(\vp)e^{-i\w_2(p)t}\right)v_1^s(\vp)e^{i\w_1(p)t}\right)=\\
     			 	=\frac{\theta}{2}\left(
     			 	 \sum_{r=1,2}\left(u^{s\dagger}_1(\vp)u^r_2(\vp)+v^{r\dagger}_2(\vp)v^s_1(\vp)\right)e^{i\w_-(p)t}a_2^r(\vp)+\right.\\
     			 	\left.+\sum_{r=1,2}\left(u^{s\dagger}_1(\vp)v^r_2(-\vp)+u_2^{r\dagger}(-\vp)v^s_1(\vp)\right)e^{i\w_+(p)t}a_2^{r\dagger}(-\vp)\right)=\\
     			 	=\frac{\theta}{2}\sum_{r=1,2}\left(W^{sr}_{12}(\vp)e^{i\w_-(p)t}a^r_2(\vp)+Y^{sr}_{12}(\vp)e^{i\w_+(p)t}a^{r\dagger}_2(-\vp)\right)
\end{multline}
with
\be\label{Wdefinition}
W^{sr}_{12}(\vp)\equiv u_1^{s\dagger}(\vp)u^r_2(\vp)+v^{r\dagger}_2(\vp)v^s_1(\vp)
\ee
\be\label{Ydefinition}
Y^{sr}_{12}(\vp)\equiv u_1^{s\dagger}(\vp)v_2^r(-\vp)+u_2^{r\dagger}(-\vp)v^s_1(\vp)
\ee
Regarding $X[a^s_2(x)]$ we simply have
\begin{multline}\label{Xaf2}
\left[X,a_2^s(\vp)\right]=-\left[-X,a_2^s(\vp)\right]\via{formulaquattro}-\left[\otto{X},a_2^s(\vp)\right]=\\
						=-\left[\otto{X},\otto{a_1^s(\vp)}\right]=-\otto{\left[X,a_1^s(\vp)\right]}=\\
	=-\frac{\theta}{2}\sum_{r=1,2}\left(W^{sr}_{21}(\vp)e^{-i\w_-(p)t}a_1^r(\vp)+Y^{sr}_{21}(\vp)e^{i\w_+(p)t}a^{r\dagger}_1(-\vp)\right)
\end{multline}
To evaluate now $X^2[a^s_i(x)]$ we need to know $X[a^{s\dagger}_i(x)]$, since $X[\slot]$ acting on the ladder operators mixes
up creation and annihilation operators. Therefore, since
\begin{multline}
\com{X}{a_i^{s\dagger}(-\vp)}=-\com{-X}{a^{s\dagger}_i(-\vp)}=-\com{X^\dagger}{a_i^{s\dagger}(-\vp)}=\\
=-\left( X^\dagger a^{s\dagger}_i(-\vp)-a_i^{s\dagger}(-\vp)X^\dagger \right)=-\left(	\left(a_i^s(-\vp)X\right)^\dagger-\left(Xa_i^s(-\vp)\right)^\dagger\right)=\\
			=-\left(\com{a_i^s(-\vp)}{X}\right)^\dagger=\left(\com{X}{a_i^s(-\vp)}\right)^\dagger
\end{multline}
we have that
\begin{multline}
\com{X}{a_1^{s\dagger}(-\vp)}=\frac{\theta}{2}\sum_{r=1,2}\left(W^{sr\dagger}_{12}(-\vp)e^{-i\w_-(p)t}a_2^{r\dagger}(-\vp)+\right.\\
				\left.+Y^{sr\dagger}_{12}(-\vp)e^{-i\w_+(p)t}a_2^r(\vp)\right)
\end{multline}
\begin{multline}
\com{X}{a_2^{s\dagger}(-\vp)}=\frac{\theta}{2}\sum_{r=1,2}\left(W^{sr\dagger}_{21}(-\vp)e^{i\w_-(p)t}a_1^{r\dagger}(-\vp)+\right.\\
				\left.+Y^{sr\dagger}_{21}(-\vp)e^{-i\w_+(p)t}a_1^r(\vp)\right)
\end{multline}
Finally, we can write
\begin{multline}\label{XXaf1}
\com{X}{\com{X}{a_1^s(\vp)}}=\\
	=\frac{\theta}{2}\sum_{t=1,2}\left(W^{st}_{12}(\vp)e^{i\w_-(p)t}\com{X}{a_2^t(\vp)}+Y^{st}_{12}(\vp)\com{X}{a_2^{t\dagger}(-\vp)}\right)=\nonumber
\end{multline}
\begin{multline}
	=\frac{\theta}{2}\sum_{t=1,2}\Bigg[W^{st}_{12}(\vp)e^{i\w_-(p)t}\left(-\frac{\theta}{2}\right)\times\\
	\times\sum_{r=1,2}\left(W^{tr}_{21}e^{-i\w_-(p)t}a^r_1(\vp)+Y^{tr}_{21}(\vp)e^{i\w_+(p)t}a^{r\dagger}_1(-\vp)\right)+\\
				+Y^{st}_{12}(\vp)e^{i\w_+(p)t}\left(-\frac{\theta}{2}\right)\times\\
	\times\sum_{r=1,2}\left(W^{tr\dagger}_{21}(-\vp)e^{i\w_-(p)t}a_1^{r\dagger}(-\vp)+Y^{tr\dagger}_{21}(-\vp)e^{-\w_+(p)t}a^r_1(\vp)\right)\Bigg]=\nonumber
\end{multline}
\begin{multline}	
										=-\left(\frac{\theta}{2}\right)^2\sum_{r,t=1,2}\left(
										\left(W^{st}_{12}(\vp)W^{tr}_{21}(\vp)+Y^{st}_{12}(\vp)Y^{tr\dagger}_{21}(-\vp)\right)a_1^r(\vp)+\right.\\
					\left.+\left(W^{st}_{12}(\vp)Y^{tr}_{12}(\vp)+Y^{st}_{12}(\vp)W^{tr\dagger}_{21}(-\vp)\right)e^{21\w_1(p)t}a_1^{r\dagger}(-\vp)\right)=\\
							\viam
							-\frac{\theta^2}{4}\sum_{r=1,2}\left( 4 \;\delta^{sr}a_1^r(\vp)+0\;e^{2i\w_1(p)t}a_1^{r\dagger}(-\vp)\right)=-\theta^2a^s_1(\vp)
\end{multline}
and
\begin{multline}\label{XXaf2}
X^2[a_2^s(\vp)]=[X,[X,a_2^s(\vp)]]=\com{-\otto{X}}{\com{-\otto{X}}{\otto{a_1^s(\vp)}}}=\\
							 =\com{\otto{X}}{\com{\otto{X}}{\otto{a_1^s(\vp)}}}=\otto{\com{X}{\com{X}{a_1^s(\vp)}}}=\\
							 =\otto{X^2[a_1^s(\vp)]}=-\theta^2a_2^s(\vp)
\end{multline}

Collecting (\ref{Xaf1}), (\ref{Xaf2}), (\ref{XXaf1}) and (\ref{XXaf2}), we can now appply (\ref{BCHsc}) that leads to
\begin{multline}\label{proprioqui1}
a_A^r(\vk)=\ct a_1^r(\vk)+\\
		-\st \frac{1}{2}\sum_{s=1,2}
			\left( W^{rs}_{12}(\vk)e^{i\w_-(k)t}a_2^s(\vk) +Y^{rs}_{12}e^{i\w_+(k)t}a_2^{s\dagger}(-\vk)\right)
\end{multline}
\begin{multline}\label{proprioqui2}
a_B^r(\vk)=\ct a_2^r(\vk)+\\
	+\st \frac{1}{2}\sum_{s=1,2}\left(W^{rs}_{21}e^{-i\w_-(k)t}a_1^s(\vk)+Y^{rs}_{21}(\vk)e^{i\w_+(k)t}a_1^{s\dagger}(-\vk)\right)
\end{multline}
We can write those latter expression in a simpler way, by noticing that
\begin{equation}\begin{split}
W^{rs}_{21}(\vk)\via{Wdefinition}&u^{r\dagger}_2(\vk)u^s_1{\vk}+v^{s\dagger}_2(\vk)v^r_1(\vk)=\left(W^{sr}_{12}(\vk)\right)^*\\
Y^{rs}_{21}(\vk)\via{Ydefinition}&u^{r\dagger}_2(\vk)v^s_1(-\vk)+u^{s\dagger}_1(-\vk)v^r_2(\vk)=Y^{sr}_{12}(-\vk)
\end{split}\end{equation}
and therefore we can define
\be\begin{array}{lr	}
W^{rs}(\vk,t)\equiv W^{rs}_{12}(\vk)e^{i\w_-(k)t}/2&Y^{rs}(\vk)\equiv Y^{rs}_{12}(\vk)e^{i\w_+(k)t}/2
\end{array}\ee
and from that formulae (\ref{proprioqui1}) and (\ref{proprioqui2}) are written as
\begin{equation}\begin{split}\label{bogomajo}
a_A^r(\vk)=&\ct a_1^r(\vk)-\st \sum_{s=1,2}\left(W^{rs}(\vk,t)a^s_2(\vk)+Y^{rs}(\vk,t)a_2^{s\dagger}(-\vk)\right)\\
a_B^r(\vk)=&\ct a_2^r(\vk)+\st \sum_{s=1,2}\left(W^{sr*}(\vk,t)a_1^s(\vk)+Y^{sr}(-\vk,t)a_1^{s\dagger}(-\vk)\right)\;\;\;
\end{split}\end{equation}
and
\begin{multline}\label{bogomajomp1}
a_A^{r\dagger}(-\vk)=\ct a_1^{r\dagger}(-\vk)+\\
		-\st \sum_{s=1,2}\left(W^{rs*}(-\vk,t)a^{s\dagger}_2(-\vk)+Y^{rs*}(-\vk,t)a_2^{s}(\vk)\right)
\end{multline}
\begin{multline}\label{bogomajomp2}
a_B^{r\dagger}(-\vk)=\ct a_2^{r\dagger}(-\vk)+\\
		+\st \sum_{s=1,2}\left(W^{sr}(-\vk,t)a_1^{s\dagger}(-\vk)+Y^{sr*}(\vk,t)a_1^{s}(\vk)\right).
\end{multline}

\subsubsection{Fermionic MLO}\label{mlof}

In last section, we derived simple expressions (formulae (\ref{bogomajo}), (\ref{bogomajomp1}), and (\ref{bogomajomp2}))
for FLO as function of MLO, not involving any infinite series as in (\ref{flomlof}).
We want now to use them to write the MLO in terms of the FLO.

Combining (\ref{bogomajo}) and (\ref{bogomajomp1})-(\ref{bogomajomp2}) it follows that
\begin{multline}
\ct a_1^t(\vk)=a^t_A(\vk)+\\
+\st \sum_{s=1,2}\left(W^{ts}(\vk,t)a_1^s(\vk)+Y^{ts}(\vk,t)a^{s\dagger}_2(-\vk)\right)
\end{multline}
\be\label{daggereaste}
\ct a^{t\mp}_1=a_A^{t\mp}+\st \sum_{s=1,2}\left( W^{ts\dagger}(-\vk)a_2^{s\dagger}+Y^{ts\dagger}(-\vk)a^s_2\right)
\ee
\be
\ct a_2^r =a_B^r-\st \sum_{t=1,2}\left(W^{tr\dagger}(\vk)a_1^t+Y^{tr}(-\vk)a_1^{t\mp}\right)
\ee
and using these three formulae it follows that
\begin{multline}\label{a2r}
\ct a_2^r=a_B^r+\\
	-\frac{\st}{\ct}\sum_{t}
			\Bigg[ W^{tr\dagger}(\vk)\left(a_A^t+\st\sum_s\left(W^{ts}(\vk)a_2^s+Y^{ts}(vk)a_2^{s\mp}\right)\right)+\\
	+Y^{tr}(-\vk)\left( a_A^{t\mp}+\st \sum_s\left( W^{ts\dagger}(-\vk)a_2^{s\mp}+Y^{ts\mp}(-\vk)a^s_2 \right) \right)\Bigg]=\nonumber
\end{multline}
\begin{multline}
=a_B^r-\tan \theta \sum_t \Big[W^{tr\dagger}(\vk)a_A^t+Y^{tr}(-\vk)a_A^{\dagger\mp}+\\
			+\st\sum_s\left(W^{tr\dagger}(\vk)W^{ts}(\vk)+Y^{tr}(-\vk)Y^{ts\dagger}(-\vk)\right)a_2^s+\\
		+\st \sum_s\left(W^{tr\dagger}(\vk)Y^{tr}(-\vk)W^{ts\dagger}(-\vk)\right)a^{s\mp}_2\Big]=\nonumber
\end{multline}
\begin{multline}
		=a_B^r-\tan \theta \sum_t \left( W^{tr\dagger}(\vk)a^t_A+Y^{tr}(-\vk)a_A^{t\mp}  \right)+\\
		-\tan \theta \st \sum_{t,s}\left( W^{tr\dagger}(\vk)W^{ts}(\vk)+Y^{tr}(-\vk)Y^{ts\dagger}(-\vk) \right)a_2^s+\\
		-\tan \theta \st \sum_{t,s}\left(W^{tr\dagger}(\vk)Y^{ts}(\vk)+Y^{tr}(-\vk)W^{ts\dagger}(-\vk)\right)a_2^{s\mp}=\nonumber
\end{multline}
\begin{multline}
		\via{4nb}a_B^r-\tan \theta \sum_t \left(W^{tr\dagger}(\vk)a_A^t+Y^{tr}(-\vk)a_A^{t\mp}\right)-\tan \theta \st a_2^r
\end{multline}
being
\begin{equation}\begin{split}\label{4nb}
\sum_t\left(W^{tr\dagger}(\vk)W^{ts}(\vk)+Y^{tr}(-\vk)Y^{ts\dagger}(-\vk)\right)\viam&\delta^{rs}\\
\sum_t\left(W^{tr\dagger}(\vk)Y^{ts}(\vk)+Y^{tr}(-\vk)W^{ts\dagger}(-\vk)\right)\viam&0.
\end{split}\end{equation} 
From (\ref{a2r}) it follows that
\begin{multline}
a_2^r(\ct+\tan \theta \st)=\\
		=a_B^r-\tan \theta \sum_t\left(W^{tr\dagger}(\vk)a_A^t+Y^{tr}(-\vk)a_A^{t\mp}\right)
\end{multline}
\be
\Rightarrow a_2^r=\ct a^r_B-\st \sum_t \left(W^{tr\dagger}(\vk)a_A^t+Y^{tr}(-\vk)a_A^{t\mp}\right)
\ee
since $\ct+(\st/\ct)\st=1/\ct$. Therefore we have
\be
a^{r\mp}_2=\ct a_B^{r\mp}-st \sum_t\left(W^{tr}(-\vk)a_A^{t\mp}+Y^{tr\dagger}(\vk)a_A^t\right)
\ee
Similarly
\begin{multline}
\ct a_1^r=a_A^r+\st \sum_s\left(W^{rs}(\vk)a_2^s+Y^{rs}(\vk)a_2^{s\mp}\right)=\\
	=a_A^r+\st \sum_s \left(W^{rs}(\vk)\ct a_B^s+\right.\\
			-\st \sum_t\left(W^{rs}(\vk)W^{ts\dagger}(\vk)a_A^t+W^{rs}(\vk)Y^{ts}(-\vk)a^{t\mp}_A\right)+\\
		\left.+Y^{rs}(\vk)\ct a^{s\mp}_B-\st\sum_t\left(Y^{rs}(\vk)W^{ts}(-\vk)a^{t\mp}_A+Y^{rs}(\vk)Y^{ts\dagger}(\vk)a_A^t\right)\right)=\\
		=a_A^r+\st \ct \sum_s\left(W^{rs}(\vk)a_B^s+Y^{rs}(\vk)a_B^{s\mp}\right)+\\
		-\sst \sum_{st}\left(W^{rs}(\vk)W^{ts\dagger}(\vk)+Y^{rs}(\vk)Y^{ts\dagger}(\vk)\right)a_A^t+\\
		-\sst \sum_{st} \left(W^{rs}(\vk)Y^{ts}(-\vk)+Y^{rs}(\vk)W^{ts}(-\vk)\right)a_A^{t\mp}=\\
		\via{4nbbis}a_A^r+\st \sum_s\left(W^{rs}(\vk)a_B^s+Y^{rs}(\vk)a^{s\mp}_B\right)-\sst a_A^r=\\
		=\ct\left(\ct a_A^r +\st \sum_s\left(W^{rs}(\vk)a_B^s+Y^{rs}(\vk)a_B^{s\mp}\right)\right)
\end{multline}
using
\begin{equation}\begin{split}\label{4nbbis}
\sum_s\left(W^{rs}(\vk)W^{ts\dagger}(\vk)+Y^{rs}(\vk)Y^{ts\dagger}(\vk)\right)\viam&\delta^{rt}\\
\sum_s\left(W^{rs}(\vk)Y^{ts\dagger}(-\vk)+Y^{rs}(\vk)W^{ts\dagger}(-\vk)\right)\viam&0
\end{split}\end{equation}
and therefore
\be
a_1^r=\ct a^r_A+\st \sum_s\left(W^{rs}(\vk)a_B^s+Y^{rs}(\vk)a_B^{s\mp}\right)
\ee
and
\be
a_1^{r\mp}=\ct a^{r\mp}_A+\st \sum_s\left(W^{rs\dagger}(-\vk)a_B^{s\mp}+Y^{rs\dagger}(-\vk)a_B^{s}\right)
\ee

\subsubsection{Fermionic Quantum Algebra}\label{qaf}

Once we have expressed the MLO as function of the FLO, we are able to evaluate objects in the form
\be\label{fvmlofv}
\lfv a_i^{r\dagger}(\vp)a_j^{s\dagger}(\vq)\rfv.
\ee
We start by acting with single MLO on $\lfv$ and $\rfv$ separately:
\be
a_1^r(\vq)\rfv=\st \sum_s Y^{rs}(\vq)a_B^{s\dagger}(-\vq)\rfv
\ee
\be
a^{r\dagger}_1(\vq)\rfv=\left(\ct a_A^{r\dagger}(\vq)+\st \sum_s W^{rs\dagger}(\vq)a_B^{s\dagger}(\vq)\right)\rfv
\ee
\be
a^r_2(\vq)\rfv=-\st\sum_s Y^{sr}(-\vq)a_A^{s\dagger}(-\vq)\rfv
\ee
\be
a_2^{r\dagger}(\vq)\rfv=\left(\ct a_B^{r\dagger}(\vq)-\st \sum_s W^{sr}(\vq)a_A^{s\dagger}(\vq)\right)\rfv
\ee
\be
\lfv a_1^r(\vp)=\lfv \left(\ct a_A^r(\vp)+\st \sum_s W^{rs}(\vp)a_B^s(\vp)\right)
\ee
\be
\lfv a_1^{r\dagger}(\vp)=\lfv \left(\st \sum_s Y^{rs \dagger}(\vp)a_B^s(-\vp)\right)
\ee
\be
\lfv a^r_2(\vp)=\lfv \left(\ct a_b^r(\vp)-\st\sum_s W^{rs\dagger}(\vp) a_A^s(\vp)\right)
\ee
\be
\lfv a^{r\dagger}_2(\vp)=\lfv\left(-\sst \sum_s Y^{sr\dagger}(-\vp)a_A^s(-\vp)\right)
\ee
We can now evaluate (\ref{fvmlofv}):
\begin{equation}\begin{split}
\lfv a^{r\dagger}_1(\vp)a_1^{r'}(\vq)\rfv=&\sst \sum_{ss'}Y^{rs\dagger}(\vp)Y^{r's'}(\vq)\lfv a_B^s(-\vp)a^{s'\dagger}_B(-\vq)\rfv=\\
		=&\sst \sum_{ss'}Y^{rs\dagger}(\vp)Y^{r's'}(\vq)\delta_{ss'}\delta^3(\vp-\vq)=\\
		=&\sst \sum_s Y^{rs\dagger}(\vp)Y^{r's}(\vp)\delta^3(\vp-\vq)=\\
		=&\sst \delta^{rr'}\delta^3(\vp-\vq)Y(\vp)
\end{split}\end{equation}
in which
\be
\sum_s Y^{rs\dagger}(\vp)Y^{r's}(\vp)\viam Y(\vp)\delta^{rr'}
\ee
with
\begin{multline}
Y(\vp)\equiv\frac{1}{2 \w_1 \w_2 (k_y+\w_1) (k_y-\w_2)}\times\\
			\times\Big[-k^4-k^2 \left(k_y^2-\w_2 (2 k_y+\w_1)+2 k_y \w_1+m_1^2+m_2^2\right)+\\
		+k_y^2 (\w_1 \w_2-m_1 m_2)+k_y (m_1+m_2) (m_1 \w_2-m_2 \w_1)+\\
		+m_1 m_2 (\w_1 \w_2-m_1 m_2) \Big].
\end{multline}
\begin{equation}\begin{split}
\lfv a^{r\dagger}_2(\vp)a^{r'}_2(\vq)\rfv=&\sst \sum_{ss'}Y^{sr\dagger}(-\vp)Y^{s'r'}(-\vq)\delta_{ss'}\delta^3(\vp-\vq)=\\
					=&\sst \sum_s Y^{sr\dagger}(-\vp)Y^{sr'}(-\vp)\delta^3(\vp-\vq)=\\
					=&\sst \delta^{rr'}\delta^3(\vp-\vq)Y(\vp)
\end{split}\end{equation}
in which
\be
\sum_s Y^{sr\dagger}(-\vp)Y^{sr'}(-\vp)\viam Y(\vp)\delta^{rr'}.
\ee
\be
\lfv a^{r\dagger}_1(\vp)a_1^{r'\dagger}(\vq)\rfv =\sst \sum_{s} Y^{rs\dagger}(\vp)W^{r's\dagger}(-\vp) \delta^3(\vp+\vq)
\ee
\be
\lfv a^{r\dagger}_2(\vp)a_2^{r'\dagger}(\vq)\rfv=\sst \sum_{s}Y^{sr\dagger}(-\vp)W^{sr'}(-\vp) \delta^3(\vp+\vq)
\ee
\be
\lfv a^{r}_1(\vp)a_1^{r'}(\vq)\rfv =\sst \sum_{s}W^{rs}(\vp)Y^{r's}(-\vp) \delta^3(\vp+\vq)
\ee
\be
\lfv a^{r}_2(\vp)a_2^{r'}(\vq)\rfv= \sst \sum_{s}W^{rs\dagger}(\vp)Y^{sr'}(\vp) \delta^3(\vp+\vq)
\ee
\begin{multline}
\lfv a^r_1(\vp)a^{r'\dagger}_1(\vq)\rfv=\lfv\left(\ct a_A^r(\vp)+\st \sum W^{rs}(\vp)a_B^s(\vp)\right)\times \\
				\times\left(\ct a_A^{r'\dagger}(\vq)+\st \sum_{s'} W^{r's'\dagger}(\vq)a_B^{s'\dagger}(\vq)\right)\rfv=\\
				=\cct \delta^{rr'}\delta^3(\vp-\vq)+\sst \sum_{ss'}W^{rs}(\vp)W^{r's'\dagger}(\vq)\delta^{ss'}\delta^3(\vp-\vq)=\\
				=\delta^3(\vp-\vq)\left(\delta^{rr'}\cct+\sst \sum_{s}W^{rs}(\vp)W^{r's\dagger}(\vp)\right)=\\
				=\delta^3(\vp-\vq)\delta^{rr'}\left(\cct+\sst W(\vp)\right)
\end{multline}
in which
\be
\sum_s W^{rs}(\vp)W^{r's\dagger}(\vp)\viam \delta^{rr'}W(\vp).
\ee
with
\begin{multline}
W(\vp)\equiv\frac{1}{2 \w_1 \w_2 (k_y+\w_1) (k_y-\w_2)}\times\\
	\times\Big[k^4+k^2 \left(k_y^2-\w_2 (2 k_y+\w_1)+2 k_y \w_1+m_1^2+m_2^2\right)+\\
			+k_y^2 (m_1  m_2-\w_1 \w_2)-k_y (m_1+m_2) (m_1 \w_2-m_2 \w_1)+\\
			+2 \w_1 \w_2 (k_y+\w_1) (k_y-\w_2)+m_1 m_2 (m_1 m_2-\w_1 \w_2)\Big].
\end{multline}
\begin{multline}
\lfv a^r_2(\vp)a_2^{r'\dagger}(\vq)\rfv=\lfv \left(\ct a_B^r(\vp)-\st \sum_s W^{rs\dagger}(\vp)a_A^s(\vp)\right)\times\\
				\times\left(\ct a^{r'\dagger}_B(\vq)-\st \sum_{s'}W^{s'r'}(\vq)a_A^{s'\dagger}(\vq)\right)\rfv=\\
				=\cct \delta^{rr'}\delta^3(\vp-\vq)+\sst \sum_s W^{rs\dagger}(\vp)W^{sr'}(\vp)\delta^3(\vp-\vq)=\\
				=\delta^{rr'}\delta^3(\vp-\vq)\left(\cct+\sst W(\vp)\right)
\end{multline}
in which
\be
\sum_s W^{rs\dagger}(\vp)W^{sr'}(\vp)\viam \delta^{rr'}W(\vp).
\ee

\subsubsection{Fermionic Flavour Vacuum}\label{fvevf}
Since
\be
T_{00}^{majorana}=i\bpsi \gamma_0\partial_0 \psi
\ee
we then have
\begin{multline}\label{emajoranac}
\lfv T^{f}_{00}\rfv=\!\!\!\sum_{i=1,2}\sum_{rs}\int d\vp d\vq \left(L^{rs}_{00}(\vp,\vq,m_i)\lfv a_i^{r\dagger}(\vp)a^s_i(\vq)\rfv+\right.\\
			+M^{rs}_{00}(\vp,\vq,m_i)\lfv a_i^{r\dagger}(\vp)a_i^{s\dagger}(\vq)\rfv+N^{rs}_{00}(\vp,\vq,m_i)\lfv a^r_i(\vp)a^s_i(\vq)\rfv+\\
			\left.+K^{rs}_{00}(\vp,\vq,m_i)\lfv a^r_i(\vp)a_i^{s\dagger}(\vq)\rfv\right)=\\
								\viam\sum_r\int \frac{d\vp }{(2\pi)^3}\left(\w_1(p)+\w_2(p)\right)\left(\sst Y(\vp)-\cct-\sst W(\vp)\right)=\\
								=\sum_r\int \frac{d\vp }{(2\pi)^3}\left(\w_1(p)+\w_2(p)\right)\left(\sst -\cct -\sst(W(\vp)-Y(\vp))\right)=\\
								=\sum_r\int \frac{d\vp }{(2\pi)^3}\left(\w_1(p)+\w_2(p)\right)\left(-\cct -\sst(1-2 Y(\vp))\right)=\\
								=\int \frac{d\vp }{(2\pi)^3}2\left(\w_1(p)+\w_2(p)\right)\left(-1+\sst 2 Y(\vp)\right)
\end{multline}
in which we used
the relation
\be
W(\vp)\viam 1-Y(\vp)
\ee
and previous relations on the quantum algebra.
It is possible to recover expressions of \cite{Blasone:2003hh} in
\be
Y(0,0,p)\viam Y(0,p,0)\viam Y(p,0,0)\viam |V_{\;P\&B}|^2.
\ee
Since
\begin{multline}
\int_0^{2\pi}Y(p \sin \alpha, \cos \phi,p\st \sin \phi,p\ct)d\phi=\\
\viam \pi \frac{-m_1 m_2 \w_1 \w_2+m_1^2\w_2^2+p^2(m^2_2+p^2-\w_1 \w_2)}{\w_1^2\w_2^2}
\end{multline}
and therefore
\begin{multline}
\int d\Omega Y(\vp)=\int_0^{\pi}\st \int^{2 \pi}_0 Y(\vp)d\phi d\theta=\\
	=2\pi\frac{\w_1(p)\w_2(p)-p^2-m_1 m_2}{\w_1(p)\w_2(p)}
\end{multline}
we finally can write
\begin{multline}\label{emajoranac2}
\lfv T^{f}_{00}\rfv=\int \frac{d p }{2\pi^2}p^2\left(\w_1(p)+\w_2(p)\right)\times\\
		\times\Big(-2+2\sst  \frac{\w_1(p)\w_2(p)-p^2-m_1 m_2}{\w_1(p)\w_2(p)})\Big).
\end{multline}

For the pressure we have:
\begin{multline}
T_{jj}^{majorana}=i\bpsi \gamma_j\partial_j \psi=\\
	=\sum_{rs}\int\frac{d^3pd^3q}{(2\pi)^3}i\left(u^{r\dagger}(\vp) e^{ipx}a^{r\dagger}(\vp)+v^{r\dagger}(\vp)e^{-ipx}a^r(\vp)\right)\times\\
	\times\gamma_0\gamma_j\left(-iq_j u^s(\vq)e^{-iqx}a^s(\vq)+iq_jv^s(\vq)e^{iqx}a^{s\dagger}(\vq)\right)=\\
 =\sum_{rs}\int\frac{d^3pd^3q}{(2\pi)^3}i\left(u^{r\dagger}(\vp)\gamma_0\gamma_j u^s(\vq)q_je^{i(p-q)x}a^{r\dagger}(\vp)a^s(\vq)+\right.\\
 -u^{r\dagger}(\vp)\gamma_0\gamma_j v^s(\vq)q_je^{i(p+q)x}a^{r\dagger}(\vp)a^{s\dagger}(\vq)+\\
  -v^{r\dagger}(\vp)\gamma_0\gamma_j v^s(\vq)q_je^{-i(p-q)x}a^{r}(\vp)a^{s\dagger}(\vq)+\\
 \left.+v^{r\dagger}(\vp)\gamma_0\gamma_j u^s(\vq)q_je^{-i(p+q)x}a^r(\vp)a^{s}(\vq)\right)	
\end{multline}
Using the quantum algebra, we get to
\begin{multline}
\lfv T_{jj}^{f}(x)\rfv=\\
	\sum_{r,s}\int \frac{d^3p}{(2\pi)^3}\Bigg[
		\left(u^{r\dagger}_1(\vp)\gamma_0\gamma_ju^s_1(\vp)p_j\sst\sum_{r'}Y^{rr'\dagger}(\vp)Y^{sr'}(\vp)\right)+\\
		+\left(u^{r\dagger}_2(\vp)\gamma_0\gamma_ju^s_2(\vp)p_j\sst\sum_{r'}Y^{r'r\dagger}(-\vp)Y^{r's\dagger}(-\vp)\right)+\\
		+\left(u^{r\dagger}_1(\vp)\gamma_0\gamma_jv^s_1(-\vp)p_je^{i2\w_1(p)t}\sst\sum_{r'}Y^{rr'\dagger}(\vp)W^{sr'\dagger}(-\vp)\right)+\\
		+\left(u_2^{r\dagger}(\vp)\gamma_0\gamma_jv_2^s(-\vp)p_je^{i2\w_2(p)t}\sst\sum_{r'}Y^{r'r\dagger}(-\vp)W^{r's}(-\vp)\right)+\\
		+\left(-v_1^{r\dagger}(\vp)\gamma_0\gamma_ju^s_1(-\vp)p_je^{-i2\w_1(p)t}\sst\sum_{r'}W^{rr'}(\vp)Y^{sr'}(-\vp)\right)+\\
		+\left(-v_2^{r\dagger}(\vp)\gamma_0\gamma_ju^s_2(-\vp)p_je^{-2i\w_2(p)t}\sst\sum_{r'}W^{rr'\dagger}(\vp)Y^{r's}(\vp)\right)+\\
		+\left(-v^{r\dagger}_1(\vp)\gamma_0\gamma_jv^s_1(\vp)p_j\sst\sum_{r'}W^{rr'}(\vp)W^{sr'\dagger}(\vp)\right)+\\
		+\left(-v_2^{r\dagger}(\vp)\gamma_0\gamma_jv_2^s(\vp)p_j\sst\sum_{r'}W^{rr'\dagger}(\vp)W^{r's}(\vp)\right)+\\
		+\left(-v_1^{r\dagger}(\vp)\gamma_0\gamma_jv^r_1(\vp)p_j\cct\delta^{rs}
					-v^{r\dagger}_2(\vp)\gamma_0\gamma_jv^r_2(\vp)p_j\cct\delta^{rs}\right)\Bigg]
\end{multline}
This expression simplifies to 
\be\label{majoranapre}
\lfv T_{jj}^f(x)\rfv\viam\int\frac{d^3p}{(2\pi)^3}\left(-\frac{8\pi}{3}p^4\left(\frac{1}{\w_1(p)}+\frac{1}{\w_2(p)}\right)\right).
\ee

\label{ffvevc}
\subsectionItalic{WZ Flavour Vacuum}

We can now combine the results for the bosonic contribution to the condensate and the fermionic one.

About the total energy, recalling (\ref{enebosV}), (\ref{enebos1}), (\ref{emajoranac}) and (\ref{emajoranac2}), we have
\begin{multline}
\lfv T^{WZ}_{00}\rfv=2\lfv T^{b}_{00}\rfv+\lfv T^{f}_{00}\rfv=\\
				=\int \frac{d\vp}{(2\pi)^3}(\w_1(p)+\w_2(p))\sst 4\left(|V|^2+Y\right)
\end{multline}
and
\begin{multline}
\int d\vp \left(Y(\vp)+|V(p)|^2\right)(\w_1+\w_2)=\\
	=\int dp \;p^2(\w_1+\w_2)\left(2\pi\frac{\w_1\w_2-p^2-m_1 m_2}{\w_1\w_2}+4\pi\frac{(\w_1 -\w_2)^2}{4\w_1 \w_2}\right)=\\
 =\int dp\;p^2(\w_1+\w_2)\frac{\pi}{\w_1\w_2}\left(\w_1^2+\w_2^2-2p^2-2m_1m_2\right)=\\
 =\int dp \;p^2(\w_1+\w_2)\frac{\pi}{\w_1\w_2}\left(\w_1^2-p^2+\w_2^2-p^2-2m_1m_2\right)=\\
 =\int dp \; p^2(\w_1+\w_2)\frac{\pi}{\w_1\w_2}\left(m_1^2+m_2^2-2m_1m_2\right)=\\
 =\int dp \; p^2(\w_1+\w_2)\frac{\pi}{\w_1\w_2}\left(m_1-m_2\right)^2=\\
 =\pi(m_1-m_2)^2 \int_0^K dp\;p^2 \frac{\w_1+\w_2}{\w_1 \w_2}\equiv f(K)
\end{multline}
with
\begin{multline}
f(K)=\frac{\pi}{2} (m_1-m_2)^2 \Big[K (\w_1(K)+\w_2(K))+\\
	-m_1^2 \log \left(\frac{K+\w_1(K)}{m_1}\right)-m_2^2 \log \left(\frac{K+\w_2(K)}{m_2}\right)\Big]
\end{multline}
\bea
f(K)\approx K^2\;\;\;\;\;&\mbox{when}&\;\;\;\;\;K\rightarrow +\infty\\
f(K)\approx \frac{K^3}{3}\left(\frac{1}{m_1}+\frac{1}{m_2}\right)\;\;\;\;\;&\mbox{when}&\;\;\;\;\;K\rightarrow0
\eea
and
\be
f'(K)=K^2\left(\frac{1}{\w_1(K)}+\frac{1}{\w_2(K)}\right)
\ee

For the pressure, holding (\ref{prebos}) and (\ref{majoranapre}), we can write
\begin{equation*}
\lfv T^{WZ}_{jj}(x)\rfv=2\lfv T_{jj}^b\rfv+\lfv T_{jj}^f\rfv=
\end{equation*}
\begin{multline}
=\int dp\Bigg[\frac{2p^4}{6\pi^2}\left(\frac{1}{\w_1(p)}+\frac{1}{\w_2(p)}\right)+\\
		-\sst\frac{2}{(2\pi)^2}(m_1^2-m_2^2)p^2\frac{(\w_1(p)-\w_2(p))}{\w_1(p)\w_2(p)}\Bigg]+\\
			+\int dp\frac{2p^4}{6\pi^2}\left(-\frac{1}{\w_1(p)}-\frac{1}{\w_2(p)}\right)=\nonumber
\end{multline}
\begin{multline}
			=-\sst \frac{1}{2\pi^2}(m_1^2+m_2^2)\int_0^K dp \;p^2\frac{\w_1(p)-\w_2(p)}{\w_1(p)\w_2(p)}=\\
			\equiv-\sst \frac{1}{2\pi^2}(m_1^2+m_2^2) g(K)
\end{multline}
Solving the integral we have
\begin{multline}
g(K)=\frac{1}{2}\Big[K(\w_2(K)-\w_1(K))+\\
		+m_1^2\log\left(\frac{K+\w_1(K)}{m_1}\right)-m_2^2\log\left(\frac{K+\w_2(K)}{m_2}\right)\Big]
\end{multline}
and therefore
\be
g(K\rightarrow \infty)\approx\frac{1}{2}(m_1^2-m_2^2)\log (K).
\ee

\newpage
\section{APPENDIX TO CHAPTER \ref{cANMOF}}\label{aC5}
\markright{\textit{\thesection \quad APPENDIX TO CHAPTER \ref{cANMOF}}}
\subsectionItalic{Bosonic Energy}\label{aC5be}

Within the context of a theory for two free scalars with flavour mixing (we follow the convention of Appendix \ref{aC4},
which can also be used as reference for all quantities not explicitly defined in this Appendix),
we want to evaluate the quantity
\be
\lfv T_{00}^{b}(x)\rfv.
\ee
A preliminary remark about the operator $G_\theta(t)$:
\be\label{Gvarious}
G^{\dagger}_\theta(t)=G^{-1}_\theta(t)=G_{-\theta}(t)
\ee
that follows straightforwardly from formula (Section \ref{G1}), here recalled:
\be
G_\theta (t)\equiv e^{i\theta \int d^3 x \left(\pi_2 (x)\phi_1 (x)-\pi_1(x)\phi_2(x) \right)}.
\ee
Then, combining the definition of the flavour vacuum 
\be
\rfv\equiv G^{\dagger}_{\theta}(t)\rmv
\ee
and the expression of the stress-energy tensor (Section \ref{bsetc})
in terms of the fields $\phi_i(x)$ and its conjugate momentum $\pi_i(x)=\dot{\phi}_i(x)$ we have:
\begin{multline}\label{gtg}
\lfv T_{00}^{b}(x)\rfv=\\
		=\lmv G_{\theta}(t)\sum_i\left(\pi_i^2(x)+\left(\vec{\nabla}\phi_i(x)\right)^2+m_i^2\phi_i^2(x)\right)G^\dagger_{\theta}(t) \rmv=\\
		=\lmv
		  \sum_i\left(\left(G_{\theta}(t)\pi_i(x)G^\dagger_{\theta}(t)\right)^2
		 +\left(\vec{\nabla}G_{\theta}(t)\phi_i(x)G^\dagger_{\theta}(t)\right)^2+\right	.\\
		 \left.+m_i^2\left(G_{\theta}(t)\phi_i(x)G^\dagger_{\theta}(t)\right)^2\right)\rmv
\end{multline}
since $G_{\theta}(t) \slot^2 G_{\theta}^\dagger(t)\via{Gvarious}
     (G_{\theta}(t)\slot G^\dagger_{\theta}(t))(G_{\theta}(t)\slot G^\dagger_{\theta}(t))$ and $\vec{\nabla}G^{(\dagger)}_\theta(t)=0$.

In the expression (\ref{gtg}) we can recognize two kinds of term, according to the appearance of the index $i$:
a first type in which the only objects carrying the index $i$ are the \textit{fields}:
\be
\lmv \sum_i\left(G_{\theta}(t)\pi_i(x)G^\dagger_{\theta}(t)\right)^2\rmv
\ee
\be
\lmv \sum_i \left(\vec{\nabla}G_{\theta}(t)\phi_i(x)G^\dagger_{\theta}(t)\right)^2 \rmv
\ee
and a second one, in which also \textit{masses} appear explicitly: 
\be\label{secondkind}
\lmv m_i^2\left(G_{\theta}(t)\phi_i(x)G^\dagger_{\theta}(t)\right)^2\rmv.
\ee
As we will see, all information about the condensate are encoded in this latter kind of terms,
since the operator $G_\theta(t)$ induces on the terms of the first type a trivial transformation.
This sort of distinction is rather general and can be applied to many cases in which
the expectation value of an observable with respect to the flavour vacuum is considered.
It also applies to different theories: in the fermionic case (Section (\ref{fecc}) and (\ref{fpcc}))
we will show how terms of the latter kind will play the analogous r\^ole of  (\ref{secondkind}).

To see how the two types of terms are transformed differently by the operator $G_\theta(t)$
in this specific case, we start by noticing that, similarly to (\ref{phiAB}), we have
\begin{equation}\begin{split}
G_{\theta}^\dagger(t)\pi_1(x)G_{\theta}(t)=&\pi_1(x) \ct+\pi_2(x) \st\\
G_{\theta}^\dagger(t)\pi_2(x)G_{\theta}(t)=&-\pi_1(x) \st+\pi_2(x) \ct
\end{split}\end{equation}
being
\be
[X,\pi_1] =i\theta \int d^3x (-\pi_2\com{\phi_1,\pi_1})=i\theta(-\pi_2i)=\theta\pi_2
\ee
\be
[X,\pi_2] =i\theta \int d^3x (\phi_1\com{\phi_2,\pi_2})=i\theta\phi_1i=-\theta\pi_1
\ee
from (\ref{G1}) and (\ref{ccrphipi}), and then applying (\ref{BCHsc}).
Then, because of (\ref{Gvarious}), we can write
\begin{multline}
G_{\theta}(t)\pi_1(x)G^\dagger_{\theta}(t)=G_{-\theta}^\dagger(t)\pi_1(x)G_{-\theta}(t)=\\
		=\pi_1(x) \cos (-\theta)+\pi_2(x) \sin (-\theta)=\pi_1(x) \ct -\pi_2(x) \st
\end{multline}
\begin{multline}
G_{\theta}(t)\pi_2(x)G^\dagger_{\theta}(t)=G_{-\theta}^\dagger(t)\pi_1(x)G_{-\theta}(t)=\\
		=-\pi_1(x) \sin (-\theta)+\pi_2(x)\cos (-\theta)=\pi_1(x) \st+\pi_2(x)\ct
\end{multline}
and analogously
\begin{equation}\begin{split}
G_{\theta}(t)\phi_1(x)G^\dagger_{\theta}(t)=&\phi_1(x) \ct -\phi_2(x) \st \\
G_{\theta}(t)\phi_2(x)G^\dagger_{\theta}(t)=&\phi_1(x) \st+\phi_2(x)\ct
\end{split}\end{equation}
Furthermore, we have
\begin{equation*}
\left(G_{\theta}(t)\pi_1(x)G^\dagger_{\theta}(t)\right)^2+\left(G_{\theta}(t)\pi_2(x)G^\dagger_{\theta}(t)\right)^2=
\end{equation*}
\begin{equation*}
=\left(\pi_1(x) \ct -\pi_2(x) \st\right)^2+\left(\pi_1(x) \st+\pi_2(x)\right)^2=
\end{equation*}
\begin{equation*}
=\pi_1^2(x)\cct+\pi_2^2(x)\sst-2\pi_1(x)\pi_2(x)\st \ct+
\end{equation*}
\begin{equation*}
+\pi_1^2(x)\sst+\pi_2^2(x)\cct+2\pi_1(x)\pi_2(x)\st \ct=
\end{equation*}
\be\label{pisquare}
=\pi_1^2(x)+\pi_2^2(x)
\ee
and similarly
\begin{equation*}
\left(\vec{\nabla}G_{\theta}(t)\phi_1(x)G^\dagger_{\theta}(t)\right)^2+\left(\vec{\nabla}G_{\theta}(t)\phi_2(x)G^\dagger_{\theta}(t)\right)^2=
\end{equation*}
\be\label{phisquare}
=\left(\vec{\nabla}\phi_1(x)\right)^2+\left(\vec{\nabla}\phi_2(x)\right)^2
\ee
On the other hand, regarding the term (\ref{secondkind}), we have 
\begin{multline}\label{msquare}
m_1^2\left(G_{\theta}(t)\phi_1(x)G^\dagger_{\theta}(t)\right)^2+m_2^2\left(G_{\theta}(t)\phi_2(x)G^\dagger_{\theta}(t)\right)^2=\\
=m_1^2\left(\phi_1(x) \ct -\phi_2(x) \st\right)^2+m_2^2\left(\phi_1(x) \st+\phi_2(x)\ct\right)^2=\\
=m_1^2\phi_1^2(x)\cct+m_1^2\phi_2^2(x)\sst-2m_1^2\phi_1(x)\phi_2(x)\st \ct+\\
+m_2^2\phi_1^2(x)\sst+m_2^2\phi_2^2(x)\cct+2m_2^2\phi_1(x)\phi_2(x)\st \ct=\\
=\left(-2m_1^2\phi_1(x)\phi_2(x)+2m_2^2\phi_1(x)\phi_2(x)\right)\st \ct+\\
+(-m_1^2\phi_1^2(x) +m_1^2\phi_2^2(x)+m_2^2\phi_1^2(x)-m_2^2\phi_2^2(x) )\sst+\\
+m_1^2\phi_1^2(x)+m_2^2\phi_2^2(x).
\end{multline}
It is easy to be convinced that
\be\label{phiiphij}
\lmv \phi_i(x)\phi_j(x)\rmv\propto\delta_{ij}
\ee
since
\be
\lmv \phi_i\sim \lmv a_i\;\;\;\;\;\mbox{and}\;\;\;\;\;\phi_j \rmv \sim  a_j^{\dagger} \rmv
\ee
and therefore if $i\neq j$
\be
\lmv \phi_i(x)\phi_j(x)\rmv\sim\lmv a_j^{\dagger} a_i  \rmv=0.
\ee
Finally, collecting all these information, we have
\begin{multline}
\lfv T_{00}^{b}(x)\rfv= \lmv \sum_i\Big[\left(G_{\theta}(t)\pi_i(x)G^\dagger_{\theta}(t)\right)^2+\\
+\left(\vec{\nabla}G_{\theta}(t)\phi_i(x)G^\dagger_{\theta}(t)\right)^2+m_i^2\left(G_{\theta}(t)\phi_i(x)G^\dagger_{\theta}(t)\right)^2\Big]\rmv=\nonumber
\end{multline}
\begin{multline}
	 \triplevia{pisquare}{phisquare}{msquare} \lmv \sum_i\left( \pi_i^2(x)+\left(\vec{\nabla}\phi_i^2(x)\right)^2+m_i^2\phi_i^2(x)\right)\\
		  +\left(-2m_1^2\phi_1(x)\phi_2(x)+2m_2^2\phi_1(x)\phi_2(x)\right)\st \ct\\
		  +(-m_1^2\phi_1^2(x) +m_1^2\phi_2^2(x)+m_2^2\phi_1^2(x)-m_2^2\phi_2^2(x) )\sst\rmv=\nonumber
\end{multline}
\begin{multline}
		 \via{phiiphij} \lmv \sum_i\left( \pi_i^2(x)+\left(\vec{\nabla}\phi_i^2(x)\right)^2+m_i^2\phi_i^2(x)\right)\\
		  +(-m_1^2\phi_1^2(x) +m_1^2\phi_2^2(x)+m_2^2\phi_1^2(x)-m_2^2\phi_2^2(x) )\sst\rmv
\end{multline}
that can be written as
\begin{multline}\label{energyscalarcc}
\lfv T_{00}^{b}(x)\rfv=\lmv T^{b}_{00}(x)\rmv+\\
+\sst \lmv \left(\phi_2^2(x)-\phi_1^2(x)\right)\left(m_1^2-m_2^2\right) \rmv.
\end{multline}
This last expression allows us to easily apply the normal ordering with respect to the usual vacuum:
\be\label{escalarcc}
\lfv : T_{00}^{b}(x):\rfv=\sst \lmv \left(\phi_2^2(x)-\phi_1^2(x)\right)\left(m_1^2-m_2^2\right) \rmv.
\ee
As we can see, once the normal ordering is performed, only the term of second kind (\ref{secondkind}) contributes
to the final expression. We have now a very simple formula for the (normal ordered) energy of the flavour vacuum,
in terms of the usual vacuum $\rmv$, the field $\phi(x)_i$ and the masses $m_i$ ($i=1,2$).

Moreover, we deduced this expression
without referring to the mode decomposition of the field itself and just using the algebra of $\phi_i(x)$ and $\pi_i(x)$.
Because of this, such an approach might be crucial in the study of the interactive theory and 
in curved spacetime, at a non-perturbative level.

We can finally show the equivalence of (\ref{escalarcc}) and (\ref{phienergyc}) by considering that
\begin{multline}\label{phiphivev}
\Big(\lmv \phi_i(x)\Big)\Big( \phi_i(x)\rmv\Big)= \\
		=\left(\lmv 
		\int \frac{d^3p}{(2 \pi)^{3/2}}\frac{a_i(\vp)}{\sqrt{2 \w_i(p)}}e^{-i p x}
		\right)\left(\int \frac{d^3q}{(2 \pi)^{3/2}}\frac{a^\dagger_i(\vq)}{\sqrt{2 \w_i(q)}}e^{ i q x}
		\rmv\right)=\\
	= \int \frac{d^3p}{(2 \pi)^{3}}\frac{1}{2 \w_i(p)}
\end{multline}
and therefore
\begin{equation}\begin{split}
\lfv : T_{00}^{b}(x):\rfv=&\sst \lmv \left(\phi_2^2(x)-\phi_1^2(x)\right)\rmv \left(m_1^2-m_2^2\right) =\\
				=&\sst \frac{m_1^2-m_2^2}{2}\int \frac{d^3p}{(2 \pi)^{3}}\left(\frac{1}{ \w_2(p)}-\frac{1}{ \w_1(p)}\right).
\end{split}\end{equation}
\label{becc}
\subsectionItalic{Bosonic Pressure}\label{aC5bp}

Following the path traced in the previous Section, we have (Section \ref{bsetc})
\begin{multline}
\lfv T^{b}_{jj}(x)\rfv=\\
= \lfv \sum_{i=1,2}\left(2(\partial_j \phi_i(x))^2+\pi_i^2(x)-\left(\vec{\nabla}\phi_i(x)\right)^2-m_i^2\phi_i^2(x)\right)\rfv=\\
      = \lmv G_{\theta}(t)      
      \sum_{i=1,2}\left(
      2(\partial_j \phi_i(x))^2+\pi_i^2(x)-\left(\vec{\nabla}\phi_i(x)\right)^2-m_i^2\phi_i^2(x)\right)
      G_{\theta}^\dagger (t)\rmv=\\
      = \lmv   
      				\sum_{i=1,2} \left(2\left(\partial_j G_{\theta}(t)\phi_i(x)G_{\theta}^\dagger(t)\right)^2+
      				\left( G_{\theta}(t) \pi_i(x) G_{\theta}^\dagger(t)\right)^2+\right.\\
       \left.-\left(\vec{\nabla}G_{\theta}(t)\phi_i(x)G_{\theta}^\dagger(t)\right)^2
      	-m_i^2\left(G_{\theta}(t)\phi_i(x)G_{\theta}^\dagger (t)\right)^2\right)
      \rmv
\end{multline}
again here we recognize one kind of terms that transforms trivially (thanks to (\ref{pisquare}) and (\ref{phisquare}))
\be
\sum_{i=1,2} 2\left(\partial_j G_{\theta}(t)\phi_i(x)G_{\theta}^\dagger(t)\right)^2=
				\sum_{i=1,2} 2\left(\partial_j \phi_i(x) \right)^2
\ee
\be
\sum_{i=1,2} \left( G_{\theta}(t) \pi_i(x) G_{\theta}^\dagger(t)\right)^2=
		\sum_{i=1,2} \left( \pi_i(x) \right)^2
\ee
\be
-\sum_{i=1,2}\left(\vec{\nabla}G_{\theta}(t)\phi_i(x)G_{\theta}^\dagger(t)\right)^2=
		-\sum_{i=1,2}\left(\vec{\nabla}\phi_i(x)\right)^2
\ee
and the second kind, in the exact form of the previous Section, up to a minus sign:
\begin{multline}
 \lmv \sum_{1,2} \left(-m_i^2\left(G_{\theta}(t)\phi_iG_{\theta}^\dagger (t)\right)^2(x)\right)\rmv=
 			=\lmv \sum_{1,2} \left(-m_i^2\phi_i^2(x)\right)\rmv+\\
		-\sst \lmv \left(\phi_2^2(x)-\phi_1^2(x)\right)\left(m_1^2-m_2^2\right) \rmv.
\end{multline}
Hence
\be\label{pressurescalarcc}
\lfv T^{b}_{jj}(x)\rfv=\lmv T^{b}_{jj}(x)\rmv-\sst \lmv \left(\phi_2^2(x)-\phi_1^2(x)\right)\left(m_1^2-m_2^2\right) \rmv
\ee
and
\be
\lfv :T^{b}_{jj}(x):\rfv=-\sst \lmv \left(\phi_2^2(x)-\phi_1^2(x)\right)\left(m_1^2-m_2^2\right) \rmv
\ee
from which
\be
\eqbox{\lfv :T^{b}_{jj}(x):\rfv=-\lfv :T^{b}_{00}(x):\rfv.}
\ee\label{bpcc}
\subsectionItalic{Fermionic Energy}\label{aC5fe}

In analogy with (\ref{gtg}), we write
\be
\lfv T_{00}^{f}(x)\rfv=\lmv G_{\theta}(t)\sum_i\left(i\bpsi_i(x)\gamma_0\partial_0 \psi_i(x)\right)G^\dagger_{\theta}(t)\rmv
\ee
This time, the operator $G_\theta(t)$ can not pass through the derivative operator without being affected, neither
$i\bpsi \gamma_0\partial_0 \psi$ transforms trivially under $G_{\theta}(t) \slot G^\dagger_{\theta}(t)$.
However, recalling the equation of motion
\be
(i\diracpartial -m_i)\psi_i(x)=0
\ee
we can write
\be
i\gamma_0\partial_0 \psi_i(x)=i\sum_{j}\gamma_j \partial_j \psi_i(x)+m_i \psi_i(x)
\ee
and therefore
\begin{multline}\label{tgpsig}
\lfv T_{00}^{f}(x)\rfv=\lmv G_{\theta}(t)\sum_i\bpsi_i(x)
				\left(i\sum_{j}\gamma_j \partial_j+m_i\right)\psi_i(x)G^\dagger_{\theta}(t)\rmv=\\
				=\lmv \sum_i\left(G_{\theta}(t) \bpsi_i(x)G_{\theta}^\dagger(t) \right)
				\left(i\sum_{j}\gamma_j \partial_j+m_i\right)\left(G_{\theta}(t)\psi_i(x)G^\dagger_{\theta}(t)\right)\rmv.\;\;\;
\end{multline}
Again, using
\be\label{Gvariousf}
G^{\dagger}_\theta(t)=G^{-1}_\theta(t)=G_{-\theta}(t)
\ee
that can be easily checked from (\ref{Gferm}) we have
\begin{equation}\begin{split}\label{gpsi1gd}
G_{\theta}(t)\psi_1(x)G^\dagger_{\theta}(t)=&G^\dagger_{-\theta}(t)\psi_1(x)G_{-\theta}(t)=\\
																					 =&\psi_1(x)\cos (-\theta)+\psi_2(x)\sin (-\theta)=\\
																					 =&\psi_1(x)\ct-\psi_2(x)\st
\end{split}\end{equation}
\begin{equation}\begin{split}
G_{\theta}(t)\psi_2(x)G^\dagger_{\theta}(t)=&G^\dagger_{-\theta}(t)\psi_2(x)G_{-\theta}(t)=\\
																					 =&\psi_2(x)\cos (-\theta)-\psi_1(x)\sin (-\theta)=\\
																					 =&\psi_2(x)\ct+\psi_1(x)\st
\end{split}\end{equation}
and
\begin{equation}\begin{split}
G_{\theta}(t)\psi_1^\dagger(x)G^\dagger_{\theta}(t)=&\left(G_{\theta}(t)\psi_1(x)G^\dagger_{\theta}(t)\right)^\dagger=\\
																						\via{gpsi1gd}&\left(\psi_1(x)\ct-\psi_2(x)\st\right)^\dagger	=\\
																						=&\psi_1^\dagger(x)\ct-\psi_2^\dagger(x)\st
\end{split}\end{equation}
\begin{equation}\begin{split}
G_{\theta}(t)\psi_2^\dagger(x)G^\dagger_{\theta}(t)=&\left(G_{\theta}(t)\psi_2(x)G^\dagger_{\theta}(t)\right)^\dagger=\\
																						\via{gpsi1gd}&\left(\psi_2(x)\ct+\psi_1(x)\st\right)^\dagger	=\\
																						=&\psi_2^\dagger(x)\ct+\psi_1^\dagger(x)\st.
\end{split}\end{equation}
Coming back to (\ref{tgpsig}), we distinguish the two different terms
\be
\lmv \sum_i\left(G_{\theta}(t) \bpsi_i(x)G_{\theta}^\dagger(t) \right)
				\left(i\sum_{j}\gamma_j \partial_j\right)\left(G_{\theta}(t)\psi_i(x)G^\dagger_{\theta}(t)\right)\rmv
\ee
and
\be
\lmv \sum_i\left(G_{\theta}(t) \bpsi_i(x)G_{\theta}^\dagger(t) \right)
				m_i\left(G_{\theta}(t)\psi_i(x)G^\dagger_{\theta}(t)\right)\rmv
\ee
Looking at the former, we have ($\vec{\gamma}\cdot\vec{\partial}\equiv \sum_{j}\gamma_j \partial_j$)
\begin{multline}\label{pressureinenergy}
\lmv \sum_i\left(G_{\theta}(t) \bpsi_i(x)G_{\theta}^\dagger(t) \right)
				i\vec{\gamma}\cdot\vec{\partial}\left(G_{\theta}(t)\psi_i(x)G^\dagger_{\theta}(t)\right)\rmv=\\
=\lmv \left(\bpsi_1(x)\ct-\bpsi_2(x)\st \right) i\vec{\gamma}\cdot\vec{\partial} \left(\psi_1(x)\ct-\psi_2(x)\st\right) \rmv+\\
+ \lmv \left(\bpsi_2(x)\ct+\bpsi_1(x)\st \right) i\vec{\gamma}\cdot\vec{\partial} \left(\psi_2(x)\ct+\psi_1(x)\st\right) \rmv=\\
=i\lmv \left(\bpsi_1(x) \vec{\gamma}\cdot\vec{\partial}\psi_1(x)\cct-\bpsi_1(x) \vec{\gamma}\cdot\vec{\partial} \psi_2(x)\ct \st \right.+\\
-\bpsi_2(x) \vec{\gamma}\cdot\vec{\partial} \psi_1(x)\ct \st+\bpsi_2(x) \vec{\gamma}\cdot\vec{\partial} \psi_2(x)\sst+\\
+\bpsi_2(x) \vec{\gamma}\cdot\vec{\partial}\psi_2(x)\cct+\bpsi_2(x) \vec{\gamma}\cdot\vec{\partial}\psi_1(x)\ct \st+\\
\left.+\bpsi_1(x) \vec{\gamma}\cdot\vec{\partial}\psi_2(x)\ct \st+\bpsi_1(x)\vec{\gamma}\cdot\vec{\partial}\psi_1(x)\sst\right)\rmv=\\
=i\lmv 
\left(\bpsi_1(x) \vec{\gamma}\cdot\vec{\partial}\psi_1(x)+\bpsi_2(x) \vec{\gamma}\cdot\vec{\partial}\psi_2(x)\right)\rmv=\\
=\sum_i \lmv \left(\bpsi_i(x) i\vec{\gamma}\cdot\vec{\partial}\psi_i(x)\right)\rmv
\end{multline}
in which we used
\be
\lmv \bpsi_i (x)i\vec{\gamma}\cdot\vec{\partial} \psi_j(x)\rmv \propto \delta_{ij}.
\ee
On the other hand
\begin{multline}
\lmv \sum_i\left(G_{\theta}(t) \bpsi_i(x)G_{\theta}^\dagger(t) \right) m_i\left(G_{\theta}(t)\psi_i(x)G^\dagger_{\theta}(t)\right)\rmv=\\
=\lmv \left(\bpsi_1(x)\ct-\bpsi_2(x)\st \right) m_1 \left(\psi_1(x)\ct-\psi_2(x)\st\right) \rmv+\\
+ \lmv \left(\bpsi_2(x)\ct+\bpsi_1(x)\st \right) m_2 \left(\psi_2(x)\ct+\psi_1(x)\st\right) \rmv=\\
=\lmv \bpsi_1(x) \psi_1(x)\rmv\left(m_1\cct+m_2\sst\right)+\\
=\lmv \bpsi_2(x) \psi_2(x)\rmv\left(m_2\cct+m_1\sst\right)=
\end{multline}
\begin{multline}
=\sum_i\lmv m_i \bpsi_i(x) \psi_i(x)\rmv+\\
			+\lmv\left( \bpsi_2(x) \psi_2(x)-\bpsi_1(x) \psi_1(x)\right)\rmv\left(m_1-m_2\right)\sst
\end{multline}
since
\be
\lmv \bpsi_i (x) \psi_j(x)\rmv \propto \delta_{ij}.
\ee
Considering both terms together we have
\begin{multline}
\lfv T_{00}^{f}(x)\rfv=\sum_i \lmv \left(\bpsi_i(x) \left(i\sum_{j}\gamma_j \partial_j+m_j\right)\psi_i(x)\right)\rmv+\\
		+\sst\left(m_1-m_2\right)\lmv\left( \bpsi_2(x) \psi_2(x)-\bpsi_1(x) \psi_1(x)\right)\rmv
\end{multline}
that can be written as
\begin{multline}\label{energymajoranacc}
\lfv T_{00}^{f}(x)\rfv=\lmv T_{00}^{f}(x)\rmv +\\
	+\sst\left(m_1-m_2\right)\lmv\left( \bpsi_2(x) \psi_2(x)-\bpsi_1(x) \psi_1(x)\right)\rmv
\end{multline}
and therefore
\be\label{setfnormord}
\lfv: T_{00}^{f}(x):\rfv =\sst\left(m_1-m_2\right)\lmv\left( \bpsi_2(x) \psi_2(x)-\bpsi_1(x) \psi_1(x)\right)\rmv.
\ee
By comparing this last formula and (\ref{escalarcc}), the analogy between the fermionic and the bosonic condensate it is now clear.

Furthermore, Supersymmetry enables us to rewrite this result in terms of the bosonic fields only.
This is convenient firstly because it clearly shows the independence of the formulae from a specific form of the spinors,
unlike formula (\ref{emajoranac2}) derived in the previous chapter; secondarily, it will allow us to write the total energy
for the Wess-Zumino model in a much simpler form.
We start recalling the equation of motion for $\psi(x)$:
\be
(i\diracpartial-m)\psi(x)=i\gamma_0\dot{\psi}(x)-i\vec{\gamma}\cdot\vec{\nabla}\psi(x)-m\psi(x)=0
\ee
and from that
\be\label{psidot}
i\gamma_0\dot{\psi}(x)=i\vec{\gamma}\cdot\vec{\nabla}\psi(x)+m\psi(x)
\ee
Recalling also that 
\be\label{energyzero}
\langle T_{00}^{WZ} \rangle=0
\ee
we therefore have
\be\label{tmunuzero}
\langle T_{\mu\nu}^{WZ} \rangle=0
\ee
being $\langle T_{\mu\nu}^{WZ} \rangle \propto \e_{\mu\nu}$.
Using (\ref{fsetfset}) and (\ref{bsetbset}), from the former we have
\begin{multline}
0=\langle 2T^{scalar}_{00}+T^{majorana}_{00}\rangle=\\
 =\langle 4\dot{S}+i\bpsi \gamma_0\dot{\psi}-\left(2\dot{S}^2-2(\vec{\nabla}S)^2-2m^2S^2\right)\rangle=\\
 =\langle i\bpsi \gamma_0 \dot{\psi}+2\dot{S}^2+2(\vec{\nabla}S)^2+2m^2 S^2\rangle =\\
 \via{psidot}\langle 	i\bpsi \vec{\gamma}\cdot \vec{\nabla}\psi+m\bpsi \psi+2\dot{S}^2+2(\vec{\nabla}S)^2+2m^2 S^2	\rangle
\end{multline}
\be\label{gammapsivec1}
\Rightarrow m \langle \bpsi \psi \rangle =\langle -i\bpsi \vec{\gamma}\cdot \vec{\nabla}\psi-2\dot{S}^2-2(\vec{\nabla}S)^2-2m^2 S^2	\rangle
\ee
while from the latter
\begin{multline}
0=\langle 2T^{scalar}_{jj}+T^{majorana}_{jj}\rangle=\\
 =\langle 4(\partial_j S)^2+i\bpsi \gamma_j \partial_j \psi+(2\dot{S}^2-2(\vec{\nabla}S)^2-2m^2 S^2)		\rangle
\end{multline}
\begin{multline}\label{gammapsivec2}
\Rightarrow \langle i \bpsi \vec{\gamma}\cdot \vec{\nabla}\psi\rangle =
		\langle -4 (\vec{\nabla}S)^2-6 \dot{S}^2+6 (\vec{\nabla}S)^2+6 m^2 S^2\rangle \\
			=\langle -6 \dot{S}^2+2 (\vec{\nabla}S)^2+6  m^2 S^2\rangle
\end{multline}
Combining (\ref{gammapsivec1}) and (\ref{gammapsivec2}) together we have
\begin{multline}\label{psipsiphiphi}
m \langle \bpsi \psi \rangle =\langle 6 \dot{S}^2-2 (\vec{\nabla}S)^2-6  m^2 S^2-2\dot{S}^2-2(\vec{\nabla}S)^2-2m^2 S^2	\rangle=\\
	=\langle 4( \dot{S}^2- (\vec{\nabla}S)^2-m^2 S^2)-4 m^2 S^2	\rangle
\end{multline}
Since, as in (\ref{phiphivev}), 
\begin{multline}
\langle \dot{S}^2- (\vec{\nabla}S)^2-m^2 S^2 \rangle =\int \frac{d^3k}{(2 \pi)^3}\left(\frac{\w(k)}{2}-\frac{k^2}{2 \w(k)}-\frac{m^2}{2\w(k)}\right)=\\
		=\int \frac{d^3k}{(2 \pi)^3}\frac{1}{2\w(k)}\left(\w^2(k)-k^2-m^2\right)=0
\end{multline}
we have
\be\label{psipsivev}
\langle \bpsi \psi \rangle=-4 m\langle  S^2	\rangle=-2m\int \frac{d^3k}{(2 \pi)^3}\frac{1}{\w(k)}.
\ee
Finally we can rewrite formula (\ref{setfnormord}) as
\begin{multline}
\lfv: T_{00}^{f}(x):\rfv \via{psipsiphiphi}\\
	=\sst\left(m_1-m_2\right)\left(-4m_2 \lmv S_2^2(x)\rmv+4m_1 \lmv S_1^2(x)\rmv\right)=\\
												 =\sst4\left(m_1-m_2\right)\left(m_1 \lmv S_1^2(x)\rmv-m_2 \lmv S_2^2(x)\rmv\right)
\end{multline}
or
\begin{multline}
\lfv: T_{00}^{f}(x):\rfv \via{psipsivev}\\
	=\sst\left(m_1-m_2\right)\left( 2m_1 \int \frac{d^3k}{(2 \pi)^3}\frac{1}{\w_1(k)} -2m_2 \int \frac{d^3k}{(2 \pi)^3}\frac{1}{\w_2(k)}\right)=\\
				=\sst \int \frac{d^3k}{(2 \pi)^3}2(m_1-m_2)\left(\frac{m_1}{\w_1(k)}-\frac{m_2}{\w_2(k)}\right)=\\
							 \via{intf}\sst \int \frac{dk}{\pi^2}k^2(m_1-m_2)\left(\frac{m_1}{\w_1(k)}-\frac{m_2}{\w_2(k)}\right)=\\
									=\sst \int \frac{dk}{\pi^2}k^2(m_1-m_2)\left(\frac{\w_2(k)m_1-\w_1(k)m_2}{\w_1(k)\w_2(k)}\right)=\nonumber
\end{multline}
\begin{multline}
=\sst \int\!\!\frac{dk}{\pi^2}k^2\left(\frac{\w_2(k)m_1^2-\w_2(k)m_1m_2-\w_1(k)m_2m_1+\w_1(k)m_2^2}{\w_1(k)\w_2(k)}\right)=\\
					=\sst\int\frac{dk}{\pi^2}k^2\frac{1}{\w_1(k)\w_2(k)}\left(\w_2(k)\w_1(k)^2-\w_2(k)k^2+\right.\\
					\left.-(\w_2(k)+\w_1(k))m_1m_2+\w_1(k)\w_2(k)^2-\w_1(k)k^2\right)=\\
			=\sst\int\frac{dk}{\pi^2}k^2\left(\w_1(k)+\w_2(k)\right)\left(\frac{\w_2(k)\w_1(k)-k^2-m_1m_2}{\w_1(k)\w_2(k)}\right)
\end{multline}
in accordance with (\ref{emajoranac2}).\label{fecc}
\subsectionItalic{Fermionic Pressure}\label{aC5fp}

The evaluation of the fermionic pressure is implicit in the calculations of the previous section.
Since
\be
\lfv T_{jj}^{majorana}(x)\rfv=\lfv\sum_i\left(i\bpsi_i(x)\gamma_j\partial_j \psi_i(x)\right)\rfv
\ee
we can use formula (\ref{pressureinenergy}) to write
\be
\lfv T_{jj}^{majorana}(x)\rfv=\lmv\sum_i\left(i\bpsi_i(x)\gamma_j\partial_j \psi_i(x)\right)\rmv
\ee
and therefore
\be\label{pressuremajoranacc}
\eqbox{\lfv T_{jj}^{majorana}(x)\rfv=\lmv T_{jj}^{majorana}(x)\rmv}
\ee
or
\be
\lfv :T_{jj}^{majorana}(x):\rfv=0
\ee
in agreement with the analogous result of Section \ref{fvevf}.\label{fpcc}
\subsectionItalic{WZ Flavour Vacuum}
Collecting the results of last sections (\ref{energyscalarcc}), (\ref{pressurescalarcc}),
(\ref{energymajoranacc}) and (\ref{pressuremajoranacc}), we have
\begin{multline}
\lfv T^{WZ}_{00}(x)\rfv=2\sst\left((m_1^2-m_2^2)\left(\langle S_2^2(x)\rangle -\langle S_1^2(x)\rangle\right)+\right.\\
				\left.+(m_1-m_2)(\langle \bpsi_2(x) \psi_2(x)\rangle -\langle \bpsi_1(x) \psi_1(x)\rangle) \right)
\end{multline}
that, according to (\ref{psipsivev}), can be written as
\be
\eqbox{\lfv T^{WZ}_{00}(x)\rfv=2\sst (m_1-m_2)^2\big(\langle S_1^2(x)\rangle+\langle S_2^2(x)\rangle\big)}
\ee
and
\be
\eqbox{\lfv T^{WZ}_{jj}(x)\rfv=-2\sst(m_1^2-m_2^2)\big(\langle S_2^2(x)\rangle -\langle S_1^2(x)\rangle\big).}
\ee

\pagestyle{headings}

\newpage
\listoffigures\addcontentsline{toc}{section}{\textit{List of Figures}}
\chapter*{Acronyms and Recurrent~Symbols\hfill} \addcontentsline{toc}{section}{\textit{Acronyms and Recurrent Symbols}}
\pagestyle{myheadings}
\markright{\textit{LIST OF SYMBOLS}}
\begin{tabbing}\label{listofsymbols}

$AQFT$~~~~~~~~~~~~~~\=\parbox{4in}{Algebraic Quantum Field Theory  \hfill \textit{\pageref{symbol:AQFT}}}\\
\addsymbol BV: {Blasone-Vitiello}{symbol:BV}
\addsymbol CAR: {Canonical Anti-commutation Relations}{symbol:CAR}
\addsymbol CCR: {Canonical Commutation Relations}{symbol:CCR}
\addsymbol CP: {Charge-parity}{symbol:CP}
\addsymbol DE: {Dark Energy}{symbol:DE}
\addsymbol DM: {Dark Matter}{symbol:DM}
\addsymbol FLO: {Flavour ladder operators}{symbol:FLO}
\addsymbol FRW: {Friedman-Robertson-Walker}{symbol:FRW}
\addsymbol MLO: {Massive ladder operators}{symbol:MLO}
\addsymbol MSSM: {Minimal Supersymmetric Standard Model}{symbol:MSSM}
\addsymbol MSW: {Mikheyev-Smirnov-Wolfenstein}{symbol:MSW}
\addsymbol NFO: {Neutrino flavour oscillations}{symbol:NFO}
\addsymbol QFT: {Quantum Field Theory}{symbol:QFT}
\addsymbol SM: {Standard Model}{symbol:SM}
\addsymbol SUSY: {Supersymmetry}{symbol:SUSY}
\\
\addsymbolnoref \mbox{iff}: {if and only if}
\addsymbolnoref c.c.: {complex conjugate (of the previous term)}
\addsymbolnoref h.c.: {hermitian conjugate  (of the previous term)}
\addsymbolnoref	\mbox{vev}: {vacuum expectation value}
\addsymbolnoref	\mbox{\textit{f}-vev}: {flavour vacuum expectation value}
\\
\addsymbol \slot: {empty slot}{symbol:slot}
\addsymbolnoref \slot^{T}: {transpose}
\addsymbol \slot^*: {complex/hermitian conjugate}{symbol:cc}
\addsymbol \slot^{\dagger}: {$(\slot^*)^T$}{symbol:dagger}
\addsymbolnoref ;:	{such that}
\addsymbolnoref \sim:	{of the order of}
\addsymbolnoref \approx:	{approximately equal to}
\addsymbolnoref \equiv:	{defined as}
\addsymbolnoref \propto:	{proportional to}
\addsymbolnoref \propto\!\!\!\!\!/:	{not proportional to}
\addsymbolnoref	\field{N}_0: {Nonnegative Integer Numbers}
\addsymbolnoref diag\{\slot\}: {diagonal matrix}
\addsymbolnoref \{X_1 \dotsc X_n \mid Y\}: {set of elements $X_1 \dotsc X_n $, with condition $Y$}
\\
\addsymbol \mathfrak{H}: {non-separable Hilbert space for physical states}{symbol:nonseparable}
\addsymbol \mathfrak{F}_{\slot}: {separable Hilbert space (Fock space)}{symbol:separable}
\addsymbol \mathfrak{F_0}: {Fock space for massive states}{symbol:massfock}
\addsymbol \mathfrak{F_f}: {Fock space for flavour states}{symbol:flavourfock}
\addsymbolnoref \vk,\vp,\vq: {momentum}
\addsymbolnoref K: {momentum cutoff}
\addsymbolnoref \w_i(k):	{energy of a particle state: $\sqrt{\vk^2+m_i^2}$}
\addsymbolnoref 	\w_\pm:	{$\w_1\pm\w_2$}
\addsymbolnoref a^{(\dagger)}_{\slot}(\vk): {annihilation (creation) operator}
\addsymbolnoref a^{(\dagger)}_i(\vk): {MLO ($i=1,2$)}
\addsymbolnoref a^{(\dagger)}_\iota(\vk): {FLO ($\iota=A,B$)}
\addsymbol a^\mp_{\slot}:	{$a^\dagger_{\slot} (-\vk)$}{symbol:aplusminus}
\addsymbol \rmv: {vacuum for MLO}{symbol:rmv}
\addsymbol \rfv: {vacuum for FLO (\textit{flavour vacuum})}{symbol:rfv}
\addsymbolnoref \langle \slot \rangle:	{vev: $\lmv \slot \rmv$}
\addsymbolnoref _f\langle \slot \rangle_f:	{\textit{f}-vev: $\lfv \slot \rfv$}
\addsymbolnoref {:\slot:}:	{normal ordering: $\slot-\lmv \slot \rmv$}
\addsymbol \fnormal{f(\varphi_1,\varphi_2)}: {\small $f(\ct \varphi_1-\st \varphi_2,\st\varphi_1+\ct\varphi_2)-%
f(\varphi_1,\varphi_2)$\normalsize}{symbol:fnormal}
\addsymbol G_{\theta}: {mapping operator between massive and flavour states}{symbol:G}
\\
\addsymbolnoref \lag: {Lagrangian}
\addsymbolnoref \varphi: {generic field}
\addsymbolnoref \phi,S: {scalar field}
\addsymbolnoref \pi_{\slot}: {conjugate momentum of $\phi_{\slot}$: $\dot{\phi}_{\slot}$}
\addsymbolnoref \psi: {spin-\nicefrac{1}{2} field}
\addsymbolnoref \gamma^{\mu}:	{gamma matrices}
\addsymbolnoref	\diracpartial: {Dirac notation: $\gamma_\mu \partial^\mu$}
\addsymbolnoref x p:	{$x_\mu p^\mu=x^\mu p^\nu \e_{\mu\nu}$}
\\
\addsymbol T_{\mu\nu}: {stress-energy tensor}{symbol:set}
\addsymbolnoref T_{jj}: {space-space components: $T_{11},T_{22},T_{33}$}
\addsymbol \rho_{\slot}: {energy density}{symbol:energy}
\addsymbol \pressure_{\slot}: {pressure}{symbol:pressure}
\addsymbol w: {equation of state}{symbol:eqofstate}
\addsymbol \Lambda: {cosmological constant}{symbol:cosmcos}
\addsymbol G:	{Newton's gravitational constant}{symbol:newton}
\addsymbol M_{\odot}:	{solar mass}{symbol:solarmass}
\\
\addsymbolnoref \e_{\mu\nu}: {metric in flat spacetime (Minkowski)}
\addsymbolnoref g_{\mu\nu}: {metric in curved spacetime}
\addsymbolnoref \slot_\flat:	{expression in \textit{flat} spacetime}
\addsymbolnoref D_{\mu}: {covariant derivative}
\addsymbol \slot_{;\nu}:	{$D_\nu \slot$}{symbol:;cov}
\addsymbol a(t): {scale factor}{symbol:scalefactor}
\addsymbol \e: {conformal time}{symbol:conftime}
\addsymbol \Ce: {conformal scale factor}{symbol:confscalefactor}
\addsymbol \we:	{$\sqrt{\vp^2+m^2\Ce}$}{symbol:we}
\\
\addsymbol X^n[Y]:	{$\underbrace{[X,[X,[X,...}_{n},Y\underbrace{]]...]}_{n}$}{symbol:Xtothenth}
\addsymbolnoref \otto{\slot}:	{interchange of the indices 1 and 2}
\addsymbolnoref A\Big|_{x\rightarrow y}:	{$x$ is replaced by $y$ in the expression $A$}
\addsymbol A\stackrel{\mbox{\tiny{M}}}{\slot }B:	{$A$ has been reduced to $B$ via \textit{Mathematica}}{symbol:overscoreM}
\addsymbol A\stackrel{\mbox{\tiny{(n)}}}{=}B:	{$A=B$ follows from equation ($n$)}{symbol:overscore}
\end{tabbing} \clearpage
\pagestyle{headings}
\end{document}